\documentclass{sotonthesis}

\usepackage [english]{babel}

\usepackage[usenames,dvipsnames,svgnames,table]{xcolor}
\usepackage{graphicx}
\usepackage{multirow}
\usepackage{tikz}
\usetikzlibrary{patterns}

\usepackage{caption}
\usepackage{subcaption}
\numberwithin{figure}{section}
\numberwithin{equation}{section}

\usepackage{amsmath, amssymb, amscd, amsbsy}
\usepackage{setspace}
\usepackage{framed}
\usepackage{color}
\usepackage{verbatim}
\usepackage{graphicx}
\usepackage{epstopdf}
\usepackage{mathrsfs}
\usepackage [english]{babel}
\usepackage [autostyle, english = american]{csquotes}
\usepackage{bm}
\usepackage{cite}

\usepackage[makeroom]{cancel}
\usepackage{colortbl}

\usepackage{epstopdf}
\usepackage [english]{babel}
\usepackage{float}
\usepackage{array,booktabs}
\usepackage{hyperref} 

\newcommand{\be}{\begin{equation}}
\newcommand{\ee}{\end{equation}}
\newcommand{\bea}{\begin{eqnarray}}
\newcommand{\eea}{\end{eqnarray}}
\newcommand{\ie}{\textit{i.e.}\ }

\newcommand{\aka}{{a.k.a.}\ }

\newcommand{\eps}{\varepsilon}
\newcommand{\ph}{\varphi}
\newcommand{\hG}[2]{\hat{\Gamma}_{#1}^{#2}}
\newcommand{\hS}[1]{\hat{S}^{#1}}
\newcommand{\hbS}[1]{\hat{\cal S}^{#1}}
\newcommand{\bS}{{\cal S}^\Lambda}
\newcommand{\Cl}{\tilde{C}^\Lambda}
\newcommand{\Clp}{\mathring{C}^{\mathring{\Lambda}}}

\newcommand{\Clpp}{{C}^{\mathring{\Lambda}}_k}

\newcommand{\Lp}{{\mathring{\Lambda}}}
\newcommand{\php}{\mathring{\ph}}

\newcommand{\vp}{\varphi} 
\newcommand{\bp}{\bar \varphi} 
\newcommand{\bc}{\bar \chi} 
\newcommand{\bV}{\bar V} 

\newcommand{\hV}{\hat V} 
\newcommand{\hp}{\hat \phi} 
\newcommand{\hk}{\hat k} 
\newcommand{\hc}{\hat \chi} 
\newcommand{\hatt}{\hat t} 

\newcommand{\dclnf}{\,\partial_\chi\! \ln\! f \,}

\usepackage[normalem]{ulem}

\newcommand{\bi}{\begin{itemize}}
\newcommand{\ei}{\end{itemize}}
\def\beal#1\eeal{\begin{align}#1\end{align}}

\newcommand{\nfrac}[2]{\,#1/#2}

\def\b{\beta}

\def\e{\epsilon}
\def\k{\kappa}

\def\l{\lambda}

\def\bp{\bar{\vp}}

\def\e{\epsilon}

\usepackage{chngcntr}
\newcounter{para}

\title{Fundamental Aspects of Asymptotic Safety in Quantum Gravity}
\author{Zo{\"e} Helen Slade}
\date{January 2017}
\supervisor{Tim R.\ Morris}
\degree{Doctor of Philosophy}
\university{University of Southampton}
\faculty{Faculty of Science and Engineering}
\academicunit{Physics and Astronomy}

\begin{document}
\pagestyle{empty}
\frontmatter

\maketitle

\begin{abstract}
This thesis is devoted to exploring various fundamental issues within asymptotic safety.
Firstly, we study the reconstruction problem and present two ways in which to solve it within the context of scalar field theory, by utilising a duality relation between an effective average action and a Wilsonian effective action.
Along the way we also prove a duality relation between two effective average actions computed with different UV cutoff profiles.
Next we investigate the requirement of background independence within the derivative expansion of conformally reduced gravity. We show that modified Ward identities are compatible with the flow equations if and only if either the anomalous dimension vanishes or the cutoff profile is chosen to be power law, and furthermore show that no solutions exist if the Ward identities are incompatible.
In the compatible case, a clear reason is found why Ward identities can still forbid the existence of fixed points.
By expanding in vertices, we also demonstrate that the combined equations generically become either over-constrained or highly redundant at the six-point level.
Finally, we consider the asymptotic behaviour of fixed point solutions in the $f(R)$ approximation and explain in detail how to construct them.
We find that quantum fluctuations do not decouple at large $R$, typically leading to elaborate asymptotic solutions containing several free parameters.
Depending on the value of the endomorphism parameter, we find many other asymptotic solutions and fixed point spaces of differing dimension.

\end{abstract}

\pagestyle{fancy}

\phantomsection{}
\tableofcontents
\cleardoublepage{}

  \phantomsection{}
  \addcontentsline{toc}{chapter}{List of Figures and Tables}  
  \listoffigures
  \cleardoublepage{}

\pagestyle{empty}

\authorshipdeclaration{\begin{itemize}
\item Tim R.\ Morris, Zo\"e H.\ Slade, \textit{Solutions to the reconstruction problem in asymptotic safety}. Journal of High Energy Physics  \textbf{11} (2015) 904. \texttt{arXiv:1507.08657 [hep-th]}\cite{Morris:2015oca}.

\item Peter Labus, Tim R.\ Morris, Zo\"e H.\ Slade, \textit{Background independence in a background dependent renormalization group}. Phys.\ Rev.\ D \textbf{94} (2016) 024007. \texttt{arXiv:1603.04772 [hep-th]}~\cite{Labus:2016lkh}.

\item Sergio Gonzalez-Martin, Tim R.\ Morris, Zo\"e H.\ Slade \textit{Asymptotic solutions in asymptotic safety}. Phys.\ Rev.\ D \textbf{95} (2017) 106010.
  \texttt{arXiv:1704.08873 [hep-th]}~\cite{Gonzalez-Martin:2017gza}.
\end{itemize}

}

\acknowledgements{First of all, I would like to thank my supervisor Tim Morris for all his help and support over the past four years, without which none of this would have been possible.

I would like to acknowledge the Women's Physics Network, and the support it receives from the department, as well as all the opportunities I have had to take part in outreach and public engagement whilst in Southampton. All of this has undoubtedly enriched my PhD experience.

Thank you to all my physics friends, past and present, who have provided stimulating conversation, endless entertainment and companionship throughout the years. Thank you also to the other friends I have made over the course of my studies here for all your support and for truly enhancing my time in Southampton.

A special thank you goes to Peter Jones for always being there for me and filling these past 4 years with so much fun and laughter. I am really grateful I was able to share this adventure with you.

Last, but by no means least, I would like to thank my family for their endless love, support and encouragement, without which I would not be where I am today. Thank you for all the opportunities you have given me.

}

\pagestyle{fancy}



\mainmatter{}

\setlength{\parindent}{5mm}
\setlength{\parskip}{0em}

\raggedbottom{}


\chapter{Introduction}
\label{cha:introduction}

General relativity (GR) is one of the towering achievements of twentieth-century physics. Its predictions have received spectacular experimental confirmation time and time again since its publication over one hundred years ago \cite{Einstein:1915ca}.
However, GR is not the end of the story as far as gravity is concerned. Singularities appearing in the theory provide internal evidence that it is somehow incomplete, and furthermore GR is a classical description of gravity whilst nature at a fundamental level behaves quantum mechanically.
At scales approaching the Planck length quantum effects are expected to become important and it is believed that a theory of quantum gravity is needed in order to describe nature at the Planck scale and beyond.\footnote{Even though probing Planck-scale physics may require energies far above those accessible at current particle accelerators, there are ways to study quantum gravitational effects e.g.\ from the finger prints of the very early universe left on the CMB. See chapter \ref{cha:outlook} for more discussions on experimental searches for quantum gravity.}
Such a theory promises to bring a deeper understanding of fascinating phenomena such as black holes and the big bang, and its discovery remains one of the biggest open challenges in fundamental physics.

Actually, there is nothing preventing us from quantising GR using the standard perturbative techniques that have been successfully applied to nature's other fundamental fields.
The resulting quantum field theory (QFT) can be used to make testable predictions, for example in the form of corrections to the Newtonian potential \cite{Donoghue:1994dn}.\footnote{Although these effects are very small and therefore not likely to be measured any time soon.} However, if we wish to describe gravity at distances approaching the Planck length predicitivity is lost.
It turns out that an infinite number of measurements need to be performed by experiment in order to determine the parameters required to cancel the divergences of the theory i.e.\ the theory is perturbatively non-renormalizable \cite{tHooft:1974toh, Goroff:1985sz, Goroff:1985th, vandeVen:1991gw}.

Perturbative quantisation of GR therefore only provides an effective description of the graviton. Still, effective field theories are commonplace in physics and some of the most successful field theories of the last century come under this umbrella. The Standard Model for example can be considered an effective description of the interactions of fundamental particles. Likewise, Newton's theory of gravity is a low energy approximation to Einstein's GR, which in turn must be an effective description of some higher-energy theory of the gravitational field (whether this be a QFT or something more exotic).

The shortcomings of perturbative approaches\footnote{Another example comes from \cite{Stelle:1976gc} in which adding higher derivative operators to the Einstein-Hilbert action leads to a perturbatively renormalizable quantum theory of gravity, but which does not respect unitary.} do not mean that QFT and gravity are incompatible however. A well-behaved quantum theory of gravity might be recovered by taking the dynamics of the non-perturbative regime into account. One such non-perturbative route, which retains the fields and symmetries of GR, is asymptotic safety. Asymptotic safety posits the existence of a non-Gaussian UV fixed point of the gravitational renormalization group flow to control the behaviour of the theory at high energies and thereby keep physical quantities safe from unphysical divergences.
This idea was first put forward by Steven Weinberg \cite{Weinberg:1980} and has since been the focus of many searches for quantum gravity, the majority of which offer encouraging signs that an appropriate high-energy fixed point could indeed exist.

It may well turn out that we have to go beyond conventional QFT in order to describe gravity at the Planck scale and in the process introduce additional degrees of freedom and symmetries like those of string theory or additional spacetime structure as in loop quantum gravity, or perhaps something else is required altogether.
However, whether or not the asymptotic safety hypothesis turns out to be ultimately correct, it is important to make progress with fundamental aspects of the approach, a collection of which provide the focus of this thesis.

In the sections that follow we give the necessary background for the research presented in chapters \ref{cha:recon}, \ref{cha:back ind} and \ref{cha:asymp}. We begin with a review of the renormalization group as understood by Kenneth Wilson in section \ref{sec:Wilson RG}, before introducing the theory space on which the renormalization group flows play out in section \ref{sec:theory space}.
In the following section, \ref{sec:EAA}, we review the specific application of the renormalization group in the asymptotic safety approach to quantum gravity. Section \ref{sec:approxs} contains a discussion on popular approximation schemes employed in asymptotic safety, many of which are then used in the chapters that follow. Finally, we conclude this introductory chapter with an outline of the rest of the thesis.


\section{The Wilsonian renormalization group}
\label{sec:Wilson RG}

Naturally the scale at which we observe the world determines how we describe it.
We construct theories in terms of variables appropriate for the viewing scale and in fact we need not worry about what goes on at shorter distances (or equivalently, higher energies) in order to make successful predictions.
For example, to describe water flowing in a stream we do not need an understanding of water at the molecular level, instead the physics of fluid mechanics is enough.

However, by the very nature of their construction our theories are often blind to UV dynamics; they are effective theories with limited descriptive power and a finite realm of validity.
The scale at which a theory ceases to be applicable is aptly named the cutoff scale. It indicates the point at which our knowledge breaks down and beyond which new physics lies. As we have already mentioned, in the case of perturbative quantum gravity this is the Planck scale.
How then are we able to gain access to a high-energy (short-distance) description of nature?

An answer comes from the understanding of the renormalization group (RG) owed to Kenneth Wilson \cite{Wilson:1971bg}.
The RG is the mathematical formalism that enables us to systematically generate and relate descriptions of a system befitting different viewing scales, and for this reason is often said to be analogous to a microscope with varying magnification. The basic idea is that a system's microscopic degrees of freedom can be replaced by effective ones, together with appropriate rescaled interaction strengths, to give a different description of the system but which produces the same predictions for physical observables.
RG methods are at the heart of the asymptotic safety approach to quantum gravity and as such provide the focus of this section.

Wilson's RG has its origins in the study of condensed matter systems and so we begin this section by introducing key RG concepts through a discussion on Kadanoff blocking.
We then move on to review the continuum description of the RG due to Wilson and visit Polchinski's flow equation.
We end this section with a comparison between the renormalization of perturbation theory and the modern view of renormalizability that Wilson's ideas brought about.

\subsection{Kadanoff blocking}

\begin{figure}[t]
\centering
\begin{tikzpicture}
\foreach \j in {0, 2.4}{
\foreach \i in {0,0.4,1.6,2.4, 3.2}{
\draw[->] (0+\i,0+\j) -- (0+\i,0.25+\j);}}
\foreach \j in {0,2.4}{
\foreach \i in {0.8,1.2, 2, 2.8}{
\draw[<-] (0+\i,0+\j) -- (0+\i,0.25+\j);}}

\foreach \j in {0.4, 1.2, 2, 2.8}{
\foreach \i in {0.4,0.8,1.6,2.4}{
\draw[->] (0+\i,0+\j) -- (0+\i,0.25+\j);}}
\foreach \j in {0.4,1.2,2, 2.8}{
\foreach \i in {0,1.2,2,2.8, 3.2}{
\draw[<-] (0+\i,0+\j) -- (0+\i,0.25+\j);}}

\foreach \j in {0.8, 1.6, 3.2}{
\foreach \i in {0, 1.2, 2, 2.8, 3.2}{
\draw[->] (0+\i,0+\j) -- (0+\i,0.25+\j);}}
\foreach \j in {0.8, 1.6, 3.2}{
\foreach \i in {0.4, 0.8, 1.6, 2.4}{
\draw[<-] (0+\i,0+\j) -- (0+\i,0.25+\j);}}

\draw (2.2, 2.3) rectangle (3.4, 3.5);
\node at (1.6,-0.75) {(a)};

\draw[thick, ->] (3.45, 3.55) .. controls (4.7, 3.85) .. (7.55, 2.9);

\draw (1,  3.9) node {$\delta$};
\draw[thick] (0.8, 3.5) -- (0.8, 3.9);
\draw[thick, <-] (0.79, 3.9 ) -- (0.92, 3.9 );
\draw[thick] (1.2, 3.5 ) -- (1.2, 3.89);
\draw[thick, ->] (1.08, 3.9 ) -- (1.2, 3.9);

\foreach \k in {0.1,2.5}{
\foreach \l in {0.1,2.5}{
\draw[->] (5.25+\l,0.26+\k) -- (5.25+\l,0.51+\k);}}
\foreach \k in {0.1, 2.5}{
\foreach \l in {1.3}{
\draw[<-] (5.25+\l,0.26+\k) -- (5.25+\l,0.51+\k);}}
\foreach \k in {1.3}{
\foreach \l in {0.1, 1.3, 2.5}{
\draw[<-] (5.25+\l,0.26+\k) -- (5.25+\l,0.51+\k);}}

\draw[dashed] (5.15,0.25) rectangle (7.95, 3.2);
\node at (6.56,-0.75) {(b)};

\foreach \j in {0, 2.4, 3.2}{
\foreach \i in {0,0.4,1.6,2.4, 3.2}{
\draw[->] (10+\i,0+\j) -- (10+\i,0.25+\j);}}
\foreach \j in {0,2.4, 3.2}{
\foreach \i in {0.8,1.2, 2, 2.8}{
\draw[<-] (10+\i,0+\j) -- (10+\i,0.25+\j);}}

\foreach \j in {0.4, 1.2, 2}{
\foreach \i in {0.4,0.8,1.6,2.4}{
\draw[->] (10+\i,0+\j) -- (10+\i,0.25+\j);}}
\foreach \j in {0.4,1.2,2}{
\foreach \i in {0,1.2,2,2.8, 3.2}{
\draw[<-] (10+\i,0+\j) -- (10+\i,0.25+\j);}}

\foreach \j in {0.8, 1.6}{
\foreach \i in {0, 1.2, 2, 2.8, 3.2}{
\draw[->] (10+\i,0+\j) -- (10+\i,0.25+\j);}}
\foreach \j in {0.8, 1.6}{
\foreach \i in {0.4, 0.8, 1.6, 2.4}{
\draw[<-] (10+\i,0+\j) -- (10+\i,0.25+\j);}}

\foreach \j in {2.8}{
\foreach \i in {0.8, 1.6, 2}{
\draw[->] (10+\i,0+\j) -- (10+\i,0.25+\j);}}
\foreach \j in {2.8}{
\foreach \i in {0, 0.4, 1.2, 2.4, 2.8, 3.2}{
\draw[<-] (10+\i,0+\j) -- (10+\i,0.25+\j);}}

\draw[dashed] (12.2, 2.3) rectangle (13.4, 3.5);

\draw[thick, ->] (8, 3.25) .. controls (10.5, 4) .. (12.65, 3.64);

\node at (11.6,-0.75) {(c)};

\end{tikzpicture}
\vspace{5pt}
\caption{Illustration of a block-spin RG transformation in a 2-dimensional lattice of spins. Coarse graining proceeds from (a) to (b), followed by rescaling from (b) to (c). In (c) lattice sites previously outside the picture have been pulled in.}
\label{fig:blockspin}
\end{figure}
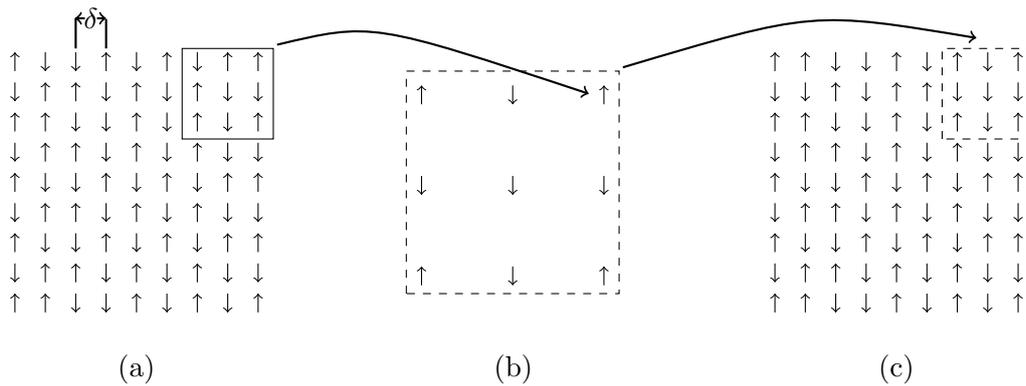
Consider a two-dimensional lattice of atoms each possessing two spin degrees of freedom, up or down, and with nearest-neighbour interactions, as shown in figure \ref{fig:blockspin}(a).
In this example the cutoff scale is given by the lattice spacing $\delta$.
Now suppose we average over a group of neighbouring spins and replace them by a single ``blocked'' spin at the centre. For example, a 3 by 3 block of spins containing mostly up spins is replaced by a single spin-up degree of freedom, and vice versa for down spins. The resulting picture is one with fewer degrees of freedom at an increased separation, see figure \ref{fig:blockspin}(b). This procedure is known as blocking or more generally as coarse graining.

In order to compare the coarse grained description of the system to the original microscopic picture, a second step is performed -- a rescaling -- to reduce the lattice spacing back to its original size, see figure \ref{fig:blockspin}(c).
This two-step process of coarse graining and rescaling is known as block-spin renormalization and was introduced by Leo Kadanoff in 1966 \cite{Kadanoff:1966wm}. Together the two steps make up a renormalization group transformation.

Block-spin renormalization gives us an alternative way of describing the lattice of spins i.e.\ in terms of coarse grained variables with appropriately scaled interaction strengths as opposed to the original microscopic degrees of freedom.
In this sense, a RG transformation is like a reorganization of what we already know.
In fact, not only does the block-spin procedure modify the spin-spin interactions, but it also gives rise to new ones. In the original lattice there are only nearest-neighbour interactions but the block-spin transformation generates next-to-nearest neighbour interactions, next-to-next-to-nearest neighbour interactions and so on.

Crucially though, these different pictures of the system still predict the same values for physical observables, so long as we consider physics at length scales much greater than the cutoff.
Performing an RG transformation changes the couplings in such a way so as to leave observables unchanged.
Indeed it seems reasonable to expect that when describing some long-distance phenomena, far away from the cutoff scale, predictions for observables should be insensitive to changes in it.
In the case of a lattice of spins such an observable would be the resistivity of a metal, which is independent of the precise inter-atomic spacing.


\subsection{Wilsonian renormalization}
\label{Wilsonian renormalization}
The RG transformations of Kadanoff's blocking procedure are concerned with discrete changes in the cutoff scale. In 1971 Kenneth Wilson introduced a version of the RG adapted to continuous changes in the cutoff which could be implemented through the path integral formulation of quantum field theory \cite{Wilson:1971bg}.

To illustrate this approach let us consider a single-component scalar field $\phi(x)$ with bare action $\hat{\mathcal{S}}[\phi]$. In the language of path integrals, physical observables of the field are then given by derivatives of the generating functional
\begin{equation}
\label{PI scalar}
Z[J]=\int^{\Lambda}\!\!\mathcal{D}\phi\,\mathrm{e}^{-\hat{\mathcal{S}}[\phi] +J\cdot\phi} \,,
\end{equation}
with respect to the external current $J(x)$.
We will use a dot notation to denote integration over position or momentum space: 
\be
\label{dot1}
J\cdot\phi\equiv J_{x}\phi_{x} \equiv \int d^{d}x J(x)\phi(x)=\int \frac{d^{d}p}{(2\pi)^{d}} J(p)\phi(-p)\,.
\ee 
For bilinear terms we regard the kernel as a matrix, thus the following forms are equivalent:
\be
\label{dot2}
\phi\cdot\Delta^{\!-1}\!\!\cdot\phi \equiv \phi_{x}\Delta^{\!-1}_{xy}\phi_{y} \equiv\int \!\!d^{d}xd^{d}y\,\phi(x)\Delta^{\!-1}(x,y)\phi(y)=\int \frac{d^{d}p}{(2\pi)^{d}}\phi(p)\Delta^{\!-1}(p^{2})\phi(-p)\,.
\ee 
Note that when transforming to momentum space, Green's functions $G(p_{1},\cdots,p_{n})$ come with momentum conserving delta functions such that they are only defined for $p_{1}+\cdots+p_{n}=0$. Thus two-point functions are functions of just a single momentum $p=p_{1}=-p_{2}$.
The integral \eqref{PI scalar} is endowed with a sharp UV cutoff $\Lambda$ such that only those modes propagating with momentum $|p| \equiv \sqrt{p^2}\leq\Lambda$ are integrated over.
Here and throughout the rest of this thesis we will now deal with energy scale cutoffs as opposed to length scale cutoffs $\delta=1/\Lambda$.
Note the Euclidean signature of the functional integral needed in order to take proper account of modes with nearly light-like four momenta. (In gravitational theories the Euclidean signature gives rise to the well-known conformal factor problem which has profound consequences for the RG properties of the theory in question. We discuss this in more detail in section \ref{subsec:CRG}.)
Finally, the requirement for physics to be independent of the cutoff in the context of the path integral means for the generator of Green's functions $Z[J]$ to be independent of $\Lambda$:
\begin{equation}
\label{phys ind}
\Lambda\frac{d Z[J]}{d\Lambda} = 0\,.
\end{equation}

\begin{figure}[t]
\centering
\includegraphics[scale=1.2]{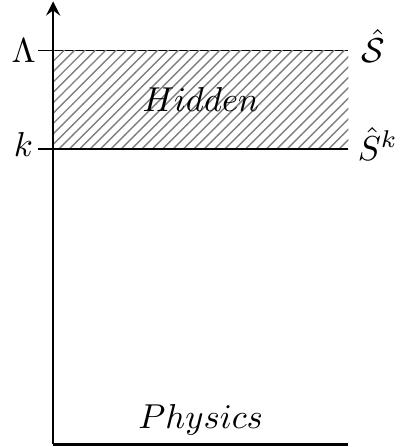}
\caption{Energy spectrum over which modes are integrated out.}
\label{fig:scales}
\end{figure}
Here coarse graining corresponds to lowering the cutoff by integrating out the high-energy degrees of freedom between $\Lambda$ and some lower energy scale $k$, cf.\ figure \ref{fig:scales}.
For simplicity, let us only consider observables with momenta less than the lower cutoff $k$ so that $J(p)=0$ for $|p|>k$. Splitting the modes into two sets, those with momenta $|p|>k$ denoted by $\phi_>$ and those with $|p|\leq k$ denoted $\phi_<$, we can rewrite \eqref{PI scalar} as
\begin{align}
Z[J]=&\int^{0<|p|\leq k}\!\!\mathcal{D}\phi_<
\int^{k<|p|\leq \Lambda}\!\!\mathcal{D}\phi_>\,\mathrm{e}^{-\hat{\mathcal{S}}[\phi_<+\phi_>] +J\cdot\phi_<}\\
\label{effective S}
=&\int^{0<|p|\leq k}\!\!\mathcal{D}\phi_<\,\mathrm{e}^{-\hat{S}^k[\phi_<] +J\cdot\phi_<}
\end{align}
where the result of integrating over a shell of momenta $k<|p|\leq \Lambda$ has been re-expressed in terms of a new, effective action\footnote{See chapter \ref{cha:recon} for a more comprehensive example and further discussions.}
\begin{equation}
\label{Seff}
\hat{S}^k[\phi] =  - \ln \int^{k<|p|\leq \Lambda}\!\!\mathcal{D}\phi_>\,\text{e}^{-\hat{\mathcal{S}}[\phi_<+\phi_>]}\ \,.
\end{equation}
This effective action predicts exactly the same low-energy ($E\ll k$) physics as the original bare action $\hat{\mathcal{S}}$.
It contains new interactions arising from the coarse graining procedure (just like we saw appear in block-spin renormalization).
As the cutoff is lowered, modes are removed from the propagator and ``hidden away" in the effective action, manifesting themselves as changes in the couplings, cf.\ figure \ref{fig:scales}.
These changes compensate for the change in the cutoff, meaning that $Z[J]$ and its functional derivatives remain unchanged i.e.\ they obey \eqref{phys ind}.
It follows that a simple Lagrangian at the cutoff scale $\Lambda$
will become more complicated as the the cutoff is lowered, growing new interactions, including contributions from irrelevant operators.\footnote{With this in mind, it no longer makes sense to insist that Lagrangians only contain relevant operators. Indeed, in the application of the Wilsonian RG to asymptotic safety we allow for all possible operators consistent with symmetry constraints.}

We still need to perform the rescaling step. This can be most easily achieved by making all quantities (fields and their couplings) dimensionless by dividing by the effective scale $k$ raised to the power of their scaling dimension.
This is equivalent to rescaling distances and momenta, and sends the cutoff back to its original size.
Thus writing everything in terms of dimensionless quantities, in addition to the coarse graining step as described above, completes a RG transformation in the Wilsonian approach.
Applying successive RG transformations gives a series of effective actions:
\begin{equation}
\hat{S}\quad \rightarrow \quad \hat{S}'\quad \rightarrow\quad \hat{S}''\quad \rightarrow\quad \cdots
\end{equation}
describing a system up to successively decreasing cutoff scales.

Joseph Polchinski adapted Wilson's RG by introducing a smooth momentum scale cutoff in a more direct way \cite{Polchinski:1983gv}.
This is achieved by incorporating a smoothly varying cutoff-dependent function $f$ into the propagator like so\footnote{where the mass term is contained within the interactions.}
\begin{equation}
\frac{1}{p^2} \rightarrow \frac{f(p^2/k^2)}{p^2} \equiv \Delta^k\,.
\end{equation}
The function $f$ has the property that for $|p|<k$, $f \approx 1$ and mostly leaves modes unaffected whilst for $|p|>k$, $f$ suppresses modes, vanishing rapidly at infinity.
Using the modified propagator, \eqref{effective S} instead becomes
\begin{equation}
\label{pol path int}
Z[J]=\int\!\!\mathcal{D}\phi\,\mathrm{e}^{-\frac{1}{2}\phi\cdot(\Delta^k)^{\!-1}\cdot\phi - S^k[\phi]+J\cdot\phi} \,,
\end{equation}
where for the sake of neatness we have made the replacement $\phi_<\rightarrow\phi$ and where the effective action $\hat{S}^k[\phi]$ has been split into a kinetic part and interactions $S^k[\phi]$.
The path integral is smoothly regulated in the UV by the cutoff function $f$.
Polchinski showed that if the effective interactions $S^k[\phi]$ satisfy the following integro-differential equation \cite{Polchinski:1983gv}
\begin{equation}
\label{pol flow}
\frac{\partial}{\partial k}S^k[\phi]=\frac{1}{2}\frac{\delta S^k}{\delta\phi}\cdot \frac{\partial\Delta^k}{\partial k}\cdot	\frac{\delta S^k}{\delta\phi}-\frac{1}{2}\text{Tr}\bigg(\frac{\partial\Delta^k}{\partial k}\cdot \frac{\delta^{2}S^k}{\delta\phi\delta\phi}\bigg)\,
\end{equation}
then \eqref{phys ind} (with the replacement $\Lambda\rightarrow k$ for the case at hand) follows.
This is Polchinski's version of Wilson's flow equation \cite{Morris:1993,Morris:1998}\footnote{For a more careful comparison between Wilson's and Polchinski's versions see reference \cite{Bervillier:2013kda}.} which we will see again shortly in chapter \ref{cha:recon}.
It expresses how the effective interactions must change as the cutoff is lowered in order to keep $Z[J]$ constant.
It is commonly referred to as an exact RG equation (ERGE) as no approximation is used in its derivation. In particular, it does not rely on small couplings.

\subsection{The Wilsonian perspective}
Wilson's approach brought about a new understanding of renormalizability in quantum field theory.
In the old view of renormalization a cutoff is introduced to loop integrals to enable their computation on the way to calculating scattering amplitudes and is nothing more than a mathematical trick.
Physical quantities are then made independent of the cutoff (they are ``renormalized") such that its value can be safely taken to infinity at the end of the calculation with physical quantities remaining finite.
This is the familiar renormalization of perturbation theory, implemented for example by redefining bare couplings in terms of renormalized ones or subtracting divergences with a finite number of counter terms.

From the modern Wilsonian perspective the cutoff should be viewed as physically meaningful and all quantum field theories in possession of one should be treated as effective theories only valid up to the cutoff scale.
As already mentioned, the cutoff represents the scale at which our knowledge breaks down and therefore we cannot justify taking the limit $\Lambda\rightarrow\infty$, at least not before knowing the high-energy behaviour of a theory.\footnote{Indeed from this point of view the action of sending $\Lambda\rightarrow\infty$ in perturbation theory is misleading. For example, QED can be renormalized perturbatively -- at low energy when the couplings are small -- but at high enough energies ($\approx 10^{300}$ GeV) it still develops divergences in spite of the limit $\Lambda\rightarrow\infty$ having already been taken.}
For a theory to be renormalizable from the Wilsonian perspective means that it is truly free from divergences at \emph{all} scales: no divergences appear, no matter how high we take the cutoff.
Technically, this is achieved by arranging for the theory to emanate from a UV fixed point, the subject of the next section.

Unlike perturbation theory, the Wilsonian RG does not rely on couplings being small and therefore represents a non-perturbative approach to renormalization.
This is one of its chief advantages as it opens the door to exploring the non-perturbative regime of quantum theories such as gravity.

In summary, in both the perturbative and nonperturbative regimes, the word ``renormalization'' refers to a way of dealing with divergences, but the methods by which this is done are conceptually and technically different. From the Wilsonian viewpoint, theories such as QED which appear renormalizable where perturbation theory is valid, are not truly renormalizable in the full non-perturbative sense of the word.
Wherever we use the term renormalization we will mean it in the sense of the Wilsonian renormalization group, also known in the continuum as the exact renormalization group (ERG), the functional renormalization group (FRG) and the continuous renormalization group.


\section{Fixed points and theory space}
\label{sec:theory space}

Now that we have reviewed the Wilsonian RG and seen an example of an exact RG equation, we are ready to examine the space on which its solutions live: theory space.
In this section we introduce the concept of theory space and discuss its key features, namely fixed points, as well as highlighting the properties they must exhibit in order for asymptotic safety to be realised.
We continue to use the scalar field throughout for illustrative purposes.

Theory space by definition is the space containing all possible actions that can be built from a given set of fields obeying certain symmetry constraints.
An action in the space is assumed to have the form:
\begin{equation}
\label{S sum}
S^k[\phi] = \sum g_i(k) \mathcal{O}_i(\phi)\,,
\end{equation}
where $g_i$ are the dimensionless, $k$-dependent couplings and $\mathcal{O}_i$ are operators made up of products of the dimensionless fields and their derivatives.
Furthermore, the $g_i$s do not include redundant (a.k.a.\ inessential) couplings i.e.\ those which can be eliminated from the action by a field redefinition.
The operators form the basis of the theory space whilst the couplings play the role of coordinates.
In this way, each point in the space represents a different possible action.
A priori the sum \eqref{S sum} is infinite as we allow for all possible couplings and therefore so is the dimension of the theory space.
In a later section, \ref{sec:approxs}, we will discuss reducing the dimension by making approximations.

\begin{figure}
\centering
\includegraphics[scale=0.75]{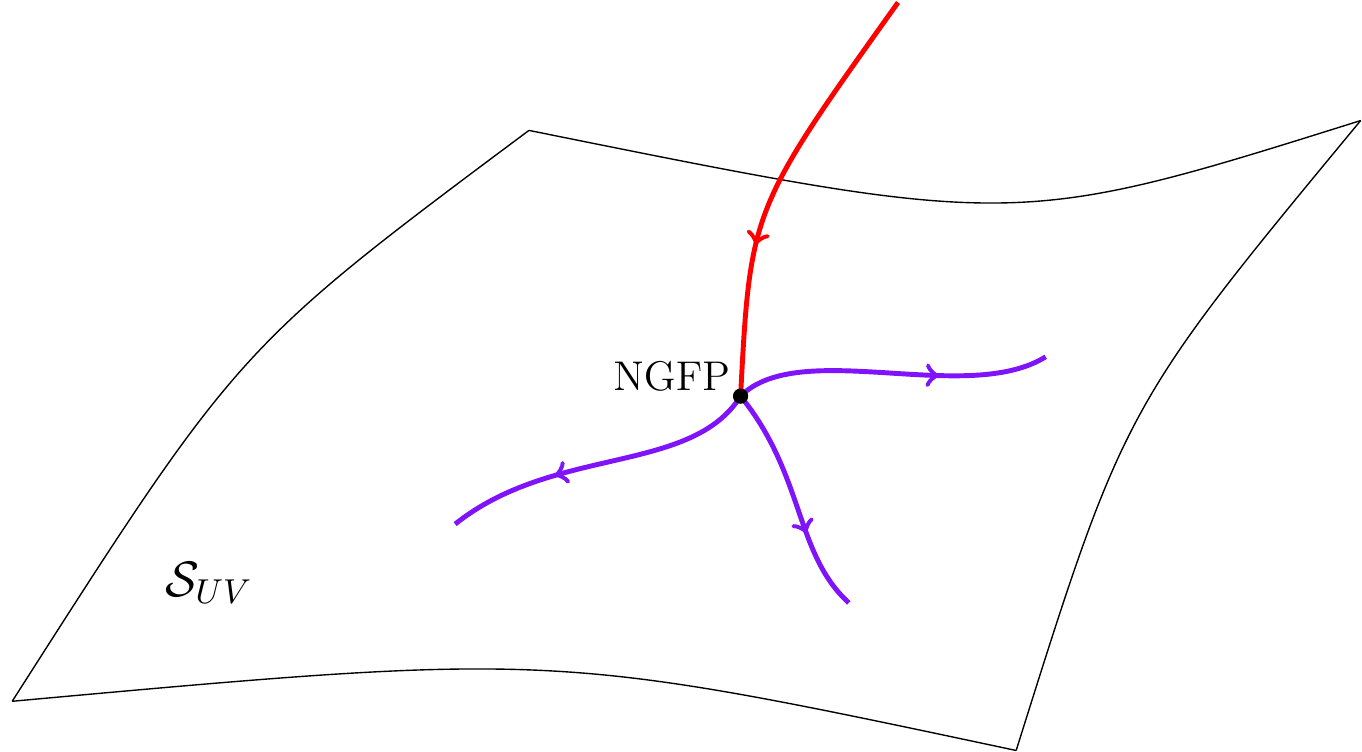}
\caption{Schematic picture of UV critical surface $\mathcal{S}_{UV}$ in theory space and RG trajectories flowing from high to low $k$ in the direction of coarse graining. The surface contains a non-Gaussian fixed point (NGFP) supporting renormalized trajectories (purple lines). There is also a trajectory coming from outside the surface and flowing into the fixed point (red line); for this trajectory, the fixed point is IR.}\label{fig:RGflows}
\end{figure}

Performing a RG transformation corresponds to moving between effective actions in theory space along a RG trajectory or flow line.
In geometrical terms, these RG trajectories are the induced integral curves of the vector field defined by a RG equation, such as Polchinski's, \eqref{pol flow}.
Thus the trajectory gives a way of visualizing the evolution of a theory with changes in the cutoff scale as described by the RG.
By convention we flow from high to low energy, in the direction of increasing coarse graining\footnote{Coarse graining can only be performed in one direction - we can only integrate out modes, we cannot ``integrate them in'' - but once the trajectory is defined, we can flow in either direction.} as indicated by the arrows in figure \ref{fig:RGflows}.
It is important to point out here that it is the trajectory itself that we identify with a theory, not the individual actions.

Features of theory space of particular interest are fixed points.
These are sources and sinks of RG flows and are home to scale-invariant theories $S^*$, i.e.\
\begin{equation}
k \frac{\partial}{\partial k} S^*[\phi] = 0 \,.
\end{equation}
Recall that all variables have been made dimensionless using $k$ and so independence of $k$ implies that $S^*$ depends on no scale at all.
It follows that fixed point theories are massless.
This scale independence also makes fixed point theories trivially renormalizable as we can trivially send $k\rightarrow\infty$.
This limit is referred to as the continuum limit and theories which have one are said to be UV complete.
A fixed point action therefore describes physics at the Planck scale and beyond.

For a given UV fixed point, there exists a submanifold called the critical surface $\mathcal{S}_{UV}$, as shown in figure \ref{fig:RGflows}. By definition, any point in theory space -- i.e.\ any action -- on this surface is pulled towards the fixed point under the reverse RG flow (against the directions of the arrows).
The portion of the critical surface local to the fixed point, is spanned by so-called relevant operators\footnote{These also include marginally relevant operators.} -- those whose coefficients in the action increase as we move out from the fixed point i.e.\ as $k\rightarrow 0$.
Perturbing the fixed-point action along the relevant directions gives rise to a ``renormalized trajectory", indicated by the purple lines in the figure.
The trajectory represents a renormalizable theory as its high-energy behaviour is controlled by a fixed point, i.e.\ as we take the limit $k\rightarrow\infty$ and approach the UV fixed point, the couplings of the theory tend to fixed finite values and are protected from blowing up.
Since observable quantities can be expressed as functions of the couplings, this means that they will also remain finite when the continuum limit is taken.

The effective actions sitting on a renormalized trajectory are called ``perfect actions" \cite{Hasenfratz:1993sp}. All their scale dependence is carried through the couplings and the anomalous dimension $\eta(k)$:  $S^k[\phi] = S[\phi]\left(g_1(k),...,g_n(k), \eta(k)\right)$. This means that the actions undergo a self-similar evolution under RG transformations. We return to these perfect actions in the next chapter.

The number of relevant operators spanning the fixed point (a.k.a.\ eigenperturbations or eigenoperators) gives the dimension $d_{UV}$ of $\mathcal{S}_{UV}$, which will therefore contain a $d_{UV}$-parameter set of trajectories.
Which trajectory is realized in nature will be decided by experiment.
For asymptotic safety we require $d_{UV}$ to be finite otherwise we lose predictivity -- we would have to take an infinite number of measurements to fix the infinite number of couplings.
Consequently, the smaller the dimension of $\mathcal{S}_{UV}$, the more predictive the theory will be.
In the asymptotic safety literature, usually fixed points with a finite number of relevant directions (typically three) are found (see e.g.\ reviews \cite{Reuter:2012, Percacci:2011fr, Niedermaier:2006wt, Nagy:2012ef, Litim:2011cp}), however there are also examples of fixed points which support a continuous spectra of eigenperturbations, see e.g.\ \cite{Dietz:2016gzg}.

Whether a fixed point is classified as UV or IR will depend on the trajectory under consideration.
If instead as we flow in the direction of coarse graining, we are pulled \textit{into} a fixed point then, as far as this trajectory is concerned, it is an IR fixed point.
Furthermore, what is a UV fixed point for one trajectory may be an IR fixed point for another.
Hence, in addition to the renormalized trajectory flowing out of the fixed point in figure \ref{fig:RGflows}, there may also be trajectories flowing into the fixed point, indicated by the red line.
Interestingly, this implies that very different physical systems described by very distinct theories can exhibit the same low-energy behaviour.
The observation that the macroscopic description of a phenomenon is independent of the microscopic details is known as universality.
Indeed this situation could be realised in a UV complete theory of quantum gravity if it supported more than one high-energy fixed point.

The UV fixed points required for non-perturbative renormalizability can be Gaussian or non-Gaussian, home to either free or interacting theories respectively.
For example, the theory of QCD possesses a Gaussian fixed point in the UV supporting interacting relevant directions: a free theory at the fixed points grows into a theory of interacting quarks as we flow to the infrared.
Theories exhibiting such fixed points are said to be asymptotically free and are naturally renormalizable since again the UV dynamics are controlled by a fixed point.
Of course the UV fixed points of most interest to quantum gravity searches are non-Gaussian. (Asymptotic safety at a Gaussian fixed point would be equivalent to perturbative renormalizability plus asymptotic freedom, but as noted at the start of the chapter, perturbative quantisation of gravity fails.) Theories emanating from such fixed points exhibit asymptotic safety. For this reason renormalized trajectories are also called asymptotically safe.

Preferably we want the theory space to support only one non-Gaussian fixed point (NGFP), or at least a finite number, otherwise again we lose predictivity.
However, there are examples in the literature in which lines and planes of fixed points have been uncovered, see \cite{Dietz:2016gzg, DietzMorris:2013-1, Dietz:2013sba, Gonzalez-Martin:2017gza}.
On the contrary, it might turn out that the theory space contains no fixed points.
One reason found for this in gravitational theories is that background independence has not been properly taken care of \cite{Dietz:2015owa, Labus:2016lkh} as discussed in chapter \ref{cha:back ind}.
The number of fixed points supported by the theory space is determined by counting up the number of independent parameters and constraints coming from the RG equation at the fixed point and its asymptotic solutions. This is the subject of chapter \ref{cha:asymp}.

In summary, we have seen that for the asymptotic safety scenario to be realised, a theory space must contain NGFPs (and preferably only one) with a finite number of relevant directions. Further to this, fixed points must support a renormalized trajectory that reproduces the behaviour of classical gravity at low energies.


\section{The effective average action and its flow}
\label{sec:EAA}

Having introduced the concept of the RG and the space on which its flows play out, in this section we review the specific application of the RG to asymptotic safety.
In the first part we introduce the central tools of the field -- namely the effective average action and its flow equation -- whilst continuing to work within the setting of scalar field theory so as to illustrate the key concepts in the simplest way possible.
The purpose of the proceeding subsection is then to review the necessary modifications when applying these ideas to gravity.
The final subsection contains a discussion on background independence, an important requirement for any theory of gravitation, which will be of particular relevance to chapter \ref{cha:back ind}.

Historically the first hints of asymptotically safe gravity came from applying Wilson's ideas in $2+\epsilon$ dimensions \cite{Weinberg:1980}.
Nowadays proponents of the field use a reformulation of Wilson's exact RG given in terms of the effective average action $\Gamma_k$, a scale dependent version of the usual effective action $\Gamma$ i.e.\ the generator of one-particle irreducible Green's functions. For a scalar field the effective average action is defined via the Legendre transform of a functional integral of the following form
\begin{equation}
\label{scalar func}
Z[J] = \int^\Lambda \!\! \mathcal{D} \phi \, \mathrm{e}^{-\hat{\mathcal{S}}^[\phi]-\Delta S_k[\phi] + J\cdot\phi} 
\equiv\mathrm{e}^{W^\Lambda_k[J]} \,,
\end{equation}
which is related in the usual way to the generator of connected Green's functions $W^\Lambda_k[J]$.
Just as in section \ref{Wilsonian renormalization}, the integral is subject to an overall UV cutoff that is required to make sense of the integration. Here it is implemented by a sharp cutoff at $\Lambda$, but it could equally well be of a different type (see chapter \ref{cha:recon} for examples).
The functional integral also depends on another cutoff scale, $k$.
When working with the Wilsonian action in the previous chapter, $k$ denoted the effective UV cutoff scale, whereas here it represents an IR cutoff.
This might seem like an unnecessary complication, however the reason for this choice becomes clear in chapter \ref{cha:recon} where a relationship between the effective average action and the Wilsonian effective action is derived.
To avoid confusion, we will denote any UV cutoff parameter with a superscript and any IR cutoff parameter with a subscript and use this pictorial guide throughout.

The dependence on $k$ is introduced via the IR cutoff operator $R_k$ which lives inside the cutoff action:
\begin{equation}
\Delta S_k[\phi] = \frac{1}{2} \phi \cdot  R_k \cdot \phi \,.
\end{equation}
The cutoff operator is a function of the Laplacian: $R_k=R_k(-\nabla^2)$, and acts on the field $\phi$ to turn $\Delta S_k$ into a mass-like term.
Roughly speaking, $R_k$ suppresses modes propagating with momentum $p^2<k^2$, otherwise leaving them unaffected.
The precise way in which it does this is unimportant but it must satisfy the two limits
\begin{equation}
\lim_{p^2/k^2\rightarrow 0} R_k(p^2) = k^2 \quad\text{and}\quad \lim_{p^2/k^2 \rightarrow \infty} R_k(p^2) = 0 \,.
\end{equation}
Popular choices for $R_k$ include the optimized cutoff $R_k(p^2) = (k^2 -p^2)\Theta(k^2 - p^2)$ \cite{opt1, Litim:2001, opt3} and the exponential cutoff $R_k(p^2)=(p^2/k^2)[\text{exp}(p^2/k^2)-1]^{-1}$.

The effective average action $\Gamma^\Lambda_k$ is obtained by subtracting the cutoff action $\Delta_k S$ (as a functional of the classical fields) from the Legendre transform of \eqref{scalar func}:
\begin{equation}
\label{EAA 2 cutoffs}
\Gamma^\Lambda_k[\varphi] \equiv \tilde{\Gamma}^\Lambda_k[\varphi]
- \frac{1}{2}\varphi\cdot R_k \cdot\varphi \,,
\end{equation}
where $\tilde{\Gamma}^\Lambda_k= -W^\Lambda_k[J] + J\cdot\varphi$ is the Legendre transform and $\varphi(x)\equiv\langle\phi(x)\rangle$ is the expectation value a.k.a.\ classical field.

The flow equation for the effective average action is obtained by taking the derivative of \eqref{EAA 2 cutoffs} with respect to $k$:\footnote{The steps are given in chapter \ref{cha:back ind}.}
\begin{equation}
\label{flow 2 cutoffs}
 \partial_k \Gamma^\Lambda_k[\varphi] = \frac{1}{2}\text{Tr}^\Lambda\!\left[\left(\frac{\delta^2\Gamma^\Lambda_k}{\delta\varphi\delta\varphi} + R_k \right)^{\!\!-1} \partial_k R_k \right] \,.
\end{equation}
The trace is taken over position (or momentum) space coordinates and here is restricted to only those modes propagating with momentum $|p|\leq\Lambda$.
Note that at this point both the effective action $\Gamma^\Lambda_k$ and the flow equation depend on \textit{two} scales: the IR cutoff scale $k$ and the UV cutoff scale $\Lambda$.
However, the derivative $\partial_k R_k$ is sharply peaked around $p^2 = k^2$, dying off rapidly for $p^2\gg k^2$, and so the left-hand side of the flow equation only receives contributions from modes near (or below) $k$.
This means that the trace is prevented from blowing up in the limit $\Lambda\rightarrow\infty$ and the UV cutoff can be safely removed. Doing this yields the following RG equation \cite{Wetterich:1992, Morris:1993}\footnote{Dimensionless RG time $t=\ln(k/\mu)$ where $\mu$ is a fixed reference scale is also commonly used instead of $k$.}
\begin{equation}
\label{FRGE}
\partial_k \Gamma_k[\varphi] = \frac{1}{2}\text{Tr}\left[\left(\frac{\delta^2\Gamma_k}{\delta\varphi\delta\varphi} + R_k \right)^{\!\!-1} \partial_k R_k \right] \,.
\end{equation}
It is this ``$\Lambda$-free" flow equation which is employed in current investigations into asymptotically safe gravity.
In contrast to \eqref{flow 2 cutoffs}, its solutions $\Gamma_k$ depend only on a single scale, $k$. This is crucial to its use as it allows us to express everything in terms of dimensionless couplings $g_i(k)$ with respect to the single dimensionful parameter $k$, i.e.\ to recover the power of the Wilsonian RG. From now on when referring to the flow equation and its solutions we will mean the $\Lambda$-free versions.

Now let us comment on some key features of the flow equation \eqref{FRGE}. First of all, a solution $\Gamma_k$ of \eqref{FRGE} represents an action for a system in which the high-energy modes (with respect to $k$) have been integrated out and provides a natural effective action for processes occurring at energies $E \approx k$. A complete set of well-behaved solutions to the flow equation $\{\Gamma_k, 0\leq k< \infty\}$ corresponds to a complete RG trajectory, free from divergences in both the IR and UV. As we saw in the previous section, the latter condition is realized by arranging the trajectory to originate from a high-energy fixed point.\footnote{Note that all theory space concepts described in the previous section apply equally well to the effective average action.}

Secondly, just like Polchinski's, \eqref{FRGE} is an exact RG equation suitable for the non-perturbative regime. However, despite the flow equation itself being exact, in practice it is not possible to solve it exactly and an approximation to the effective average action has to be made. These approximations are the subject of section \ref{sec:approxs}.

Thirdly, since the infrared cutoff $k$ is introduced by hand, it is an artificial quantity that must not feature in physical observables. The physical part of the effective action is therefore only recovered when the cutoff is removed. This is done by taking the limit $k\rightarrow 0$ whilst holding all physical, i.e.\ unscaled, quantities fixed.
It is in this limit that we recover the information contained in the full path integral.

There is a further property of this set up which is important to recognise: given a solution $\Gamma_k$ of the flow equation, is it not possible to exactly recover the path integral \eqref{scalar func} from which it was derived. Or more specifically, there is no exact way to reconstruct the bare action $\hat{\mathcal{S}}$ from an effective average action $\Gamma_k$.
In short, the reason for this is that a UV regulated path integral cannot through the Legendre transform procedure return an effective action $\Gamma_k$ but instead necessarily gives rise to an effective action $\Gamma_k^\Lambda$ which depends explicitly on two cutoffs.
At the point of defining $\Gamma_k$, all reference to the UV cutoff is lost and so there is no way to gain access to the bare action in the original UV regularized functional integral from the solutions of the $\Lambda$-free flow equation.

Conceptually there is nothing wrong with simply working with the flow equation \eqref{FRGE} and forgoing defining a path integral representation of the theory.
In this way, we dispense with the need to define a bare action at the overall cutoff scale and concomitant tuning required to reach the continuum limit.
One of the main advantages of working with the effective average action over the path integral is that it lends itself to more powerful approximation techniques. 
Being able to work with approximations is crucial as solving the flow equation is equivalent to, and practically as difficult as, solving the original path integral from which it came.
Furthermore, since $\Gamma_k$ is the $k$-dependent generator of one-particle irreducible Green's functions, it is directly related to scattering amplitudes which means that once we have found a complete trajectory, taking consecutive functional derivatives of $\Gamma_k$ give us all the Green's functions of the theory and in the limit $k\rightarrow 0$ they coincide with those of the standard effective action $\Gamma \equiv \Gamma_{0}$.

Despite the advantages of using the effective average action, there are still reasons for wanting a path integral formulation of the theory.
For example, to more easily understand certain properties of the QFT such as constraints and symmetries and to compare with other approaches to quantum gravity.
The challenge of obtaining a path integral representation is called the reconstruction problem and is the subject of chapter \ref{cha:recon}.

Even though we cannot directly obtain the bare action from the effective average action as emphaszied above, a simple and exact relationship between $\Gamma_k$ and the Wilsonian effective action $\hat{S}^k$ (introduced in \eqref{Seff}) does exist \cite{Morris:1993, Morris:2015oca}.
Referring back to figure \ref{fig:scales}, it need not seem so surprising that there is such a relationship \cite{Morris:1998}. 
In the discussions on the Wilsonian RG in section \ref{Wilsonian renormalization}, we saw that integrating out degrees of freedom between $\Lambda$ and some lower cutoff scale $k$ resulted in a Wilsonian effective action $S^k$ with the scale $k$ acting as a UV cutoff for the unintergrated modes. On the other hand, $k$ can also be regarded as an infrared cutoff for the modes which have already been integrated out (those which reside in the shaded area of figure \ref{fig:scales}). From this perspective we see that the Wilsonian effective action is almost equivalent to the original functional integral, but modified by an infrared cutoff $k$, which in turn is straightforwardly related to $\Gamma_k$ in the continuum limit (cf.\ equation \eqref{partition to S} in chapter \ref{cha:recon}).
In chapter \ref{cha:recon} we derive this relationship and show how $\hat{S}^k$ can play the role of a perfect bare action which lives inside a fully UV regularised functional integral.


\subsection{The effective average action for gravity}
\label{sub:EAA grav}
Up to this point we have been using a scalar field to introduce key concepts in functional RG methods, but of course we need to go beyond scalar theory to study quantum gravity.
Instead of quantizing some field living on some predetermined spacetime background, in quantum gravity spacetime itself becomes the dynamical variable we wish to quantise, and with this give meaning to the path integral over all metrics
\begin{equation}
\label{PI grav}
\int\!\! \mathcal{D} \tilde{g}_{\mu\nu}\, \mathrm{e}^{-\hat{\mathcal{S}}[\tilde{g}_{\mu\nu}]}
\end{equation}
and its associated effective average action.
This brings with it new challenges, both conceptual and technical in nature.
In the following we give an overview of the construction of the effective average action for gravity and its flow equation. The derivation is more involved than for the case of the scalar field but the procedure follows the same general pattern.

To deal with the obstacles arising when applying the functional RG to gravity, a technique called the background field method is used.
It consists of decomposing the full metric $\tilde{g}_{\mu\nu}$ into a background metric $\bar{g}_{\mu\nu}$ and a fluctuation field $\tilde{h}_{\mu\nu}$ like so
\begin{equation}
\tilde{g}_{\mu\nu} = \bar{g}_{\mu\nu} + \tilde{h}_{\mu\nu}\,.
\end{equation}
The background metric is fixed but left completely arbitrary. 
The split shifts the integration \eqref{PI grav} over the total metric to one over the fluctuation field $\tilde{h}_{\mu\nu}$ i.e.\ it is the fluctuation field that is quantised in the path integral. Note that the fluctuation $\tilde{h}_{\mu\nu}$ is not restricted to being small here like in perturbation theory.

The bare action $\hat{\mathcal{S}}[\tilde{g}_{\mu\nu}]$ is invariant under diffeomorphisms,
\begin{equation}
\delta \tilde{g}_{\mu\nu} = \mathcal{L}_v \tilde{g}_{\mu\nu} \equiv v^{\rho}\partial_\rho\tilde{g}_{\mu\nu} + \partial_\mu v^\rho\tilde{g}_{\rho\nu} +\partial_\nu v^\rho\tilde{g}_{\rho\mu} \,,
\end{equation}
which after performing the background split can be written as
\begin{equation}
\delta\tilde{h}_{\mu\nu}= \mathcal{L}_v\tilde{g}_{\mu\nu} \quad\text{and}\quad \delta\bar{g}_{\mu\nu} =0 \,.
\end{equation}
Here $\mathcal{L}_v$ is the Lie derivative along the vector field $v^\mu\partial_\mu$.
These gauge transformations must be gauge-fixed to avoid over-counting seemingly distinct but physically indistinguishable metric configurations.
A gauge-fixing condition $F_\mu[\tilde{h};\bar{g}]=0$ is introduced into the path integral via the Fadeev-Popov procedure.
This results in a ghost action which then appears alongside the bare action.
The broken gauge symmetry of the path integral will eventually be communicated to the effective action via the generator of connected Green's functions, however we can restore diffeomorphism invariance to the effective action if we insist that it is invariant under the so-called background gauge transformations:
\begin{equation}
\label{back gauge trans}
\delta\bar{g}_{\mu\nu} = \mathcal{L}_v\bar{g}_{\mu\nu} \quad\text{and}\quad \delta\tilde{h}_{\mu\nu}=\mathcal{L}_v\tilde{h}_{\mu\nu} \,.
\end{equation}
These extra gauge choices are made possible thanks to the background field method.

Another key advantage of the this method is that it allows the construction of a covariant IR cutoff.
In this gravitational context, the IR cutoff operator becomes a function of the covariant Laplacian associated with the background field: $R_k(-\bar{\nabla}^2)=R_k(-\bar{g}^{\mu\nu}\bar{\nabla}_{\mu}\bar{\nabla}_{\nu})$. It is then with respect to the spectrum of $-\bar{\nabla}^2$ that fluctuation modes are compared to the cutoff scale $k$ and are either integrated out or suppressed.
A Laplacian of the total metric cannot be used as it would not preserve the structure of the flow equation as represented in \eqref{FRGE}.
This fact actually turns out to be of key significance in the quest for background independence, an important issue which we shall return to shortly. Note that the ghost fields also come with their own IR cutoff.

Once the gauge fixing, ghost and cutoff terms have all been included in the functional integral alongside the bare action and source terms for all the fields, the effective average action is obtained by following the analogous steps described in section \ref{sec:EAA} and which are explicitly laid out in \cite{Reuter:1996}. The result is the effective average action for gravity \cite{Reuter:1996}:
\begin{equation}
\Gamma_k [h,\bar{g},\xi,\bar{\xi}] \,,
\end{equation}
where $h$ is the classical fluctuation field, $\bar{g}$ is the background metric as before and $\xi, \bar{\xi}$ are the classical ghost fields.
The crucial observation here is that the effective action depends \textit{separately} on the background metric $\bar{g}$. This is due to the extra background field dependence of the ghost, gauge fixing and cutoff terms, which is in turn a consequence of using the background field method. 
As mentioned above, as long as the background gauge transformations \eqref{back gauge trans} are obeyed, i.e.\ the background metric transforms as an ordinary tensor field $\delta\bar{g}_{\mu\nu}=\mathcal{L}_\nu \bar{g}_{\mu\nu}$, the effective action is a diffeomorphism invariant functional of its fields: $\Gamma_k[\Phi +\mathcal{L}_\nu\Phi] = \Gamma_k[\Phi]$ where $\Phi=\{h_{\mu\nu},\bar{g}_{\mu\nu},\xi^{\mu},\bar{\xi}_{\mu}\}$.

The derivation of the flow equation for gravity goes through in much the same way as in the case of the scalar field (the explicit steps can be found in \cite{Reuter:1996}).
The result has the same general structure as \eqref{FRGE} but with the right-hand side featuring a trace for both the fluctuation field $h$ and ghosts $\xi,\bar{\xi}$ (with an additional minus sign for the anti-commuting ghost term).
The functional derivatives in the traces are taken at fixed $\bar{g}$.
Again, the UV cutoff on the functional integral drops out at the level of the flow equation due to the protective properties on the cutoff function $R_k$.

\subsection{Background independence}
\label{sub:BI}
As pointed out already, an essential ingredient for any theory of gravity is background independence. Background independence is the requirement that a theory be free from any prior geometry; instead, the properties of the spacetime should emerge as a prediction of the theory.
With this in mind, it might seem like a misstep to introduce dependence on a background metric through the background field method.
However, by leaving the background metric completely unspecified, no background configuration plays a distinguished role in the construction of the flow equation. This means that the flow equation does not rely on the properties of any particular background field, implying that quantisation of the fluctuation $\tilde{h}$ occurs on all backgrounds simultaneously.\footnote{Even then, background independence of the formalism is not guaranteed due to the inherent background dependence of the RG scale $k$. See end of section for further discussion.}
Nevertheless, the solutions of the flow equation \textit{do} depend on the background. They are forced to carry separate dependence on the background metric $\bar{g}_{\mu\nu}$ through the cutoff operator $R_k(-\bar{\nabla}^2)$ as previously emphasized.
Physics should depend only on the full metric, and not also on the background metric that was introduced by hand as part of the background field technique. This separate background dependence means that in general each background configuration would lead to different results for physical observables.

Not only do these solutions live in an appropriately enlarged theory space, spanned by operators of both the total metric and background metric, but the separate background field dependence makes further artificial enlargement of the theory space possible. A solution of the flow equation can be modified by an arbitrary scale-independent functional of the background field $\mathcal{F}[\bar{g}]$ such that the result $\Gamma_k[h,\bar{g},\xi,\bar{\xi}] +\mathcal{F}[\bar{g}]$ is also a solution to the flow equation. This additional freedom introduced by hand through the background field method also needs to be controlled.

It is thus necessary to go beyond simply making sure the formalism does not depend on any particular background and to also somehow manage the separate background field dependence of the effective action.
In most of the literature, the requirement of background independence refers only to the construction of the flow equation about an arbitrary background, whereas background independence in the sense that we mean it here is much more than this, and is in fact a strong extra constraint.

One way of circumventing these issues is to use the single field approximation.\footnote{spoken about in more detail in section \ref{sec:approxs}.}
This approximation consists of neglecting the evolution of the gauge-fixing and ghost sectors and setting $\bar{g}=g$ (equivalently, $h=0$) in $\Gamma_k[h,\bar{g}]$ such that the effective action becomes a functional of only one field, namely the total metric $g$.
Note that this can only be done once the functional derivatives in the trace have been performed as they are taken at fixed $\bar{g}$.
With the solutions of the flow equation then just depending on the total metric, the aforementioned issues are bypassed.

The single field approximation has been employed in the majority of works in asymptotic safety to date.
A severe drawback of this approximation however is that it cannot be used to explore the effects of background dependence as of course dependence on the background metric becomes invisible.
This can lead to unphysical results as has been seen in the LPA for scalar field theory \cite{Bridle:2013sra} and obscures the significance of fixed point solutions at large field in the $f(R)$ approximation as emphasized in \cite{Dietz:2016gzg}.
Instead, background dependence can only be investigated in bi-metric truncations in which dependence on both the full metric and the background metric is retained.
For studies going beyond the single field approximation in different ways see \cite{Manrique:2010am, Manrique:2009uh, Becker:2014qya, Manrique:2010mq, Codello2013, Christiansen:2014raa, Groh:2010ta, Eichhorn:2010tb, Dona:2013qba, Dona:2014pla}.

Working within bi-metric truncations, and therefore being able to take full account of the effects of background dependence, requires us to find an alternative way to manage the separate background dependence of the effective action.
This can be achieved by imposing an additional constraint alongside the flow equation known as a modified split Ward Identity (msWI). (See equation \eqref{equ:sWiGamma} for an example of what the msWI looks like in the context of conformally reduced gravity.)
Even though for all $k>0$ background independence will inevitably be lost due to the cutoff, imposing the msWI in addition to the flow equation ensures that exact\footnote{By exact we mean background independence in the strict sense defined previously.} background independence is recovered in the limit $k\rightarrow 0$ (the limit in which $R_k$ drops out) after going ``on-shell".
This is imperative for the attainment of background independent physical observables as it is in this limit that the physical part of the effective action is recovered, as already explained at the start of this section.
Furthermore, solutions of the flow equation do not automatically satisfy the msWI and in this way the msWI also controls the arbitrary enlargement of the theory space manufactured by the background split.

It is important to note that the msWI constraint is not an optional extra: it is derived from the same functional integral as the flow equation and therefore any set of (exact) solutions to the flow equation must also satisfy the msWI. In other words, the flow equation and msWI must be \textit{compatible}. 
In chapter \ref{cha:back ind} we prove that this is indeed true at the exact level, before any approximation to the effective action has been made. For approximate solutions, compatibility is not automatically guaranteed.
We show that in the case of approximation, namely a derivative expansion up to $\mathcal{O}(\partial^2)$ for conformally reduced gravity, extra conditions must be placed on the form of the cutoff or the anomalous dimension in order to achieve compatibility.

An unsettling conclusion from the research reported in \cite{Dietz:2015owa} was that fixed points with respect to the RG scale $k$ are in general forbidden by the msWIs that are enforcing background independence.
With hindsight, this can be seen as a useful signal that a background dependent description of quantum gravity does not make sense and a hint that there might be some deeper understanding of the meaning of RG in quantum gravity to be unearthed.
For scalar field theory at the level of the LPA in \cite{Bridle:2013sra}
and later in the setting of conformally reduced gravity \cite{Dietz:2015owa}, it was discovered that it is possible to combine the msWI and flow equation to uncover a background independent description of the entire flow, written in terms of background independent variables, including a background independent notion of the RG scale.

The employment of the msWI thus also remedies the issue of the ambiguity in the meaning of the scale $k$ in a gravitational setting.
Since it is the metric that provides us with the definition of length, the RG scale $k$ (which can be equally thought of as some inverse length $1/k$) is inherently dependent on it.
But moreover, in a quantum gravity theory, length scales fluctuate and so it is not clear what meaning should be ascribed to $k$ or indeed scale dependence as expressed through the RG.
Using the background field method alone does not resolve this issue since then $k$ is defined with respect to modes of the covariant background field Laplacian $-\bar{\nabla}^2$ and becomes inherently dependent on the background metric instead.


\section{Approximations}
\label{sec:approxs}

As emphasized in section \ref{sec:EAA}, it is usually impossible to solve the flow equation exactly and in order to actually make any progress we need to make an approximation for the effective average action.
Making an approximation corresponds to truncating the theory space to some lower dimensional subspace and evaluating the flow equation there.\footnote{One option is to do this by expanding the trace with respect to a small coupling, but of course this would only then allow us to explore the perturbative regime.}
The subspace should be chosen in such a way that it is small enough to make calculations feasible but yet still big enough to capture the essential physics. Despite not retaining all the information within the full effective action\footnote{or equivalently, the path integral}, approximations make computations manageable and prove a fruitful way to gain insights into important foundational issues in asymptotic safety.
The purpose of this section is to introduce well-known and much-used approximation schemes, the majority of which are employed in the chapters to come.


\subsection{The Einstein-Hilbert truncation}

The earliest truncation for which RG flows have been found is the Einstein-Hilbert truncation \cite{Reuter:1996}:
\begin{equation}
\label{EH truncation}
\Gamma_k[h,\bar{g},\xi,\bar{\xi}] = \frac{1}{16\pi G_k}\int d^4x \sqrt{g} \left( - R + 2\Lambda_k \right) + S_{\text{gf}}[h,\bar{g}] + S_{\text{gh}}[h,\bar{g},\xi,\bar{\xi}] \,,
\end{equation}
where the classical gauge fixing $S_{\text{gf}}$ and ghost actions $S_{\text{gh}}$ are chosen to be independent of $k$.
This ansatz utilizes the single field approximation which, now stated more precisely, means that the evolution of the ghosts is neglected, it features no $k$-dependent piece for which $\bar{g}\neq g$ and as before, we set $h=0$ once the Hessians have been computed.
Most notably, \eqref{EH truncation} contains two parameters which are allowed to run with energy: the cosmological constant $\Lambda_k$ and Newton's coupling $G_k$.

By inserting the ansatz into the flow equation, RG flows for the dimensionless Newton's coupling $\tilde{G}_k = k^2G_k$ and dimensionless cosmological constant $\tilde{\Lambda}_k = k^{-2}\Lambda_k$ can be determined.
This requires projecting the flow on to the chosen subspace of theory space.
Let us briefly review how this is done in the general case of a theory space comprised of functionals of the form $\Gamma_k[\varphi] = \sum_{i=1} g_i(k)\mathcal{O}_i(\varphi)$.
An approximation $\check{\Gamma}_k[\varphi]$ is made up of operators (perhaps infinitely many of them) belonging to the subspace only, for example $\check{\Gamma}_k[\varphi] = \sum^N_{j=1} g_j(k)\mathcal{O}_j(\varphi)$.
The general idea is to expand the trace on the right-hand side of the flow equation with the ansatz $\check{\Gamma}_k$ inserted on the basis $\{\mathcal{O}_i\}$ of the full theory space i.e.\
\begin{equation}
\label{projection}
\frac{1}{2}\text{Tr}[\cdots] = \sum^{\infty}_{i=1}\beta_i \mathcal{O}_i(\varphi) = \sum^N_{j=1} \beta_j \mathcal{O}_j(\varphi) + \text{rest}
\end{equation}
and retaining only those terms contained within the subspace i.e.\ neglecting the ``rest".
Here $\beta=\beta(g_1,g_2,\cdots)$ are the beta functions for the couplings which
unlike the beta functions of perturbation theory are not restricted to be functions of only small couplings.
Equating \eqref{projection} to the left-hand side of the flow equation, $\partial_k\check{\Gamma}_k = \sum^N_{j=1} \beta_j\mathcal{O}_j$, yields a system of $N$ coupled ODEs for the couplings.
Once these equations are solved, we say that the RG flow in the space of all couplings has been projected onto the $N$-dimensional subspace. Here we have used an approximation of the polynomial type as an example, but the same ideas apply to approximations involving full functionals as well; then instead of having coupled differential equations we obtain an evolution equation for the functional.\footnote{In fact this highlights a computational advantage of polynomial truncations over those retaining a full functional: the flow equation for a polynomial truncation is simply an ODE in $k$ yielding a finite number of relations for the couplings, whereas working with a full functional results in a partial differential equation which is technically more involved.}

\begin{figure}[t]
\centering
\begin{tikzpicture}
\node {\includegraphics[width=0.8\textwidth]{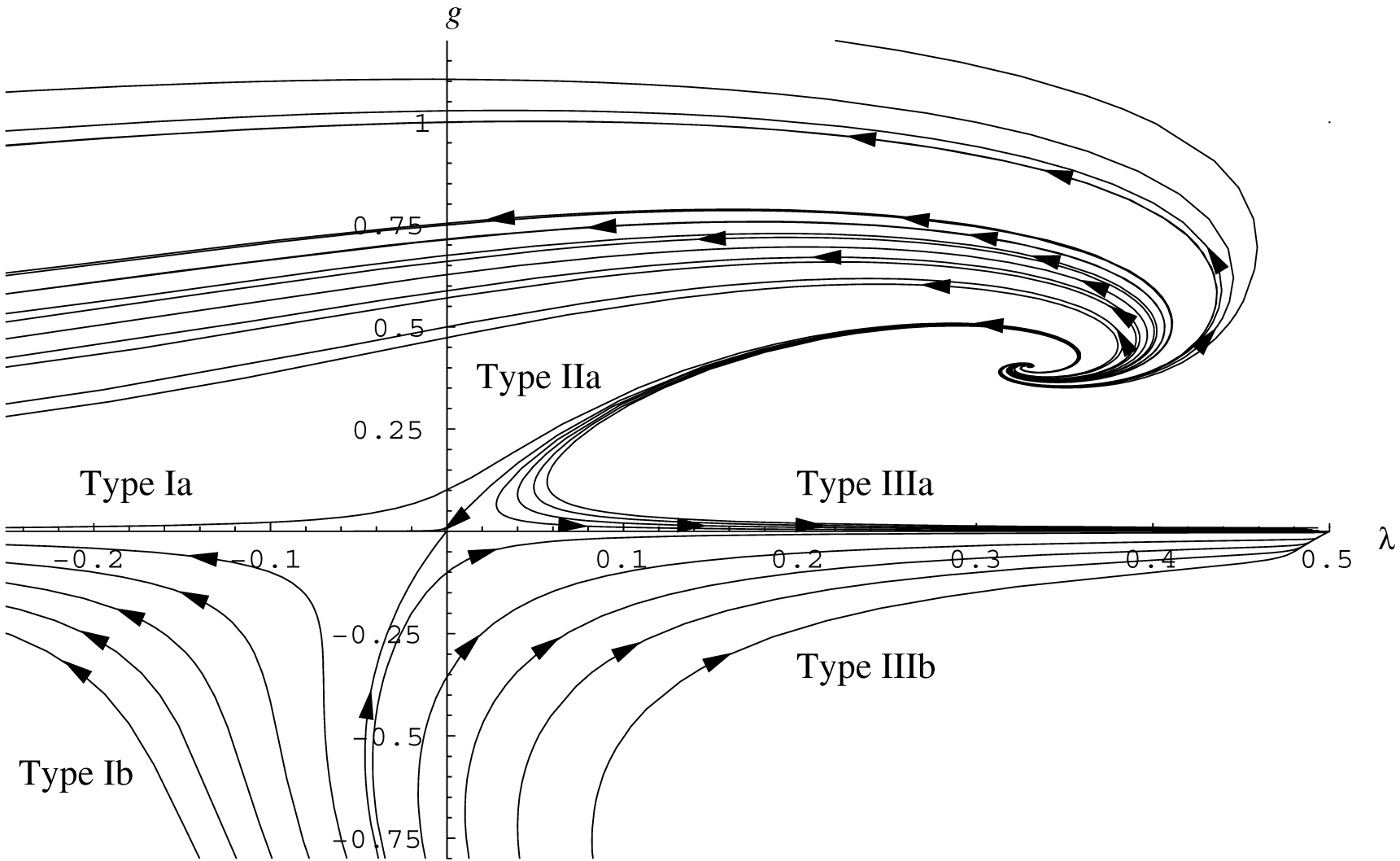}};
\fill[white] (-3, 3.4) rectangle (-1, 3.8);
\draw (-2.1, 3.75) node {$\tilde{G}_k$};
\fill[white] (5.7,-1.2) rectangle (6.1, 0.2);
\draw (5.9, -0.7) node {$\tilde{\Lambda}_k$};
\fill[white] (-6,-3.2) rectangle (-4.85, -2.65);
\fill[white] (-5.7,-0.65) rectangle (-4.2, -0.2);
\fill[white] (-2, 0.28) rectangle (-0.75, 0.65);
\fill[white] (0.7,-0.65) rectangle (2.3, -0.2);
\fill[white] (0.7,-2.2) rectangle (2.3, -1.9);
\end{tikzpicture}
\caption{Plot of RG flows from the Einstein-Hilbert truncation \eqref{EH truncation} in the $\tilde{G}_k-\tilde{\Lambda}_k$ plane \cite{Reuter:2001ag}.}
\label{fig:EHflows}
\end{figure}
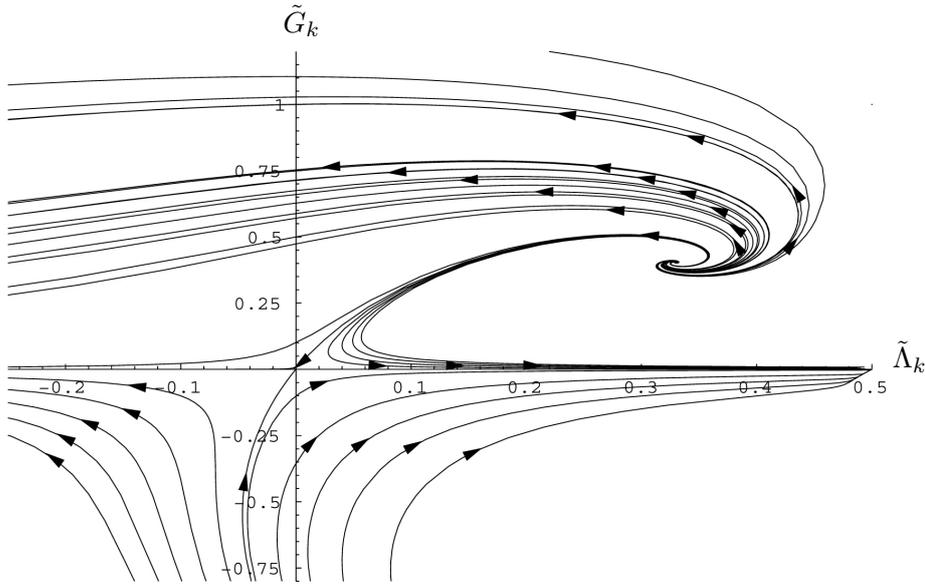
Carrying out this procedure for the Einstein-Hilbert truncation gives rise to the flows displayed in figure \ref{fig:EHflows}.
Notably the figure features two fixed points: a Gaussian one at the origin and a NGFP at positive values for both couplings.
Whilst the employment of different cutoff types shifts the position of the NGFP, it continues to be present for all cutoffs tested to date \cite{Reuter:2012}.
Furthermore, it is always found in the quadrant of positive $\tilde{G}_k$ and $\tilde{\Lambda}_k$ and is UV attractive for both couplings i.e.\ has two relevant directions.
The Einstein-Hilbert truncation \eqref{EH truncation} predicts a NGFP with the  desired properties for asymptotic safety and has been the subject of many studies within the community \cite{Reuter:1996, Lauscher:2001ya, Reuter:2001ag, Souma:1999at, Litim:2003vp, Fischer:2006fz, Codello:2008, Benedetti:2010nr}.
However, to be sure that a reputed fixed point is not just an artifact of insufficient approximation, we must go beyond the Einstein-Hilbert truncation.

\subsection{Polynomial truncations}
The natural next step is to explore less severe truncations. These so-called polynomial truncations keep successively higher powers of the scalar curvature $R$ and have to date included all powers up to $R^{34}$ \cite{Falls:2013bv, Falls:2014tra, Falls:2016wsa}.
In all cases asymptotically safe fixed points have been found.
This is encouraging, but it is easy to be misled into thinking fixed points exist as past studies have shown.
For example, in the local potential approximation for a single component scalar field, spurious fixed points have been found to persist in polynomial truncations of the potential to very high order. These fixed points can then be shown to disappear when the full potential is considered \cite{Morris:1994ki}.
Another example is given by \cite{Morris:1995he} which analysed the RG properties of U(1) theory in three dimensions using the approximation $f(F^2_{\mu\nu})$. There again non-Gaussian fixed points were found for $f(F^2_{\mu\nu})$ truncated to a polynomial, whereas using the full function resulted in no such fixed points.

We find that even though careful treatment of polynomial approximations taken to high order can allow extraction of convergent results, one does not see in this way the singularities at finite field or asymptotic behaviour at diverging field which are actually responsible for determining their high order behaviour. Indeed such large field effects can invalidate deductions from polynomial truncations \cite{Morris:1994ki,Morris:1995he,Morris:1996nx} and/or restrict or even exclude the existence of global solutions \cite{Morris:1994ie,Morris:1994jc,Morris:1997xj,Morris:1998}.
Another good example is provided by some of the most impressive evidence for asymptotic safety to date, (the polynomial expansions up to $R^{34}$),
which are however derived from a differential equation for an $f(R)$ fixed point Lagrangian \cite{Codello:2008} which was shown in \cite{Dietz:2012ic} to have no global solutions as a consequence of fixed singularities at finite field.

Furthermore, since fixed points are effectively the solutions of polynomial equations in the couplings, they only allow for discrete solutions.
But physical systems exist with lines or even higher dimensional surfaces of fixed points, parameterised by exactly marginal couplings (in supersymmetric theories these are common and called moduli). 
Moreover, lines and planes of fixed points have been found in other approximations within asymptotic safety \cite{Dietz:2012ic,Dietz:2013sba,Dietz:2016gzg}\footnote{and in a perhaps related approximation in scalar-tensor gravity \cite{Benedetti:2013nya}}.

As well as this, by construction polynomial truncations only deal with small curvatures, which has to be the case for an expansion in powers of $R$ to make sense. This means that polynomial truncations are insensitive to strong curvature effects and the deep non-perturbative regime of quantum gravity that we are ultimately interested in.

\subsection{The $f(R)$ approximation}
\label{sub:f(R)}
In order to have confidence that asymptotically safe fixed points exist we must therefore go beyond even polynomial truncations to approximations that keep an infinite number of couplings.
Arguably the simplest such approximation is to keep all powers of the scalar curvature, making the ansatz
\begin{equation}
\Gamma_k[g] = \int \!\!d^4x \sqrt{g}\, f_k(R)\,.
\end{equation}
This is called the $f(R)$ approximation and has been investigated in many works \cite{Machado:2007, Codello:2008, Benedetti:2012, Demmel:2012ub, Demmel:2013myx, Benedetti:2013jk, Demmel:2014hla ,Demmel2015b, Ohta:2015efa, Ohta2016, Percacci:2016arh, Morris:2016spn, Falls:2016msz, Ohta:2017dsq}.\footnote{In fact, to date this is the only such approximation that has been investigated, together with some closely related approximations in scalar-tensor 
\cite{Percacci:2015wwa,Labus:2015ska} and unimodular \cite{Eichhorn:2015bna} gravity, and in three space-time dimensions \cite{Demmel:2014fk}.}
Inserting such an approximation into the flow equation results in a non-linear partial differential equation which governs the evolution of $f_k(R)$ with changes in the RG scale $k$.
At fixed points, where the $k$-dependence drops out, it reduces to an ODE of either second or third order (depending on the cutoff scheme used). See equation \eqref{fp} in chapter \ref{cha:asymp} for an example written in terms of scaled variables $\varphi(r) :=k^4 f(R k^{-2})$.

In the $f(R)$ approximation we are no longer restricted to small curvatures, however this then raises the question: what significance should we attach to the behaviour of $f_k(R)$ for $R \gg 1$? Since then the size of the spacetime is much smaller than the cutoff $1/k$.
This puzzle is addressed and resolved in \cite{Morris:2016spn} and also discussed in more detail in the introduction to chapter \ref{cha:asymp}.

Finally, as already hinted at above in Polynomial truncations, in order to ascertain the true nature of fixed points it is crucial to explore the regime of large scaled curvature: $r\rightarrow \infty$, i.e.\ to develop the asymptotic solutions.
We could have already guessed that the behaviour of solutions in this limit is important to understand since for fixed background curvature\footnote{Here we commit a slight abuse of notation as, at the level of the projected flow equation, $R$ now represents the background curvature equation which emerges from employing the single field approximation.} $R$ it corresponds to the limit in which the physical effective action is recovered, $k \rightarrow 0$.
These asymptotic solutions are the central topic of chapter \ref{cha:asymp}.

\subsection{Conformally reduced gravity}
\label{subsec:CRG}
Conformally reduced gravity is the regime in which only fluctuations of the conformal factor of the metric are quantised. A small number of works have studied it using the exact RG, starting with reference \cite{Reuter:2008wj}.
To arrive at conformally reduced gravity we only consider a subset of metrics that are conformally equivalent to some fixed reference metric $\hat{g}_{\mu\nu}$:
\begin{equation}
\tilde{g}_{\mu\nu} = f(\tilde{\phi}) \hat{g}_{\mu\nu}\,.
\end{equation}
Here $\tilde{\phi}$ is the total conformal factor field and $f$ is some choice of parameterisation. It is then the fluctuation field $\tilde{\phi}$ that is integrated over in the path integral.
This leads to a scalar-like theory and a simpler model than say $f(R)$ for investigating the effects of background dependence and is of particular relevance to chapter \ref{cha:back ind}. 

Recent investigations in conformally reduced gravity have shed light on important foundational issues in asymptotic safety which deserve some comment.
Even though conformally reduced gravity and standard 4-dimensional scalar theory are very similar in structure (after all the conformal factor is a single component scalar field), the flow equation for the former comes with an additional minus sign, a result of the conformal factor problem already briefly alluded to below equation \eqref{PI scalar}.
As is well-known, the Euclidean signature functional integral for the Einstein-Hilbert action suffers from this problem \cite{Gibbons:1978ac}, which is that the negative sign for the kinetic term of the conformal factor yields a wrong-sign Gaussian destroying convergence of the integral. At first sight, providing the cutoff is adapted, the change in sign seems not to pose any special problem for the exact RG equation \cite{Reuter:1996}.
However as is shown in \cite{Dietz:2016gzg}, this one sign change has profound consequences for the RG properties of the solutions, broadly resulting in a continuum of fixed points supporting both a discrete and a continuous eigenoperator spectrum.

The conclusions reached in \cite{Dietz:2016gzg} seem to be strongly at variance with the asymptotic safety literature 
where a single fixed point with a handful of relevant directions (typically three) is found.\footnote{Actually a continuum of fixed points supporting a continuous spectra of eigenoperators has been found for the $f(R)$ approximation already in \cite{Dietz:2012ic}.}
The great majority of work in the literature however focuses on the single field approximation and/or polynomial truncations which can obscure the effects of the conformal factor problem;
whereas, in \cite{Dietz:2016gzg} use of these type of approximations was avoided -- the only approximations used were that of conformally reduced gravity itself and the slow field limit for the background field -- and furthermore, background independence was incorporated.
Further work is needed to understand whether this picture persists when working with the full metric; perhaps this might qualitatively alter the results.

\subsection{The derivative expansion}
The derivative expansion is an approximation originally developed for scalar field theory \cite{Morris:1994ie} and as such can be straightforwardly applied to conformally reduced gravity.
It consists of expanding an action in powers of derivatives of the field.
For standard scalar field theory, an expansion of the effective average action up to the third order looks like
\begin{align}
\Gamma_k[\varphi] = \int d^dx \big\{ &V(\varphi,t) + \frac{1}{2}K(\varphi,t)(\partial_\mu \varphi)^2 + H_1(\varphi,t)(\partial_\mu \varphi)^4 \nonumber\\
&+ H_2(\varphi,t)(\Box\varphi)^2+ H_3(\varphi,t)(\partial_\mu \varphi)^2(\Box\varphi) + \cdots \big\}\,,
\end{align}
which in momentum space amounts to an expansion in powers of momenta.

The leading order of the derivative expansion is called the Local Potential Approximation (LPA) introduced in \cite{Nicoll:1974zz} and since rediscovered by many authors e.g.\ \cite{Morris:1994ki,Hasenfratz:1985dm, Morris:1995af}.
This functional truncation keeps a general potential $V(\varphi)$ for the field and therefore incorporates infinitely many operators.
When keeping all components of the metric tensor, the $f(R)$-approximation is as close to the LPA as one can get, as emphasized in \cite{Benedetti:2012}.
We make use of the LPA, and more generally the derivative expansion, in chapter \ref{cha:back ind} in the setting of conformally reduced gravity.
\newline

\noindent Let us close this section by remarking that in practice expanding the trace and extracting the terms belonging to the subspace of an approximation is a rather involved technical process.
The background metric is often fixed to be that of a four-sphere to simplify calculations.\footnote{But note that there is no conceptual necessity for this and final results should be independent of the choice of background metric.}
A transverse-traceless decomposition of the fluctuation field $\tilde{h}_{\mu\nu}$ is performed to facilitate the computation of the inverse Hessian on the right-hand side of the flow equation and this introduces new fields.
Also to facilitate computation, different types of cutoffs are used, e.g.\ a type I cutoff where $R_k$ is just a function of $-\bar{\nabla}^2$ as in section \ref{sec:EAA}, or a type II cutoff, $R_k=R_k(-\bar{\nabla}^2 + E)$, where $E$ is a non-trivial endomorphism \cite{Codello:2008}.
Type II cutoffs allow flexibility in how different modes are integrated out and will appear again in chapter \ref{cha:asymp}.
The spacetime trace in the flow equation itself is evaluated by a type of heat kernel expansion.
Finally, solving the differential equations resulting from the projection often entails a combination of analytical and numerical methods.

\section{Thesis outline}

Each of the following three chapters focuses on a different fundamental aspect of asymptotic safety.
In chapter \ref{cha:recon} we consider the reconstruction problem. As explained in section \ref{sec:EAA}, this is the problem of how to recovery a path integral formulation of a theory from the effective average action.
Presenting the research of \cite{Morris:2015oca}, we provide two exact solutions to this problem and understand how they compare to a one-loop approximate solution in the existing literature. 
In chapter \ref{cha:back ind} we present the work of \cite{Labus:2016lkh} in which the fundamental requirement of background independence in quantum gravity is addressed.
Working within the derivative expansion of conformally reduced gravity, we explore the notion of compatibility (introduced in section \ref{sub:BI}) and uncover the underlying reasons for background dependence generically forbidding fixed points in such models, extending the work of \cite{Dietz:2015owa}.
As emphasized in section \ref{sub:f(R)}, in order to understand the true nature of fixed point solutions, it is necessary to study their asymptotic behaviour.
Chapter \ref{cha:asymp} presents the work of \cite{Gonzalez-Martin:2017gza} in which we explain how to find the asymptotic form of fixed point solutions in the $f(R)$ approximation.
In the fifth and final chapter we give a brief summary of the research presented in chapters \ref{cha:recon}, \ref{cha:back ind} and \ref{cha:asymp}, discussing the significance of the key findings and commenting on useful extensions of the work.
We finish by considering the need to incorporate matter into the formalism in a compatible way and touch upon potential opportunities to test asymptotic safety in the future.



\chapter{Solutions to the reconstruction problem}
\label{cha:recon}

\section{Introduction}

In this chapter we return to a foundational issue raised in section \ref{sec:EAA} called the reconstruction problem. In the discussions that follow we phrase all arguments in terms of a single-component scalar field so that none of the extra structure that comes along when dealing with metric degrees of freedom plays a role.
However, it is straightforward to adapt the equations to fields with more indices and/or different statistics as required. We make more comments on this in the conclusions.

Recall that the flow equation for the scalar field is derived from a functional integral \eqref{scalar func}, which is subject to some overall UV cutoff.
However, providing the IR cutoff profile $R_k(p^2)$ varies sufficiently fast, the flow itself only receives support from finite $|p|/k$ and thus is well defined in the limit that the UV cutoff is removed, $\Lambda\to\infty$.
The resulting flow equation for the effective average action $\hat{\Gamma}_k$ is repeated here for ease of reference:
\begin{equation}
\label{FRGE2}
 \partial_k \hat{\Gamma}_k[\varphi] = \frac{1}{2}\text{Tr}\left[\left(\frac{\delta^2\hat{\Gamma}_k}{\delta\varphi\delta\varphi} + R_k \right)^{\!\!-1} \!\partial_k R_k \right] \,.
\end{equation}
Note that we have introduced a change of notation here as the effective average action is now denoted with a hat. As emphasized in \ref{sec:EAA}, the removal of the UV cutoff $\Lambda$ from the flow equation is crucial to its use. Only in this way can we find fixed points with respect to $k$ (implying the absence of any other dimensionful parameter), and construct the continuum limit in the standard way envisaged in asymptotic safety literature, namely via the a renormalised trajectory emanating from a UV fixed point. In other words, we can solve for the flow equations ``directly in the continuum" (as already emphasized in \cite{Morris:1993}).

However, as emphasised by Manrique and Reuter in \cite{Manrique:2008zw,Manrique:2009tj} this leaves us with a problem, dubbed by them ``the reconstruction problem".\footnote{See also reference \cite{Vacca:2011fx}.}
Recall that this is the problem of how to reconstruct a path integral representation of the theory, or more specifically, how to reconstruct the bare action from solutions of the flow equation \eqref{FRGE2}.
Such a formulation is desirable since potentially we need access to some bare action to obtain the microscopic degrees of freedom and from there study possible Hamiltonian formulations; understand more directly properties of the constructed quantum field theory such as constraints and local symmetries; make more direct contact with perturbative approaches; and finally more directly compare to other approaches that are formulated at the microscopic level, such as canonical quantisation, loop quantum gravity or Monte Carlo simulations \cite{Ashtekar:2004eh,Thiemann:2007zz,Rovelli:2008zza,Ambjorn:2012jv,Ashtekar:2014kba,Hamber:2015jja,Liu:2015bwa}.

In order to make the matter more concrete, Manrique and Reuter consider the following situation \cite{Manrique:2008zw,Manrique:2009tj}. They regulate the functional integral by using a sharp cutoff $\Lambda$, such that the integration is restricted to only those modes propagating with momentum $|p|\leq \Lambda$, and consider either a generic IR cutoff profile $R_k$ or the optimised cutoff profile: $R_{k}(p)=(k^2-p^2)\theta(k^2-p^2)$\cite{opt1,Litim:2001,opt3}.
Now there are two issues to confront.
Firstly, as already emphasised in section \ref{sec:EAA}, and proved in appendix \ref{app:UV}, a partition function regularised by some finite UV cutoff $\Lambda$ \emph{cannot} through the standard Legendre transform procedure yield the continuum Legendre effective action $\hG{k}{}$.\footnote{Throughout this chapter we will often refer to $\Gamma_k$ by its alternative name, the Legendre effective action, in accordance with reference \cite{Morris:1993}.} Instead it must give an effective average action $\hG{k}\Lambda$ that now also depends explicitly on $\Lambda$ as well as $k$.
Likewise, the resulting UV regulated flow equation also depends on the two cutoffs, cf.\ \eqref{flow 2 cutoffs}, and is not that of \eqref{FRGE2}. 
In reference \cite{Manrique:2008zw} it is claimed that for the optimised cutoff this dependence on $\Lambda$ disappears in the sense that providing we restrict flows to $k\le\Lambda$, we can consistently set $\hG{k}\Lambda[\ph] = \hG{k}{}[\ph]$. In fact this is not correct, as explained in appendix \ref{app:UV}.

The second issue is that even if we take the particular case of $k=\Lambda$ like the authors did in \cite{Manrique:2008zw}, there is still a functional integral $Z$ separating the bare action $\hat{\mathcal{S}}^{ \Lambda}$ from the effective action $\hat{\Gamma}^\Lambda_{k=\Lambda}$ (the Legendre transform of ln$Z$), albeit threshold-like, being only over modes with an effective mass of order the overall cutoff $\Lambda$.
This means the effective action is related to a bare action  in a way which cannot in practice be calculated exactly, and moreover we then need to invert this relation in order to find $\hat{\mathcal{S}}^{ \Lambda}$ in terms of $\hat{\Gamma}^\Lambda_{k=\Lambda}$. At the one-loop level, the partition function can be evaluated by steepest descents \cite{Manrique:2008zw} to yield:\footnote{All momenta  should be understood to be cutoff from above by $\Lambda$, including that in the momentum integral implied by the space-time trace. The mass parameter $M$ introduced in reference \cite{Manrique:2008zw} will play no significant role here so will be neglected. Also in contrast to reference \cite{Manrique:2008zw}, we will not make the momenta discrete by compactifying on a torus.}
\begin{equation}
	\label{MandR.soln}
	\hat{\Gamma}^\Lambda_{k=\Lambda}[\varphi]-\hat{\mathcal{S}}^{ \Lambda}[\varphi]=\frac{1}{2}\text{Tr}\,\text{ln}			\Big\{\hat{\mathcal{S}}^{ \Lambda (2)}[\varphi]+R_{\Lambda}\Big\}\,,
	\end{equation}
where $\hat{\mathcal{S}}^{ \Lambda (2)} = \delta^2 \hat{\mathcal{S}}^{ \Lambda}/\delta\varphi\delta\varphi$ is the Hessian of the bare action. Unfortunately in the interesting case of asymptotic safety the theory is strongly interacting at these scales, with all couplings ${\cal O}(1)$ times the appropriate power of $\Lambda$, and thus one loop is not a good approximation. Furthermore even with this approximation it is not straightforward to invert the relation to find $\hat{\mathcal{S}}^{ \Lambda}$ in terms of $\hat{\Gamma}^\Lambda_{k=\Lambda}$.

The research conducted in \cite{Morris:2015oca} and presented in this chapter provides two solutions to the reconstruction problem.
The first solves both of the issue given above for a wide range of cutoffs by utilising a kind of duality relation between a Wilsonian effective action $\hat{S}^k$ and the effective average action $\hG{k}{}$, and is closely based on results from reference \cite{Morris:1993}. (Aspects of reconstruction were already treated there at the end of section 3 and in the conclusions.)
In particular it also involves a map between an effective multiplicative UV cutoff $C^k(p^2)$, which regulates $\hat{S}^k$, and the IR cutoff $R_k(p^2)$.
The central point is that since the Wilsonian effective action is already an action, fully regularised in the UV by $C^k$, it can be used as a bare action. 
As described in section \ref{sec:theory space}, since $\hS{k}$ depends on only one scale, namely $k$, it can also display all the required RG properties:
in the continuum limit the full trajectory $\hS{k}$ is then again the renormalised trajectory starting from  the UV fixed point\footnote{Here we commit a slight abuse of notation. Strictly in order for the action to reach a fixed point, we should change to the appropriate dimensionless variables. By $\hS*$ we actually mean the action such that it takes the fixed point form after such a transformation.}
$\hS*$ in the far UV ($k\to\infty$) and extending down to $k\to0$.
It follows that such an $\hS{k}$ is a continuum version of the ``perfect bare actions" mentioned previously in section \ref{sec:theory space} and explored e.g.\ in reference \cite{Hasenfratz:1993sp}, since, as we review in section \ref{sec:SandGamma}, setting $\hbS\Lambda=\hS{k=\Lambda}$ to be the bare action (together with UV cutoff $C^{k=\Lambda}$) results in a partition function that is actually \emph{independent} of $\Lambda$ and thus in particular equal to the partition function obtained in the continuum limit $\Lambda\to\infty$.  

Unlike the map described in \eqref{MandR.soln}, the map between $\hS{k}$ and $\hG{k}{}$ is exact. Unlike the map \eqref{MandR.soln}, it is straightforward to explicitly construct it in either direction, via a tree-diagram expansion which can be developed vertex by vertex, as we will see in section \ref{sec:vertices}.\footnote{It is also possible to solve the relation explicitly in approximations that go beyond an expansion in vertices. For example the duality relation remains exact in the Local Potential Approximation and thus at this level can be analysed exactly, both analytically and numerically \cite{Morris:2005ck, Bridle:2016nsu}.}
Constructing $\hbS{\Lambda}=\hS\Lambda$ in this way, already constitutes a practical solution to the reconstruction problem, since it provides a bare action that expresses the same asymptotically safe renormalised trajectory as $\hG{k}{}$. 

This still leaves a puzzle however, since it is not immediately clear how this solution should be related to the one-loop expression \eqref{MandR.soln}. 
Since we cannot obtain $\hat{\Gamma}_k$ directly from $Z$ through the standard Legendre transform procedure, if we want $\hbS\Lambda$ to be associated to $\hG{k}{}$, then the best we can hope to achieve is to find a map from the continuum $\hG{k}{}$ to a pair $\{\hbS\Lambda, \hG{k}\Lambda\}$ consisting of a bare action and the resulting effective average action, such that $\hG{k}\Lambda\to\hG{k}{}$ as $\Lambda\to\infty$.
In sections \ref{sec:details} and \ref{sec:solving} we set out exactly how to compose such a map, again explicitly constructable vertex by vertex, and show how it is consistent with the one-loop formula \eqref{MandR.soln}.
This map provides an alternative solution to the reconstruction problem.
It requires using what we term ``compatible cutoffs", defined in the following section.

The structure of the chapter is then as follows. In the next section we give the definitions necessary to set out precisely our two prescriptions for reconstructing a bare action and give a detailed explanation on why they constitute solutions.
For the second prescription we use a special case of a remarkable relation proved in section \ref{sec:duality-Gamma}. There we prove another Legendre transform (\aka duality) relation between two effective average actions, or simply two Legendre effective actions, with different overall UV cutoff profiles but the same associated Wilsonian effective action. In section \ref{sec:SandGamma} we derive the main Legendre transform relation between Wilsonian effective actions and effective average actions, and show how these are in turn derived from the partition function, extending the results of reference \cite{Morris:1993} to more general cutoff profiles. In section \ref{sec:vertices} we compute the vertices of the Wilsonian effective action $\hat{S}^k$ from $\hat{\Gamma}_k$ through the tree-level expansion implied by the duality relation. This expansion can also be used in the other direction and for the other duality relations simply by renaming propagators and vertices. In section \ref{sec:solving} we provide more detail on our second solution to the reconstruction problem and show how it is related to \eqref{MandR.soln}.
In section \ref{sec:compatible} we give some examples of compatible cutoff profiles, and finally in section \ref{sec:conclusions} we summarise and draw our conclusions.

\section{Detailed prescription for reconstruction}
\label{sec:details}

Here we set out in detail the definitions we need in order to precisely give our two prescriptions for reconstructing a bare action as sketched in the introduction.

\begin{figure}
\centering
\begin{minipage}{0.45\textwidth}
\includegraphics[scale=0.8]{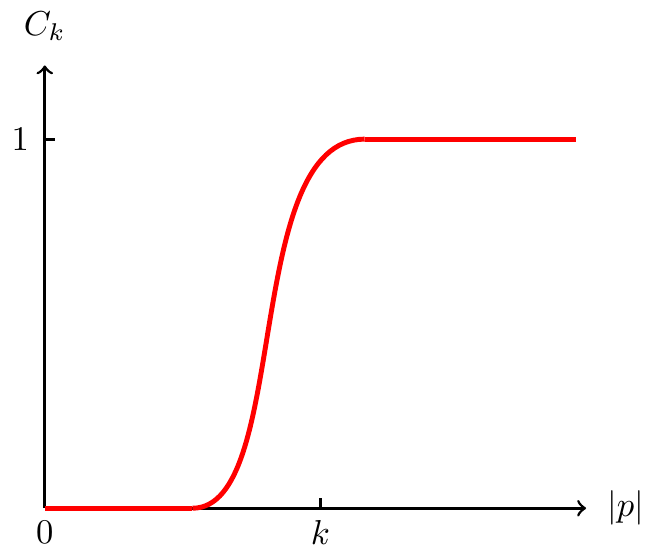}
\end{minipage}
\hspace{3pt}
\begin{minipage}{0.45\textwidth}
\includegraphics[scale=0.8]{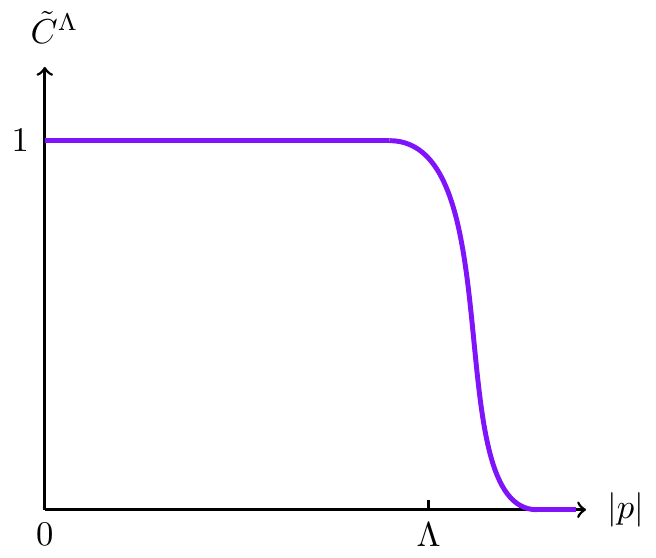}
\end{minipage}
\caption{Example behaviour of multiplicative IR cutoff function $C_k$ \eqref{IR-only} and UV cutoff function $\tilde{C}^\Lambda$ \eqref{DeltaTotalUV} respectively. The other UV cutoff, $C^k$ \eqref{DeltaUV}, displays similar behaviour to $\tilde{C}^\Lambda$ but regulates at $k$ instead of $\Lambda$.}\label{fig:cutoffsplots1}
\end{figure}

Let us choose to define the interaction part of the effective average action $\Gamma_k$ to be the part obtained by splitting off a normalised massless kinetic term:
\be
\label{interactions}
\hat{\Gamma}_{k}[\ph]= \frac{1}{2}\varphi\cdot p^{2}\cdot \varphi+\Gamma_{k}[\ph]\,.
\ee
Note that, as before, we regard a mass term $\frac{1}{2}m^{2}\ph^{2}$ as contained within the interactions. The total effective action contains also the additive infrared cutoff:
\bea 
\label{LEAA-con}
\Gamma^{\text{tot}}_{k}[\ph] &=& \hat{\Gamma}_{k}[\ph]+\frac{1}{2}\varphi\cdot R_{k}\cdot \varphi\,, \\
\label{LEAA-alt}
&=& \Gamma_{k}[\ph]+
	\frac{1}{2}\varphi\cdot \left(\Delta_{k}\right)^{\!-1}\!\!\cdot \varphi\,,
\eea
where in the second line we have combined the massless kinetic term with the additive cutoff to form a  propagator with a multiplicative cutoff:
\be
\label{DeltaIR}
\Delta_k = \frac{C_k(p)}{p^2}\,,
\ee
such that
\be 
\label{IR-only}
C_k(p) = \frac{p^2}{p^2+R_k(p)}\,.
\ee
This provides the translation between multiplicative IR cutoff profiles and additive IR cutoff profiles, but explicitly uses the fact that the overall UV cutoff has been removed. Note that $C_k$ inherits from $R_k$ the properties that for $|p|<k$ it suppresses modes, and in particular $C_{k}(p)\to 0$ as $|p|/k\to0$, while for $|p|>k$, $C_k(p)\approx1$ and mostly leaves the modes unaffected and in particular  $C_{k}(p) \rightarrow1$ as $|p|/k\to\infty$.
The behaviour of $C_k$ is represented by the red line in figure \ref{fig:cutoffsplots1}.

As already mentioned in the introduction, Manrique and Reuter choose to regularise the functional integral in the UV with an overall sharp cutoff such that all momenta $|p|\le\Lambda$ \cite{Manrique:2008zw}. This is equivalent to ensuring that the internal momentum running through any propagator is cut off so that this propagator vanishes for $|p|>\Lambda$. Both the ultraviolet regularisation and the infrared regularisation can therefore be carried by a multiplicative cutoff 
\be
\label{UVIRMR}
C^\Lambda_k(p) = \frac{p^2\theta(\Lambda-|p|)}{p^2+R_k(p)}\,,
\ee
which appears in the resulting effective action like so:
\be
\label{total-Gamma}
\Gamma^{\text{tot},\, \Lambda}_{k}[\ph]= \Gamma^\Lambda_{k}[\ph]+
	\frac{1}{2}\varphi\cdot \left(\Delta_{k}^\Lambda\right)^{\!-1}\!\!\cdot \varphi\,,
\ee
where 
\be
\label{DeltaUVIR}
\Delta_k^\Lambda = \frac{C^\Lambda_k(p)}{p^2}\,.
\ee
As we have previously noted, the effective average action now depends also on the overall UV cutoff $\Lambda$. We recover the previous case when the UV cutoff is removed: $C_k(p)\equiv C^\infty_k(p)$. 

As already emphasised in the introduction to this chapter, our constructions go through for much more general UV cutoffs, providing that the UV and IR cutoffs are always implemented together, multiplicatively, as defined via the above relations \eqref{total-Gamma} and \eqref{DeltaUVIR}.
As we recall in section \ref{sec:SandGamma}, the flow equation for the interactions then takes the general form
\begin{equation}
	\label{Gamma.flow}
	\frac{\partial}{\partial k}\Gamma^\Lambda_{k}[\varphi]=-\frac{1}{2}\text{Tr}\bigg[\bigg(1+\Delta^\Lambda_k\cdot 						\frac{\delta^{2}\Gamma^\Lambda_{k}}{\delta\varphi\delta\varphi}\bigg)^{\!-1}\frac{1}									{\Delta^\Lambda_k}\frac{\partial\Delta^\Lambda_k}{\partial k}\bigg]\,.
	\end{equation}
By recasting the right-hand side in terms of $\left(\Delta^\Lambda_k\right)^{\!-1}$, and using $1/$\eqref{DeltaIR}, $1/$\eqref{IR-only}, and \eqref{interactions}, it is easy to see that in the limit $\Lambda\to\infty$ this flow equation gives back \eqref{FRGE2}.

Now we define in precisely the same way both the bare interactions $\mathcal{S}^\Lambda[\phi]$ and Wilsonian interactions $S^k[\Phi]$:
\be
\label{other-interactions}
\hbS\Lambda[\phi]= \frac{1}{2}\phi\cdot p^{2}\cdot \phi+\mathcal{S}^\Lambda[\phi]\,,\qquad 
\hat{S}^{k}[\Phi]= \frac{1}{2}\Phi\cdot p^{2}\cdot \Phi+S^{k}[\Phi]\,.
\ee
(We choose different symbols for the fields in each case for convenience as will become clear later.) We define the total bare action to include also the UV cutoff profile and thus
\be 
\label{total-bare}
\mathcal{S}^{\mathrm{tot},\Lambda}[\phi] = \mathcal{S}^\Lambda[\phi]+\frac{1}{2}\phi\cdot \left(\tilde{\Delta}^{\Lambda}\right)^{-1}\!\!\cdot \phi\,,
\ee
where 
\be 
\label{DeltaTotalUV}
\tilde{\Delta}^\Lambda = \frac{\Cl(p)}{p^2}\,.
\ee
For the sharp cutoff case 
\be
\label{sharpUV}
\Cl(p)=\theta(\Lambda-|p|)\,,
\ee 
but again we emphasise that the UV cutoff profile can be more general and we will in general take it to be so. All we then require is that for $|p|<\Lambda$, $\Cl(p)\approx1$ and mostly leaves the modes unaffected and in particular $\Cl(p) \to 1$ for $|p|/\Lambda\to0$, while for $|p|>\Lambda$ it suppresses modes, and in particular for $|p|/\Lambda\to\infty$, $\Cl(p) \to0$  sufficiently fast to ensure that all momentum integrals are regulated in the ultraviolet.
This cutoff is represented by the purple line in figure \ref{fig:cutoffsplots1}.
Finally, the total Wilsonian effective action can be written
\be 
\label{total-Wilsonian}
S^{\mathrm{tot},k}[\Phi] = S^k[\Phi] + \frac{1}{2}\Phi\cdot (\Delta^{k})^{\!-1}\!\!\cdot \Phi\,,
\ee
where 
\be 
\label{DeltaUV}
\Delta^k = \frac{C^k(p)}{p^2}\,,
\ee
and $C^k(p)$ is an ultraviolet cutoff profile for this effective action and effective partition function, which regularises at scale $k$. $C^k(p)$ has to satisfy the same conditions as $\Cl(p)$ above (with the replacement $\Lambda\mapsto k$ of course). Since the functional integral with this action $S^{\mathrm{tot},k}$ is therefore already completely regularised in the ultraviolet, there is no need for any dependence on the overall UV cutoff $\Lambda$. We will therefore choose $C^k(p)$ to depend only on the one cutoff scale $k$ as already indicated, and indeed apart from obeying the same general conditions,  the profiles $C^k$ and $\Cl$ will otherwise be unrelated. However, we will require one ``sum rule" relation between these three profiles:\footnote{This goes beyond the sum rule introduced in reference \cite{Morris:1993} since we now allow $\Cl$ to be unrelated to $C^k$.}
\be 
\label{sum rule}
C^\Lambda_k(p)+C^k(p) = \tilde{C}^\Lambda(p)\,.
\ee
For example, from \eqref{UVIRMR}, \eqref{sharpUV}, and \eqref{sum rule}, we can deduce the  UV cutoff profile for the Wilsonian effective action which is implied by the regularisation used in reference \cite{Manrique:2008zw}:
\be
\label{MR-Wilson}
C^k(p) = \left(1-\frac{p^2}{k^2}\right)\theta(k-|p|)
\ee
(where $k<\Lambda$).
We see that it behaves sensibly as a UV cutoff profile and actually depends only on the one cutoff scale as required. 
Thus \eqref{UVIRMR}, \eqref{sharpUV} and \eqref{MR-Wilson} provide an example of a consistent set of cutoffs satisfying the sum rule \eqref{sum rule}.

Additionally, cutoffs may or may not be what we call ``compatible". If the UV and IR cutoffs are compatible, it means that $C^\Lambda_k$ vanishes at $k=\Lambda$ i.e.\ when the IR and UV cutoffs meet. 
The set of cutoffs listed previously (effectively those used in \cite{Manrique:2008zw}) are therefore not compatible. This in short explains the difference in nature between our second solution and the one-loop approximate solution \eqref{MandR.soln} as we specialise to compatible cutoffs when constructing our second solution to the reconstruction problem.
Examples of cutoff profiles satisfying \eqref{sum rule} that also satisfy this compatibility condition are given in section \ref{sec:compatible}.

In general we can use \eqref{sum rule} to \emph{define} $C^\Lambda_k(p)=\Cl(p)-C^k(p)$. Since  $\Lambda>k$, the general properties given above for $C^k$ and $\Cl$ ensure that it behaves as a multiplicative UV cutoff at $\Lambda$ and multiplicative IR cutoff at $k$ as  required. Thus for $|p|>\Lambda$ modes are suppressed such that as $|p|/\Lambda\to\infty$, $C^\Lambda_k(p)\to0$ sufficiently fast that all momentum integrals are UV regulated. For $k<|p|<\Lambda$, $C^k(p)$ is small (vanishingly so for $|p|\gg k$) while $\Cl(p)\approx1$, and thus $C^\Lambda_k(p)\approx1$ and mostly leaves modes unaffected. For $k\ll |p|\ll \Lambda$, $C^\Lambda_k(p)$ will be very close to one. Finally for $|p|<k$, $C^k(p)\approx1$ and $\Cl(p)$ is close to one (very close for $k\ll\Lambda$) and thus $C^\Lambda_k(p)\approx0$ suppresses modes, while for $|p|/k\to0$, since both $C^k(p)\to1$ and $\Cl(p)\to1$, we have that $C^\Lambda_k(p)\to0$ thus providing the expected IR cutoff $k$.
\begin{figure}
\begin{center}
\includegraphics[scale=0.8]{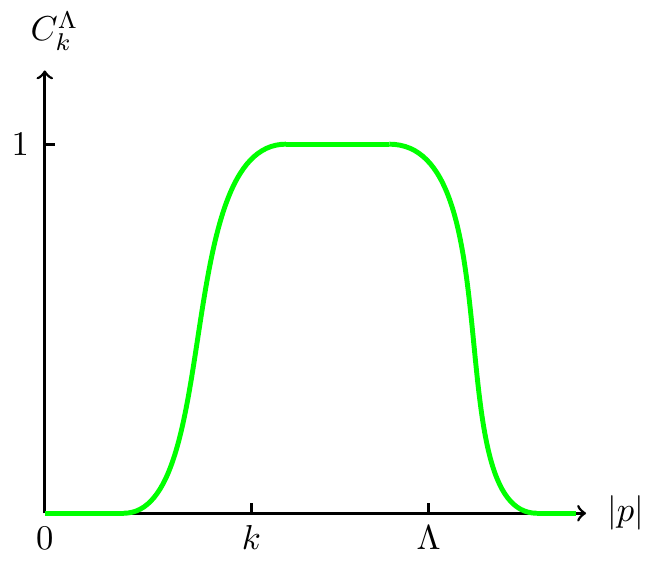}
\caption{Example behaviour of multiplicative cutoff function $C_k^\Lambda$ \eqref{DeltaUVIR} which regulates in both the IR and UV.}\label{fig:IRUV}
\end{center}
\end{figure}
This cutoff is represented by the green line in figure \ref{fig:IRUV}.

By adding the infrared cutoff profile to the bare action in order to generate the effective average action in the usual way, we equivalently change the multiplicative cutoff profile $\Cl$ into one that depends on both $\Lambda$ and $k$. We have already anticipated in our discussion of $\Gamma^\Lambda_k$ that the new multiplicative cutoff profile is $C^\Lambda_k$. Thus the bare action becomes
\be 
\label{total-bare-k}
\mathcal{S}^{\mathrm{tot},\Lambda}_k[\phi] = \mathcal{S}^\Lambda[\phi]+\frac{1}{2}\phi\cdot \left({\Delta}^{\Lambda}_k\right)^{\!-1}\!\!\cdot \phi\,.
\ee
It is this bare action that generates \eqref{total-Gamma} in the usual way and leads to the UV modified flow equation \eqref{Gamma.flow}. (Note that the total bare action then necessarily depends on both cutoffs. The bare interactions $\bS$ however do not, and indeed consistent with the usual philosophy of renormalisation they should be taken to depend only on the UV modification.)

\vspace{40pt}
\noindent\textbf{Solution 1}
\vspace{3pt}

\noindent Now we can state the duality in its general form:
\begin{equation}
	\label{duality-gen}
	S^{k}[\Phi]=\Gamma_{k}^\Lambda[\varphi]+\frac{1}{2}(\varphi-\Phi)\cdot \left(\Delta_{k}^{\Lambda}\right)^{\!-1}\!\!\cdot (\varphi-\Phi)\,.
\end{equation}
This is a Legendre transform relation that maps between two apparently very different pictures of the exact RG \cite{Morris:1993}. On the one hand we have the effective average action which flows with respect to an IR cutoff $k$ as in \eqref{Gamma.flow} (or in the limit $\Lambda\to\infty$, as in  \eqref{FRGE2}) and on the other hand we have a Wilsonian effective action whose interactions flow with respect to an effective UV cutoff $k$ according to Polchinski's flow equation\cite{Polchinski:1983gv}, restated here using the current notation:
	\begin{equation}
	\label{S.flow}
	\frac{\partial}{\partial k}S^k[\Phi]=\frac{1}{2}\frac{\delta S^k}{\delta\Phi}\cdot \frac{\partial\Delta^k}{\partial k}\cdot			\frac{\delta S^k}{\delta\Phi}-\frac{1}{2}\text{Tr}\bigg(\frac{\partial\Delta^k}{\partial k}\cdot \frac{\delta^{2}S^k}			{\delta\Phi\delta\Phi}\bigg)\,.
	\end{equation}
As we review in section \ref{sec:SandGamma}, and outlined in the introduction, the original partition function with bare action \eqref{total-bare} can be exactly re-expressed as a partition function with the bare action replaced with the Wilsonian one \eqref{total-Wilsonian}, which is thus a so-called perfect action. 

In particular if we have an effective average action solution $\Gamma_k$  to the continuum flow equation \eqref{FRGE2} such that it exists for all $0< k<\infty$, we can construct $S^k$ by using \eqref{duality-gen} with the identifications $\Gamma_k\equiv\Gamma^\infty_k$, and $\Delta_k \equiv \Delta^\infty_k$ as in \eqref{DeltaIR} and \eqref{IR-only}:
\begin{equation}
	\label{duality-con}
	S^{k}[\Phi]=\Gamma_{k}[\varphi]+\frac{1}{2}(\varphi-\Phi)\cdot \left(\Delta_{k}\right)^{\!-1}\!\!\cdot (\varphi-\Phi)\,.
\end{equation}
$S^k$ can then be constructed from this for example vertex by vertex as in section \ref{sec:vertices}.  

We can then also reconstruct the partition function Z even in this continuum limit, by using the perfect bare action \eqref{total-Wilsonian} with $k$ set to some initial upper scale of our choice, $k=\Lambda$ for example. Note that as required such an action has the same structure as the general form of the bare action \eqref{total-bare}, and indeed just involves the replacements $\Cl\mapsto C^\Lambda$, and $\bS\mapsto S^\Lambda$. The new UV cutoff profile $C^k(p)=1-C_k(p)$ as follows from \eqref{sum rule} with $\tilde{C}^\infty \mapsto 1$. This then provides our first solution to the reconstruction problem.

Note that such a bare action, and thus partition function, does not incorporate an infrared cutoff $R_k$ and thus there is no connection to the effective average action $\hat{\Gamma}_k$ through the standard route of taking a Legendre transform of ln$Z$. If we add the infrared cutoff term to this bare action, we still do not recover $\hat{\Gamma}_k$ this way.  As emphasised previously and appendix \ref{app:UV}, it is impossible to recover the continuum effective average action this way since the result is a $\hat{\Gamma}^\Lambda_k$ that necessarily now depends on both cutoffs. It is possible however to construct a map from the continuum solution $\hat{\Gamma}_k$ to a pair $\{\hat{\bS}, \hat{\Gamma}_k^\Lambda\}$, where $\hat{\Gamma}^\Lambda_k$ is related to $\hat{\bS}$ in the usual way, and such that as $\Lambda\to\infty$ we have $\hat{\Gamma}_{k}^\Lambda\to\hat{\Gamma}_k$. This is our second solution to the reconstruction problem which we now proceed to describe in detail.\\

\noindent\textbf{Solution 2}
\vspace{3pt}

\begin{figure}
\begin{center}
\begin{tikzpicture}\hspace{0pt}

\tikzstyle{b} = [rectangle, draw, fill=none, node distance=4.8cm, text width=6.5em, text centered, rounded corners, 				    minimum height=3.4em, thick]
\tikzstyle{c} = [rectangle, draw, fill=blue!20, node distance=4.8cm, text width=4.8em, text centered, rounded corners, 				    minimum height=2.3em, thick]
\tikzstyle{d} = [rectangle, draw, fill=blue!20, node distance=4.8cm, text width=3.5em, text centered, rounded corners, 				    minimum height=2.3em, thick]
\tikzstyle{e} = [rectangle, draw, fill=blue!20, node distance=4.8cm, text width=3.5em, text centered, rounded corners, 				    minimum height=3.4em, thick]
\tikzstyle{l} = [draw, line width=0.5mm, ->]

\node [b] (first) {$\hat{\Gamma}_k\equiv\hat{\Gamma}^\infty_k$};

\node [b, right of=first, node distance=5.5cm] (second){$\hat{S}^{k}$\\$C^k$};		
\node [b, right of=second, node distance=5.5cm] (third) {$\hat{S}^{k=\Lambda}=\mathcal{\hat{S}}^\Lambda$
								      \\$C^{k=\Lambda}=\tilde{C}^\Lambda$};
\node [b, below of=third, node distance=3cm] (fourth) {$\mathcal{\hat{S}}^\Lambda$\\
								         $C^\Lambda_k =C^\Lambda - C^k$};
\node [b, left of=fourth, node distance=5.5cm](fifth){$Z^\Lambda_k$};
\node [b, left of=fifth, node distance=5.5cm](sixth){$\hat{\Gamma}^\Lambda_k$};
  
\path [l] (first) -- (second);
\path [l] (second) -- (third);
\path [l] (third) -- (fourth);
\path [l](fourth)--(fifth);
\path [l](fifth)--(sixth);

\draw (2.65,0.3) node{Duality};
\draw (2.65, -0.3) node{relation};
\draw (8.25,0.72) node{Compatible};
\draw (8.15, 0.3) node{cutoffs};
\draw (8.15,-0.3) node{$k=\Lambda$};
\draw (10.95, -1.45) node{Sum \hspace{1.75pt}rule};
\draw (2.75, -2.7) node{Legendre};
\draw (2.775,-3.3) node{transform};
\end{tikzpicture}
\end{center}
\caption{Summary of second solution to the reconstruction problem.}
\label{fig:map}
\end{figure}
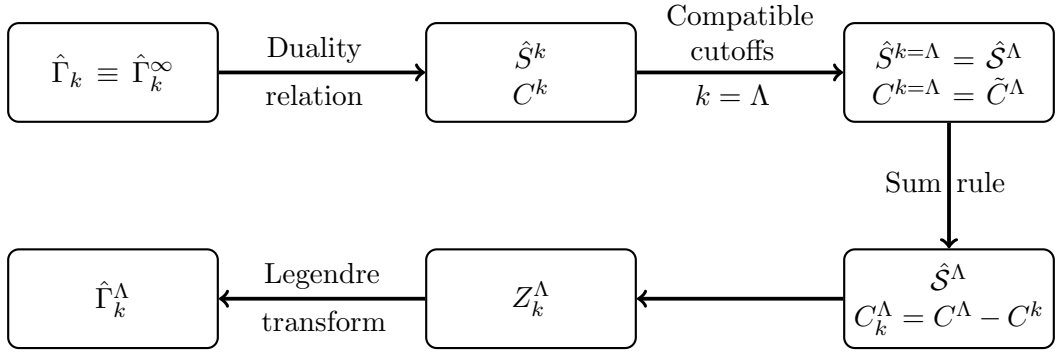

\noindent Assume we have found the appropriate renormalised trajectory $\hG{k}{}\equiv\hG{k}\infty$ of \eqref{FRGE2}, where we emphasise that this solution corresponds to the case where the overall UV cutoff has been removed. Using the duality relation we construct the corresponding Wilson effective action $\hS{k}$ together with its associated effective UV cutoff $C^k$.
Next we specialise to compatible cutoffs. Recall that this means that the overall UV cutoff $\Cl = C^{k=\Lambda}$ is identical to the effective UV cutoff set at scale $k=\Lambda$.
Thus we set the effective action to be a bare action at $k=\Lambda$, \ie $\hat{\mathcal{S}}^\Lambda = \hS{k=\Lambda}$, regulated in the UV with $C^\Lambda$. 
Now we replace the multiplicative cutoff $C^\Lambda$ with  $C^\Lambda_k = C^\Lambda - C^k$. 
which regularises both in the IR and the UV.
The corresponding partition function yields by the standard procedure an effective average action $\hG{k}\Lambda$.
This $\hG{k}\Lambda$ satisfies a UV regularised version of the flow equation \eqref{FRGE2} with the property that as $\Lambda\to\infty$ it goes back to the original flow equation \eqref{FRGE2}.
Thus we have constructed an exact, explicit and calculable map from any continuum solution $\hG{k}{}\equiv\hG{k}\infty$ with its associated IR cutoff $R_k$, to the pair, $\hat{\mathcal{S}}^\Lambda$ and $\hG{k}\Lambda$, related in the standard way through a functional integral regularised in the UV and IR by $C^\Lambda_k$.
This pair has the property that as $\Lambda\to\infty$, the regularised solution $\hG{k}\Lambda\to\hG{k}{}$. Since, given $\hG{k}{}$, everything is explicitly calculable, we see that this provides an alternative solution to the reconstruction problem.
This map is summarised in figure \ref{fig:map}.

Note that $C^\Lambda_\Lambda(p)$ vanishes for all $p$, by the sum rule formula \eqref{sum rule}. Thus \eqref{duality-gen} implies that 
\be
\label{initial}
\hat{\Gamma}^\Lambda_\Lambda[\ph]=\hat{S}^\Lambda[\ph] 
\ee
(and $\ph=\Phi$) as can be seen either directly from the fact that $1/\Delta_{\Lambda}^{\Lambda}(p)$ is infinite for all $p$, or more carefully by first solving the Legendre transform relation as done for the continuum version in \eqref{soln2}. The UV boundary condition \eqref{initial} for the flow \eqref{Gamma.flow} is therefore particularly simple, and is furthermore a triple equality since the right-hand side is also the bare interactions $\mathcal{S}^\Lambda$. Moreover, the right-hand side is also dual to the original continuum solution $\hG{k}{}$ evaluated at $k=\Lambda$. More details are given in section \ref{sec:solving}.

We do not need to compute the functional integral or solve the flow \eqref{Gamma.flow} to find $\Gamma_k^\Lambda$ in this map.
It can also be constructed vertex by vertex from the original continuum $\Gamma_k$ using the same recipe as in section \ref{sec:vertices}. The clue is hidden in a remarkable property of the duality relation \eqref{duality-gen}. Note that by construction $S^k$ need have no dependence on $\Lambda$. (It is just a solution to \eqref{S.flow} which also has no dependence on $\Lambda$.) Therefore if we choose to keep $S^k$ fixed, the duality relation \eqref{duality-gen} actually implies that the right-hand side is independent of the choice of overall UV cutoff $\Cl$, and in particular that it is independent of $\Lambda$. As we show in the next section, this implies that the interaction parts of the two $\hat{\Gamma}$s are related by
\be
\label{duality-Gamma-con}
\Gamma^\Lambda_k[\Phi] = \Gamma_k[\ph]+\frac{1}{2}(\ph-\Phi)\cdot \left(\Delta_{\Lambda}\right)^{\!-1}\!\!\cdot (\ph-\Phi)\,,
\ee
where the notation for the inverse propagator on the right-hand side indicates that it is regularised in the infrared by $C_\Lambda:=C_{k=\Lambda}$. Comparing \eqref{duality-Gamma-con} to \eqref{duality-con}, we see that the vertices of $\Gamma^\Lambda_k[\Phi]$ are thus given by those of $S^k$ in the recipe set out in section \ref{sec:vertices},
providing we make the replacement $\Delta_k\mapsto\Delta_\Lambda$. Of course it then follows that the same tree-diagram expansion illustrated in fig. \ref{fig:vertices} is  also correct for $\Gamma^\Lambda_k[\Phi]$ after this replacement.

In section \ref{sec:solving} we show how our second solution is consistent with the one-loop formula \eqref{MandR.soln}.  On the one hand, by construction the multiplicative cutoff $C^\Lambda_{k}$ vanishes at $k=\Lambda$, which means effectively that the modified $R_k$ diverges at $k=\Lambda$. As a consequence, apart from a field independent piece, \eqref{MandR.soln} implies that $\hG\Lambda\Lambda = \hat{\mathcal{S}}^\Lambda$, recovering our result. On the other hand if the UV and IR cutoffs are not compatible, there is still a functional integral to do at $k=\Lambda$. Then the formula \eqref{MandR.soln} supplies the approximate relation, valid to one loop. As we review in sections \ref{sec:SandGamma} and \ref{sec:solving}, the Wilsonian effective action $\hS{k}$ can also be derived from the bare action $\hat{\mathcal{S}}^\Lambda$ via  a functional integral. In section \ref{sec:solving}, we show directly by the method of steepest descents that in the non-compatible case this functional integral yields at one loop an $\hS\Lambda$ which is precisely the one which is dual to the effective action given by the formula \eqref{MandR.soln}, proving consistency also in the non-compatible case.

\section{Proof of a duality relation between effective actions with different UV regularisations}
\label{sec:duality-Gamma}
We will consider a more general case and then specialise to \eqref{duality-Gamma-con}, since the proof is just as simple. We thus go back to the UV regularised form \eqref{duality-gen} of the duality relation between the Wilsonian interactions $S^k$ and the effective average action $\Gamma^\Lambda_k$. Now consider an alternative overall UV cutoff $\Clp(p)$ in place of $\Cl(p)$, where for generality we change both the profile form $\mathring{C}$ and the magnitude $\Lp$. Without loss of generality we can assume $\Lp>\Lambda$ however. We choose to keep the same effective UV cutoff $C^k$ and therefore through the sum rule relation \eqref{sum rule} we define an alternative joint regulator profile $\Clpp = \Clp - C^k$. Again, providing $\Clp$ is chosen to behave sensibly as a UV cutoff, as discussed below \eqref{sharpUV}, $\Clpp$ will behave correctly in the UV and infrared, as discussed below \eqref{MR-Wilson}.
Relabelling \eqref{duality-gen} in the obvious way, we evidently therefore have the alternative duality relation:
\begin{equation}
	\label{duality-gen-2}
	S^{k}[\Phi]=\Gamma_{k}^\Lp[\php]+\frac{1}{2}(\php-\Phi)\cdot \left(\Delta_{k}^{\Lp}\right)^{\!-1}\!\!\cdot (\php-\Phi)\,.
\end{equation}
As observed in the previous section, $S^k$ is not forced to have any dependence on these overall cutoffs. Since $S^k$ satisfies a flow equation \eqref{S.flow} which itself is independent of these cutoffs we can choose to keep the same solution $S^k$ after these changes.
Eliminating the left-hand side we thus have the relation
\be
\label{eliminated-S}
\Gamma_{k}^\Lambda[\varphi]+\frac{1}{2}(\varphi-\Phi)\cdot \left(\Delta_{k}^{\Lambda}\right)^{\!-1}\!\!\cdot (\varphi-\Phi)
=\Gamma_{k}^\Lp[\php]+\frac{1}{2}(\php-\Phi)\cdot \left(\Delta_{k}^{\Lp}\right)^{\!-1}\!\!\cdot (\php-\Phi)\,.
\ee
This is a Legendre transform relation in which all three fields can be varied independently. Varying $\Phi$ we thus have
\be 
\label{Phi-elimination}
\left[\left(\Delta_{k}^{\Lp}\right)^{\!-1}-\left(\Delta_{k}^{\Lambda}\right)^{\!-1}\right]\Phi = \left(\Delta_{k}^{\Lp}\right)^{\!-1}\php-
\left(\Delta_{k}^{\Lambda}\right)^{\!-1}\ph\,.
\ee
Define $C^\Lp_\Lambda = \Clp-\Cl = \Clpp-C^\Lambda_k$, where the second equality follows from the sum rule \eqref{sum rule}. Given the general behaviour of its component parts, $C^\Lp_\Lambda$ is a multiplicative cutoff profile that is cutoff in the UV by $\Lp$ and in the IR by $\Lambda$, with properties discussed below \eqref{MR-Wilson}. Thus also define $\Delta^\Lp_\Lambda(p)=C^\Lp_\Lambda(p)/p^2$. Then \eqref{Phi-elimination} can be rearranged to give
\bea
\ph-\Phi &=& \frac{\Delta^\Lambda_k}{\Delta^\Lp_\Lambda}\cdot(\php-\ph)\,, \\
\php-\Phi &=& \frac{\Delta^\Lp_k}{\Delta^\Lp_\Lambda}\cdot(\php-\ph)\,.
\eea
Substituting these back into \eqref{eliminated-S} gives us the desired general duality relation between effective average actions with different UV cutoffs:
\be 
\label{duality-Gamma-gen}
\Gamma^\Lambda_k[\ph] = \Gamma^\Lp_k[\php] + \frac{1}{2}(\ph-\php)\cdot \left(\Delta_\Lambda^{\Lp}\right)^{\!-1}\!\!\cdot (\ph-\php)\,.
\ee
An alternative proof of this relation is given in reference \cite{Morris:1993}, by demonstrating directly that this transformation turns the flow equation \eqref{Gamma.flow} into the equivalent one for $\Gamma^\Lp_k$. Reference \cite{Morris:1993} however specialised to the case where only the scale $\Lambda\mapsto\Lp$ changes. As we see here the relation is more general including also the option to change the form of the cutoff profile. 

It is remarkable that such a generalised Legendre transformation relationship exists between two effective average actions regularised in the UV with different cutoff profiles, $\Cl$ versus $\Clp$. To drive the point home, note that we can take the limit $k\to0$ and then this is a Legendre transform relation between two standard Legendre effective actions regularised in different ways in UV of our choosing. This latter result is therefore significant in general, not just within the context of functional RG. As we see explicitly in section \ref{sec:vertices}, it implies that the vertices of two effective actions are related by tree diagram expansions which can be constructed exactly. 

Since a change in regularisation obviously affects the loop integrals in the quantum corrections, this result looks surprising at first sight. However note that the key to the relation is that the Wilsonian effective action \eqref{total-Wilsonian} is unchanged. Since $S^{{\rm tot},k}$ is ultimately derived from a functional integral that depends on the bare action \eqref{total-bare-k} 
(see \eqref{partition to S}),
which most certainly does depend on the form of the overall UV cutoff, the change from $\Cl$ to $\Clp$ implies a change of bare interactions $\bS\mapsto \mathring{{\cal S}}^\Lp$ sufficient to completely compensate for this when computing $S^{{\rm tot},k}$. We make further comments on this in the conclusions. Although it makes no change to the Wilsonian effective action computed with these methods it leaves a remnant change to the Legendre effective action (with or without an IR cutoff $k$) which is summarised in the duality relation \eqref{duality-Gamma-gen}. 

In the special case where $\Cl = C^{k=\Lambda}$ and $\Clp = C^{k=\Lp}$, i.e.\ where the UV scale changes but not the form of the cutoff, which is furthermore fixed to be the Wilsonian one, we have the situation already analysed in reference \cite{Morris:1993}. Then the bare interactions change only trivially in that in each case ($k=\Lambda,\Lp$) the bare interactions are just equal to the Wilsonian interactions at that scale ${\cal S}^k = S^k$ as determined through the flow equation \eqref{S.flow}.

Finally, let us choose $\Cl=C^{k=\Lambda}$ and send $\Lp\to\infty$. Then $\Gamma^\Lp_k\to\Gamma_k$ and $C_\Lambda^{\Lp}\to1-C^\Lambda= C_\Lambda$, where we have used $\Clp\to1$ and \eqref{sum rule}. Thus with these changes, \eqref{duality-Gamma-gen} becomes the equation \eqref{duality-Gamma-con} we set out to prove.

\section{The Wilsonian Effective Action versus the Legendre Effective Action}
\label{sec:SandGamma}

In this section we recall most of the steps that give rise to the exact relationship \eqref{duality-gen} between the Wilsonian effective action and the Legendre effective action. They are adapted here from reference \cite{Morris:1993} both because the relationship goes marginally beyond what was proven there and also because they underpin the claims in the rest of the paper.

We consider the functional integral for a scalar field $\phi(x)$ in a $d$-dimensional Euclidean spacetime:
	\begin{equation}
	\label{func.int}
	Z^\Lambda[J]=\int\!\!\mathcal{D}\phi\,\mathrm{e}^{
	-\mathcal{S}^{\mathrm{tot},\Lambda}[\phi] +J\cdot\phi}
	=\int\!\!\mathcal{D}\phi\,\mathrm{e}^{-\frac{1}{2}\phi\cdot\left(\tilde{\Delta}^\Lambda\right)^{\!-1}\!\!\cdot\phi-\mathcal{S}^{\Lambda}[\phi] +J\cdot\phi}\,,
	\end{equation}
where the UV regulated bare action was introduced in \eqref{total-bare} and where now we will include superscripts and subscripts on $Z$ to indicate the regularisation within the functional integral.
We introduce an intermediate cutoff scale $k$ 
by re-expressing the propagator as:
	\begin{equation}
	\label{Delta-partition}
	\tilde{\Delta}^\Lambda=\Delta^\Lambda_k+\Delta^k\,,
	\end{equation}
where $\tilde{\Delta}^\Lambda$, $\Delta^\Lambda_k$ and $\Delta^k$ are defined in \eqref{DeltaTotalUV}, \eqref{DeltaUVIR} and \eqref{DeltaUV} respectively, and the split above follows from the sum rule relation \eqref{sum rule}. The integral can identically be rewritten as\footnote{up to a constant of proportionality. We ignore these from now on.}
	\begin{equation}
	\label{mod.int}
	Z^\Lambda[J]=\int\!\! \mathcal{D}\phi_{>}\mathcal{D}\phi_{<}\,\mathrm{e}^{-\frac{1}{2}\phi_{>}\cdot\left(\Delta^\Lambda_k\right)^{\!-1}\!\!\cdot\phi_{>}-
	\frac{1}{2}\phi_{<}\cdot\left(\Delta^k\right)^{\!-1}\!\!\cdot\phi_{<}-\mathcal{S}^{\Lambda}[\phi_{>}+\phi_{<}]+J\cdot(\phi_{>}+\phi_{<})}\,.
	\end{equation}
To see that this is true perturbatively, note that as a consequence of the sum form of the interactions, every Feynman diagram constructed from \eqref{mod.int} now appears twice for every internal propagator it contains: once with $\tilde{\Delta}^\Lambda$ replaced by $\Delta^\Lambda_k$ and once with $\tilde{\Delta}^\Lambda$ replaced by $\Delta^k$. Thus for every propagator line, what actually counts is the sum, which however is just $\tilde{\Delta}^\Lambda$ again by \eqref{Delta-partition} \cite{Morris:1998}. To prove the identity non-perturbatively, make the change of variables to $\phi=\phi_{>}+\phi_{<}$, for example by eliminating $\phi_>$. Evidently in \eqref{mod.int}, the action then has only up to quadratic dependence on $\phi_<$. Making the change of variables $\phi_<=\phi'_<+(\Delta^k/\tilde{\Delta}^\Lambda)\cdot\phi$, and using \eqref{Delta-partition}, results in the partition function factorising into a decoupled Gaussian integral over $\phi'_<$ (the constant of proportionality) and \eqref{func.int}, as required \cite{Morris:1993}.

Clearly, $\phi_{>}$ and $\phi_{<}$ beg to be regarded as the modes with momenta above and below $k$ respectively. This distinction is however only precise in the limit that the cutoff functions $C^\Lambda_k$ and $C^k$ become sharp. In general, modes in $\phi_{>}$ with $|p|<k$ and those in $\phi_{<}$ with $|p|>k$ will only be damped by the relevant cutoff functions. Even so, from now on we refer to $\phi_{>}$ ($\phi_{<}$) as high (low) momentum modes.

Consider computing the integral over the high momentum modes only in \eqref{mod.int}:
	\begin{equation}
	\label{high.int}
	Z^\Lambda_k[J, \phi_{<}]\equiv\int\!\! \mathcal{D}\phi_{>}\,\mathrm{e}^{-\frac{1}{2}\phi_{>}\cdot\left(\Delta^\Lambda_k\right)^{\!-1}\!\!\cdot\phi_{>}
	-\mathcal{S}^{\Lambda}[\phi_{>}+\phi_{<}]+J\cdot(\phi_{>}+\phi_{<})}
	\end{equation}
where $\phi_{<}$ now plays the role of a background field. Indeed, setting $\phi_{<}=0$ gives back the standard construction from which we can define the (UV and IR regulated) Legendre effective action,  \aka effective average action, as we will recall later:
	\begin{equation}
	\label{high.int-standard}
	Z^\Lambda_k[J]:=Z^\Lambda_k[J, 0]\equiv\int\!\! \mathcal{D}\phi_{>}\,\mathrm{e}^{-\frac{1}{2}\phi_{>}\cdot\left(\Delta^\Lambda_k\right)^{\!-1}\!\!\cdot\phi_{>}
	-\mathcal{S}^{\Lambda}[\phi_{>}]+J\cdot\phi_{>}}\,.
	\end{equation}
From \eqref{high.int}, performing the linear shift $\phi_{>}=\phi-\phi_{<}$ and rewriting the interaction $\mathcal{S}^{\Lambda}$ as a function of $\delta/\delta J$ gives
	\begin{equation}
	Z^\Lambda_k[J, \phi_{<}]=\mathrm{e}^{-\frac{1}{2}\phi_{<}\cdot\left(\Delta^\Lambda_k\right)^{\!-1}\!\!\cdot\phi_{<}}\,\mathrm{e}^{-\mathcal{S}^{\Lambda}			[\frac{\delta}{\delta J}]}\int\!\!\mathcal{D}\phi \,\mathrm{e}^{-\frac{1}{2}\phi\cdot\left(\Delta^\Lambda_k\right)^{\!-1}\!\!\cdot\phi+
	\phi\cdot(J+\left(\Delta^\Lambda_k\right)^{\!-1} \cdot	\phi_{<})}\,.
	\end{equation}
Following another change of variables $\phi'=\phi-\Delta^\Lambda_k\cdot J-\phi_{<}$, the remaining integral is a decoupled Gaussian in $\phi'$ and, after some rearranging, we obtain
	\begin{align}
	Z^\Lambda_k[J, \phi_{<}]= &\,\mathrm{e}^{\frac{1}{2}J\cdot\Delta^\Lambda_k\cdot J+J\cdot\phi_{<}}\,\mathrm{e}^{-\frac{1}{2}(J+
	\left(\Delta^\Lambda_k\right)^{\!-1}\!\!\cdot\phi_{<})\cdot{\Delta^\Lambda_k}\cdot(J+\left(\Delta^\Lambda_k\right)^{\!-1}\!\!\cdot\phi_{<})}\times\nonumber\\
	&\quad\mathrm{e}^{-\mathcal{S}^{\Lambda}[\frac{\delta}{\delta J}]}\,\mathrm{e}^{\frac{1}{2}(J+
	\left(\Delta^\Lambda_k\right)^{\!-1}\!\!\cdot\phi_{<})\cdot{\Delta^\Lambda_k}\cdot(J+\left(\Delta^\Lambda_k\right)^{\!-1}\!\!\cdot\phi_{<})}\,.
	\end{align}
Performing all derivatives in $\mathcal{S}^{\Lambda}[\delta/\delta J]$, we find
	\begin{equation}
	\label{high.Z}
	Z^\Lambda_k[J,\phi_{<}]=\,\mathrm{e}^{\frac{1}{2}J\cdot{\Delta^\Lambda_k}\cdot J+J\cdot\phi_{<}-S^k[{\Delta^\Lambda_k}\cdot J+\phi_{<}]}
	\end{equation}
for some functional $S^k$. Substituting the above expression into \eqref{mod.int}, we have another identity \cite{Morris:1993} for the original partition function \eqref{func.int}:
	\begin{equation}
	\label{low.int}
	Z^\Lambda[J]=\int\!\!\mathcal{D}\phi_{<}\,\,\mathrm{e}^{-\frac{1}{2}\phi_{<}\cdot \left(\Delta^k\right)^{\!-1}\!\!\cdot \phi_{<}+
	\frac{1}{2}J\cdot {\Delta^\Lambda_k}\cdot J+J\cdot \phi_{<}-S^k[{\Delta^\Lambda_k}\cdot J+\phi_{<}]}\,.
	\end{equation}
All the high modes have been integrated out. Consider for the moment the case where $J$ couples only to low energy modes i.e.\ so that ${\Delta^\Lambda_k}\cdot J=0$. Such is the case for example if the cutoff is of compact support so that $C^\Lambda_k(p)=0$ for $|p|<k$, and we choose $J$ to vanish for high energy modes, i.e.\ $J(p)=0$ for $|p|>k$. Choosing $J(p)=0$ for $|p|>k$ of course just means not considering Green's functions with momenta greater than this effective cutoff. Then $Z^\Lambda[J]$ simplifies to
	\begin{equation}
	\label{low.int.simplified}
	Z^\Lambda[J]=\int\!\!\mathcal{D}\phi_{<}\,\,\mathrm{e}^{-\frac{1}{2}\phi_{<}\cdot \left(\Delta^k\right)^{\!-1}\!\!\cdot \phi_{<}-S^k[\phi_{<}]
	+J\cdot \phi_{<}}\,.
	\end{equation}
It is now straightforward to recognize the functional $S^k$ as the interaction part of the total Wilsonian effective action 
\eqref{total-Wilsonian}
regulated in the UV at $k$. 

Since \eqref{low.int} is nothing but the original partition function \eqref{func.int}, it gives Green's functions which are all actually independent of $k$, despite appearances to the contrary. Note also that from \eqref{high.Z} and \eqref{high.int}, we obtain a prescription for computing the Wilsonian effective action from the bare action via a functional integral. We will return to this in section \ref{sec:solving}. 

The identification as a Wilsonian action, is still valid if we let $J$ couple to all modes. We just have to recognise that it then also enters non-linearly with the precise prescription given in \eqref{low.int}, i.e.\ as well as being the source it also plays the part of a space-time dependent coupling. Alternatively, we can use \eqref{low.int.simplified} even if ${\Delta^\Lambda_k}\cdot J\ne0$. In this case it is no longer true that \eqref{low.int.simplified} is independent of $k$, since we are missing the terms in \eqref{low.int} that contribute to making $Z^\Lambda[J]$ and thus all Green's functions independent of $k$. However for Green's functions all of whose (external) momenta $|p|\ll k$, we have $\Delta^\Lambda_k(p)=0$ to very good approximation. Furthermore $\Delta^\Lambda_k(p)\to0$ as $|p|/k\to0$, implying that in this limit \eqref{low.int.simplified} becomes exactly independent of $k$. 

The flow equation for $S^k$ is found by first differentiating \eqref{high.int} with respect to $k$ to obtain the flow equation for $Z^\Lambda_k[J,\phi_{<}]$:
	\begin{equation}
	\label{flow.Z}
	\frac{\partial}{\partial k}Z^\Lambda_k[J,\phi_{<}]=-\frac{1}{2}\bigg(\frac{\delta}{\delta J}-\phi_{<}\bigg)\cdot						\bigg(\frac{\partial}{\partial k}\left(\Delta^\Lambda_k\right)^{\!-1}\bigg)\cdot \bigg(\frac{\delta}{\delta J}-\phi_{<}\bigg)Z^\Lambda_k[J,\phi_{<}]\,.
	\end{equation}
Then by inserting \eqref{high.Z} into the above expression 
and defining $\Phi\equiv{\Delta^\Lambda_k}\cdot J+\phi_{<}$, we obtain the Polchinski flow equation given in \eqref{S.flow}.

Turning our attention to \eqref{high.Z} once more, we can recognise it as being related to the generator of connected Green's functions $W^\Lambda_k$ with IR cutoff $k$:
	\begin{equation}
	\label{W}
	\mathrm{e}^{W^\Lambda_k[J,\phi_{<}]}\equiv Z^\Lambda_k[J,\phi_{<}]=\,\mathrm{e}^{\frac{1}{2}J\cdot {\Delta^\Lambda_k}\cdot J+J\cdot \phi_{<}
	-S^k[{\Delta^\Lambda_k}\cdot J+\phi_{<}]}
	\end{equation}
and in taking the limit $k\rightarrow 0$, we recover the standard Green's functions (regulated in the UV through $\tilde{\Delta}^\Lambda$). The Legendre transform of $W^\Lambda_k$ gives the  Legendre effective action $\Gamma^{\text{tot},\Lambda}_{k}$: 
	\begin{align}
	\label{Legendre}
	\Gamma^{\text{tot},\Lambda}_{k}[\varphi,\phi_{<}]&=-W^\Lambda_k[J,\phi_{<}]+J\cdot \varphi\\ \label{Legendre-interactions}
	&=\frac{1}{2}(\varphi-\phi_{<})\cdot \left(\Delta^\Lambda_k\right)^{\!-1}\!\!\cdot (\varphi-\phi_{<})+\Gamma^\Lambda_k[\varphi]
	\end{align}
where $\varphi\equiv\delta W^\Lambda_k/\delta J$ is the classical field and $\Gamma^\Lambda_k$ is the interaction part which carries no $\phi_{<}$ dependence \cite{Morris:1993}, as follows from 
\be 
\label{no-phi<}
\frac{\delta}{\delta\phi_<}\Gamma^{\text{tot},\Lambda}_{k}[\varphi,\phi_{<}] = -\frac{\delta}{\delta\phi_<}W^\Lambda_k[J,\phi_{<}] 
 =-\left(\Delta^\Lambda_k\right)^{\!-1}\!\!\cdot\left(\frac{\delta W^\Lambda_k}{\delta J}-\phi_<\right) = \left(\Delta^\Lambda_k\right)^{\!-1}\!\!\cdot\left(\phi_<-\varphi\right)\,,
\ee
where we have used \eqref{Legendre} and then \eqref{W}. 

Notice that when $\phi_<=0$, we have the standard definition of the partition function \eqref{high.int-standard} and from it the standard definition of $W^\Lambda_k[J]$ in \eqref{W} and thus from \eqref{Legendre} the standard definition of the (IR and UV regulated) Legendre effective action. Thus from \eqref{Legendre-interactions} with $\phi_<=0$,   it follows that 
$\Gamma^\Lambda_k[\varphi]$ is the same interactions part of the effective average action as defined in \eqref{total-Gamma}. See also the discussion in section \ref{sec:details} leading up to  \eqref{total-Gamma}.  Recall that $\Gamma^\Lambda_k[\varphi]$ is thus equivalently the interactions part of the generator of one particle irreducible (1PI) Green's functions, cutoff in the IR at $k$, and coincides with the interactions part of the standard effective action $\Gamma$ in the limit $k\rightarrow0$. 
 
Substituting the Legendre transform equation \eqref{Legendre} into \eqref{flow.Z}, we obtain the flow equation for $\Gamma^\Lambda_k$ already stated in \eqref{Gamma.flow} . From equation \eqref{Legendre} follows the well-known fact that connected Green's functions can be expressed as a tree level sum of 1PI vertices (in this case connected by IR cutoff propagators). Thus equation \eqref{W} implies that the vertices of $S^k$ will also have a similar expansion (see section \ref{sec:vertices}). Indeed, we can find a direct relationship between $S^k$ and $\Gamma^\Lambda_k$ by substituting \eqref{W} into \eqref{Legendre}, using \eqref{Legendre-interactions} and recalling that $\Phi={\Delta^\Lambda_k}\cdot J+\phi_{<}$. The result is the duality equation \eqref{duality-gen} we have been aiming for.
 
To reiterate, \eqref{duality-gen} is an exact relationship between the interaction part of the Wilsonian effective action, $S^k$, regulated in the UV at $k$ and the interaction part of the Legendre effective action, $\Gamma^\Lambda_k$ regulated in the UV at $\Lambda$ and regulated in the IR at $k$ (\aka effective average action). It gives rise to a duality between the flow equations \eqref{S.flow} and \eqref{Gamma.flow}. If we have a complete RG trajectory for $\Gamma_k$, that is a solution of \eqref{FRGE2} where the UV cutoff $\Lambda$ has been removed, and where by complete we mean that it extends from a UV fixed point as $k\rightarrow \infty$ down to $k\rightarrow 0$, then we can take the continuum limit of the key equations given in this section simply by replacing $\Delta^\Lambda_k$ with $\Delta_k$. In this way we equivalently have a solution to \eqref{Gamma.flow}, the duality relation now reads \eqref{duality-con}, which allows us to compute the equivalent RG trajectory for $S^k$ with the equivalent fixed point solution, and the continuum limit of the effective partition functions can then be computed directly from \eqref{low.int} and \eqref{low.int.simplified}.

\section{Vertices of the Wilsonian Effective Action}
\label{sec:vertices}

In this section we use result \eqref{duality-con} to derive explicit expressions for the vertices of $S^k$ in terms of those of $\Gamma_{k}$. Clearly \eqref{duality-con} is symmetric under the map: $S_k\leftrightarrow\Gamma_k$ with $\Delta_k\mapsto-\Delta_k$, so by relabelling in this way we can also use the expressions below to derive the vertices of $\Gamma_k$ from $S^k$.
Clearly these expressions can therefore also be used after some renaming to give the vertices of one action in terms of another for any of the alternative expressions of duality, namely  \eqref{duality-gen},  \eqref{duality-Gamma-con}, \eqref{duality-gen-2} and \eqref{duality-Gamma-gen}. For example to obtain the vertices of $\Gamma^\Lambda_k$ in terms of those of $\Gamma_k$ using \eqref{duality-Gamma-con} (part of our second solution to the reconstruction problem) it is only necessary to replace $S^k$ with $\Gamma^\Lambda_k$ and $\Delta_k$ with $\Delta_\Lambda$ in the following expressions.

Extracting the momentum conserving Dirac delta-function in what follows, vertices of $S^k$ will be denoted by
	\begin{equation}
	(2\pi)^d\delta(p_1+\cdots+p_n)\,S^{(n)}(p_{1},\cdots,p_{n};k)\equiv\frac{\delta^{n} S^k[\Phi]}{\delta\Phi(p_{1})\cdots\delta\Phi(p_{n})}\bigg|				_{\Phi=0}
	\end{equation}
and the vertices of $\Gamma_{k}$ by
	\begin{equation}
	\label{Gamma-vertices}
	(2\pi)^d\delta(p_1+\cdots+p_n)\,\Gamma^{(n)}(p_{1},\cdots,p_{n};k)\equiv\frac{\delta^{n} \Gamma_{k}[\Phi]}{\delta\Phi(p_{1})\cdots						\delta\Phi(p_{n})}\bigg|_{\Phi=0}
	\end{equation}
with the exception of its 2-point function which we write as $\Sigma(p^{2};k)$. We often omit the momentum arguments of the vertices for neatness. For simplicity we impose a $Z_{2}$ symmetry $\phi \leftrightarrow -\phi$ on ${\cal S}^\Lambda$ so that it only contains even powers of $\phi$ and hence $S^{(n)}(p_{1},\cdots,p_{n};k)$ and $\Gamma^{(n)}(p_{1},\cdots,p_{n};k)$ vanish for odd $n$.

We start by writing \eqref{duality-con} more conveniently as
	\begin{equation}
	\label{soln2}
	S^k[\Phi]=\Gamma_{k}[\Phi-{\Delta_k}\cdot \frac{\delta S^k}{\delta\Phi}]+\frac{1}{2}\frac{\delta S^k}				{\delta\Phi}\cdot {\Delta_k}\cdot \frac{\delta S^k}{\delta\Phi}
	\end{equation}
by recognising that $\varphi=\Phi-{\Delta_k}\cdot (\delta S^k/\delta\Phi)$.  Taylor expanding the right-hand side, keeping only bilinear terms in $\Phi$ and rearranging, we find the following expression for the 2-point function:
	\begin{equation}
	\label{two.point}
	S^{(2)}(p^{2};k)=\Sigma(p^{2};k)\left(1+{\Delta_k(p)}\Sigma(p^{2};k)\right)^{\!-1}\,.
	\end{equation}
Expanding the right-hand side perturbatively in $\Sigma$ gives the expected expansion of $S^{(2)}$ in terms of 1PI vertices, connected by IR cutoff propagators. Note that in obtaining this result, it is only necessary to expand to second order in the Taylor series as the $Z_{2}$ symmetry kills the cross-terms from one-point and three-point vertices that would otherwise appear.

\begin{figure}
	\begin{center}
	\setlength{\unitlength}{2mm}
	\begin{picture}(75,35)
	\includegraphics[scale=0.9]{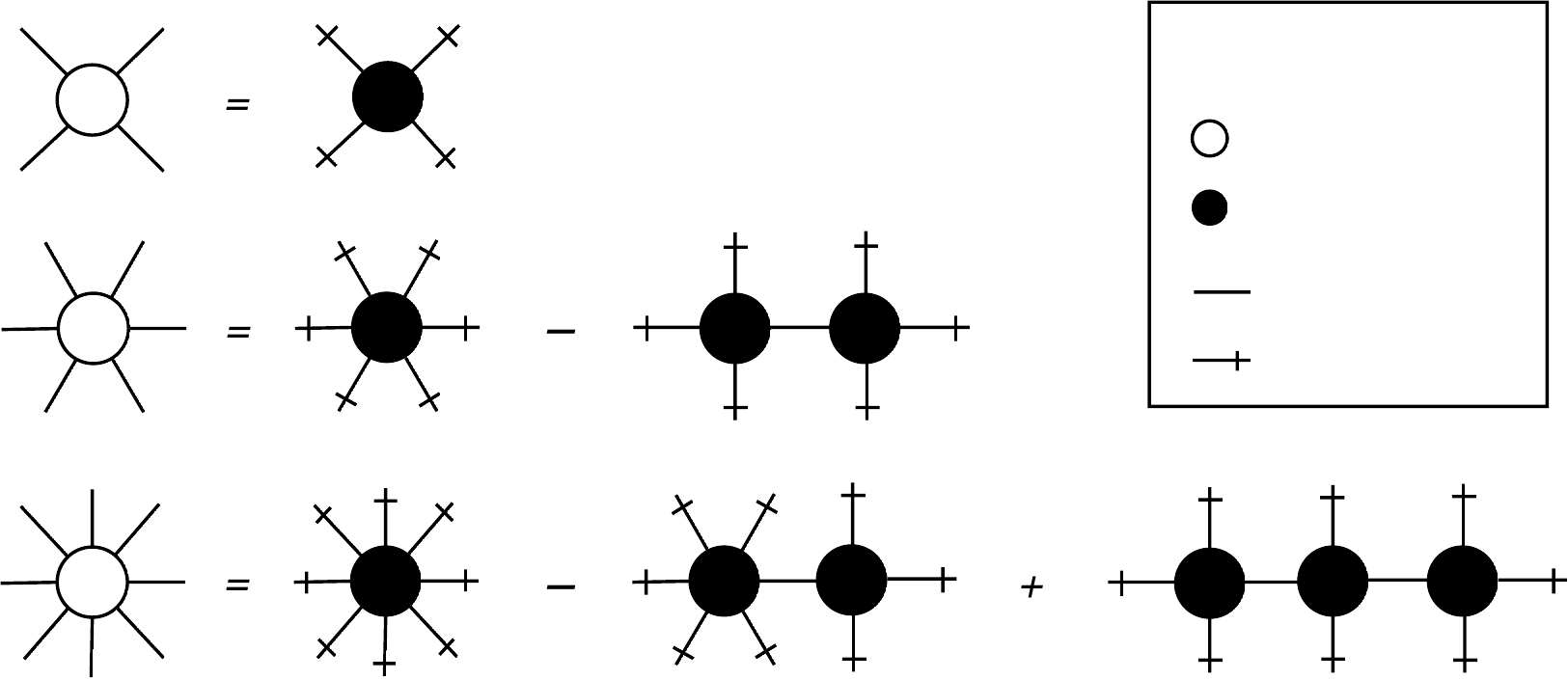}
	\put(-18.5,29.5){Key}
	\put(-13,25.25){vertex of $S^k$}
	\put(-13, 22){vertex of $\Gamma_{k}$}
	\put(-11,18){$\frac{\Delta_kS^{(2)}}{\Sigma}$}
	\put(-10, 14.5){$\frac{S^{(2)}}{\Sigma}$}
	\put(-46, 16.25){$\frac{1}{2}$}
	\put(-23.5,4.2){$\frac{1}{2}$}
	\end{picture}
	\end{center}
	\caption{Vertices of the Wilsonian effective interaction $S^k$ for $n=4,6$ and 8 respectively from top to bottom. Each diagram containing more than one vertex represents a sum over disjoint subsets of momenta corresponding to the number and type of vertices in each diagram e.g.\ the final diagram in the expansion of $S^{(8)}$ stands for a sum over partitions of $\{p_{1},\cdots,p_{8}\}$ into 2 sets of 3 momenta and 1 set of 2 momenta.\label{fig:vertices}}
\end{figure}
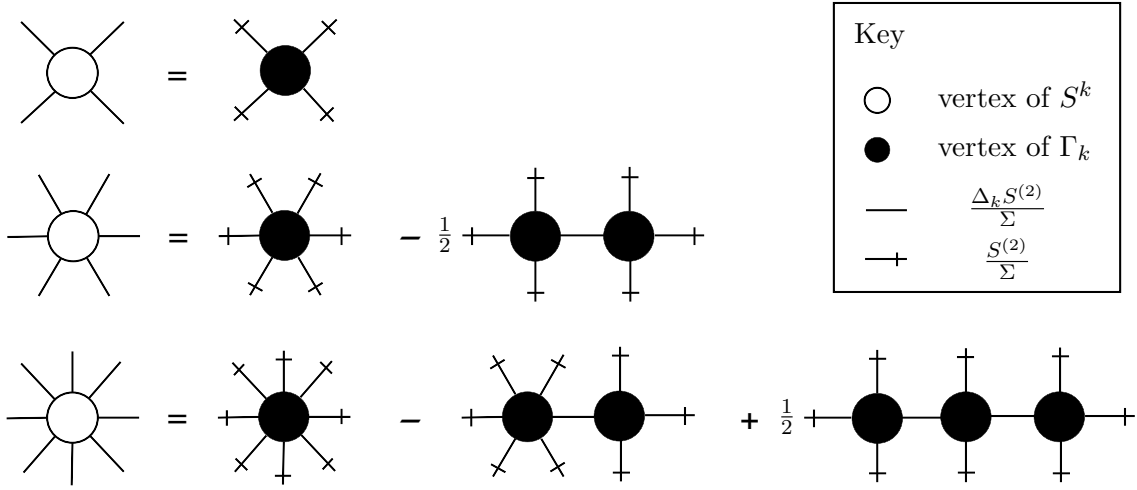

To compute expressions for vertices for $n>2$, we need to isolate the 2-point pieces from $S^k$ and $\Gamma_{k}$, like so
	\begin{align}
	S^k[\Phi]=\frac{1}{2}\Phi\cdot S^{(2)}\cdot \Phi + S'^k[\Phi]&&
	\Gamma_{k}[\varphi]=\frac{1}{2}\varphi\cdot \Sigma\cdot \varphi + \Gamma'_{k}[\varphi]
	\end{align}
such that all terms but those quadratic in the fields are contained in $S'^k$ and $\Gamma'_{k}$. Upon substituting the above into \eqref{soln2} and using \eqref{two.point}, we have
	\begin{equation}
	\label{soln3}
	S'^k[\Phi]=\Gamma_{k}'[\frac{S^{(2)}}{\Sigma}\cdot \Phi-{\Delta_k}\cdot \frac{\delta S'^k}{\delta\Phi}]+\frac{1}				{2}\frac{\delta S'^k}{\delta\Phi}\cdot \frac{\Delta_k\Sigma}{S^{(2)}}\cdot \frac{\delta S'^k}{\delta\Phi}\,.
	\end{equation}
Again, by Taylor expanding the right-hand side to the desired order, we obtain our vertex of choice. In general, for an $n$-point function, we only have to keep terms in the Taylor series up to and including the $n$th order: higher order terms vanish either from the $Z_{2}$ symmetry or because they then contain too many $\Phi$s. For the 4-point function we have
	\begin{equation}
	\label{4.point}
	S^{(4)}(p_{1},p_{2},p_{3},p_{4};k)=\Gamma^{(4)}(p_{1},p_{2},p_{3},p_{4};k)\prod_{i=1}^{4}\frac{S^{(2)}			(p^{2}_{i};k)}{\Sigma(p^{2}_{i};k)}\,.
	\end{equation}
Likewise, the 6-point function is given by
	\begin{align}
	S^{(6)}(p_{1},\cdots,p_{6};k)=&\Gamma^{(6)}(p_{1},\cdots,p_{6};k)\prod_{i=1}^{6}\frac{S^{(2)}(p^{2}_{i};k)}				{\Sigma(p^{2}_{i};k)}\nonumber\\
	&\quad-\frac{1}{2}\sum_{\{I_{1},I_{2}\}}\bigg\{\Gamma^{(4)}(I_{1},q;k)\prod_{p_{i}\in I_{1}}
	\frac{S^{(2)}(p_{i}^{2};k)}{\Sigma(p_{i}^{2};k)}\nonumber\\
	&\qquad\times{\Delta_k}(q^{2})\frac{S^{(2)}(q^{2};k)}{\Sigma(q^{2};k)}\Gamma^{(4)}(-q,I_{2};k)\prod_{p_{j}
	 \in I_{2}}\frac{S^{(2)}(p_{j}^{2};k)}{\Sigma(p_{j}^{2};k)}\bigg\}
	\end{align}
where $I_{1}$ and $I_{2}$ are disjoint subsets of 3 momenta such that $I_{1}\cup I_{2}=\{p_{1},\cdots,p_{6}\}$. The sum over $\{I_{1},I_{2}\}$ means sum over all such subsets. By momentum conservation, the momentum $q$ carried by certain 2-point functions is equivalent to a partial sum i.e.\ $q=\sum_{p_{i} \in I} p_{i}$ where $I$ is a subset of the total set of external momenta. Graphical representations of these expressions, as well as one for the 8-point function, are given in figure \ref{fig:vertices} and are much easier to interpret. Of course the expansion can be continued to higher orders.

\section{Second solution to the reconstruction problem}
\label{sec:solving}

In this section we provide more detail on our second solution to the reconstruction problem and show how it is related to the one-loop approximate solution \eqref{MandR.soln} provided in reference \cite{Manrique:2008zw}. As explained in the introduction and section \ref{sec:details}, given a complete RG trajectory for $\Gamma_k[\varphi]$, \eqref{duality-con} then provides us with $S^k[\Phi]$ which is the interaction part of a perfect bare action. This already provides us with an acceptable solution to the reconstruction problem, but as we emphasised in appendix \ref{app:UV} it cannot give us back $\Gamma_k$ via the standard path integral route \eqref{high.int-standard} since such a UV regulated path integral necessarily leaves its imprint on the Legendre effective action such that it now depends on both cutoffs: $\Gamma\equiv\Gamma^\Lambda_k$. However what can be done is to use $\Gamma_k[\varphi]$ to construct a pair 
$\{\bS, \Gamma_{k}^\Lambda\}$, where $\Gamma^\Lambda_k$ is related to $\bS$ in the usual way, and such that as $\Lambda\to\infty$ we have $\Gamma_{k}^\Lambda\to\Gamma_k$. This is our second solution. The question then is how this solution is to be compared with the one-loop approximate relation \eqref{MandR.soln}.

Let us first note that in \eqref{MandR.soln} we can split off the bare interactions and effective average action interactions as defined in \eqref{other-interactions} and \eqref{interactions} respectively. For the left-hand side of \eqref{MandR.soln} that just means dropping the hats, 
but in the right-hand side we recognise that as in the shift from \eqref{LEAA-con} to \eqref{LEAA-alt} we can incorporate the infrared cutoff through a multiplicative profile \eqref{IR-only} and then make explicit the UV sharp cutoff by replacing this by \eqref{UVIRMR}. The net result is that we re-express equation \eqref{MandR.soln} as
\begin{equation}
	\label{descents.soln.1}
	\Gamma_{k=\Lambda}^\Lambda[\varphi]=\mathcal{S}^{\Lambda}[\varphi]+
	\frac{1}{2}\text{Tr}\,\text{ln}\Big\{\mathcal{S}^{\Lambda (2)}[\varphi]+\left(\Delta^\Lambda_\Lambda\right)^{\!-1}\Big\}\,.
	\end{equation}
This has two advantages. Firstly it makes the overall UV sharp cutoff explicit, and secondly actually this formula is valid as a one-loop approximation in general, whatever the precise profile of IR and UV cutoff we implement via $C^\Lambda_k(p)$. 
The total effective average action is then in general given as in \eqref{total-Gamma} and the total bare action as in \eqref{total-bare-k}. As already reviewed below \eqref{no-phi<}, the standard construction using the partition function \eqref{high.int-standard} yields of course this $\Gamma^\Lambda_k[\ph]$. 

Our second solution to the reconstruction problem follows  from employing compatible cutoffs. Recall that by compatible cutoffs we mean that $C^\Lambda_\Lambda(p)=0$ for all $p$, i.e.\ such that when the IR cutoff meets the UV cutoff the result completely kills the propagator: $\Delta^\Lambda_\Lambda \equiv 0$. Up to a (divergent but irrelevant) constant we then have $\Gamma_{k=\Lambda}^\Lambda[\varphi]=\mathcal{S}^{\Lambda}[\varphi]$ as is clear from \eqref{descents.soln.1} if we note that
\be 
\label{one-loop-compatible}
\text{Tr}\,\text{ln}\Big\{\mathcal{S}^{\Lambda (2)}[\varphi]+\left(\Delta^\Lambda_\Lambda\right)^{\!-1}\Big\} = -\text{Tr}\,\text{ln}\left\{\Delta^\Lambda_\Lambda\right\} +\text{Tr}\,\text{ln}\Big\{1+\Delta^\Lambda_\Lambda\cdot\mathcal{S}^{\Lambda (2)}[\varphi]\Big\}\,.
\ee
Indeed the fact that in \eqref{high.int-standard}, $\left(\Delta^\Lambda_k\right)^{\!-1}\to\infty$ as $k\to\Lambda$, turns the steepest descents calculation that gives \eqref{descents.soln.1} into an exact statement. For the same reason, from the most direct expression relating the Wilsonian interactions $S^k$ to the bare interactions, obtained by setting $J=0$ in \eqref{high.int} and \eqref{high.Z}:
	\begin{equation}
	\label{partition to S}
	Z^\Lambda_k[0,\phi_{<}]=\,\mathrm{e}^{-S^k[\phi_{<}]}=\int\!\!\mathcal{D}\phi_{>}
	\,\,\mathrm{e}^{-\frac{1}{2}\phi_{>}\cdot \left(\Delta^\Lambda_k\right)^{\!-1}\!\!\cdot \phi_{>}-
	\mathcal{S}^{\Lambda}[\phi_{>}+\phi_{<}]}\,,
	\end{equation}
we see that we have no choice but to have the equality $S^{\Lambda}[\varphi]=\mathcal{S}^{\Lambda}[\varphi] $. To make the map from the continuum solution $\Gamma_k$ to this system, we insist that the Wilsonian interactions $S^k$ and thus also the effective Wilsonian cutoff $C^k(p)$, are still  the continuum ones. Then this fixes via \eqref{sum rule} the overall bare cutoff to be the continuum Wilsonian one: $\Cl=C^\Lambda$, and as we see already the bare interactions must taken to be $\mathcal{S}^{\Lambda}[\varphi] = S^{\Lambda}[\varphi]$. Then the map \eqref{duality-Gamma-con} from $\Gamma_k$ to $\Gamma^\Lambda_k$ follows, as proved in section \ref{sec:duality-Gamma}, and worked out in detail in section \ref{sec:vertices}. We thus have all the elements of our second solution.

If the UV and IR cutoff imposed by ${\Delta^\Lambda_k}$ are not compatible, then $\Delta^\Lambda_\Lambda\ne0$ and in both \eqref{high.int-standard} and \eqref{partition to S} there is still a non-trivial functional integral to compute in the limit $k\to\Lambda$. To one loop, the result for $\Gamma^\Lambda_\Lambda$ is the one given in \eqref{descents.soln.1}.
 In analogy with \cite{Manrique:2008zw}, let us also compute the integral in \eqref{partition to S} to one loop, using the method of steepest descents. The exponent is at a minimum when
	\begin{equation}
	\label{min}
	\phi_{>}=-{\Delta^\Lambda_k}\cdot \frac{\delta \mathcal{S}^{\Lambda}[\phi_>+\phi_<]}{\delta\phi_{>}}\equiv\phi^{0}_{>}\,.
	\end{equation}
We define $\phi_{>}\equiv\phi_{>}^{0}+\tilde{\phi}_{>}$ 
and expand about $\tilde{\phi}_{>}=0$, keeping only up to second derivatives of $\mathcal{S}^{\Lambda}$:
	\begin{equation}
	\mathrm{e}^{-S^k[\phi_{<}]}=\,\mathrm{e}^{-\frac{1}{2}\phi^{0}_{>}\cdot \left(\Delta^\Lambda_k\right)^{\!-1}\!\!\cdot \phi^{0}_{>}}
	\,\mathrm{e}^{-\mathcal{S}^{\Lambda}[\phi_{>}^{0}+\phi_{<}]}
	\int\!\!\mathcal{D}\tilde{\phi}_{>}\,\,\mathrm{e}^{-\frac{1}{2}\tilde{\phi}_{>}\cdot \left(\Delta^\Lambda_k\right)^{\!-1}\!\!\cdot \tilde{\phi}_{>}
	-\frac{1}{2}\tilde{\phi}_{>}\cdot \frac{\delta^{2}\mathcal{S}^{\Lambda}}{\delta\phi_{>}\delta\phi_{>}}\cdot \tilde{\phi}_{>}}\,.
	\end{equation}	
The terms linear in $\tilde{\phi}_{>}$ cancel by \eqref{min}. Performing the Gaussian integral over $\tilde{\phi}_{>}$, we find
	\begin{equation}
	\label{one-loop-Wilsonian}
	S^k[\phi_{<}]-\frac{1}{2}\phi_{>}^{0}\cdot \left(\Delta^\Lambda_k\right)^{\!-1}\!\!\cdot \phi_{>}^{0}
	=\mathcal{S}^{\Lambda}[\phi^{0}_{>}\!+\!\phi_{<}]+
	\frac{1}{2}\text{Tr}\,\text{ln}
	\Big\{\frac{\delta^{2}\mathcal{S}^{\Lambda}[\phi^{0}_{>}\!+\!\phi_{<}]}									{\delta\phi_{>}\delta\phi_{>}}+\left(\Delta^\Lambda_k\right)^{\!-1}\Big\}\,.
	\end{equation}
Introducing $\varphi\equiv\phi_{>}^{0}+\phi_{<}$, we thus have
	\begin{equation}
	\label{descents.soln-S}
	S^k[\phi_{<}]-\frac{1}{2}(\varphi-\phi_<)\cdot \left(\Delta^\Lambda_k\right)^{\!-1}\!\!\cdot (\varphi-\phi_<)=\mathcal{S}^{\Lambda}[\varphi]+
	\frac{1}{2}\text{Tr}\,\text{ln}\Big\{\mathcal{S}^{\Lambda (2)}[\varphi]+\left(\Delta^\Lambda_k\right)^{\!-1}\Big\}\,.
	\end{equation}
Comparing \eqref{descents.soln.1} we recognise that the right-hand side is nothing but the one-loop approximation to the effective average action at a general value of $k$:
	\begin{equation}
	\label{descents.soln}
	\Gamma_{k}^\Lambda[\varphi]=\mathcal{S}^{\Lambda}[\varphi]+
	\frac{1}{2}\text{Tr}\,\text{ln}\Big\{\mathcal{S}^{\Lambda (2)}[\varphi]+\left(\Delta^\Lambda_k\right)^{\!-1}\Big\}\,.
	\end{equation}
Finally comparing \eqref{descents.soln} and \eqref{descents.soln-S}, we see that we recover  the duality relation \eqref{duality-gen} in section \ref{sec:SandGamma}.\footnote{It can also be shown that this is consistent to one loop with the solution \eqref{min}.} We have therefore explicitly confirmed the duality relation to one loop via the steepest descents method.
Through the above demonstration and also our discussion of the compatible case, cf.\ \eqref{one-loop-compatible}, we have also comprehensively explored how our solution is related to the one-loop result \eqref{MandR.soln}.

\section{Some compatible cutoffs}
\label{sec:compatible}

In this section we briefly explore some possible forms of compatible cutoffs, i.e.\ such that $C^\Lambda_k(p)$ vanishes identically when $k\to\Lambda$. We also insist that the effective Wilsonian UV cutoff $C^k(p)$ depends only on the one cutoff scale $k$ as indicated. Through the sum rule \eqref{sum rule} it follows that we take the overall UV cutoff to be the Wilsonian one at scale $\Lambda$: $\Cl=C^{k=\Lambda}$. 

There are various possibilities for compatible cutoffs. One straightforward option is to make all the cutoff functions sharp:
	\begin{equation}
	\label{sharp}
	C^\Lambda= \left\{
  	\begin{array}{l l}
   	 0 & \quad |p|\geq\Lambda\vspace{3pt}\\
    	1 & \quad |p|<\Lambda\vspace{3pt}
 	 \end{array} 
 	 \right., \qquad
	C^k= \left\{
  	\begin{array}{l l}
   	 0 & \quad |p|>k\vspace{3pt}\\
    	 1& \quad  |p|\leq k\vspace{3pt}
 	 \end{array} 
 	 \right.,\qquad
	C^\Lambda_k= \left\{
  	 \begin{array}{l l}
   	 0 & \quad |p|\geq\Lambda \vspace{3pt}\\
    	 1 & \quad k<|p|<\Lambda\vspace{3pt}\\
	 0& \quad |p|\leq k\vspace{3pt}
 	 \end{array} 
 	 \right.\,.
	\end{equation}
Another choice of compatible cutoffs is:
	\begin{equation}
	\label{choice.2}
	C^\Lambda=\! \left\{
  	\begin{array}{l l}
   	 0 &  |p|\geq\Lambda\vspace{3pt}\\
    	1-\frac{p^{2}}{\Lambda^{2}} &  |p|<\Lambda\vspace{3pt}
 	 \end{array} 
 	 \right.\!,\quad
	\!C^k=\! \left\{
  	\begin{array}{l l}
   	 0 &  |p|\geq k\vspace{3pt}\\
    	 1-\frac{p^{2}}{k^{2}}&   |p|<k\vspace{3pt}
 	 \end{array} 
 	 \right.\!,\quad
	\!C^\Lambda_k=\!\left\{
  	\begin{array}{l l}
   	 0 &  |p|\geq\Lambda \vspace{3pt}\\
    	1-\frac{p^{2}}{\Lambda^{2}} &  k\leq |p|<\Lambda\vspace{3pt}\\
	\frac{p^{2}}{k^{2}}-\frac{p^{2}}{\Lambda^{2}}&  |p|<k\vspace{3pt}
 	 \end{array} 
 	 \right.\!\!.
 	\end{equation}
It can be easily checked that all cutoff functions have the desired regulating behaviour and that for $k=\Lambda$, we have ${\Delta^\Lambda_k}=0$.
The cutoff functions \eqref{choice.2} have been obtained by first of all using \eqref{IR-only} to find the multiplicative IR cutoff function $C_k$ corresponding to the optimized cutoff. In order to ensure that the effective Wilsonian UV cutoff depends only on the one cutoff scale $k$, we   define it as $C^k=1-C_k$, i.e.\ via \eqref{sum rule} but with the overall cutoff $\Lambda\to\infty$, and thus $\Cl\to1$.  (The result agrees with \eqref{MR-Wilson} since we already found that cutoff  $C^k$ to be dual to the optimised IR cutoff and also to be independent of $\Lambda$.) 
As we have seen, compatibility for finite overall cutoff then requires $\Cl\equiv C^\Lambda$. This however forces us to change the IR profile via \eqref{sum rule} to one, $C^\Lambda_k = C^\Lambda-C^k$, that includes both cutoffs. The resulting choices \eqref{choice.2} thus also have the property that as $\Lambda\to\infty$, $C^\Lambda_k$ returns to the (multiplicative form \eqref{IR-only} of the) optimised cutoff.

Another choice of additive IR regulator from which we can define compatible cutoff functions following these steps is
	\begin{equation}
	\label{alt-additive}
	\tilde{R}_{k}(p^{2})=\frac{1}{\mathrm{e}^{\frac{p^{2}}{k^{2}}}-1}\,.
	\end{equation}
This corresponds to the following choice of cutoffs:
	\begin{equation}
	C^k=\frac{1}{1+p^{2}\big(\mathrm{e}^{\frac{p^{2}}{k^{2}}}-1\big)}\,,
	\end{equation}
again the overall UV cutoff is just $C^\Lambda$, and
	\begin{equation}	
	C^\Lambda_k=\frac{p^{2}\big(\mathrm{e}^{\frac{p^{2}}{k^{2}}}-\mathrm{e}^{\frac{p^2}{\Lambda^2}}\big)}							{\big(1+p^2\big(\mathrm{e}^{\frac{p^2}{\Lambda^2}}-1\big)\big)\big(1+p^2\big(\mathrm{e}^{\frac{p^2}{k^2}}-1\big)\big)}\,.
	\end{equation}
These cutoffs regulate as required, exhibiting the behaviour described below \eqref{IR-only}, below \eqref{sharpUV} and below \eqref{MR-Wilson} respectively. Again we have defined the cutoffs so that  ${\Delta^\Lambda_k}=0$ when $k=\Lambda$,  whilst as $\Lambda\to\infty$, $C^\Lambda_k$ returns to the multiplicative version of \eqref{alt-additive}.
In summary, we have seen how we can formulate compatible cutoff functions
using a sharp cutoff, or based closely on the optimised cutoff $R_{k}$ or $\tilde{R}_{k}$.

\section{Conclusions}
\label{sec:conclusions}

Let us start by briefly summarising our main conclusions. We set out two solutions to the reconstruction problem, giving the recipes in detail in section \ref{sec:details}. Starting from a full renormalised trajectory for the effective average action \eqref{interactions}, whose interactions are given by $\Gamma_k[\varphi]$, we can reconstruct a suitable bare action by using the corresponding Wilsonian interactions $S^k[\Phi]$. This also describes the full renormalised trajectory, but in the Wilsonian language. $S^k[\Phi]$ is computed through the duality relation \eqref{duality-con}. The vertices are then related via a tree expansion to the vertices of $\Gamma_k$ and these are worked out in detail in section \ref{sec:vertices}. 
The full Wilsonian effective action $S^{\mathrm{tot},k}[\Phi]$ is given by \eqref{total-Wilsonian}, where the effective multiplicative UV cutoff profile $C^k(p)=1-C_k(p)$, and $C_k$ is the multiplicative version of the additive IR cutoff via the translation \eqref{IR-only}.
The partition function constructed using $S^{\mathrm{tot},k}[\Phi]$ is actually independent of $k$, and thus this bare action is an example of a perfect bare action. Written in the form \eqref{low.int.simplified} (where the superscript $\Lambda=\infty$ since we have taken the continuum limit), the independence with respect to $k$ is only approximate, becoming exact when we compute Green's functions with momenta $|p|\ll k$, unless the source $J$ obeys some restrictions, as discussed around \eqref{low.int.simplified}. Alternatively we can embed the source inside the action as well, as in \eqref{low.int}, and then the independence with respect to $k$ is indeed exact.

A potential problem with this first solution to the reconstruction problem is that we have only the one cutoff $k$ involved which now plays the role of a UV cutoff for this perfect bare action. For some purposes we may want to investigate a system where a suitable bare action with UV regularisation set at some scale $\Lambda$ gives back the effective average action through the usual procedure. In other words, we insert an infrared cutoff $k$ into the bare action to give \eqref{total-bare-k}, where the overall multiplicative UV cutoff has been replaced by $C^\Lambda_k$ incorporating also the IR cutoff, and then form the partition function \eqref{high.int-standard}.
As we emphasised in appendix \ref{app:UV}, we cannot get the continuum $\Gamma_k$ in such a way, since it is then guaranteed that the effective average action $\hat{\Gamma}^\Lambda_k$, bilinear part and interactions, now depends on both cutoffs, as displayed in \eqref{total-Gamma}.  What we can do however is again to take the bare interactions to be the perfect Wilsonian ones computed from $\Gamma_k$, thus $\bS = S^{k=\Lambda}$, and then the above procedure gives us a $\Gamma^\Lambda_k[\varphi]$, such that as $\Lambda\to\infty$, $\Gamma^\Lambda_k\to\Gamma_k$. The UV boundary conditions on the flow equation \eqref{Gamma.flow} for this effective average action are just $\Gamma^\Lambda_{k=\Lambda} = S^\Lambda = \bS$. 
We do not need to compute the functional integral, or the flow equation, to find $\Gamma^\Lambda_k[\varphi]$ however, since it is also directly related to the original continuum $\Gamma_k$ via a duality relation \eqref{duality-Gamma-con}, which may also be solved vertex by vertex as in section \ref{sec:vertices}. This is our second solution to the reconstruction problem, summarized in figure \ref{fig:map}.

We proved the latter duality relation by first proving an even more remarkable duality relation in section \ref{sec:duality-Gamma}, namely \eqref{duality-Gamma-gen}. This is a tree-level relation between two effective average actions computed with different overall cutoff profiles $\Cl$ and $\Clp$, but whose corresponding effective Wilsonian actions $S^{\mathrm{tot},k}$ actually coincide.  As we explain in section \ref{sec:duality-Gamma}, this assumes that the bare interactions $\bS$ and $\mathring{{\cal S}}^\Lp$ can be chosen precisely to ensure this. If we choose a solution $S^k$ of the flow equation \eqref{S.flow} that does not correspond to a full renormalised trajectory, then clearly this is not always possible, for example it is then not possible to raise the overall cutoff $\Lambda$ or $\Lp$ all the way to infinity. Even if we choose $S^k$ to be a renormalised trajectory, it still may not be possible to change the bare cutoff arbitrarily in such a way.
The ability to do this is a statement of universality, but universal behaviour typically has a basin of attraction, so it should be expected that $\Cl$ cannot be changed completely arbitrarily. However these limitations do not apply to the required duality relation \eqref{duality-Gamma-con} since as we saw in section \ref{sec:duality-Gamma},
this corresponds to the special case where the form of the overall cutoff profile $C^\Lambda$ does not change, only the overall scale $\Lambda\mapsto\Lp$, and furthermore the bare interactions are perfect Wilsonian ones corresponding to a full renormalised trajectory, and thus exist at any scale.

In section \ref{sec:solving} we fully explored how our solutions to the reconstruction problem are related to the one-loop formula \eqref{MandR.soln} derived in reference \cite{Manrique:2008zw}. The key was to recognise that in our second solution we employed compatible cutoffs such that when the IR cutoff meets the UV cutoff, $k\to\Lambda$, the propagator is forced to vanish identically. In section \ref{sec:compatible} we set out a recipe for constructing such cutoff combinations.

Although we phrased all relations in terms of a single scalar field, it is a straightforward generalisation to write the relations for multiple fields including fields with indices and those with fermionic statistics. It is therefore straightforward to generalise these relations to the case of full quantum gravity for instance. At various stages we discarded additive constant terms, but these would become background dependent. Their functional form can be determined however, and thus this would be a useful extension of this work.
Finally, since $S^{\Lambda}$ are perfect bare interactions, or equivalently since they are made via a tree-diagram expansion using the vertices of $\Gamma_{k=\Lambda}{}$, we can expect them to be as complicated as $\hG{k}{}$, arguably more so. 
For any large but finite $\Lambda$, we can however use $S^\Lambda$ as the starting point for constructing equally valid alternative bare actions based on either of our solutions of the reconstruction problem. 
We have already seen a small example of this in that using $S^\Lambda$ together with the standard coupling between source and fields as in \eqref{low.int.simplified} only yields a perfect action lying on a renormalised trajectory in the limit of infinite $\Lambda$, unless we impose restrictions on the source (cf.\ section \ref{sec:SandGamma}).
In fact we have an infinite dimensional space of possible bare actions to choose from (a reflection of universality). In general we can choose $\hat{\mathcal{S}}^\Lambda$ to be any action close to any point on the (infinite dimensional) critical surface containing the UV (asymptotically safe) fixed point $\hS*$, such that after appropriate tuning back into the critical surface in the limit $\Lambda\to\infty$, we again construct the renormalised trajectory (see e.g.\ \cite{Morris:1998}). In practice for example we can choose $\hbS\Lambda= \hS\Lambda+\sum_{i\notin{\cal R}} \alpha_i(\Lambda) {\cal O}_i$, where the sum is over the integrated irrelevant operators and $\alpha_i(\Lambda)$ are arbitrary functions of $\Lambda$ providing they remain small enough for the linearised approximation to be valid as $\Lambda\to\infty$.
\appsection{}
\section{Why a UV regulated effective average action must depend on the UV regulator}
\label{app:UV}

It is clear that at least for a general form of UV cutoff,  the effective average action $\hat{\Gamma}_{k}^\Lambda[\varphi]$ must depend on the UV regulator $\Lambda$ as indicated. Indeed if we embed the UV cutoff in the free propagator as done in \eqref{Gamma.flow} then the Feynman diagrams that follow from its perturbative expansion will evidently have all free propagators $1/p^2$ replaced by $\Delta^\Lambda_k(p)$. The fact that $\hat{\Gamma}_{k}^\Lambda[\varphi]$ thus depends on two scales, means that a bare action cannot be reconstructed which would directly give the continuum version $\hat{\Gamma}_k$ in the usual way. This is the first ``severe issue''  outlined above \eqref{MandR.soln}. 

Following reference \cite{Manrique:2008zw}, a sharp UV cutoff and infrared optimised cutoff would appear to provide an exception however. With  a sharp UV cutoff in place, \eqref{FRGE2} can alternatively be written
\begin{equation}
\label{R-flow-2}
	\frac{\partial}{\partial k}\hat{\Gamma}_{k}^\Lambda[\varphi]=\frac{1}											{2}\text{Tr}\bigg[\bigg(R_{k}+\frac{\delta^{2}\hat{\Gamma}_{k}^\Lambda}										{\delta\varphi\delta\varphi}\bigg)^{\!-1}\frac{\partial R_{k}}{\partial k}\bigg]-\frac{1}										{2}\text{Tr}\bigg[\theta(|p|-\Lambda)\bigg(R_{k}+\frac{\delta^{2}\hat{\Gamma}_{k}^\Lambda}										{\delta\varphi\delta\varphi}\bigg)^{\!-1}\frac{\partial R_{k}}{\partial k}\bigg]\,,
	\end{equation}
where the first space-time trace leads to an unrestricted momentum integral
\be 
\int \!\!\frac{d^{d}p}{(2\pi)^{d}}\, \bigg(R_{k}+\frac{\delta^{2}\hat{\Gamma}_{k}^\Lambda}										{\delta\varphi\delta\varphi}\bigg)^{\!-1}\!\!\!\!\!\!(p,-p)\,\,\frac{\partial R_{k}(p)}{\partial k}\,,
\ee
and we mean that the second  term, the ``remainder term'', has the momentum integral defining the trace restricted to $|p|>\Lambda$ as indicated. With the optimised IR cutoff profile we have $\partial R_k(p)/\partial k = 2k\theta(k^2-p^2)$ and thus, since $k\le\Lambda$, the remainder term vanishes in this case. At first sight this would appear then to allow us to consistently set $\hG{k}\Lambda[\ph] = \hG{k}{}[\ph]$ in \eqref{R-flow-2} (providing only that we restrict flows to $k\le\Lambda$), meaning that for these choice of cutoffs, the dependence of the effective average action on $\Lambda$ disappears. This is not correct however as can be seen by expanding the inverse kernel. Define the full inverse propagator as
\be 
\hat{\Delta}^{\!-1}(p) := R_k(p) +\frac{\delta^{2}\hat{\Gamma}_{k}^\Lambda}										{\delta\varphi(p)\delta\varphi(-p)}\bigg|_{\varphi=0}\,,
\ee
(temporarily suppressing the $k$ and $\Lambda$ dependence) and similarly define $\Gamma^{\prime}[\varphi]$ to be the remainder after the term quadratic in the fields is removed (which thus starts at ${\cal O}(\varphi^3)$ in a field expansion).
Then 
\begin{align}
\label{inverse-expanded}
\Bigg(R_{k}\,\, +& \,\, \frac{\delta^{2}\hat{\Gamma}_{k}^\Lambda}	{\delta\varphi\delta\varphi}\Bigg)^{\!-1}(p,-p)\\
 =&\, \left(\hat{\Delta}^{\!-1}+\frac{\delta^{2}\Gamma'}	{\delta\varphi\delta\varphi}\right)^{\!-1}(p,-p)\nonumber\\
=&\,\,\, \hat{\Delta}(p)- \hat{\Delta}(p)\frac{\delta^{2}\Gamma'	}{\delta\varphi(p)\delta\varphi(-p)}\hat{\Delta}(p)\nonumber\\
&+\int^\Lambda \!\!\!\!\frac{d^{d}q}{(2\pi)^{d}}\, \hat{\Delta}(p)\frac{\delta^{2}\Gamma'}{\delta\varphi(p)\delta\varphi(-p-q)}\hat{\Delta}(p+q)\frac{\delta^{2}\Gamma'}{\delta\varphi(p+q)\delta\varphi(-p)}\hat{\Delta}(p)-\cdots.\nonumber
\end{align}
The momentum $q$ is the external momentum injected by the fields remaining in $\Gamma'$:
\be 
\frac{\delta^{2}\Gamma'}{\delta\varphi(p)\delta\varphi(-p-q)} = \Gamma^{(3)}(p,-p-q,q;k,\Lambda)\varphi(-q)+{\cal O}(\varphi^2)\,,
\ee
where we have displayed as a simple example the 1PI three-point vertex defined as in \eqref{Gamma-vertices}. (The higher point vertices will have an integral over the field momenta with a delta-function restricting the sum to $-q$.) With a sharp UV cutoff in place, not only are the external momenta $|q|\le\Lambda$ restricted, but the momentum running through any internal line is also restricted, thus here we also have $|p+q|\le\Lambda$. This is because ultimately all the free propagators come (via Wick's theorem) from a Gaussian integral over the fields $\phi(r)$ in the path integral whose momenta $|r|\le\Lambda$ have been restricted by the sharp UV cutoff. Although the momentum $p$ already has a sharp UV cutoff $k$ provided by $\partial R_k(p)/\partial k$ which means the overall UV cutoff $\Lambda$ is invisible for it, this invisibility does not work for the other internal momenta, such as $p+q$, hidden in the construction of the inverse kernel. In other words even if the argument $p$ above is freed from its UV cutoff at $\Lambda$, this cutoff remains inside the construction in all the internal propagators, such as displayed in \eqref{inverse-expanded}, and thus despite appearances the first term on the right hand side of \eqref{R-flow-2} actually still does depend non-trivially on $\Lambda$, implying also that $\hat{\Gamma}_{k}^\Lambda[\varphi]$ is a non-trivial function of $\Lambda$.

\chapter{Background independence in a background dependent RG}
\label{cha:back ind}

\section{Introduction}
\label{sec:introduction}
In section \ref{sub:BI} we underlined the importance of background independence in quantum gravity and motivated going beyond the single field approximation to instead work within bi-metric truncations in which separate dependence on the background field is retained.
Recall that using bi-metric truncations requires imposing a modified split Ward identity (msWI) to ensure that background independence is recovered in the limit the cutoff $k$ is removed, $k \rightarrow 0$.
We also remarked in \ref{sub:BI} that fixed points can be forbidden by the very msWIs that are enforcing background independence, an unsettling conclusion from the research reported in \cite{Dietz:2015owa}.
In this chapter we present the research of \cite{Labus:2016lkh} in which we uncover the underlying reasons for fixed points being forbidden within the derivative expansion and polynomial truncations of conformally reduced gravity, extending the work of \cite{Dietz:2015owa}.

In the conformally truncated gravity model investigated in \cite{Dietz:2015owa}, fixed points are forbidden generically when the anomalous dimension $\eta$ is non-vanishing.
This can however be avoided by a careful choice of parametrisation $f$ (setting it to be a power of the background field $\chi$ determined by its scaling dimension \cite{Dietz:2015owa}).
On the other hand, it was shown in \cite{Dietz:2015owa} that the situation is saved in all cases, at least in the conformally reduced gravity model, by the existence of an alternative background independent description of the flow. This involves in particular a background independent notion of RG scale, $\hat{k}$. This background independent description exists at a deeper underlying level since in terms of these background-independent variables, the RG fixed points and corresponding flows always exist, and are manifestly independent of the choice of parametrisation $f(\chi)$.  

After approximating the exact RG flow equations and msWIs to second order in the derivative expansion (as will be reviewed later), the crucial technical insight was to notice that, just as in the scalar field theory model \cite{Bridle:2013sra}, the msWIs and RG flow equations can be combined into linear partial differential equations. It is the solution of the latter equations by the method of characteristics, that uncovers the background independent variables. And it is by comparing the description in these variables with the equivalent description in the original variables,
that we see that fixed points in the original variables are in general forbidden by  background independence.

However, in order to facilitate combining the RG flow equations and msWIs when the anomalous dimension $\eta\ne0$, the authors of \cite{Dietz:2015owa} were led to a particular form of cutoff profile $R_k$, namely a power-law cutoff profile.
In the research presented in this chapter we will show that in fact this cutoff profile plays a role that is much deeper than the convenience of a mathematical trick.
It in fact provides a condition that, if obeyed, means that the flow and msWI are compatible.
Recall from section \ref{sub:BI}, that compatibility is achieved if solutions to the flow equation also satisfies the msWI.
The notion of compatibility is of great importance as without it fixed points are forbidden to exist, as we will see in the ensuing sections.
As already argued in section \ref{sub:BI}, at the exact level the msWIs are guaranteed to be compatible with the exact RG flow equation, but this will typically not be the case once approximated.\footnote{Note that even though conformally reduced gravity is a truncation of the full theory in which we only focus on one particular mode of the metric (the conformal mode), approximation in the sense that we mean it here involves an expansion, terminated at some order, like the other approximation schemes outlined in section \ref{sec:approxs}.}
We will see that in the $O(\partial^2)$ derivative expansion approximation derived in reference \cite{Dietz:2015owa}, the msWI and flow equations are in fact compatible {if and only if} either the cutoff profile is power law\footnote{Power law cutoff profiles have nice properties in that they ensure that the derivative expansion approximation preserves the quantisation of the anomalous dimension in non-gravitational systems, e.g.\ scalar field theory \cite{Morris:1994ie,Morris:1994jc,Morris:1998}. (Although as with the optimised cutoff \cite{opt1,opt3}, they do not allow a derivative expansion to all orders \cite{Morris:2005ck,Morris1999,Morris2001}.) Nevertheless, given the unsettling nature of the conclusions in reference \cite{Dietz:2015owa}, it is important to understand  to what extent the results depend on cutoff profile.}, or we have the special case that $\eta=0$\footnote{In fact it is natural to expect $\eta$ to be \textit{non-vanishing} at the LPA level for conformally truncated gravity, as explained in reference \cite{Dietz:2016gzg}.}.

The structure of this chapter is then as follows.
In section \ref{sec:review} we quickly review the results we need from \cite{Dietz:2015owa} and their context.
In section \ref{sec:1compatmain} we provide a proof of compatibility at the exact level and investigate compatibility in the $O(\partial^2)$ derivative expansion along with deriving the requirements needed in order to achieve it.
If the msWIs are not compatible with the flow equations within the derivative expansion, it does not immediately follow that there are no simultaneous solutions to the system of equations. However, as we argue in section \ref{sec:1compatmain}, if the msWIs are not compatible, the equations are overconstrained and it is for this reason that it is hopeless to expect any solutions.
We verify this using the LPA in section \ref{sec:LPA} (see also \ref{sec:truncations}).
We also see in section \ref{sec:LPA} that when the LPA equations are compatible they can indeed be combined to give a background independent description of the flow; however, even when compatibility is achieved with power-law cutoff, we understand why the msWI can still forbid fixed points for general parameterisation $f$ and $\eta\ne0$.
Finally, in section \ref{sec:truncations}, we consider how these issues become visible in polynomial truncations and without resorting to the trick of combing the flow and msWI equations. It is instructive to do so since it seems likely that this is the only way we could investigate these issues using the exact non-perturbative flow equations. We verify very straightforwardly that generically there can be no fixed points as the equations over-constrain the solutions if the truncation is taken to a sufficiently high order.

\section{Conformally reduced gravity at order derivative-squared\!}
\label{sec:review}

In this section we give a quick review of the results we need and their context from reference \cite{Dietz:2015owa}.
Recall that we arrive at conformally reduced gravity (in Euclidean signature) by writing:
\begin{equation}
\label{conformal-reduction}
	\tilde g_{\mu\nu} = f(\tilde\phi) \hat g_{\mu\nu}= f(\chi +\tilde\varphi )\hat g_{\mu\nu} \qquad\text{and} \qquad \bar g_{\mu\nu} = f(\chi)\hat g_{\mu\nu} \,.
\end{equation}
The metric $\tilde g_{\mu\nu}$ is restricted to an overall conformal factor $f(\tilde\phi)$ times a reference metric which in fact we set to flat: $\hat g_{\mu\nu}=\delta_{\mu\nu}$.
Following the background field method, we split the total conformal factor field $\tilde\phi(x)$ into a background conformal factor field $\chi(x)$ and fluctuation conformal factor field $\tilde\vp(x)$. It is then the latter that is integrated over. Similarly, we parametrise the background metric $\bar{g}_{\mu\nu}$ in terms of the background conformal factor field $\chi$.

Examples of parametrisations used previously in the literature include  
$f(\phi) = \exp(2\phi)$ \cite{Machado:2009ph} and $f(\phi)=\phi^2$ \cite{Manrique:2009uh,Bonanno:2012dg}. However we leave the choice of parametrisation unspecified. It is important to note however that $f$ cannot depend on $k$ since it is introduced at the bare level and has no relation to the infrared cutoff (moreover if $f$ depended on $k$, the flow equation \eqref{equ:flowGamma} would no longer hold).  Later we will change to dimensionless variables using $k$ and in these variables it can be forced to depend on $k$ (see sections \ref{sec:required-cutoff} and \ref{sec:forbids}).

By keeping only the conformal factor of the metric, diffeomorphism invariance is destroyed. Therefore gauge fixing and ghosts are not required in this setup.
A remnant diffeomorphism is preserved however, a multiplicative rescaling, which constrains appearances of the background field.

Introducing the classical fluctuation field $\vp = \langle \tilde \vp \rangle$ and total classical field $\phi = \langle \tilde \phi \rangle = \chi + \vp$, the effective action  satisfies the flow equation
\be
\label{equ:flowGamma}
\partial_t \Gamma_k[\vp,\chi] = \frac{1}{2}\mathrm{Tr}\left[\frac{1}{\sqrt{\bar g}\sqrt{\bar g}}\frac{\delta^2\Gamma_k}
				  {\delta \vp \delta \vp}+ R_k[\chi]\right]^{-1} \partial_t R_k[\chi]\,.
\ee
Note that here we are using the same notation for the effective action as in chapter \ref{cha:introduction}. We have also introduced RG time 
\be
\label{time}
t=\ln(k/\mu)\,,
\ee 
with $\mu$ being a fixed reference scale, which can be thought of as being the usual arbitrary finite physical mass-scale.
Recall from section \ref{sub:EAA grav} that in the context of the background field method the cutoff operator $R_k$ itself depends on the background field $\chi$ as it becomes a function of the covariant Laplacian of the background metric $R_k\left(-\bar \nabla^2\right)$.
We see that again, now in the context of conformally reduced gravity, that the effective action possesses a separate dependence on the background field, enforced through the cutoff.

By specialising to a background metric ${\bar g}_{\mu\nu}$ that is slowly varying, so that space-time derivatives of it can be neglected, we
effectively terminate at the level of the LPA for the background conformal factor $\chi$. For the classical fluctuating conformal factor $\vp$ however,  $\mathcal{O}(\partial^2)$ in the derivative expansion approximation is fully implemented, making no other approximation.
The effective action in its most general form at this level of truncation is thus given by
\begin{equation}
\label{trunc}
	\Gamma_k[\varphi, \chi] = \int d^dx \sqrt{\bar g} \left( -\frac{1}{2}K(\varphi,\chi)
	\bar g^{\mu\nu}\partial_{\mu}\varphi\partial_{\nu}\varphi + V(\varphi,\chi)  \right)\,.
\end{equation}

The msWI originates from the observation that the introduction of the cutoff action into the functional integral violates split symmetry:
\be
\label{equ:split-symmetry}
\tilde \vp(x) \mapsto \tilde \vp(x) + \eps(x) \qquad \text{and}\qquad \chi(x) \mapsto \chi(x) -\eps(x)\,,
\ee
under which the bare action is invariant.\footnote{The source term also breaks the symmetry but does not contribute to the separate background field dependence in $\Gamma_k[\vp,\chi]$.}
It is the breaking of this symmetry that signals background independence has been lost, both at the level of the functional integral and at the level of the effective action.
The msWI encodes the extent to which the effective action violates split symmetry:\footnote{When all metric degrees of freedom are considered in full gravity, the msWI receives extra contributions to its right-hand side originating from gauge fixing and ghosts.}
\be
\label{equ:sWiGamma}
\frac{1}{\sqrt{\bar g}}\left(\frac{\delta\Gamma_k}{\delta \chi}-\frac{\delta \Gamma_k}{\delta \vp}\right)\!
      =\frac{1}{2}\mathrm{Tr}\left[\frac{1}{\sqrt{\bar g}\sqrt{\bar g}}\frac{\delta^2\Gamma_k}
				  {\delta \vp \delta \vp}+ R_k[\chi]\right]^{\!-1}\! \frac{1}{\sqrt{\bar g}}
				  \left\{\frac{\delta R_k[\chi] }{\delta \chi}+\frac{d}{2}\dclnf R_k[\chi]\right\}\,.
\ee
Exact background independence would be realised if the right-hand side of the msWI was zero, implying that the effective action is only a functional of the total field $\phi = \chi + \vp$. The presence of the cutoff operator however causes the right-hand side to be non-vanishing in general. It is only in the limit $k\rightarrow0$ (again whilst holding unscaled momenta and fields fixed) that the cutoff operator drops out and background independence can be restored exactly. We now see how imposing the msWI in addition to the flow equation \eqref{equ:flowGamma} automatically ensures exact background independence in the limit $k\rightarrow0$. The observation we further explore in the work presented in this chapter is that restricting flows to satisfy \eqref{equ:sWiGamma} then has consequences for RG properties, in particular fixed point behaviour, that follow from \eqref{equ:flowGamma}.

Computing the flow equation and msWI in the derivative expansion \eqref{trunc} results in flow equations and modified split Ward identities\footnote{Although we always mean these modified identities, we will sometimes refer to them  simply as Ward identities.},  for the potential $V$:
\begin{align}
	\label{flowV}
	\partial_t V(\varphi,\chi) &= f(\chi)^{-\frac{d}{2}}\int dp\, p^{d-1} Q_p\dot R_p \,,\\
	\label{msWIV}
	\partial_\chi V - \partial_\varphi V +\frac{d}{2}\partial_\chi \text{ln} f V 
	&= f(\chi)^{-\frac{d}{2}}\int dp\, p^{d-1} Q_p\left(\partial_\chi R_p + \frac{d}{2}\partial_\chi \text{ln} f  R_p\right),
\end{align}
and for $K$:
\begin{align}
	\label{flowK}
	f^{-1}\partial_t K(\varphi,\chi) &= 2 f^{-\frac{d}{2}}\int dp\,p^{d-1} P_p(\varphi,\chi)\dot R_p \,,\\
	\label{msWIK}
	 f^{-1}\!\left(\partial_\chi K- \partial_\varphi K + \frac{d-2}{2}\partial_\chi\text{ln}fK\right)
	 &\!= 2 f^{-\frac{d}{2}}\!\!\int dp\,p^{d-1} P_p(\varphi,\chi) \left(\partial_{\chi}R_p 
	 + \frac{d}{2}\partial_\chi \text{ln} f   R_p\right).
\end{align}
The $p$ subscripts denote the momentum dependence of $Q_p, P_p$ and the cutoff $R_p$ and as usual RG time derivatives are denoted also by a dot on top. $Q_p$ is defined as
\begin{equation}
	\label{Q}
	Q_p=\left(\partial^2_\varphi V - p^2\frac{K}{f} + R_p \right)^{-1}.
\end{equation}
and $P_p$ is given by
\begin{align}
	P_p=&-\frac{1}{2}\frac{\partial_\varphi K}{f}Q_p^2
	+\frac{\partial_\varphi K}{f}\left(2\partial_\varphi^3 V 
	- \frac{2d+1}{d}\frac{\partial_{\varphi}K}{f}p^2\right)Q_p^3\nonumber\\
	&-\left[\left\{\frac{4+d}{d}\frac{\partial_\varphi K}{f}p^2 - \partial^3_\varphi V\right\}
	\left(\partial_{p^2}R_p - \frac{K}{f}\right) 
	+\frac{2}{d}p^2\partial^2_{p^2} R_p\left(\frac{\partial_\varphi K}{f} - \partial_\varphi^3V\right)\right]\nonumber\\
	&\quad\times\left(\partial_\varphi^3 V - \frac{\partial_\varphi K}{f}p^2\right)Q_p^4\nonumber\\
	&-\frac{4}{d}p^2\left(\partial_{p^2}R_p-\frac{K}{f}\right)^2
	\left(\partial_\varphi^3 V - \frac{\partial_\varphi K}{f} p^2\right)^2 Q_p^5 \,.
\end{align}

\section{Compatibility of the msWI with the flow equation}\label{sec:1compatmain}

Compatibility of the msWI with the flow equation can be phrased in the following way. Write the msWI in the form $\mathcal{W}=0$ and assume that this holds at some scale $k$. Computing $\dot{\mathcal{W}}$ by using the flow equation, we say that the msWI is compatible if $\dot{\mathcal{W}}=0$ then follows at scale $k$ without further constraints. 

In the first part of this section we rederive the flow equation and msWI for conformally reduced gravity but organised in a different way from reference \cite{Dietz:2015owa} so as to make the next derivation more transparent. We then prove that they are compatible with one another. As previously emphasized, this is naturally to be expected since both are derived from the same partition function. For completeness we include it here in order to  fully understand the issues once we consider derivative expansions. (For a proof of the exact case in a more general context see reference \cite{Safari:2015dva}.)
In the second part we study the notion of compatibility for conformally reduced gravity in the truncation \eqref{trunc}. Asking for compatibility in the derivative expansion is actually non-trivial. We derive the requirements necessary to achieve it.

\subsection{Compatibility at the exact level}\label{sec:exact}

The proof of compatibility of the un-truncated system consists of demonstrating that the RG time derivative of the msWI is proportional to the msWI itself \cite{Litim1998,Litim1999}. In analogy with references \cite{Litim1998,Litim1999}, we expect to find that this RG time derivative is, more specifically, proportional to a second functional derivative with respect to $\varphi$ acting on the msWI and it is with this in mind that we proceed (see also reference\cite{Safari:2015dva}).

We begin by considering the following Euclidean functional integral over the fluctuation field $\tilde\varphi$
\begin{equation}
	\label{Z}
	\text{e}^{W_k}=\int\!\mathcal{D}\tilde\varphi \, \text{exp}\left(-S[\chi+\tilde\varphi]-\Delta S_k[\tilde\varphi,\bar g]
	+ S_{\text{src}}[\tilde\varphi,\bar g]\right).
\end{equation}
This integral is regulated in the UV (as it must be), however we leave this regularisation implicit in what follows. Compatibility can be shown most easily by presenting both the flow equation and the msWI as matrix expressions. Thus we begin by rewriting the source term using matrix notation, introduced in chapter \ref{cha:introduction}:
\begin{equation}
	S_{\rm src}[\tilde\varphi,\bar g]=\int\!d^d x\sqrt{\bar g(x)}\,\tilde\varphi(x) J(x)\equiv 
	\tilde\varphi_xT_{xy}J_y\equiv \tilde\varphi\cdot T\cdot J \,,
\end{equation}
where $T_{xy}\equiv T(x,y)\equiv\sqrt{\bar g(x)}\delta(x-y)$.  Similarly, we write the cutoff action as
\begin{equation}
\label{cutoff-action}
	\Delta S_k[\tilde\varphi, \bar g]=\frac{1}{2}\int\!d^d x\sqrt{\bar g(x)}\,\tilde\varphi(x)R_k[\bar g]\tilde\varphi(x)
	\equiv\frac{1}{2}\tilde\varphi_x r_{xy} \tilde\varphi_{y}\equiv\frac{1}{2}\,\tilde\varphi\cdot r \cdot\tilde\varphi \,,
\end{equation}
where 
\be 
\label{odd-r}
r_{xy}\equiv r(x,y)\equiv\sqrt{\bar g(x)}\sqrt{\bar g(y)}R_{k}(x,y)\,,
\ee 
and where the cutoff operator and its kernel are related according to
\begin{equation}
	R_k(x,y) = R_{k,x}\frac{\delta(x-y)}{\sqrt{\bar g(y)}} \,.
\end{equation}
We refrain from putting a $k$ subscript on $r_{xy}$ to avoid clutter with indices, but note that it still has $k$-dependence. Also note that now the factors of $\sqrt{\bar g}$ are no longer part of the integration; this is to enable all $\chi$-dependent quantities to be easily accounted for when acting with $\delta/\delta\chi$ later on.

With these definitions in place, the RG time derivative of \eqref{Z} gives
\begin{equation}
	\label{W_flow}
	\dot W_k=-\frac{1}{2}\dot r_{xy} \left<\tilde\varphi_x \tilde\varphi_y \right>.
\end{equation}
In the usual way, we take the Legendre transform of $W_k$:
\begin{equation}
	\label{LTR}
	\tilde\Gamma_{k}=J\cdot T \cdot \varphi - W_k \quad \text{with} \quad T\cdot\varphi=\frac{\delta W_k}{\delta J}
\end{equation}
and from this we define the effective average action
\begin{equation}
	\label{EAA}
 	\Gamma_k[\varphi,\bar g]=\tilde\Gamma_k[\varphi,\bar g]-\Delta S_k[\varphi,\bar g] \,.
\end{equation}
From \eqref{LTR}, it also follows that
\begin{equation}
	\label{2pt}
	\langle\tilde\varphi_x \tilde\varphi_y \rangle =
	\left(\frac{\delta^{2}\tilde\Gamma_k}{\delta\varphi_x \delta\varphi_y}\right)^{\!-1}+\,\varphi_x \varphi_y \,.
\end{equation}
Finally substituting \eqref{LTR} and \eqref{2pt} into \eqref{W_flow}, together with \eqref{EAA}, we obtain the flow equation for the effective average action
\begin{equation}
	\label{flow1}
	\dot\Gamma_k=\frac{1}{2}\text{Tr}\left[\left(\frac{\delta^{2}\Gamma_k}{\delta\varphi \delta\varphi}
	+r\right)^{\!-1}\dot r\right]
	\equiv \frac{1}{2}\text{Tr}\,\Delta\, \dot r \,,
\end{equation}
where
\begin{equation}
\label{new-inverse}
	\Delta_{xy}\equiv\left(\frac{\delta^{2}\Gamma_k}{\delta\varphi_x \delta\varphi_y}+r_{xy}\right)^{-1} \,.
\end{equation}

The msWI is derived by applying the split symmetry transformations \eqref{equ:split-symmetry}, with infinitesimal $\varepsilon(x)$, to the functional integral \eqref{Z}.
Applying these shifts we obtain
\begin{equation}
	\label{Wshift}
	-\frac{\delta W_k}{\delta\chi}\cdot\varepsilon=\left< \varepsilon\cdot T\cdot J-
	\tilde\varphi\cdot\left(\frac{\delta T}{\delta\chi}\cdot\varepsilon\right)\cdot J 
	- \varepsilon\cdot r\cdot\tilde\varphi
	+\frac{1}{2}\tilde\varphi\cdot\left(\frac{\delta r}{\delta\chi}\cdot\varepsilon\right)\cdot\tilde\varphi\right>.
\end{equation}
Under these same shifts, the Legendre transformation \eqref{LTR} gives
\begin{equation}
	\frac{\delta W_k}{\delta\chi}\cdot\varepsilon
	= J\cdot\left(\frac{\delta T}{\delta\chi}\cdot\varepsilon\right)\cdot\varphi
	- \frac{\delta\tilde\Gamma_k}{\delta\chi}\cdot\varepsilon \,.
\end{equation} 
Substituting the above relation into \eqref{Wshift} together with \eqref{EAA}, we obtain the msWI:
\begin{equation}
	\label{msWIuntrunc}
	\frac{\delta \Gamma_k}{\delta\chi_\omega}-\frac{\delta \Gamma_k}{\delta\varphi_\omega}=
	\frac{1}{2}\,\Delta_{xy}\,\frac{\delta r_{yx}}{\delta\chi_\omega} \,,
\end{equation}
where we have used the fact that the identity must hold for arbitrary $\varepsilon(\omega)$. Note that in deriving \eqref{msWIuntrunc} the contribution of the source term to the separate background field dependence of $\Gamma_k[\varphi,\chi]$ drops out.

The equations just derived, \eqref{flow1} and \eqref{msWIuntrunc}, appear at first sight to be in conflict with \eqref{equ:flowGamma} and \eqref{equ:sWiGamma} respectively.
In particular  factors of $\sqrt{\bar g}$ are apparently missing. This is because the $\sqrt{\bar g}$ factors are absorbed in a different definition of the inverse kernel. Indeed the inverse kernel \eqref{new-inverse} now satisfies
\be 
\left(\frac{\delta^{2}\Gamma_k}{\delta\varphi_x \delta\varphi_y}+r_{xy}\right) \Delta_{yz} =\delta_{xz}
\ee
without a $\sqrt{\bar g}(y)$  included in the integration over $y$.


Now that we have derived the flow equation and msWI written in a convenient notation, we are ready to prove that they are compatible. We begin by defining
\begin{equation}
\label{cal-W}
	\mathcal{W}_\omega\equiv\frac{\delta \Gamma_k}{\delta\chi_\omega}-\frac{\delta \Gamma_k}{\delta\varphi_\omega}
	-\frac{1}{2}\Delta_{xy}\frac{\delta r_{yx}}{\delta\chi_\omega} =0 \,.
\end{equation}
Taking the RG time derivative of $\mathcal{W_\omega}$ then gives
\begin{equation}
	\label{WIdot}
	\mathcal{\dot W}_\omega= \frac{\delta \dot\Gamma_k}{\delta\chi_\omega}
	-\frac{\delta \dot\Gamma_k}{\delta\varphi_\omega}+
	\frac{1}{2}\left[\Delta\left(\frac{\delta^{2}\dot{\Gamma}_k}{\delta\varphi \delta\varphi}
	+\dot r\right)\Delta\right]_{\!xy}\frac{\delta r_{yx}}{\delta\chi_\omega}
	-\frac{1}{2}\Delta_{xy}\frac{\delta \dot r_{yx}}{\delta\chi_\omega}
\end{equation}
and upon substituting the flow equation \eqref{flow1} into the right-hand side, we have
\begin{align}
\label{WIdot-2}
	\mathcal{\dot W}_{\omega}=&-\frac{1}{2}\Delta_{xz}
	\frac{\delta^{3}\Gamma_k}{\delta\varphi_{z}\delta\varphi_{z'}\delta\chi_{\omega}}\Delta_{z'y}\dot r_{yx}
	 + \frac{1}{2}\Delta_{xz}\frac{\delta^{3}\Gamma_k}{\delta\varphi_{z}
	 \varphi_{z'}\varphi_\omega}\Delta_{z'y}\dot{r}_{yx}
	 \nonumber\\ &\quad
	 +\frac{1}{4}\Delta_{xz}\left(\frac{\delta^{2}}{\delta\varphi_{z}\delta\varphi_{z'}}\Delta_{uu'}\right)
	 \dot r_{u'u}\Delta_{z'y}\frac{\delta r_{yx}}{\delta\chi_{\omega}}\nonumber\\
	=&-\frac{1}{2}(\Delta \dot{r} \Delta)_{zz'}\frac{\delta^{2}}{\delta\varphi_{z'}\delta\varphi_{z}}
	 \left(\frac{\delta\Gamma}{\delta\chi_\omega}-\frac{\delta\Gamma}{\delta\varphi_\omega}\right)
	  +\frac{1}{4}\left(\frac{\delta^{2}}{\delta\varphi_{z}\delta\varphi_{z'}}\Delta_{uu'}\right)
	 \dot r_{u'u}\,\Delta_{z'y}\frac{\delta r_{yx}}{\delta\chi_{\omega}}\Delta_{xz} \,.
\end{align}
The first term in the last equality is in the form we want: a differential operator acting on (part of) $\mathcal{W_\omega}$. We now expand out the second term with the aim of also putting it into the desired form. For the sake of neatness let us define
\begin{equation}
		\Gamma_{x_1...x_n}\equiv \frac{\delta^n\Gamma_k}{\delta\varphi_{x_1}...\delta\varphi_{x_n}} \,.
\end{equation}
Expanding out the second term then gives
\begin{align}
\label{step1}
\!\!\left(\! \frac{\delta^2}{\delta\varphi_z\delta\varphi_{z'}}\Delta_{uu'}\!\right)\! 
	\dot r_{u'u}\,\Delta_{z'y}\frac{\delta r_{yx}}{\delta\chi_\omega}\Delta_{xz}
	\! =\!\Delta_{xz}& \bigg(\! \Delta_{uv}\Gamma_{zvs}\Delta_{sv'}\Gamma_{z'v's'}\Delta_{s'u'}
	\! +\! \Delta_{uv'}\Gamma_{v's'z'}\Delta_{s'v}\Gamma_{zvs}\Delta_{su'}\nonumber\\
	&\qquad-\Delta_{uv'}\Gamma_{v's'zz'}\Delta_{s'u'}\bigg)
	\dot r_{u'u}\Delta_{z'y}\frac{\delta r_{yx}}{\delta\chi_\omega} \,.
\end{align}
Upon exchanging factors of $\Delta$ and relabelling indices, we find
\begin{equation}
\label{step2}
	\left(\frac{\delta^2}{\delta\varphi_{z}\delta\varphi_{z'}}\Delta_{uu'}\right)
	\dot r_{u'u}\left(\Delta_{z'y}\frac{\delta r_{yx}}{\delta\chi_{\omega}}\Delta_{xz}\right)=
	(\Delta\dot r\Delta)_{s'v'}\frac{\delta^2}{\delta\varphi_{v'}\delta\varphi_{s'}} 
	\Delta_{xy}\frac{\delta r_{yx}}{\delta\chi_{\omega}} \,,
\end{equation}
which now has the structure we require. Thus we have shown that the RG time derivative of the msWI can be written as
\begin{equation}
	\mathcal{\dot W}_{\omega}=-\frac{1}{2}\text{Tr}
	\left(\Delta \dot r\Delta\frac{\delta^2}{\delta\varphi\delta\varphi}\right)\mathcal{W}_{\omega} \,,
\end{equation}
i.e.\ that it is proportional to the msWI itself. If $\Gamma_{k}$ satisfies $\mathcal{W_\omega}$ at some initial scale $k_0$, and satisfies the flow equation there, it thus follows without further restriction that $\mathcal{\dot W_\omega}|_{k_{0}}=0$ since it is proportional to $\mathcal{W_\omega}$. Thus the msWI is compatible with the flow equation. 
If $\Gamma_{k}$ continues to evolve according to the flow equation, 
it then follows that $\mathcal{W_\omega}$ and thus $\mathcal{\dot W_\omega}$ will be zero for all $k$.

\subsection{Compatibility versus derivative expansion}\label{sec:exact-vs-derivatives}
\begin{figure}[ht]
\centering
\includegraphics[scale=0.23]{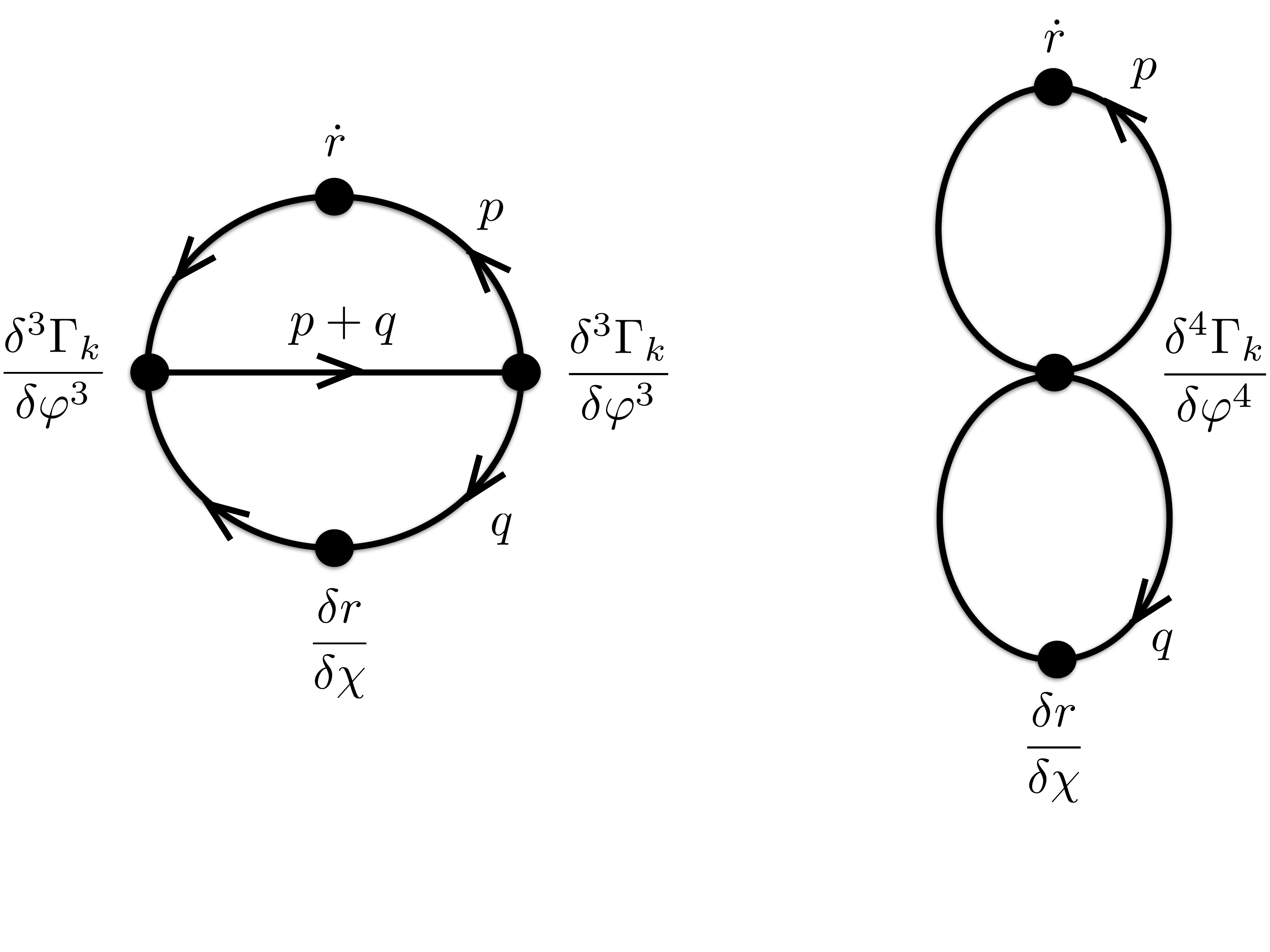}
\vskip-30pt
\caption{The two-loop diagrams in \eqref{step1}. Their symmetry immediately implies the identity \eqref{step2}. Momentum flow is indicated in the case where the fluctuation field $\varphi$ is then set to zero.}
\label{fig:two-loop}
\end{figure}
Recalling from \eqref{new-inverse} that $\Delta$ is an infrared regulated full propagator, we see from \eqref{step1} that the identity \eqref{step2} can be understood diagrammatically in terms of two-loop diagrams as sketched in figure \ref{fig:two-loop}. 
The symmetry of these diagrams means that nothing changes if we exchange $\dot{r}\leftrightarrow\delta r/\delta\chi$. This exchange immediately leads to the identity \eqref{step2}.

This identity breaks down in general in the derivative expansion. If the Ward identity  is approximated by a derivative expansion, the full propagator in the one-loop term in \eqref{cal-W} is also expanded in a derivative expansion. This full propagator
 has loop momentum $q$ say, and is then expanded in powers of momenta carried by the external fluctuation field $\varphi(p)$, i.e.\ by the external legs.
 The RG time derivative of the Ward identity yields the RG time derivative of such vertices, as can be seen from the $\delta^2\dot{\Gamma}_k/\delta\varphi^2$ term  in \eqref{WIdot}. This latter term has two internal legs given by the explicit functional derivatives, carrying the loop momentum $q$ and joining full internal propagators $\Delta$, and any number of external legs contained in the vertices of $\dot{\Gamma}_k$. Substituting the flow equation \eqref{flow1} then gives in particular the last term in eqn. \eqref{WIdot-2} in which two of these external legs are now joined to form a loop connected via $\dot{r}$. However it is momenta external to \emph{this new loop} which are Taylor expanded in the derivative expansion of the flow equation (see also \cite{Morris1999,Morris2001}). This is illustrated in the diagram displayed in fig. \ref{fig:two-loop}. In particular when the remaining external fluctuation field dependence is removed by setting $\varphi=0$, we have exactly the momentum dependence displayed in the figure. We see that a derivative expansion of the Ward identity involves Taylor expanding 
 in small $p$, while integrating over $q$. However a derivative expansion of the flow equation involves Taylor expanding 
 in small $q$, and integrating over $p$ instead. Thus the symmetry between the two loops is broken and the identity \eqref{step2} no longer follows.
 
 On the other hand we see that if $\dot{r}$ and $\delta r/\delta\chi$ have the same momentum dependence then the identity \eqref{step2} is restored because it is no longer possible to distinguish the two loops. Returning the placement of $\sqrt{\bar{g}}$ from \eqref{odd-r} to the integration measure, this in fact would give us the relation \eqref{same-p-dependence} that is necessary and sufficient for compatibility of the Ward identities within the derivative expansion, and which we will now derive directly within the derivative expansion.

\subsection{Compatibility at order derivative-squared}\label{sec:compatibility-at-d2}

We now proceed to compute the flow of the msWI for the system truncated at $\mathcal{O}(\partial^2)$ as described in section \ref{sec:review},
and investigate directly under which circumstances it vanishes. Let us start by writing the flow equations and msWIs for both $V$ and $K$ in the following form so that we can study both cases simultaneously:
\begin{align}
	\label{flowA}
	\dot A(\varphi,\chi) &= \int_p B_p\dot R_p \,,\\
	\label{msWIA}
	\mathcal{W}^{(A)}&=\bar\partial A - \gamma A + \int_p B_p(\partial_\chi R_p + \gamma R_p)=0 \,,
\end{align}
where $A$ is either $V$ or $K/f$ such that $B_p$ is either $Q_p$ or $2 P_p$ respectively. Here we have also introduced the shorthand notation 
\be \label{gamma}
\int_p \equiv f(\chi)^{-\frac{d}{2}}\int dp \,p^{d-1}\,,\qquad \gamma \equiv \frac{d}{2}\partial_\chi \text{ln}f\,,\qquad{\rm and}\qquad \bar\partial \equiv \partial_\varphi - \partial_\chi\,.
\ee 
It will also be useful to have to hand the following relations:
\begin{align}
	\label{DV}
	\left( \bar \partial + \partial_t - \gamma \right) V &=
	\mathcal W^{(V)} + \int_q Q_q (\dot R_q - \partial_{\chi} R_q - \gamma R_q)\,,\\
	\label{DK}
	\left( \bar \partial + \partial_t - \gamma \right) \frac{K}{f} &=
	\mathcal W^{(K)} + 2 \int_q P_q (\dot R_q - \partial_{\chi} R_q - \gamma R_q)\,,\\
	\label{DQ}
	\left( \bar \partial + \partial_t + n \gamma \right) Q_p^n &=
	-n \, \, Q_p^{n+1} \int_q (\partial_{\varphi}^2 Q_q - 2 \, p^2 P_q)(\dot R_q - \partial_{\chi} R_q
	 - \gamma R_q)\nonumber\\ 
	&\qquad-n \, Q_p^{n+1} (\dot R_p - \partial_{\chi} R_p - \gamma R_p)
	-n \, Q_p^{n+1} (\partial_{\varphi}^2 \mathcal W^{(V)} - p^2 \mathcal W^{(K)}).
\end{align}
The first two relations are derived by subtracting the msWI from the flow equation for $V$ and $K/f$ respectively. The last relation is then derived by using the first two relations above together with the definition of $Q_p$ given in \eqref{Q}.

We begin by taking the RG time derivative of \eqref{msWIA}. Substituting in the flow equation for $\dot A$, and remembering the power of $f(\chi)$ hidden in the integral over $p$, this gives
\begin{equation}
	\label{flowWA1}
	\dot{ \mathcal W } ^ {(A)} =
 	\int_p \dot R_p \left( \bar\partial + \partial_t + \gamma \right) B_p
 	- \int_p \dot B_p \left( \dot R_p - \partial_{\chi} R_p - \gamma R_p \right) .
\end{equation}
In order to proceed we have to assume a particular form of $B_p$ so that we can compute the result of the linear operators under the integral acting on it. A general term in $P_p$ takes the form
\begin{equation}
	\tilde B_p =
 	\left( \partial_{\varphi}^i V \right) ^ a
 	\left( \partial_{\varphi}^j \frac{K}{f} \right) ^ b
 	\left( \partial_{p^2}^k R_p \right) ^ c \left(p^2\right)^l Q_p^e \,,
\end{equation}
where $a,b,c,e,i,j,$ $k$ (not to be confused with the cutoff scale), and $l$ are non-negative integers. From the structure of the terms in $P_p$ one can read off the following sum rule for the exponents:
\begin{equation}
	a+b+c = e-1 \,.
\end{equation}
Notice that the case $B_p = Q_p$  for the potential is also included, since  $a=b=c=l=0$ and $e=1$ also satisfies the sum rule.
Taking the term under the first integral of \eqref{flowWA1}, we find
\begin{align}
\nonumber
	\left( \bar\partial + \partial_t + \gamma \right) \tilde B_p &=
	\bigg[a \left( \partial_{\varphi}^i V \right) ^ {-1} \partial_{\varphi}^i \left(  \bar\partial + \partial_t \right) V
	+ b \left( \partial_{\varphi}^j \frac{K}{f} \right) ^ {-1} \partial_{\varphi}^j
	 \left(  \bar\partial + \partial_t \right) \frac{K}{f} \\
	&\quad+ c \left( \partial_{p^2}^k R_p \right) ^ {-1} \partial_{p^2}^k \left( - \partial_{\chi} + \partial_t \right) R_p
	+ e \, Q_p^{-1} \left( \bar\partial + \partial_t \right) Q_p+ \gamma\bigg ] \tilde B_p \,.
\end{align}
Substituting equations \eqref{DV}--\eqref{DQ} into the above expression and using the sum rule, we obtain
\begin{align}
\label{DB}
	\nonumber
	\left( \bar\partial + \partial_t + \gamma \right)\! \tilde B_p &\!=\!\!
	\bigg[a \left( \partial_{\varphi}^i V \right) ^ {-1} \partial_{\varphi}^i \left(  \mathcal W^{(V)}
	 + \int_q Q_q \bar R_q \right)\\ \nonumber
	&\!\quad+ b \left( \partial_{\varphi}^j \frac{K}{f} \right) ^ {-1} \partial_{\varphi}^j \left(  \mathcal W^{(K)}
	+ 2 \int_q P_q \bar R_q \right)
	+ c \left( \partial_{p^2}^k R_p \right) ^ {-1} \partial_{p^2}^k \bar R_p \\
	&\!\quad- e\, Q_p\! \int_q\!\! \left( \partial_{\varphi}^2 Q_q - 2 p^2 P_q \right) \!\bar R_q
	- e\, Q_p \bar R_p
	- e\, Q_p \left( \partial_{\varphi}^2 \mathcal W^{(V)}\! - p^2 \mathcal W^{(K)} \right)
	\! \bar R_q\bigg] \tilde B_p
\end{align}
where we have introduced the shorthand notation 
\be 
\bar R_p = \dot R_p - \partial_\chi R_p - \gamma R_p\,.
\ee 
Turning our attention now to the second integral of \eqref{flowWA1} we take the RG time derivative of $\tilde B_p$ and again substitute in the flow equations for $V$ and $K/f$. This gives
\begin{align}
\label{Bdot}
\nonumber
	\dot{ \tilde B }_p &=
	\bigg[	a \left( \partial_{\varphi}^i V \right) ^ {-1} \partial_{\varphi}^i \int_q Q_q \dot R_q
	+ b \left( \partial_{\varphi}^j \frac{K}{f} \right) ^ {-1} \partial_{\varphi}^j \int_q 2 P_q \dot R_q \\
	&\quad+ c \left( \partial_{p^2}^k R_p \right) ^ {-1} \partial_{p^2}^k \dot R_p
	- e \, Q_p \int_q  \left( \partial_{\varphi}^2 Q_q - 2 p^2 P_q \right) \dot R_q
	- e \, Q_p  \dot R_q\bigg] \tilde B_p  \,.
\end{align}
Inserting \eqref{DB} and \eqref{Bdot} into \eqref{flowWA1} we obtain
\begin{align}
\nonumber
 	\mathcal{\dot W}^{(A)}\! &=\! \sum_{\tilde B_p} \Bigg\{a \int_{p,q} 
 	\tilde B_p ( \partial_{\varphi}^i V ) ^ {-1} \partial_{\varphi}^i
	\bigg( \dot R_p \mathcal{W}^{(V)} +Q_q [ \dot R , \partial_{\chi} R + \gamma R ]_{qp} \bigg) \\ \label{flowWA}
	&\quad+ b \int_{p,q} \tilde B_p \left( \partial_{\varphi}^j \frac{K}{f} \right) ^ {-1} \partial_{\varphi}^j
	\bigg( \dot R_p \mathcal{W}^{(K)} + 2 P_q [ \dot R , \partial_{\chi} R + \gamma R ]_{qp} \bigg) \\\nonumber
	&\quad+ c \int_p \tilde B_p \left( \partial_{p^2}^k R_p \right) ^ {-1}
	\left( \left(\partial_\chi R_p + \gamma R_p\right) \partial_{p^2}^k \dot R_p - \dot R_p \partial_{p^2}^k \left(\partial_\chi R_p + \gamma R_p\right) \right) \\ \nonumber
	&\quad-\! e \!\int_p \!\tilde B_p  Q_p  \dot R_p \bigg( \partial^2_{\varphi} \mathcal{W}^{(V)}\!\! - p^2  \mathcal{W}^{(K)}\! \bigg)\!- e\!\! \int_{p,q}\!\!\tilde B_p Q_p \bigg(  \partial_{\varphi}^2 Q_q\!\! - 2 p^2 P_q\! \bigg) [ \dot R , \partial_{\chi} R + \gamma R ]_{qp}\!\Bigg\}
\end{align}
where we have introduced the commutator-like construct $[A,B]_{qp} = A_q B_p - B_qA_p$.

When $A = V$ the above expression simplifies considerably to
\begin{equation}
\label{flowWV}
	 \mathcal{\dot W}^{(V)}=-  \int_p  Q^2_p  \dot R_p\, \bigg( \partial^2_{\varphi} \mathcal{W}^{(V)}
	  - p^2  \mathcal{W}^{(K)} \bigg) - \int_{p,q}   Q^2_p\, \bigg(  \partial_{\varphi}^2 Q_q
	  - 2 p^2 P_q \bigg) [ \dot R , \partial_{\chi} R + \gamma R ]_{qp} \,,
\end{equation}
which we see contains only terms that contain either the Ward identities or the `commutator' $[ \dot R , \partial_{\chi} R + \gamma R ]_{qp}$. On the other hand for the flow of the $K/f$ msWI, the terms do not collect, so that it remains separately dependent on the individual $\tilde B_p$. However each term either contains the Ward identities themselves, the `commutator'  $[ \dot R , \partial_{\chi} R + \gamma R ]_{qp}$, or the additional commutator-like structures:
\be
\label{k-comm}
 \left(\partial_\chi R_p + \gamma R_p\right) \partial_{p^2}^k \dot R_p - \dot R_p \partial_{p^2}^k \left(\partial_\chi R_p + \gamma R_p\right)\,.
 \ee
These appear in the third line of \eqref{flowWA}, and the integer $k$ takes values $1$ and $2$. For a general cutoff $R_p$, these two additional commutator terms neither vanish nor combine with other terms of the flow.

If $[ \dot R , \partial_{\chi} R + \gamma R ]_{qp}$ vanishes, the flow \eqref{flowWV} of the V msWI is automatically satisfied providing that both the $K$ and $V$ msWI are also satisfied. In this case we have by rearrangement that
\be
\left( \partial_\chi R_p + \gamma R_p \right)/\dot R_p = \left( \partial_\chi R_q + \gamma R_q \right)/\dot R_q\,,
\ee
which means that the ratio is independent of momentum. Equivalently
\be
\label{same-p-dependence}
   \partial_\chi R_p + \gamma R_p = F(\chi,t) \,\dot R_p \,,
\ee
where $F$ can be a function of $\chi$ and $t$ but not of $p$. 
However it is straightforward to see that \eqref{same-p-dependence} also forces the additional commutators \eqref{k-comm} to vanish.

We have therefore shown that all the commutator-like terms vanish if and only if $\dot R_p$ and $\partial_{\chi} R_p + \gamma R_p$ have the same dependence on $p$, with the consequence  that both the $\mathcal{\dot W}^{(A)}$ vanish, if the Ward identities $\mathcal{W}^{(A)}$ themselves vanish. Since for general choices of the functions, the vanishing of the `commutators' is surely necessary to achieve $\mathcal{\dot W}^{(A)}=0$ without further restriction, we have thus shown that the condition \eqref{same-p-dependence} is necessary and sufficient to ensure compatibility, as defined at the beginning of this section.

\subsection{Incompatibility implies no solutions}\label{sec:incompatibility}

However even if the commutators do not vanish, and thus the Ward identities are incompatible with the flow equations, {\it a priori} there could still be a non-empty restricted set of solutions that both satisfy the flow equations and Ward identities. In this case the equations are satisfied not by the vanishing of the commutators themselves, but by the fact that for the given solutions the sum of all these terms vanish after performing the integration over momenta. Therefore, as well as obeying the flow equations and the msWIs $\mathcal{W}^{(A)}=0$, the solutions must also separately obey two further conditions, namely the vanishing of the right-hand sides of \eqref{flowWA}. In the language of Dirac's classification of constraints \cite{Dirac-primary,Dirac1950}, the $\mathcal{W}^{(A)}=0$ provide the primary constraints. We have shown that if the `commutators' do not vanish, then the solutions are subject also to non-trivial secondary constraints $\mathcal{\dot W}^{(A)}=0$. Given the involved form of $\mathcal{\dot W}^{(K)}$ in particular,  we can be sure that the procedure does not close and that actually there is then an infinite tower of secondary constraints, $\partial^n_t \,\mathcal{W}^{(A)}=0,\;\forall \,n>0$,  all of which must be satisfied. It would therefore seem inevitable that there are in fact no non-trivial solutions in this case. We will confirm this by example in section \ref{sec:incompatible-no-solns}. We conclude that \emph{the vanishing of the `commutators', and hence condition \eqref{same-p-dependence}, is both necessary and sufficient for there to be any solutions to the flows and Ward identities in the derivative expansion approximation outlined in section \ref{sec:review}}.

The condition \eqref{same-p-dependence} was already used in reference \cite{Dietz:2015owa}, where however it was introduced as a mathematical trick to help solve the coupled system of flow equations and msWI. As we recall below, it implies either that $\eta=0$ or $R_p$ is of power-law form. We now see that the requirement for $\dot R_p$ and $\partial_{\chi} R_p + \gamma R_p$ to have the same dependence on $p$, goes much deeper: the flow equations \eqref{flowV} and \eqref{flowK}, and the Ward identities \eqref{msWIV} and \eqref{msWIK}, are incompatible without this constraint, and incompatibility forces there to be no solutions to the combined system.


\subsection{Required form of the cutoff profile}\label{sec:required-cutoff}

Note that $R_p$ must take a form that respects the scaling dimensions. Introducing dimensionless variables for use in the next section and later on, we can make these scaling dimensions explicit by employing the RG scale $k$.
We denote the new dimensionless quantities with a bar. We have
\begin{align}
\label{dim-vars}
	\varphi = k^{\eta/2}\bar\varphi, \qquad\chi = k^{\eta/2}\bar\chi, \qquad f(\chi) = k^{d_f}\bar f(\chi),\nonumber\\
	V(\varphi,\chi) = k^{d_V}\bar V(\bar\varphi,\bar\chi), \qquad K(\varphi, \chi) = k^{d_R- 2 + d_f} \bar  K(\bar\varphi,\bar\chi),
\end{align}
where 
\be 
\label{dimensions}
d_V = d(1-d_f/2)\qquad{\rm and}\qquad d_R = d_V - \eta\,,
\ee 
and thus from \eqref{cutoff-action} and \eqref{conformal-reduction}, we have by dimensions that $R_p$ must take the form
\begin{equation}
\label{equ:cutoff}
	R(p^2/f)= - k^{d_R} \,r\left(\frac{p^2}{k^{2-d_f}f}\right) = - k^{d_R} \,r(\hat p^2) \,,
\end{equation}
where $r$ is a dimensionless cutoff profile of a dimensionless argument,\footnote{The minus sign in \eqref{equ:cutoff} is necessary to work with the wrong sign kinetic term in \eqref{trunc} \cite{Dietz:2015owa}.} and we have introduced the dimensionless momentum magnitude $\hat p = p\sqrt{k^{d_f-2}/f}$. 

If $\dot R_p$ and $\partial_{\chi} R_p + \gamma R_p$ have the same dependence on $p$, i.e.\ satisfy \eqref{same-p-dependence}, then either $\eta=0$ or $R_p$ is of power-law form \cite{Dietz:2015owa}. To see this, note that from \eqref{equ:cutoff} and \eqref{gamma} we have
\be
\label{R-scaling-relation}
\gamma {\dot R}_p = d_V \left[\partial_\chi R_p+\gamma R_p\right] -\eta \gamma R_p\,.
\ee
Thus (choosing $F=\gamma/d_V$) we see that \eqref{same-p-dependence} is satisfied if $\eta=0$, without further restriction on $R$. However if $\eta\ne0$, then \eqref{R-scaling-relation} together with \eqref{same-p-dependence} implies
\be 
f \frac{\partial R_p}{\partial f} = \frac{d}{2}\left( \frac{\eta F}{ d_VF- \gamma} - 1\right) R_p\,,
\ee
and thus from \eqref{equ:cutoff}
\be 
\hat{p} \frac{d}{d\hat{p}} r(\hat{p}^2) = -d\left( \frac{\eta F}{ d_VF- \gamma} - 1\right) r(\hat{p}^2)\,.
\ee
Since the term in brackets does not depend on $p$, we see that this is only possible if in fact the term in brackets is a constant. Setting this constant to be $2n/d$ for some constant $n$, we thus also deduce that $r\propto \hat{p}^{-2n}$.

An example of a cutoff that does not satisfy \eqref{same-p-dependence} if $\eta\ne0$, and thus leads to incompatible msWIs in this case, is the optimised cutoff \cite{opt1,opt3}:
\be 
\label{optimised}
r(\hat{p}^2) = (1-\hat{p}^2)\theta(1-\hat{p}^2)\,.
\ee
It is straight-forward to confirm that this does not satisfy \eqref{same-p-dependence} if $\eta\ne0$. Using \eqref{equ:cutoff} and \eqref{R-scaling-relation} we find
\be 
\dot{R}_p \propto d_V \left[ \frac{2}{d}\,\theta(1-\hat{p}^2) + (1-\hat{p}^2)\theta(1-\hat{p}^2)\right] - \eta\, (1-\hat{p}^2)\theta(1-\hat{p}^2)\,.
\ee
In order for \eqref{optimised} to satisfy \eqref{same-p-dependence}, the right-hand side must be proportional to $\partial_\chi R_p+\gamma R_p$ i.e.\ to the term in square brackets. This is only true if $\eta=0$.

\section{LPA equations}
\label{sec:LPA}

We will now use the Local Potential Approximation to further investigate the restriction imposed by the msWI on the RG flow equation, in terms of general solutions and also on the existence of $k$-fixed points (i.e.\ RG fixed points with respect to variations in $k$). We start with a very clear example where the msWI forbids the existence of $k$-fixed points. 

Then using the concrete example of the optimised cutoff we show explicitly that compatibility forces $\eta=0$ for non-power-law cutoffs. Setting $\eta=0$ we will see that background independent variables exist, in other words they exist whenever the msWI is compatible with the flow. We will also see that such $\hat{k}$-fixed points\footnote{Recall that $\hat{k}$ is the background independent notion of RG scale.} coincide with the $k$-fixed points. The background independent variables allow us to solve for the fixed points explicitly, uncovering a line of fixed points, consistent with the findings for power-law cutoff \cite{Dietz:2016gzg}.

\subsection{Demonstration of background independence forbidding fixed \\ points in general}
\label{sec:forbids}

We use the change to dimensionless variables \eqref{dim-vars} and \eqref{equ:cutoff}. In the LPA we discard the flow and Ward identity for $K$, and set $\bar K=1$. The result, for a general cutoff profile $r(\hat{p}^2)$, is:
\begin{align}
\label{flow}
\partial_t \bar V + d_V \bar V - \frac{\eta}{2} \, \bar\varphi \, \frac{\partial \bar V}{\partial \bar\varphi} - \frac{\eta}{2} \, \bar\chi \, \frac{\partial \bar V}{\partial \bar\chi} &=
\int_0^{\infty} d\hat p \, \hat p^{d-1} \, \frac{d_R\, r - \frac{d_V}{d} \, \hat p \, r'}{\hat p^2 + r - \partial^2_{\bar\varphi}\bar V}\,,\\
\label{msWI}
\frac{\partial \bar V}{\partial \bar\chi} - \frac{\partial \bar V}{\partial \bar\varphi} + \bar \gamma \, \bar V &= \bar \gamma
\int_0^{\infty} d\hat p \, \hat p^{d-1} \, \frac{r - \frac{1}{d} \, \hat p \, r'}{\hat p^2 + r - \partial^2_{\bar\varphi}\bar V} \,,
\end{align}
where $r'$ means $dr(\hat{p}^2)/d\hat{p}$ and from the change to dimensionless variables we find:
\be 
\label{fbar-general}
\bar{\gamma} = \frac{d}{2}\frac{\partial}{\partial\bc}\ln\bar{f}\left({\rm e}^{\eta t/2}\mu^{\eta/2}\bc\right)\,.
\ee
Note that since $f$ cannot depend on $t$ (see the discussion in section \ref{sec:review}), once we go to dimensionless (i.e.\ scaled) variables, $\bar{f}$ is in general forced to depend on $t$ if $\chi$ has non-vanishing scaling dimension $\eta$.
At the ($k$-)fixed point we must have $\partial_t \bar V = 0$. We see at once why fixed points are generically forbidden by the msWI: the fixed point potential $\bar{V}$ would have to be independent of $t$, but through \eqref{msWI} and \eqref{fbar-general} this is impossible in general since $\bV$ is forced to be dependent on explicit $t$-dependence in $\bar{f}$ through the Ward identity. This is true even in the case of power-law cutoff profile\footnote{And indeed this issue was highlighted, but in a different way in reference \cite{Dietz:2015owa}.} which as we have seen allows \eqref{msWI} to be compatible with the flow \eqref{flow}.

At first sight an escape from this problem is simply to set $f$ to be power law. Indeed setting $f\propto\chi^{\rho}$
for some constant $\rho$,  \eqref{fbar-general} implies
\be 
\bar{\gamma} = \frac{d}{2} \frac{\rho}{\bc}\,,
\label{powlaw-gamma}
\ee 
and thus \eqref{msWI} no longer has explicit $t$ dependence. 
Recall that for power-law cutoff profiles $r$, it was indeed found that $k$-fixed points for $\bV$ 
are allowed if $f$ is chosen to be of power law form \cite{Dietz:2015owa}.\footnote{This is true also for $\bar{K}$. However if the dimensions of $f$ and $\chi$
do not match up, 
these fixed points do not agree with the background independent $\hat{k}$-fixed points and furthermore  the effective action $\Gamma_k$ still runs with $k$ \cite{Dietz:2015owa}.} 
However we have seen in section \ref{sec:required-cutoff} that any other cutoff profile does not allow the Ward identity to be compatible with the flow unless $\eta=0$. We argued in section \ref{sec:incompatibility} that incompatibility overconstrains the equations leading to no solutions. In the next subsection, section \ref{sec:incompatible-no-solns}, we will confirm this explicitly, choosing as a concrete example the optimised cutoff profile and space-time dimension $d=4$.

On the other hand, if we set $\eta=0$ then the msWI \eqref{msWI} is compatible with the flow \eqref{flow}, for any parametrisation $f$. Apparently $k$-fixed points are also now allowed without further restriction, since again \eqref{fbar-general} loses its explicit $t$ dependence.  Opting once more for optimised cutoff profile and $d=4$, we will see in section \ref{sec:works} that indeed they are allowed and furthermore they coincide with fixed points in a background independent description that we also uncover.

\subsection{Confirmation of no solutions if the msWI is incompatible with the flow}\label{sec:incompatible-no-solns}

Specialising to the optimised cutoff and (for simplicity) the most interesting case of spacetime dimension $d=4$, the equations read
\begin{align}
\label{flowComplete_opt}
\partial_t\bar{V}+
d_V \bar V - \frac{\eta}{2} \bar \varphi \, \partial_{\bar\varphi}\bar V -\frac{\eta}{2} \bar \chi \, \partial_{\bar\chi}\bar V &= \left(  \frac{d_R}{6} + \frac{\eta}{12} \right) \frac{1}{1 -  \partial^2_{\bar\varphi}\bar V},\\
\label{msWI_opt}
\partial_{\bar\chi}\bar V - \partial_{\bar\varphi}\bar V + \bar{\gamma} \bar V &= \frac{\bar{\gamma}}{6} \frac{1}{1 - \partial^2_{\bar\varphi}\bar V}\,.
\end{align}
Choosing power law $f$ and thus eliminating explicit dependence on $t$, cf.\ \eqref{powlaw-gamma}, apparently these equations can work together. 
Combining them 
by eliminating their right-hand sides, we get
\be
\label{pde1}
2\partial_t\bar{V}+\eta\bar{V}-\left(\eta\bar\varphi-\alpha\bc\right)\partial_{\bar{\varphi}}\bar{V}-(\eta+\alpha)\bar\chi\partial_{\bar{\chi}} \bar{V}=0\,,
\ee
where we have introduced the constant $\alpha = (d_R+\eta/2)/\rho$.
This equation can be solved by the method of characteristics (see e.g.\ the appendix in reference \cite{Dietz:2015owa}). Parametrising the characteristic curves with $t$, they are generated by the following equations:
\be 
\frac{d\bV}{dt} = -\frac{\eta}{2}\bV\,,\quad\frac{d\bc}{dt}=-\frac{\alpha+\eta}{2}\bc\,,\quad \frac{d\bp}{dt}=\frac{\alpha\bc-\eta\bp}{2}\,.
\ee
Solving the second equation before the third, it is straightforward to find the curves:
\be 
\bV = \hV {\rm e}^{-\eta t/2}\,,\quad \bc = \hc\, {\rm e}^{-(\eta+\alpha)t/2}\,,\quad \bp+\bc = \hp\, {\rm e}^{-\eta t/2}\,,
\ee
in terms of initial data $\hV,\hp,\hc$.
Thus the solution to \eqref{pde1} can be written as
\be 
\bV ={\rm e}^{-\eta t/2}\, \hV(\hp,\hc) ={\rm e}^{-\eta t/2} \,\hV\left({\rm e}^{\eta t/2}[\bp+\bc],{\rm e}^{(\eta+\alpha)t/2}\bc\right)\,,
\ee
as can be verified directly. Plugging this into either \eqref{flowComplete_opt} or \eqref{msWI_opt} gives the same equation, which in terms of the hatted variables reads
\be 
\hc \partial_{\hc}\hV+2\rho\hV = \frac{\rho}{3}\frac{1}{{\rm e}^{-\frac{\eta}{2} t} \,-\partial^2_{\hp}\hV}\,.
\ee
Since $\hV(\hp,\hc)$ is independent of $t$, we see there are no solutions unless $\eta=0$. 
We saw in section \ref{sec:required-cutoff} that this was also the necessary and sufficient condition for compatibility in this case. 

\subsection{Background independence at vanishing anomalous dimension}\label{sec:works}
We now set $\eta=0$. As recalled in section \ref{sec:required-cutoff}, the msWI is now compatible with the flow, and furthermore  from \eqref{fbar-general} the explicit $t$ dependence has dropped out.
For power-law cutoff profiles we found that $k$-fixed points exist and coincide with background independent $\hk$-fixed points for any form of $f$ with any dimension $d_f$ \cite{Dietz:2015owa}. We will see that for non-power law cutoff that the same is true. (Again we choose optimised cutoff and $d=4$ as an explicit example.) We will uncover consistent background independent variables for which the full line of fixed points is visible \cite{Dietz:2016gzg}. 

Since $\eta=0$, in the equations \eqref{flowComplete_opt} and \eqref{msWI_opt}, we also have $d_R=d_V=2(2-d_f)$ and $\bar{\gamma}=2\partial_{\bc}\ln\bar{f}(\bc)$. Note that from \eqref{equ:cutoff}, $d_f=2$ is excluded otherwise the IR cutoff no longer depends on $k$. Also note that since $\eta=0$ we can drop the bars on $\chi$ and $\vp$.
Combining the equations into a linear partial differential equation we get
\be 
\label{pde2}
\partial_t\bV +\frac{2-d_f}{\partial_{\chi}\ln\bar{f}}\left(\partial_{\vp}\bV-\partial_{\chi}\bV\right) =0\,,
\ee
whose characteristic curves satisfy
\be 
\label{eta0curves}
\frac{d\chi}{dt}=\frac{d_f-2}{\partial_{\chi}\ln\bar{f}}\,,\quad\frac{d\vp}{dt}=\frac{2-d_f}{\partial_{\chi}\ln\bar{f}}\,,\quad \frac{d\bV}{dt}=0\,.
\ee
Solving the first equation gives:
\be 
\label{hatt}
\hatt = t+\frac{\ln\bar{f}}{2-d_f}\,,
\ee
where the integration constant $\hatt$ is thus the background independent definition of RG time (see the appendix to reference \cite{Dietz:2015owa}). Exponentiating,
\be 
\hk = k \left\{\bar{f}(\chi)\right\}^{\frac{1}{2-d_f}} = k^{2\frac{1-d_f}{2-d_f}} \left\{f(\chi)\right\}^{\frac{1}{2-d_f}}\,,
\ee
where the second equality follows from \eqref{dim-vars}. The sum of the first two equations in \eqref{eta0curves} tells us that $\phi=\vp+\chi$ is an integration constant for the characteristics, and finally the last equation says that $\bV$ is also constant for characteristics. Thus we learn that the change to background independent variables is achieved by writing
\be 
\label{background-independent}
\bV = \hV(\phi,\hatt\,)\,.
\ee
It is straightforward to verify that this solves \eqref{pde2}. Substituting into either \eqref{flowComplete_opt} or \eqref{msWI_opt} gives the same flow equation:
\be 
\label{bi-flow}
\partial_{\hatt}\hV +d_V\hV = \frac{d_V}{6}\frac{1}{1-\partial^2_\phi\hV^{\vphantom{H^H}}}\,,
\ee
which is indeed now background independent, i.e.\ independent of $\chi$, and indeed independent of parametrisation $f$. There remains a dependence on the dimension of $f$ through $d_V = 2(2-d_f)$ although this disappears for $\hk$-fixed points, and can be removed entirely by a rescaling $\hatt\mapsto \hatt\, d_V$ which however changes the dimension of $\hat{k}$ to $d_V$. 

We also see from \eqref{hatt} and \eqref{background-independent} that
\be 
\partial_t \bV = \partial_{\hatt} \hV\,,
\ee
and thus fixed points in $k$ coincide with the background independent fixed points. 

Finally, the fixed points are readily found from \eqref{bi-flow} similarly to references \cite{Dietz:2016gzg,Morris:1994jc} by recognising that 
\be 
\frac{d^2\hV}{d\phi^2} = 1-\frac{1}{6\hV}
\ee
is equivalent to Newton's equation for acceleration with respect to `time' $\phi$ of a particle of unit mass at `position' $\hV$ in a potential $U=-\hV+ (\ln\hV)/6$. In this way it can be verified that there is a line of fixed points ending at the Gaussian fixed point, which is here $\hV=1/6$.
The appearance of this line of fixed points is a consequence of the conformal factor problem, discussed in section \ref{sec:approxs} of chapter \ref{cha:introduction}, and is in agreement with the findings for power-law cutoff in reference \cite{Dietz:2016gzg} in which the problem has been addressed.

\section{Polynomial truncations}
\label{sec:truncations} 

The analysis so far has used properties of conformally truncated gravity and the derivative expansion approximation method. In order to gain insight about what might happen at the non-perturbative level, and in full quantum gravity, we will consider how the issues would become visible in polynomial truncations.

The generic case treated in section \ref{sec:forbids} will be just as clear in the sense that truncations of the Ward identity will still force the effective potential (effective action in general) to be $t$-dependent if the dimensionless parametrisation \eqref{fbar-general} is similarly forced to be $t$-dependent. In general therefore, if the way the metric is parametrised forces the parametrisation to become $t$-dependent, we can expect that background independence excludes the possibility of fixed points, at least with respect to $t$.

Consider next the situation treated in section \ref{sec:incompatible-no-solns}. Expanding the dimensionless potential and the equations in a double power series in the fluctuation and the background field, we write:
\begin{equation}
\label{pot-expand}
\bar V(\bar\varphi,\bar\chi) = \sum_{n,m=0}^{\infty} a_{nm} \bar\varphi^n \bar\chi^m\,.
\end{equation}
Substituting \eqref{powlaw-gamma} into \eqref{msWI_opt} and multiplying through by $\bc$, we can read off from this and \eqref{flowComplete_opt} the zeroth level equations:
\be 
d_V a_{00} = \left(  \frac{d_R}{6} + \frac{\eta}{12} \right) \frac{1}{1 -  2a_{20}}\,,\qquad 2\rho\, a_{00} = \frac{\rho}{3} \frac{1}{1 -  2a_{20}}\,.
\ee
Since $\rho$ cannot vanish and $a_{20}$ cannot diverge, combining these equations gives $d_V = d_R+\eta/2$ which from \eqref{dimensions} implies $\eta=0$. Thus we recover already from the zeroth order level that fixed points are excluded unless $\eta=0$. (Of course the real reason, namely that the equations are incompatible, and the full consequence that there are no $t$-dependent solutions either, is maybe not so easy to see this way.)

\subsection{Counting argument}\label{sec:counting}

We remarked the Introduction that generically the solutions become over-constrained if we consider a sufficiently high truncation. We now proceed to make a careful count of the coefficients appearing in the equations and estimate the level at which this happens.

We concentrate on fixed point solutions to the LPA system \eqref{flow}, \eqref{msWI} and \eqref{fbar-general} where either $\eta=0$ or we choose power-law $f$, so that explicit $t$ dependence does not already rule out such solutions. We introduce the short-hand notation $\bar V ^{(n,m)}=\partial_{\bar \varphi}^n \partial_{\bar \chi}^m \bar V(\bp,\bc)$. 
To obtain the system at order $r$ we have to plug the expansion of the potential \eqref{pot-expand}
into both the fixed point equation and msWI, act on them with operators $\frac{\partial^{i+j}}{\partial \bar\varphi^i \, \partial \bar\chi^j}$ such that $i+j=r$, before finally setting the fields to zero. In particular, for any fixed value $r_{\star}$ we have $2 \, (r_{\star}+1)$ equations and hence up to order $r$ there are
\begin{equation}
\label{number_eqns}
n_{\text{eqn}}(r) = \sum_{i=0}^{r} 2\,(i+1) = r^2 + 3r + 2 
\end{equation}
equations. 

To count the coefficients appearing in these $n_{\text{eqn}}(r)$ equations let us start with the left-hand sides. First note that 
\begin{equation}
\label{proportionality}
\bar V^{(i,j)} \bigg|_{\bar\varphi = \bar\chi = 0}
\propto a_{ij} \,.
\end{equation}
That is, for any fixed pair $(i,j)$ the left-hand side of (\ref{msWI}) will contain the coefficients $a_{ij}$, $a_{i+1,j}$ and $a_{i,j+1}$, whereas the left-hand side of (\ref{flow}) will only contain $a_{ij}$. Up to some fixed order $r$ there will be thus coefficients $a_{ij}$ where $i$ and $j$ run from $0$ to $r+1$ and $i+j \leqslant r+1$
\begin{figure}
\centering
\begin{tikzpicture}[>=stealth]
\draw (0,0) -- (1,0) -- (1,-1) -- (0,-1) -- cycle;
\draw (1,0) -- (2,0) -- (2,-1) -- (1,-1) -- cycle;
\draw (2,0) -- (3,0) -- (3,-1) -- (2,-1) -- cycle;
\draw (3,0) -- (4,0) -- (4,-1) -- (3,-1) -- cycle;

\draw (0,-1) -- (1,-1) -- (1,-2) -- (0,-2) -- cycle;
\draw (1,-1) -- (2,-1) -- (2,-2) -- (1,-2) -- cycle;
\draw (2,-1) -- (3,-1) -- (3,-2) -- (2,-2) -- cycle;

\draw (0,-2) -- (1,-2) -- (1,-3) -- (0,-3) -- cycle;
\draw (1,-2) -- (2,-2) -- (2,-3) -- (1,-3) -- cycle;

\draw (0,-3) -- (1,-3) -- (1,-4) -- (0,-4) -- cycle;

\draw (0.5,0.5) node {0}  (1.5,0.5) node {1}  (2.5,0.5) node {$\cdots$} (3.5,0.5) node {$r+1$};
\draw (-0.5,-0.5) node {0}  (-0.5,-1.5) node {1}  (-0.5,-2.5) node {$\vdots$} (-0.5,-3.5) node {$r+1$};

\draw (0.5,-0.5) node {$a_{00}$}  (1.5,-0.5) node {$a_{01}$}
(2.5,-0.5) node {$\cdots$}  (3.5,-0.5) node {$a_{0,r+1}$};
\draw (0.5,-1.5) node {$a_{10}$}  (1.5,-1.5) node {$\ddots$};
\draw (0.5,-2.5) node {$\vdots$};
\draw (0.5,-3.5) node {$a_{r+1,0}$};

\draw (-0.25,0.5) node {$j$} (-0.5,0.25) node {$i$};
\draw (-0.15,0.15) -- (-0.55,0.55);
\end{tikzpicture}
\caption{Coefficients of the potential appearing on the left sides of the equations.}
\label{diag_lhs}
\end{figure}
\begin{equation}
\label{coeff_lhs}
\bigg\lbrace a_{00}, a_{01}, \dots , a_{0,r+1}, a_{10}, \dots , a_{1,r}, \dots , a_{2,r-1}, \dots, a_{r+1,0}   \bigg\rbrace \,,
\end{equation}
 (cf.\ figure \ref{diag_lhs}). This adds up to the following number of coefficients
\begin{equation}
n_{\text{lhs}}(r) = \sum_{i=1}^{r+2} \, i = \frac{1}{2} \, r^2 + \frac{5}{2} \, r + 3 \,.
\end{equation}

Including the coefficients on the right-hand sides, we have to be careful not to double count any coefficients that have already been taken account of on the left-hand sides. Let us suppose we have fixed the cutoff and let us assume that for the moment $\bar{\gamma} =$ const. Then all additional coefficients on the right-hand sides come from the expansion of the propagator
\begin{equation}
\label{expansion_propagator}
\frac{\partial^{i+j}}{\partial \bar\varphi^i \, \partial \bar\chi^j} \, \left( \frac{1}{1 - \bar V^{(2,0)}} \right) \bigg|_{\bar\varphi = \bar\chi = 0} = 
\frac{\partial^{j}}{\partial \bar\chi^j} \left[ \frac{\partial^{i-1}}{\partial \bar\varphi^{i-1}} \left( \frac{\bar V^{(3,0)}}{(1 - \bar V^{(2,0)})^2} \right) \right] \bigg|_{\bar\varphi = \bar\chi = 0}\, .
\end{equation}
Since we can always arrange the $\bar\varphi$--derivatives to act first, the expression in the square brackets will involve terms $\bar V^{(2,0)} \cdots \, \bar V^{(i+2,0)}$. Using (\ref{proportionality}), we see that the expression given in (\ref{expansion_propagator}) will then include terms
\begin{equation}
\bigg\lbrace a_{20}, a_{21}, \dots , a_{2i}, a_{30}, \dots , a_{3i}, \dots, a_{i+2,0}, \dots, a_{i+2,j}   \bigg\rbrace \,.
\end{equation}
Up to any fixed order $r$, $i$ and $j$ can take values between $0$ and $r$ such that $i+j = r$, and in total we will have the following coefficients on the right-hand sides 
\begin{figure}
\centering
\begin{tikzpicture}[>=stealth]
\draw (0,0) -- (1,0) -- (1,-1) -- (0,-1) -- cycle;
\draw (1,0) -- (2,0) -- (2,-1) -- (1,-1) -- cycle;
\draw (2,0) -- (3,0) -- (3,-1) -- (2,-1) -- cycle;
\draw (3,0) -- (4,0) -- (4,-1) -- (3,-1) -- cycle;

\draw (0,-1) -- (1,-1) -- (1,-2) -- (0,-2) -- cycle;
\draw (1,-1) -- (2,-1) -- (2,-2) -- (1,-2) -- cycle;
\draw (2,-1) -- (3,-1) -- (3,-2) -- (2,-2) -- cycle;

\draw (0,-2) -- (1,-2) -- (1,-3) -- (0,-3) -- cycle;
\draw (1,-2) -- (2,-2) -- (2,-3) -- (1,-3) -- cycle;

\draw (0,-3) -- (1,-3) -- (1,-4) -- (0,-4) -- cycle;

\draw (0.5,0.5) node {0}  (1.5,0.5) node {1}  (2.5,0.5) node {$\cdots$} (3.5,0.5) node {$r$};
\draw (-0.5,-0.5) node {2}  (-0.5,-1.5) node {3}  (-0.5,-2.5) node {$\vdots$} (-0.5,-3.5) node {$r+2$};

\draw (0.5,-0.5) node {$a_{20}$}  (1.5,-0.5) node {$a_{21}$}
(2.5,-0.5) node {$\cdots$}  (3.5,-0.5) node {$a_{2,r}$};
\draw (0.5,-1.5) node {$a_{30}$}  (1.5,-1.5) node {$\ddots$};
\draw (0.5,-2.5) node {$\vdots$};
\draw (0.5,-3.5) node {$a_{r+2,0}$};

\draw (-0.25,0.5) node {$j$} (-0.5,0.25) node {$i$};
\draw (-0.15,0.15) -- (-0.55,0.55);
\end{tikzpicture}
\caption{Coefficients of the potential appearing in the expansion of the propagator.}
\label{diag_rhs}
\end{figure}
\begin{equation}
\bigg\lbrace a_{20}, \dots , a_{2,r}, a_{30}, \dots , a_{3,r-1}, \dots , a_{4,r-2}, \dots, a_{r+2,0}   \bigg\rbrace \,,
\end{equation}
(cf.\ figure \ref{diag_rhs}). Most of these coefficients have however already been accounted for on the left-hand sides cf.\ (\ref{coeff_lhs}). The only ones not yet counted are
\begin{figure}
\centering
\includegraphics[]{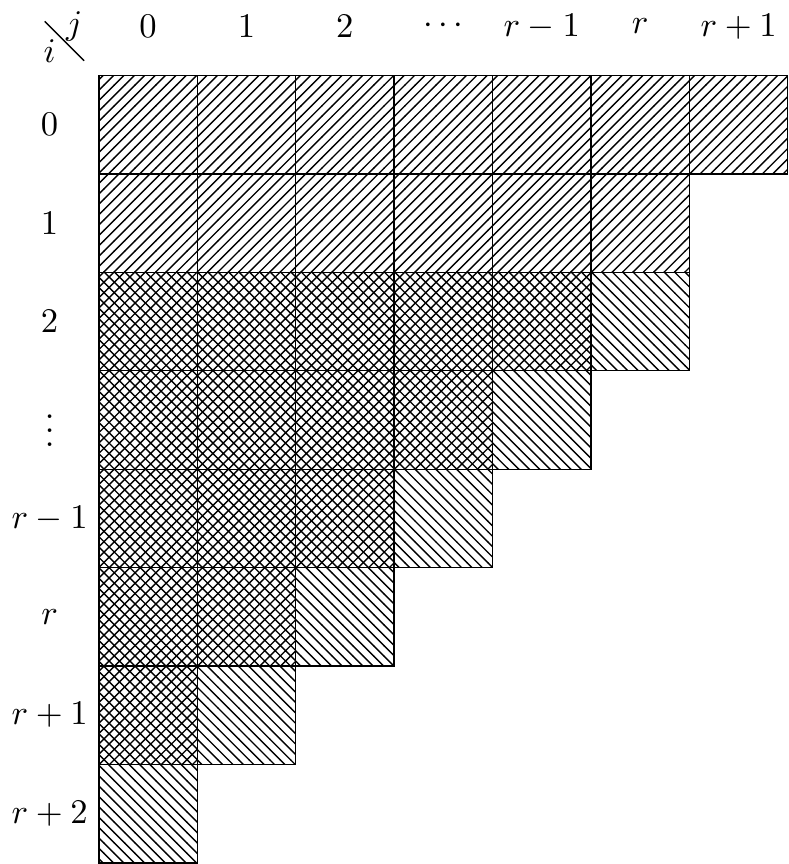}
\caption{All the coefficients of the potential appearing on both sides of the equations.}\label{diag_combined}
\end{figure}
\begin{equation}
\bigg\lbrace a_{2,r}, a_{3,r-1}, a_{4,r-2}, \dots, a_{r+2,0}   \bigg\rbrace \,,
\end{equation}
(cf.\ figure \ref{diag_combined}) which precisely add up to a further $r+1$ coefficients. We also must include another two coefficients, namely $\eta$ and $d_f$. 
Finally, since $\gamma$ is in general some function of $\chi$ it is easy to see that
\begin{equation}
\frac{d^r}{d \bar\chi^r} \, \bar{\gamma} \propto \frac{d^r}{d \bar\chi^r} \, \left( \frac{f'}{f}\right) \subseteq
\bigg\lbrace f, f', \dots, f^{(r+1)}   \bigg\rbrace \,,
\end{equation}
which gives us an additional $(r+2)$ coefficients from the Taylor expansion of $f$. The total number of coefficients from both left and right-hand sides is then given by
\begin{equation}
\label{number_coeff_general}
n_{\text{coeff}}(r) = n_{\text{lhs}} + (r+1) + (r+2) + 2  = \frac{1}{2} \, r^2 + \frac{9}{2} \, r + 8 \,.
\end{equation}

From \eqref{number_eqns} we see that for large $r$ the number of equations $\sim r^2$, while from \eqref{number_coeff_general} the number of coefficients only $\sim r^2/2$. There are therefore asymptotically twice as many equations as coefficients, as already discussed in the Introduction. 
Equating the number of equations and coefficients yields the positive solution
\begin{equation}
r = 5.3 \,.
\end{equation}
Therefore the number of equations exceeds the number of coefficients for the first time at order $r=6$. If there is to be a conflict between the existence of ($k$-)fixed points and background independence generically we would expect this to become evident at about this level.  Equally, if there is no conflict between background independence and the existence of ($k$-)fixed points then from this level onwards some equations become redundant (i.e.\ they provide constraints that are automatically satisfied once the other equations are obeyed). In the limit $r\to\infty$ fully half of the equations must become redundant if ($k$-)fixed points are to be consistent with background independence.

\section{Conclusions}
\label{sec:conclusions}

If we construct the non-perturbative flow equation for quantum gravity by introducing a cutoff defined through a background metric then independence from this artificial metric can only be achieved if the appropriate modified split Ward identity is obeyed. However even if it is obeyed, background independence is guaranteed only in the limit $k\to0$. RG properties on the other hand are defined at intermediate scales $k$. There is therefore the potential for conflict in this formulation between RG notions such as fixed points, and the requirement of background independence. Examples of such conflicts were uncovered in reference \cite{Dietz:2015owa}.

In this paper we have further investigated these issues. Our findings, together with those of reference \cite{Dietz:2015owa}, are summarised in table \ref{table:summary}.\footnote{For power law cutoff  $r(z)=z^{-n}$, $d_f = 2-\eta/(n+2)$ is excluded \cite{Dietz:2015owa}, and from section \ref{sec:required-cutoff} when $\eta=0$,  $d_f=2$ is excluded for any cutoff profile: in these cases the cutoff term is independent of $k$.} The first question that needs to be addressed is whether the msWI, $\mathcal{W}=0$, is compatible with the exact RG flow equation, i.e.\ such that $\dot{\mathcal{W}}=0$ then follows. At the exact level, compatibility is guaranteed since they are both identities derived from the partition function (see also section \ref{sec:exact}). Within the derivative expansion approximation of conformally reduced gravity considered in reference \cite{Dietz:2015owa} (reviewed in section \ref{sec:review}), we have shown  in sections \ref{sec:compatibility-at-d2} and \ref{sec:required-cutoff}, that this compatibility follows if and only if either $\eta=0$ or the cutoff profile is power law. In section \ref{sec:exact-vs-derivatives}, we saw precisely why the derivative expansion breaks compatibility in general and why these special cases restore it. We argued in section \ref{sec:incompatibility} that if the equations are incompatible they are overconstrained since there are then an infinite number of secondary constraints, and thus not even $t$-dependent solutions can exist.
We confirmed this latter conclusion by example in section \ref{sec:incompatible-no-solns} in the LPA. In section \ref{sec:truncations}, we also saw that the fixed point equations and Ward identities together generically overconstrain the system when expanded in terms of vertices beyond the six-point level.

\begin{table}[t]
\begin{center}
\begin{tabular}{|c||c|c|c||c|c|c|c|c|c|c|}
\hline
 & \multicolumn{3}{c||}{parametrisation $f$} & \multicolumn{2}{c|}{cutoff profile $r$} \\
 \hline\hline
$\eta$ & type & $d_f$ & runs & power-law & not power-law \\
\hline \hline
\multirow{3}{*}{$\ne0$} & not power-law & any & yes & \xcancel{FP}\phantom{$=$} $\widehat{\rm FP}$ & \cellcolor{red!30} \\
\cline{2-5}
 & 
   power law & $\ne\rho\eta/2$ & yes & FP $\ne$ $\widehat{\rm FP}$ &  \cellcolor{red!30} incompatible \\
\cline{3-5}
 &     $f=\chi^\rho$    & $=\rho\eta/2$ & no & FP $=$ $\widehat{\rm FP}$ & \cellcolor{red!30} \\
\hline  
\multirow{2}{*}{$=0$} & \multirow{2}{*}{any} & $\ne0$ & yes & \multicolumn{2}{c|}{\multirow{2}{*}{FP $=$ $\widehat{\rm FP}$}} 
\\ 
\cline{3-4}
                   &         & $=0$    & no & \multicolumn{2}{c|}{}  \\
\hline
 
\end{tabular}
\end{center}
\caption{RG properties of the derivative expansion for conformally reduced gravity, when the msWI is also satisfied. The results depend on whether the conformal factor develops an anomalous dimension $\eta$, on the choice of cutoff profile $r$, and on how the metric is parametrised via $f$. Depending also on its dimension $d_f$, $f$ can contain a massive parameter, and thus run with $k$ when written in dimensionless terms, as listed in the table.
$\widehat{\rm FP}$ indicates that a background independent description exists, while (\xcancel{FP}) FP  indicates that $k$-fixed points are (not) allowed; the (in)equality shows how these relate to the $\hk$-fixed points.}
\label{table:summary}
\end{table}

Even if the equations are compatible, the msWI can still forbid fixed points. In section \ref{sec:forbids} the reason was laid out particularly clearly. The Ward identity
\be 
\frac{\partial \bar V}{\partial \bar\chi} - \frac{\partial \bar V}{\partial \bar\varphi} + \bar \gamma \, \bar V = \bar \gamma
\int_0^{\infty} d\hat p \, \hat p^{d-1} \, \frac{r - \frac{1}{d} \, \hat p \, r'}{\hat p^2 + r - \partial^2_{\bar\varphi}\bar V} 
\ee
(which is compatible for power-law $r$),
forces the effective potential $\bV$ to depend on $t$ through
\be 
\bar{\gamma} = \frac{d}{2}\frac{\partial}{\partial\bc}\ln\bar{f}\left({\rm e}^{\eta t/2}\mu^{\eta/2}\bc\right)\,,
\ee
whenever this dimensionless combination is similarly forced to be $t$-dependent. For example we see that fixed points with respect to $k$ are forbidden for exponential parametrisations $f(\phi)= \exp(\phi)$  if the field grows a non-zero anomalous dimension.  It is clear that the reasons for this conflict are general and not tied to the derivative expansion of the conformally truncated model per se. Therefore this issue could provide important constraints for example on the exponential parametrisations recently advocated in the literature \cite{Demmel:2015zfa,Eichhorn:2013xr,Eichhorn:2015bna,Nink:2014yya,Percacci:2015wwa,Labus:2015ska,Ohta:2015efa,Gies:2015tca,Dona:2015tnf}.

In section \ref{sec:truncations} we considered how these issues arise in polynomials truncations.
We saw that the problem is that if the fixed point equations and msWI equations are truly independent, then they will over-constrain the solutions if carried to a sufficiently high order truncation. Indeed, expanding in powers of the fluctuation field $\vp$ to the $m^{\rm th}$ level and background field $\chi$ to the $n^{\rm th}$ level, we get one fixed point equation for each $(m,n)$-point vertex \emph{and} one msWI equation per vertex. Even though each of these equations is open (depending on yet higher-point vertices) we saw that since there are two equations for every vertex, at sufficiently high order truncation there are more equations than vertices (indeed eventually double the number) and thus either the equations become highly redundant or the vertices are constrained to the point where there are no solutions.
This analysis strongly suggests  that the full non-perturbative Ward identities would lead to important constraints on RG properties.


For full quantum gravity, such conflict between $k$-fixed points and background independence may also show up clearly in a vertex expansion, as discussed in \ref{sec:truncations}, or generically it may not become visible until the six-point level. However for full quantum gravity, if we are to follow the standard procedure, we must also fix the gauge. The original msWI, which formally expresses background independence before gauge fixing, will no longer be compatible with the flow equation. Instead we must use the appropriate version which has contributions from the background dependence of the gauge fixing and ghost terms as well as from cutoff terms for the ghost action itself. However background independence is then only restored in the limit $k\to0$ after going ``on-shell'' (assuming such an appropriate property can be defined). This last step is required to recover quantities that are independent of the gauge fixing. If we are to continue with a flow equation for a Legendre effective action \cite{Wetterich:1992,Morris:1993} then to get around this obstruction, the Vilkovisky-DeWitt covariant effective action seems called for \cite{Branchina:2003ek,Donkin:2012ud,Demmel:2014hla,Safari:2015dva}, with the msWI replaced by the corresponding modified Nielsen identities where the role of the background field is played by the ``base point'' \cite{Pawlowski:2003sk}.

Returning to the present study, it seems surely significant that  
whenever the msWI equations are actually compatible with the flow equations, it is possible to combine them and thus uncover background independent variables, including a background independent notion of scale, $\hk$. These are not only independent of $\chi$ but also independent of the parametrisation $f$. Of course such an underlying description has only been shown in this $\mathcal{O}(\partial^2)$ approximation and in conformally truncated gravity, and one might doubt that this happy circumstance could be generalised to full quantum gravity, and not only for the reasons outlined above. However we also saw in section \ref{sec:counting} that if modified Ward identities are to be compatible with the flow equations then in terms of vertices, the information they contain becomes highly redundant at sufficiently high order (the six-point level in our case).
This in itself suggests the existence of a simpler description. Finally, a formulation for non-perturbative RG has recently been proposed where computations can be made without ever introducing a background metric (or gauge fixing) \cite{Morris:2016nda}.

\chapter{Asymptotic solutions in asymptotic safety}
\label{cha:asymp}

\section{Introduction}
\label{sec:Intro}

As already hinted at in section \ref{sub:f(R)}, in order to understand fixed point solutions of the RG equation both physically and mathematically, it is necessary to study their asymptotic behaviour.
For example, the asymptotic behaviour is needed to determine the number of fixed points supported by theory space, as mentioned at the end of section \ref{sec:theory space}.
In addition, this behaviour is also important to understand because it encodes the deep non-perturbative quantum regime, contains the only physical part of the fixed point effective action and can be used to validate numerical solutions, all of which are discussed in more detail at the end of this introduction.
In this chapter we present the work of \cite{Gonzalez-Martin:2017gza} in which we explain how to find the asymptotic form of fixed point solutions in functional truncations, in particular the $f(R)$ approximation
\be
\label{ansatz EAA}
\Gamma_k[g]=\int\,d^4x\sqrt{g}\, f(R)\,,
\ee
introduced in section \ref{sec:approxs}.
It should be emphasised that \eqref{ansatz EAA} actually goes beyond keeping a countably infinite number of couplings, the Taylor expansion coefficients $g_n=f^{(n)}(0)$, because a priori the large field parts of $f(R)$ contain degrees of freedom which are unrelated to all these $g_n$.

Recall that the result of such an approximation is a fixed point equation which is either a third order or a second order, non-linear ODE for the dimensionless function $\vp(r)$, where\footnote{Again recall the abuse of notation as, at the level of the projected flow equation, $R$ now represents the background curvature associated to the background metric $\bar{g}_{\mu\nu}$.}
\be
\label{dim vars}
r \equiv {R}\, k^{-2} \qquad\quad \text{and} \qquad\quad f({R}) \equiv k^4 \vp({R}k^{-2})\,.
\ee
The asymptotic form of a solution to the ODE is determined by its behaviour at large field, which in the present case means the behaviour of $\vp(r)$ in the limit $r \rightarrow \infty$. (This is equivalent to the large background curvature $R$ limit for fixed $k$, by \eqref{dim vars} or for fixed ${R}$, $r \rightarrow \infty$ corresponds to what we more commonly refer to as the IR limit, $k\rightarrow 0$.)
We call the form of the solution in the large $r$ limit, the asymptotic solution and denote it $\vp_{asy}(r)$.
In what follows we choose to study the fixed point ODE \eqref{fp} derived in \cite{Demmel2015b} which fortuitously provides a zoo of asymptotic solutions of different types.
The equation is derived on a space of positive curvature (effectively the Euclidean four-sphere) which means that a fixed point corresponds to a smooth global solution $\vp(r)$ over the domain $r\in[0,\infty)$.\footnote{The discussion is readily adapted to negative curvatures.}

Although these ODEs are complicated, the asymptotic solution can fortunately be found analytically and in full generality \cite{Dietz:2012ic,Dietz:2016gzg} by adopting techniques developed much earlier for scalar field theories \cite{Morris:1994ki,Morris:1994ie,Morris:1994jc}. These techniques apply to any functional truncation of the exact renormalization group fixed point equations, such that the result is an ODE or coupled set of ODEs (as e.g.\ in \cite{Dietz:2016gzg}).
Perhaps because these techniques were covered only briefly and without outlining the general treatment, they have yet to be entirely adopted, meaning that the functional solution spaces for many of the formulations \cite{Percacci:2015wwa,Labus:2015ska,Eichhorn:2015bna,Demmel:2014fk,Demmel:2012ub,Demmel:2013myx,Benedetti:2013jk,Demmel:2014hla,Demmel2015b,Ohta:2015efa,Ohta2016,Percacci:2016arh,Falls:2016msz,Ohta:2017dsq} remain unexplored or at best only partially explored.
It is hoped that the work presented in this chapter improves this situation, by describing in detail and with as much clarity as possible how the techniques allow asymptotic solutions to be fully developed.

Perhaps another reason why the asymptotically large $r=R/k^2$ region may have been under-explored is that it has not been clear what meaning should be attached to this region when $1/k$ is larger than the physical size $1/\sqrt{R}$ of the manifold, despite the fact that we know that the infrared cutoff $k$ is artificial and introduced by hand and the physical effective action,
\be 
\label{phys-eff-ac}
\Gamma=\lim_{k\to0} \Gamma_k\,,
\ee
is therefore only recovered when the cutoff is removed. This puzzle was brought into sharp relief in formulations that have a gap, i.e.\ a lowest eigenvalue which is positive, so that large $r$ then corresponds to $k$ being less than any eigenvalue \cite{Demmel:2014fk,Demmel2015b,Ohta2016,Falls:2016msz}. This issue was recently resolved in reference \cite{Morris:2016spn} where it was shown to be intimately related to ensuring background independence (but in a way that can be resolved even for single-metric approximations, which is just as well since only the references \cite{Morris:2016spn,Percacci:2016arh, Ohta:2017dsq} in the list above  actually go beyond this approximation). Wilsonian renormalization group concepts do not apply to a single sphere. In particular, 
although in these formulations, $k$ can be low enough on a sphere of given curvature $R$ that there are no modes left to integrate out, the fixed point equation should be viewed as summarising the state of a continuous ensemble of spheres of different curvatures. From the point of view of the ensemble there is nothing special about the lowest mode on a particular sphere. The renormalization group should be smoothly applied to the whole ensemble, and it is for this reason that one must require that smooth solutions exist over the whole domain $0\le r<\infty$.

This chapter is structured as follows. In section \ref{sec:overview} we introduce the fixed point ODE we will study and provide a summary of our findings i.e.\ the asymptotic solutions we uncovered. We also discuss what the solutions imply for the types of fixed points supported by the theory space, for example whether discrete sets or higher-dimensional spaces of fixed point solutions exist.
In the following sections, \ref{sec:pow-laws} and \ref{sec:non-pow}, we provide the details of how the asymptotic solutions are derived and therefore layout the techniques for developing asymptotic solutions in functional truncations in general.
We conclude this chapter with a discussion of our findings and their implications in section \ref{sec:conclusions}.
We finish this introduction by giving more detail on why the asymptotic solution is so important.

\subsection{Quantum fluctuations do not decouple}
\label{sec:quantum}

In the application of the RG to scalar field theory  \cite{Morris:1994ki,Morris:1994ie,Morris:1994jc,Morris:1998}, the leading asymptotic behaviour was always found by neglecting the right-hand side of the fixed point equation (or flow equation more generally). This made physical sense since the right-hand side encodes the quantum fluctuations, and at large field one would expect that these are negligible in comparison. Therefore the asymptotic solution in the scalar setting simply encodes the passage to mean field scaling, characteristic of the classical limit.

With functional approximations to quantum gravity, the situation is radically different. The leading asymptotic solution $\vp_{asy}(r)$ intimately depends on the right-hand side of the flow equation and never on the left-hand side alone. We will see this for the large $r$ solutions of the fixed point equation, \eqref{fp}, derived in \cite{Demmel2015b}. This behaviour confirms what had already been found in a different $f(R)$ approximation in \cite{Dietz:2012ic}, and also for a conformal truncation in \cite{Dietz:2016gzg}. 
Again this makes physical sense because here the analogue of large field is large curvature which therefore shrinks the size of the space-time and thus forbids the decoupling of quantum fluctuations. In fact by Heisenberg's uncertainty principle we must expect  that the quantum fluctuations become ever wilder. We note that it is the conformal scalar contribution that is determining the leading behaviour \cite{Dietz:2012ic,Demmel2015b,Dietz:2016gzg} and appears to be related to the so-called conformal instability\cite{Gibbons:1978ac,Dietz:2012ic,Dietz:2016gzg}.
In any case, we see that for quantum gravity, the asymptotic solution $\vp_{asy}(r)$ encodes the deep non-perturbative quantum regime.

\subsection{Physical part}
\label{sec:intro-phys}

The asymptotic solution contains the only physical part of the fixed point effective average action. Recall that the infrared cutoff $k$ is added by hand and the physical Legendre effective action \eqref{phys-eff-ac} is recovered only in the limit that this cutoff is removed. In scalar field theory, the analogous object is the universal scaling equation of state, which for a constant field precisely at the fixed point takes the simple form 
\be 
\label{Vscale}
V(\vp) = A \,\vp^{d/d_\vp}\,,
\ee 
where $d$ is the space-time dimension and $d_{\vp}$ is the full scaling dimension of the field (i.e.\ incorporating also the anomalous dimension). In the present case we keep fixed the constant background scalar curvature $R$. Thus by \eqref{ansatz EAA}, the only physical part of the fixed point action in this approximation is:
\be 
\label{phys}
f(R) |_{\text{phys}} = \lim_{k\to 0} k^4\, \vp(R/k^2) = \lim_{k\to 0} k^4\, \vp_{asy}(R/k^2)\,.
\ee
The significance of this object is further discussed in section \ref{sec:phys}, in the light of the results we uncover.

\subsection{Dimensionality of the fixed point solution space}
\label{sec:dims}

For given values of the parameters, the fixed point ODEs are too complicated to solve analytically,\footnote{although special analytical solutions were found by tuning the endormorphism parameters \cite{Ohta2016}.} and challenging to solve numerically.
However, the dimension $d_{FP}$ of their solution space, namely whether the fixed points are discrete, form lines, or planar regions etc., can be found by inspecting the fixed singularities and the asymptotic solutions. 

To see this we express the fixed point ODE in normal form by solving for the highest derivative:
\be
\label{norm}
\vp^{(n)}(r) = rhs\, ,
\ee
where $n=n_{ODE}$ is the order of the ODE, and $rhs$ (right-hand side) contains only rational functions of $r$ and lower order differentials\footnote{including $\vp$ itself i.e.\ $m=0$} $\vp^{(m<n)}(r)$.
The fixed singularities are found at points $r=r_i$ where this expression develops a pole for generic $\vp^{(m<n)}(r)$. For the solution to pass through the pole requires a boundary condition relating the $\vp^{(m<n)}(r)$, one for each pole. By a fixed singularity, we will mean one of these poles.

At the same time such non-linear ODEs suffer moveable singularities, points where $rhs$ diverges as a consequence of specific values for the  $\vp^{(m<n)}(r)$. The number of these that appear in practice depends on the solution itself. However, if the solution is to exist globally then it also exists for large $r$, where we can determine it analytically in the form of its asymptotic solution $\vp_{asy}(r)$, (this being the central topic of this chapter). The number of constraints implicit in $\vp_{asy}(r)$ is equal to $n_{ODE}-n_{asy}$, where $n_{asy}$ is the number of free parameters in $\vp_{asy}(r)$. This can be seen straightforwardly by noting that the maximum possible number of free parameters is $n_{asy}=n_{ODE}$; if $\vp_{asy}(r)$ contains any less then this implies that there are $n_{ODE}-n_{asy}$ relations between the $\vp^{(0\le m\le n)}_{asy}$ at any large enough $r$, which may be used as boundary conditions. Now, the number of moveable singularities that operate for a solution with these asymptotics is also equal to $n_{ODE}-n_{asy}$, providing we have uncovered the full set of free parameters in the asymptotic solution, as has been explicitly verified by now in many cases \cite{Morris:1994ki,Morris:1994ie,Morris:1994jc,Morris:1995he,Morris:1996nx,Morris:1997xj,Morris:1998,Dietz:2012ic,Bridle:2013sra,Dietz:2016gzg}. This follows because the moveable singularities can also occur at large $r$ where they influence the form of $\vp_{asy}$. Indeed linearising the ODE about $\vp_{asy}$, the perturbations can also be solved for analytically. The missing free parameters in $\vp_{asy}$ correspond to perturbations that grow faster than $\vp_{asy}$, overwhelming it and invalidating the assumptions used to derive it in the first place. These perturbations can be understood to be the linearised expressions of these moveable singularities \cite{Morris:1994ki,Morris:1994ie,Morris:1994jc}. 

To summarise, if the number of fixed singularities operating in the solution domain is $n_{s}$, then the dimension of the solution space is simply given by
\be 
\label{counting}
d_{FP} = n_{ODE} - n_{s} - (n_{ODE}-n_{asy}) = n_{asy} - n_{s}\,,
\ee
where $d_{FP}=0$ indicates a discrete solution set which may or may not be empty, 
and $d_{FP}<0$ corresponds to being overconstrained, i.e.\ having no solutions.
In section \ref{sec:dimensionality-ef}, we will illustrate this by working out the dimension of the fixed point solutions for eleven of the possible asymptotic behaviours. We discuss their significance in section \ref{sec:which}.

The counting argument \eqref{counting} also aids in the numerical solution. For example it tells us where it is hopeless to look for global numerical solutions, namely where $d_{FP}<0$, and to improve the numerical accuracy if the numerical solution apparently enjoys more free parameters than allowed by $d_{FP}$  \cite{Dietz:2012ic}. 

\subsection{Validation of the numerical solution}
\label{sec:validate}

A priori one might think that the analytical solutions for $\vp_{asy}(r)$ can be dispensed with in favour of a thorough numerical investigation. The problem is that without knowledge of $\vp_{asy}(r)$, there is no way to tell whether the numerical solution that is found is a global one or will ultimately end at some large $r$ in a moveable singularity. In fact, if the numerical solution is accessing a regime where the number of free asymptotic parameters $n_{asy}<n_{ODE}$, it will actually prove impossible to integrate numerically out to arbitrarily large $r$. Instead the numerical integrator is guaranteed to fail at some critical value. The reason is that it requires infinite accuracy to avoid including one of the linearised perturbations that grow faster than $\vp_{asy}$ which as we said, signal that the solution is about to end in a moveable singularity. On the other hand, if one can extend the solution far enough to provide a convincing fit to the analytical form of $\vp_{asy}(r)$, then one confirms with the requisite numerical accuracy that the numerical solution has safely reached the asymptotic regime
\cite{Dietz:2012ic,Bridle:2013sra,Dietz:2016gzg}, after which its existence is established, and its form is known, over the whole domain. We will see an example of this in section \ref{sec:numer-pow} where we will see that the numerical solution found in reference \cite{Demmel2015b} matches the power-law asymptotic expansion \eqref{full-pow-beta1/6}, but such that it would need to be integrated out twice as far in order to be sure of its asymptotic fate.

\section{Overview}
\label{sec:overview}

\subsection{Fixed point equation}

The fixed point equation that we will be studying is given by \cite{Demmel2015b}:
\be
\label{fp}
4\vp-2r\vp'=\frac{\tilde{c}_1\vp'-2\tilde{c}_2 r\vp''}{3\vp-(3\alpha r + r - 3)\vp'}+\frac{c_1\vp' + c_2 \vp'' - 2c_4 r \vp'''}{(3\beta r + r -3)^2\vp''+(3-(3\beta+2)r)\vp' + 2\vp}\,,
\ee
where $\vp(r)$ is defined in \eqref{dim vars} and prime indicates differentiation with respect to $r$.
This gives the scaled dependence on the curvature of a Euclidean four-sphere.
Recall that we are searching for smooth solutions defined on the domain $0\le r<\infty$. Each such solution is a fixed point of the renormalization group flow.

The coefficients $\tilde{c}_i$ and $c_i$ depend on $r$ and the endomorphism parameters $\alpha$ and $\beta$, and are given as
\begin{align}
\tilde{c}_1 &=-\frac{5 (6 \alpha  r+r-6) \left(\left(18 \alpha^2+9 \alpha -2\right) r^2-18 (8 \alpha +1) r+126\right)}{6912 \pi ^2}\nonumber\,,\\
\tilde{c}_2 &=-\frac{5 (6 \alpha  r+r-6) ((3 \alpha +2) r-3) ((6 \alpha -1) r-6)}{6912 \pi ^2}\,,\label{coeffs-c}\\
c_1&=-\frac{((6 \beta -1) r-6) \left((6 \beta -1) \beta  r^2+(10-48 \beta ) r+42\right)}{2304 \pi^2}\,,\nonumber\\
c_2&=\!-\frac{((6 \!\beta -1) r-\! 6)\! \left(\left(54 \beta ^2-3 \beta -1\right) \beta  r^3+\left(270 \beta^2\!+42 \beta -35\right) r^2\! -\!39 (18 \beta\! +1) r+378\right)}{4608 \pi ^2}\!\,,\nonumber\\
c_4&=\frac{(\beta  r-1) ((6 \beta -1) r-6)^2 ((9 \beta +5) r-9)}{4608 \pi ^2}\,.\nonumber
\end{align}
The endomorphism parameters arise from employing a type II cutoff $R_k=R_k(-\bar{\nabla}^2 + E_{(s)})$ with endomorphisms $E_{(s)}$ for the scalar ($s=0$) and transverse-traceless tensor modes ($s=2$) resulting from the transverse-traceless decomposition used to derive the flow equation.
For a spherical background, the authors of \cite{Demmel2015b} set $E_{(0)}=\beta R$ and $E_{(2)}=\alpha R$.
These endomorphism allow flexibility in how the tensor and scalar modes are integrated out \cite{Demmel2015b}. They are further discussed in sections \ref{sec:which} and \ref{sec:conclusions}.
The authors of \cite{Demmel2015b} also set
$\alpha=\beta-2/3$ 
and we will do the same.\footnote{This sets the lowest eigenvalues equal for the scalar and tensor modes. Following reference \cite{Morris:2016spn}, see also above section \ref{sec:dims},
it is not clear what significance should be attached to this however.}

\subsection{Fixed singularities}
\label{sec:fixed-sing-count}

The ODE \eqref{fp} is third order and thus admits a three-parameter set of solutions locally.
As reviewed in \ref{sec:dims}, the fixed singularities will limit this parameter space. The positions of the fixed singularities are determined by casting the flow equation into normal form \eqref{norm} (with $n=3$). The zeroes of the coefficient $c_4$ then give the points where the flow equation develops a pole. These poles are given by \cite{Demmel2015b}:
\be
r_1=0\,,\qquad\quad r_2=\frac{9}{5+9\beta}\,, \qquad\quad r_3=\frac{1}{\beta}\,, \qquad\quad r_{4,5}=\frac{6}{6\beta-1}\,.\nonumber
\ee
Note that there is a double root $r_{4,5}$ and that the root $r_1$, which is actually there for good physical reasons \cite{Benedetti:2012,Dietz:2012ic}, is always present, whereas the positions of the last 4 roots depend on the value $\beta$ takes. Different choices for $\beta$ will result in a different number of fixed singularities being present in the range $r\geq 0$, as shown in table \ref{table: fixed sings}.
\begin{table}
\begin{center}\begin{tabular}{ c|c }
Range of $\beta$ & Singularities\\
\hline
$1/6 < \beta$ & $r_1,\,r_2,\,r_3,\,r_{4,5}$ \\ 
$0<\beta \leq 1/6$ & $r_1, \,r_2,\,r_3$ \\ 
$-5/9<\beta\leq0$ & $r_1,\, r_2$ \\ 
$\beta\leq-5/9$ & $r_1$\\
\end{tabular}
\end{center}
\caption{List of fixed singularities present for different choices for $\beta$.}
\label{table: fixed sings}
\end{table}

If no additional constraints emerge from the asymptotic behaviour of the solution then choosing $0<\beta\leq 1/6$ leads to $d_{FP} =3-n_s=0$ by \eqref{counting}, which means that isolated fixed point solutions (or no solutions) can be expected.
The authors of \cite{Demmel2015b} choose $\beta = 1/6$ for this reason. It is also noted in \cite{Demmel2015b} that in addition this choice simplifies the numerical analysis.

We will analyse the fixed point equation for general $\beta$, both to uncover the extent to which the results depend on the particular choice and to demonstrate and explain the asymptotic methods in a large variety of examples. But since $\beta=1/6$ was chosen in reference \cite{Demmel2015b}, we will pay special attention to this value.

If $\beta>1/6$ is chosen then $d_{FP}<0$, the ODE is overconstrained and global solutions do not exist. On the other hand non-positive $\beta$ give rise to continuous sets, again assuming that no extra constraints are coming from the solution at infinity. For example $-5/9<\beta\leq0$, would give rise to only 2 fixed singularities, resulting in a one-parameter set of solutions i.e.\ a line of fixed points, while $\beta\le-5/9$ gives rise to a plane of fixed points. 

\subsection{Asymptotic expansions}
\label{sec:asymptotics-overview}

We now provide a list of all the asymptotic solutions that we found.
They can have up to three parameters, which are always called $A$, $B$ and $C$.

\paragraph{(a)} As covered in section \ref{Leading behaviour}, there exists a power-law solution where the leading power is $r^0$. It takes the form
\be 
\label{n=0}
\vp(r)= A + k_1/r^2+\cdots\,,
\ee
for all $\beta\ne0$, where $k_1$ is given by \eqref{n=0 k1}. 

\paragraph{(b)} For $\beta=0$, the subleading power is altered and the solution changes to 
\be
\label{n=0no2}
\vp(r)=A -\frac{18432\,{\pi }^{2}{A}^{2}}{535\,r} +\cdots\,.
\ee
as explained in section \ref{sec:excep-numer}. At this value of $\beta$ only this asymptotic solution is allowed.

\paragraph{(c)} As covered in section \ref{Leading behaviour}, for $n$ a root of \eqref{double root soln} such that $n<2$, the asymptotic solution
\be 
\label{n general subleading}
\vp(r)= Ar^n + k_1r^{2n-2}+\cdots\,,
\ee
with $k_1$ given by \eqref{n general k1}, exists for all $\beta \notin (-0.4835,-0.4273)$, except as explained in section \ref{sec:excep-numer} for $\beta=1/6$ and $\beta=0$, and except for the values
\be
\label{beta n=2}
\beta = \beta_\pm := \frac{3}{13} \pm \frac{\sqrt{285}}{78} = 0.01433,\ 0.4472 \,,
\ee
as explained at the end of section \ref{Leading behaviour}.  
When $n$ is complex, which happens for $\beta\in(-1.326,-0.4474)$ the parameter $A$ is in general also complex and the real part of \eqref{n general subleading}
should be taken leading to $\vp(r) \sim r^{\text{Re}(n)} \sin(\text{Im}(n)\log r +B)$ type behaviour. The values $n$ are plotted in figure \ref{scenario 2}.

\paragraph{(d)} As explained at the end of section \ref{Leading behaviour}, for the values \eqref{beta n=2}, $n=2$ is a root of \eqref{double root soln}, however the asymptotic solution is not given by \eqref{n general subleading} but instead by
\be
\label{n=2sol}
\vp(r) = Ar^2+k_1r+\cdots,
\ee
where $k_1$ is given by \eqref{n=2sol k1}.

\paragraph{}The techniques set out in sections \ref{Missing parameters} and \ref{missing params 2} need to be followed to find the missing parameters in the above solutions (a) -- (d) before we can discover the dimension $d_{FP}$ of their corresponding solution space. Of course we already know from table \ref{table: fixed sings} that there are no solutions (i.e.\ $d_{FP}<0$) for $\beta>1/6$. For the following remaining asymptotic solutions we also uncover the missing parameters.

\paragraph{(e)} For generic $\beta$ the following solution:
\be
\label{solution}
\vp(r) = \vp_{pow}(r):= A\,r^{3/2} + k_1 \, r + k_2 \, r^{1/2} + k_3\, \log\left(\frac{r}{b}\right)
+k_4\frac{\log\left(\frac{r}{b}\right)}{\sqrt{r}}+ \frac{k_5}{\sqrt{r}} +\cdots\,,
\ee
where $B=\log b$ is a second free asymptotic parameter,
forms the basis for the asymptotic solutions below. As  explained in section  \ref{Exceptions} it fails to exist for $\beta=0$, 1, and $\pm1/\sqrt{27}$. The leading part is derived in sec. \ref{Leading behaviour} and the subleading parts in section \ref{Sub-leading behaviour}. The subleading coefficients are functions of $A$ and $\beta$, where $k_1$ is given in \eqref{Fval} and the others are given in appendix \ref{AppendixA}. As explained in section \ref{exceptions}, exceptions develop at poles of these subleading coefficients where the corresponding term and subleading terms then develop an extra $\log r$ piece. As shown in section \ref{Missing parameters}, the full asymptotic solution  is then one of  the following forms:
\begin{align}
&\vp_{pow}(r) \,, \!\! &&\beta\in\! (-\infty,-0.1809)\!\cup\!(0.1931,0.4042)\!\cup\!(0.8913,\infty)\backslash\{-\frac59\},\label{p3pos}\\
&\vp_{pow}(r) + C\, r^{p_3 +\frac{3}{2}}+\cdots, \! &&\beta\in(-0.1809,0.1931)\cup(0.4042,0.8913)\backslash\{\frac16\}\,,\label{p3neg}\\
&\vp_{pow}(r) + C\, r^2 e^{-\frac{r^2}{351}}+\cdots, \!&&\beta=\frac16\,,\label{full-pow-beta1/6}\\
&\vp_{pow}(r) + C\, r^4 e^{-\frac{33223}{31941}r}+\cdots, \!&&\beta=-\frac59\,,\label{powm59}
\end{align}
where $p_3<0$ is given by \eqref{p3} and the ellipses stand for further subleading terms that will mix powers of the new piece and its free parameter, $C$, with the powers of the terms in \eqref{solution}. The power $p_3$ is plotted in figure \ref{fig p3}. Since the authors of  \cite{Demmel2015b} use $\beta=1/6$, the solution \eqref{full-pow-beta1/6} is of particular interest. We show in section \ref{sec:numer-pow} that it provides a match to their numerical solution, as far as it was taken.

\paragraph{(f)} Except for $\beta=0$, and $\beta=\beta_\pm$ as in \eqref{beta n=2}, as discussed in section \ref{exceptionss}, the asymptotic series
\be 
\label{non-pow}
\vp(r) = r^2 f_{asy}\left(\log(r/A)\right)
\ee 
forms the basis for the asymptotic solutions below, where\footnote{$f_{asy}(x)$ should not be confused with the Lagrangian $f(R)$ defined via \eqref{dim vars}.}
\be 
f_{asy}(x)=k_1 x+k_2\log(x)+k_3\dfrac{\log(x)}{x}+\dfrac{k_4}{x}+k_5\dfrac{\log^2(x)}{x^2}+\cdots \,.
\label{vptotal}
\ee
For $\beta \ne -1/3$, $5/6$, the coefficient $k_1$ is derived in section \ref{sec:non-pow-leading} and is
given in \eqref{non-pow k1}, while the other $k_i$ are derived in section \ref{sec:non-pow-subleading} and are given in appendix \ref{kivalues}. As explained in section \ref{exceptionss}, for $\beta=-1/3$ and $5/6$ the coefficients take different values as given in \eqref{ki13} respectively \eqref{ki56}.
The arguments $x$ in the logs in \eqref{vptotal} can be replaced by $x/c$ as in \eqref{f} but as shown in section \ref{sec:non-pow-subleading} this is not an extra parameter and can be absorbed into the free parameter $A$ in \eqref{non-pow}. The full asymptotic solution then takes the following forms:
\beal 
f_{asy}(x) \phantom{+} &,\qquad\qquad\qquad\qquad\qquad\qquad \beta\in \left(\b_-,\b_+\right)\,,\label{non-pow-only}\\
f_{asy}(x) + &\,B e^{-\tfrac{2 }{3}\sqrt{-\tfrac{2}{h_3}}\,x^{\frac32}}\,,\qquad\qquad\quad \beta\in \left(-\infty,-\tfrac{5}{9}\right)\cup\left( -\tfrac{1}{3},\b_-\right) \cup\left(\b_+,\infty\right)\backslash\left\{0,\tfrac{5}{6}\right\}\,,\label{non-powExp32}\\
f_{asy}(x) + &\, \left\{ B \cos\left(\tfrac{2  }{3}\sqrt{\tfrac{2}{h_3}}\,x^{\frac{3}{2}}\right)+C \sin\left(\tfrac{2  }{3}\sqrt{\tfrac{2}{h_3}}\,x^{\frac{3}{2}}\right) \right\} \, e^{4h_2 x/h_3}\,,\ \
\beta\in \left(-\tfrac{5}{9},-\tfrac{1}{3}\right)\backslash\left\{-0.4111\right\}\,,
\label{non-powOsc32}\\
f_{asy}(x) + &\, B\, e^{-L_+x}\,, \qquad\qquad \qquad\qquad\beta=-\frac13\,, \label{non-powExpL}\\
f_{asy}(x) + &\,B 
e^{-\tfrac{2\sqrt{21}}{15}\,x^{\frac{3}{2}}}\,,
\qquad\qquad \qquad\quad\beta=\frac56\,,\label{non-powExp32alt}\\
f_{asy}(x) + &\,B\, e^{-\frac{23056}{22815}\,e^x}\,,\qquad\qquad \qquad\quad\beta=-\frac59\,,\label{non-powExpExp}\\
f_{asy}(x) + &\, B\, x^q \cos\left(L_I\,x^{\frac{3}{2}}\right)+C\, x^q \sin\left(L_I\,x^{\frac{3}{2}}\right)\,,\qquad\qquad\qquad \beta=-0.4111\label{non-powOsc}
\eeal
where the positive square root is taken, $h_2$ and $h_3$ are defined in \eqref{h2nonpow} and \eqref{h3nonpow},  $L_+$ in \eqref{Lpm}, $L_I=1.0648$ and $q=-2.1499$.
The top three solutions are derived in section \ref{missing params 2}. However \eqref{non-pow-only} required a separate analysis for $\beta=1/6$ and $0.3800$, in sections \ref{sec:third derivative} and \ref{k2sec} respectively.
The next two solutions are derived in section \ref{sec:altered}, \eqref{non-powExpExp} is derived  in section \ref{sec:third derivative}, and \eqref{non-powOsc} is derived in section \ref{k2sec}.

\subsection{Physical part}
\label{sec:phys}
Using \eqref{phys}, we can now extract the corresponding physical parts. As noted in section \ref{sec:intro-phys}, these give the universal equation of state precisely at the fixed point, analogous to \eqref{Vscale} in a scalar field theory. We see that except for the cases (d) and (f) discussed below, the result vanishes:
\be 
f(R) |_{\text{phys}} = 0\,.
\ee
The asymptotic solution \eqref{full-pow-beta1/6} which matches the numerical solution in 
\cite{Demmel2015b}, thus also falls in this class.
Similar results were obtained for cases in the conformal truncation model of \cite{Dietz:2015owa,Dietz:2016gzg} where also divergent results were found. Perhaps these indicate that these do not give a sensible continuum limit, although a fuller understanding is needed for example by moving away from the fixed point by including relevant couplings.

For case (d), from \eqref{n=2sol} we get (for $\beta=\beta_\pm$):
\be 
f(R) |_{\text{phys}} = A R^2\,.
\ee
This equation of state was also found in reference \cite{Dietz:2012ic} for the $f(R)$ approximation given in \cite{Benedetti:2012}, and for solutions found in reference \cite{Ohta2016}.
For any of the solutions for case (f),  we get from \eqref{non-pow} and \eqref{vptotal} that
\be 
f(R) |_{\text{phys}} = k_1 R^2 \log(R) - k_1 R^2 \log(Ak^2)\,,
\ee
and $k_1$ is given by \eqref{non-pow k1}. Since we require $k\to0$, the second term is a positive logarithmic divergence. It is perhaps a signal of the asymptotic freedom of the $R^2$ coupling in this case where thus it should be treated as in reference \cite{Falls:2016msz}. As we will see in the next section, global solutions with the asymptotics of case (f) exist only for $\beta<0$ and therefore $k_1$ is always positive.

\subsection{Dimensionality of the fixed point solution spaces}
\label{sec:dimensionality-ef}

Using the counting argument \eqref{counting} and table \ref{table: fixed sings} we can read off the dimensionality of the corresponding fixed point solution spaces for cases (e) and (f). 

\paragraph{(e)} We see that  since \eqref{solution} has two free parameters, \eqref{p3pos} only extends to global solutions for $\beta$ in the negative interval. When $\beta>-5/9$ the fixed points, if any, form a discrete set, while lines of fixed points are found for $\beta<-5/9$. The other solutions, \eqref{p3neg} -- \eqref{powm59}, all have three free parameters and thus provide no constraints on the dimension of the solution space, which thus follows the pattern discussed in section \ref{sec:fixed-sing-count}. In particular \eqref{p3neg} extends to a global solution only for $\beta<1/6$ where it can have discrete fixed points or lines of fixed points depending on the sign of $\beta$, \eqref{full-pow-beta1/6} has discrete solutions, and \eqref{powm59} has planar regions of fixed points. 

\paragraph{(f)} Using \eqref{beta n=2}, we see that  since \eqref{vptotal} has only one free parameter, solutions with asymptotic behaviour \eqref{non-pow-only} and \eqref{non-powExp32alt} do not exist since they are overconstrained by the fixed singularities, \eqref{non-powExp32} has discrete solutions for $-1/3\le\beta<0$ and lines of fixed points for $\beta\le -5/9$, \eqref{non-powExpL} has discrete solutions, and 
\eqref{non-powOsc32}, \eqref{non-powExpExp} and 
\eqref{non-powOsc} all generate lines of fixed points.

\subsection{Which fixed point?}
\label{sec:which}

As we have seen, depending on the value $\beta$ takes, either discrete or continuous sets of fixed point are produced.
As already pointed out in section \ref{sec:theory space} of the Introduction, from the point of view of the asymptotic safety programme, it would be phenomenologically preferable if the correct answer lay in only one of the discrete sets: \eqref{p3pos} for $\beta\in(-5/9,-0.1809)$, \eqref{p3neg} for $\b\in(0,\,1/6)$, \eqref{full-pow-beta1/6} (the choice made in reference \cite{Demmel2015b}), \eqref{non-powExp32} for $\b\in(-1/3,\,0]$, or \eqref{non-powExpL}. However we would still need a convincing argument for choosing one solution over the others. 

Note that no lines or planes of fixed points are found for cases (e) and (f) if $\beta>0$. This can be seen straightforwardly by recognising that the maximum number of parameters in any solution is 3 which is always less than or equal to the number of fixed singularities for positive $\beta$.
Inspection of the form of the cutoff functions used in the derivation of \eqref{fp}, see eqn. (3.13) in reference \cite{Demmel2015b}, shows that $\beta<0$ corresponds to cases where some scalar modes never get integrated out, no matter how small we take $k$. One therefore could argue that for $\beta<0$ the Wilsonian renormalization group is undermined and that solutions in this range (which means all the continuous sets) should be excluded.

Continuous sets of solutions \cite{Dietz:2012ic} were also found with another approach to the $f(R)$ approximation \cite{Benedetti:2012}, and there also there is a scalar mode that never gets integrated out. It would be very useful to know if this correlation is found for other formulations \cite{Percacci:2015wwa,Labus:2015ska,Eichhorn:2015bna,Demmel:2014fk,Demmel:2012ub,Demmel:2013myx,Benedetti:2013jk,Demmel:2014hla,Demmel2015b,Ohta:2015efa,Ohta2016,Percacci:2016arh,Falls:2016msz,Ohta:2017dsq} in the literature.

For these continuous solutions it could also be, like in reference \cite{Dietz:2012ic}, that the $f(R)$ approximation is breaking down there, such that the whole eigenspace becomes redundant \cite{Dietz:2013sba}. To check this would require developing the full numerical solutions. 

It could also be that these continuous sets are artefacts caused by the violation of background independence \cite{Bridle:2013sra}, which as we have seen in the previous chapter can be problematic. We saw there that in the LPA, providing the msWI that reunites the scalar field $\phi$ with its background counterpart $\bar{\phi}$ is satisfied, the spurious behaviour is cured.
The implementation of background scale independence in references \cite{Morris:2016spn,Percacci:2016arh,Ohta:2017dsq} is arguably the equivalent step for the $f(R)$ approximation, since it reunites the constant background curvature $\bar{R}$ with the multiplicative constant conformal factor piece of the fluctuations. The resulting formulations can be close to the single-metric approximation used to derive \eqref{fp},  in the sense that minimum changes are needed (e.g.\ setting space-time dimension to six, or choosing a pure cutoff), to convert the fixed point equation into a background scale independent version. It could therefore be promising to investigate formulations with these changes.

On the other hand, continuous solutions were found in the conformal truncation model of reference \cite{Dietz:2016gzg}, where a clear cause was found in the conformal factor problem \cite{Gibbons:1978ac} as discussed in section \ref{sec:approxs} of the chapter \ref{cha:introduction}. In \cite{Dietz:2016gzg} the cutoff implementation did not introduce fixed singularities, background independence was incorporated \cite{Dietz:2015owa}, all modes were  integrated out, and an analogous breakdown to the $f(R)$ approximation \cite{Dietz:2013sba} was either not there or not possible. Again, this suggests that the issues go deeper.

\paragraph{} In the next two sections, we provide the details of how the asymptotic solutions were discovered and developed.

\section{Asymptotic expansion of power law solutions}
\label{sec:pow-laws}

Finding asymptotic solutions initially requires a degree of guesswork. A profitable place to start is to assume that the asymptotic series expansion of the solution starts with a power, i.e.\
\be
\label{ansatz}
\vp(r)= A\,r^n + \cdots\,,
\ee
where  $A\ne0$ is an arbitrary coefficient, and subsequent terms need not be powers but are successively smaller than the leading term for large $r$.

Now, requiring that \eqref{ansatz} satisfies the fixed point equation \eqref{fp} means that at large $r$ the leading piece in this equation must itself satisfy the equation. In this way we typically determine $n$ and sometimes also $A$. The leading piece of the fixed point equation will be satisfied either because the left-hand side and the right-hand side  provide such a piece which are then equal for appropriate values of $n$ and $A$, or because only one side of the equation has such a leading piece but this can be forced to vanish by appropriate values of $n$ and $A$. In this sense we require the leading terms to `balance' in the fixed point equation. As we will see, requiring then the sub-leading pieces also to balance will determine the form of the next terms in \eqref{ansatz}.

\subsection{Leading behaviour}
\label{Leading behaviour}
We begin by finding the leading behaviour of $\vp(r)$  i.e.\ solving for the power $n$ in \eqref{ansatz}. We build the asymptotic series leaving $\beta$ unspecified for the reasons given at the end of section \ref{sec:fixed-sing-count}.
We plug the solution ansatz \eqref{ansatz} into the fixed point equation \eqref{fp}, expand about $r=\infty$ and keep only the leading terms in the large $r$ limit. Since the coefficients \eqref{coeffs-c} are expanded along with everything else and only their leading parts are kept, it is useful to introduce the following definitions:
\begin{align}
\label{coeffs d}
 \tilde{c}_1\sim\,\tilde{c}_2 &\sim\,-\frac{5\beta(2\beta-1)(6\beta-5)}{768\pi^2}
\,r^3\,\equiv \tilde{d}_1 \, r^3 \,,\nonumber\\
c_1 &\sim\, -\frac{\beta(6\beta-1)^2}{2304\pi^2}
\,r^3 \,
\equiv d_1 \, r^3 \,,\nonumber\\
c_2 &\sim \, -\frac{\beta (6\beta-1)^2(9\beta+1)}{4608\pi^2}
\,r^4 \,
\equiv d_2\,r^4 \,,\nonumber\\
c_4 &\sim\, \frac{\beta (6\beta-1)^2(9\beta+5)}{4608\pi^2}\,
r^4\,
\equiv d_4\, r^4\,,
\end{align}
where we have rewritten $\tilde{c}_1$ and $\tilde{c}_2$ using $\alpha=\beta-2/3$. 
For functions $f(r)$ and $g(r)$, $f(r) \sim g(r)$ means that $\lim_{r\rightarrow\infty}\,{f(r)}/{g(r)} = 1$. We note that for certain values of $\beta$, the leading behaviour of the coefficients will be different from those given in \eqref{coeffs d} since the leading coefficients will vanish. We discuss this in section \ref{Exceptions}, and comment there and below on the case of $\beta={1}/{6}$.

Inserting ansatz \eqref{ansatz} into the fixed point equation \eqref{fp}, we find that the leading piece on the left-hand side as $r\rightarrow\infty$ is simply
\be
\label{lhs}
(4-2n) A r^n\,,
\ee
and the leading piece on the right-hand side is given by
\be
\label{rhs}
\left\{ \frac{ n(3-2n)\, \tilde{d}_1 }{3-3  \beta  n+ n }
+\frac{ n d_1+n (n-1) d_2-2 n(n-1) (n-2)   d_4}{ n(n-1) \left(9 \beta ^2+6 \beta +1\right) -3    n\beta-2 n+2 } \right\} 
r^2\, ,
\ee
where we  substituted $\alpha=\beta-2/3$ in the first fraction. The important observation here is that the left-hand side goes like $r^n$ whereas the right-hand side goes like $r^2$, and thus which side dominates will be decided by whether $n$ is less than, greater than or equal to 2. Below we investigate these possible scenarios to determine the power $n$. We recognise that the scaling behaviour of the right-hand side could differ from $r^2$ if cancellations were to occur in either the denominators or the numerators. This is discussed in section \ref{Exceptions}.

\paragraph{Scenario (1):}
For $n>2$, the left-hand side \eqref{lhs} dominates and so we require that $n$ be chosen to set the left-hand side to zero.
However we see immediately that this is only true if $n=2$ and thus we reach a contradiction.

\paragraph{Scenario (2):}
For $n\le2$, the leading right-hand side piece, \eqref{rhs}, must vanish. For $n<2$ this is because the right-hand side dominates, whilst for $n=2$ it must happen because we have just seen that in that case the left-side vanishes on its own. Thus for this scenario, we require $n$ to be such that the coefficient of $r^2$ in \eqref{rhs} vanishes i.e.\ we require
\be
0 = \frac{ n\tilde{d}_1-2 n (n-1) \tilde{d}_2 }{3-3  \beta  n+ n }
+\frac{ n d_1\,+n (n-1) d_2\,-2 n(n-1) (n-2)   d_4 }{ n(n-1) \left(9 \beta ^2+6 \beta +1\right) -3    n\beta-2 n+2 }\, .
\ee
Solving this for generic $\beta$ we find four solutions for $n$, which we now describe.

\paragraph{} One solution is 
\be
\label{n=3/2}
n=3/2 \,.
\ee
In order to know whether this solution leads to a valid asymptotic series we must take the expansion to the next order to check whether subsequent terms are genuinely sub-leading. We pursue this solution in section \ref{Sub-leading behaviour} where we see that it does indeed allow us to build a legitimate asymptotic series, which we work out in detail to demonstrate the general method.
As we will see in section \ref{Missing parameters} there are values of $\beta$ for which this choice of scaling is not allowed. However, for generic of $\beta$, we can say that one possibility for the leading term in an asymptotic expansion \eqref{ansatz} is therefore
\be
\label{LO}
\vp(r) \sim A \, r^{3/2}\,.
\ee
This leading behaviour agrees with the quantum scaling found in \cite{Demmel2015b}, but is in conflict with the classical and balanced scaling $r^2$ that the authors ultimately use to approximate the large $r$ behaviour of their solution. In fact the full solution we find, namely \eqref{full-pow-beta1/6}, 
does not agree with that suggested in reference \cite{Demmel2015b}, 
but does match the numerical solution they found quite acceptably, as we show in section \ref{sec:numer-pow}.


Another solution is $n=0$. This is already clear from \eqref{fp} and follows from the fact that the numerators depend only on differentials of $\vp$. Setting $n=0$, we have arranged that the leading $r^2$ term vanishes, so now we turn to the subleading term. We see that the left-hand side of \eqref{fp} will provide an $r^0$ piece. For $\vp=A$ to be the beginning of an asymptotic series we will need to balance this piece with a term on the right-hand side. Whatever subleading term we add, it will generically no longer be annihilated by the numerators in \eqref{fp}. Meanwhile the denominators will go like a constant for large $r$ (from the undifferentiated $\vp$ parts). Thus, using \eqref{coeffs d}, we see by inspection that the subleading piece goes like $1/r^2$. By expanding \eqref{fp} in an asymptotic expansion and matching coefficients we thus obtain equation \eqref{n=0} with
\be 
\label{n=0 k1}
k_1 = \frac{18432\pi^2A^2}{7\beta(972\beta^3+528\beta^2-497\beta+117)}\,.
\ee
We could continue to investigate this asymptotic solution, developing further subleading terms and finding out how many parameters it ultimately contains, but  this solution should be a very poor fit to the numerical solution found in reference \cite{Demmel2015b} which numerically shows behaviour identified in reference \cite{Demmel2015b} as $\vp\sim r^2$ for large $r$.

\begin{figure}
\centering
\includegraphics[scale=1.3]{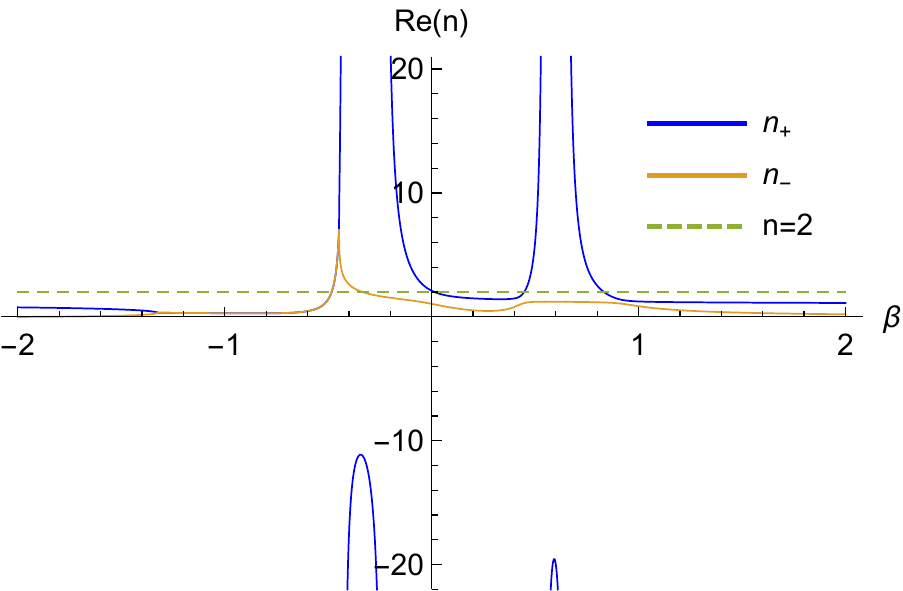}
\caption{Plot of the two solutions $n_\pm$ given by \eqref{double root soln}.} 
\label{scenario 2}
\end{figure}
The last two solutions are functions of $\beta$ and are given by the roots of the quadratic
\begin{multline}
\label{double root soln}
0=\left( 4212\,{\beta}^{4}-2268\,{\beta}^{3}-1395\,{\beta}^{2}+486\,
\beta+145 \right) {n}^{2}\\
+ \left( -4212\,{\beta}^{4}+2043\,{\beta}^{2}
+147\,\beta-458 \right) n
+972\,{\beta}^{3}+1152\,{\beta}^{2}-1185\,
\beta+321\,.
\end{multline}
The solutions \eqref{double root soln} are plotted in figure \ref{scenario 2}.
These roots take complex values when $-1.326<\b<-0.4474$ and so the allowed solutions which we want are  those for which Re$(n)\le2$. The only region where there is not a solution Re$(n)\le2$ is from the point where Re$(n)$  crosses the $n=2$ line in this range, namely at $\b=-0.4835$, through to the point where the `blue' root diverges, namely $\b=-0.4273$.

Again, in order to know whether the Re$(n)\le2$ solutions of \eqref{double root soln} really lead to valid asymptotic series we need to take the expansion to the next order. We  pursue this in a similar way to the $n=0$ case above. We know that the left-hand side of \eqref{fp} $\sim r^n$ and this term will need balancing by terms on the right-hand side. On the right-hand side the denominators also $\sim r^n$, whereas the numerators, which would have gone like $r^{n+2}$, have had the corresponding coefficient cancelled by choosing \eqref{double root soln}. Therefore the subleading term we need to add is a piece $\zeta(r)$ such that 
\be 
r^3 \zeta'(r) /r^n \sim r^n\,,
\ee
where we have trialled just the first term in the first numerator on the right-hand side of \eqref{fp} and compared it to the left-hand side. Solving this gives $\zeta\sim r^{2n-2}$. Since $2n-2 = n + (n-2)$, we see that this is genuinely subleading only if Re$(n)<2$. 
By inspection, such a power law solution then works out for the full numerators on the right-hand side of \eqref{fp}. Substituting these first two terms of the asymptotic series into \eqref{fp} and expanding for large $r$ we find the coefficient
\be
\label{n general k1}
k_1 = \frac{6912{A}^{2}{\pi }^{2} ( n-1 ) ^{2}{n}^{3}}{36{n}^{2} ( 16n-19)  ( n-1) ^{2}{\beta}^{2}-6n ( n-1 )  ( 76{n}^{3}-196{n}^{2}+140n-15) \beta+ c(n) }\,,
\ee
where we have set
\be
c(n) = ( n-3 )  ( 36{n}^{4}-139{n}^{3}-
210{n}^{2}+713n-420)\,.
\ee
We thus obtain our last two power-law asymptotic solutions \eqref{n general subleading}.
When \eqref{double root soln} has real roots, those with $n<2$ are taken. When complex, the real part of \eqref{n general subleading} should be taken, leading to $\vp(r) \sim r^{\text{Re}(n)} \sin(\text{Im}(n)\log r +B)$ type behaviour.

For $n=2$, both the left and right-hand side scale as $r^2$ and therefore could be expected to balance. As we have already seen however, if $n=2$ the left-hand side is identically zero and so again we require the right-hand side \eqref{rhs} to vanish, but now with $n$ set to 2.
As we will see in section \ref{sec:non-pow} this impasse gives a clue however to a non-power law asymptotic solution. 
Pursuing for now the power law case \eqref{ansatz} but with fixed $n=2$, we find that this equation is satisfied either for $\beta=\beta_\pm$ where $\beta_\pm$ is defined in \eqref{beta n=2},
or apparently for $\beta=0$, since then all the $\tilde{d}_i$ and $d_i$ vanish. The exceptional case of $\beta=0$ is discussed in section \ref{Exceptions}. The solutions \eqref{beta n=2} are just values of $\beta$ such that one of the roots of \eqref{double root soln} is indeed $n=2$. Substituting $n=2$ in \eqref{double root soln} gives a quartic in $\b$, but the other two roots, $\b=-1/3$, $5/6$, are cancelled at $n=2$ by the denominator in \eqref{rhs}. The remaining quadratic is in fact the one that appears in the denominator of coefficients \eqref{h2nonpow} and \eqref{h3nonpow} that we will come across later.

Again, to validate the $n=2$ solution \eqref{beta n=2}, we need to show that the expansion can be taken to the next order. We know the expansion for general $n$ given in \eqref{n general subleading} breaks down for $n=2$. In fact in this case, since the left-hand side of \eqref{fp} already vanishes, the subleading term comes from the next term on the right-hand side in a large $r$ expansion. In this way we see that the asymptotic solution is \eqref{n=2sol} where 
\be 
\label{n=2sol k1}
k_1={\frac {312\,A \left( -21353+363048\,\beta \right) }{17166809088\,A{
\pi }^{2}\beta-14017536\,A{\pi }^{2}+10800590\,\beta-110555}}\,,
\ee
and $\beta$ is either root in \eqref{beta n=2}.

We see that overall there are in general four types of power-law asymptotic solutions given by the power being one of the roots \eqref{double root soln} providing $\beta$ is such that Re$(n)\le$ 2, or two $\beta$ independent cases: $n=0$ and $n=3/2$.

\subsection{Exceptions}
\label{Exceptions}
As mentioned previously, we recognise that certain choices for $\beta$ alter the scaling behaviour of the right-hand side of the fixed point equation such that it differs from $r^2$ in the limit $r\to\infty$. 

\subsubsection{Exceptions from the denominators}
\label{excep-denom}
For instance, the leading behaviour could increase to $r^3$ if the $r^n$ terms in one or both of the denominators were to cancel amongst themselves. These cancellations occur in the first and second fractions respectively when
\be
\label{except1}
n=\frac{3}{3\beta-1}
\ee
or
\be
\label{except2}
n=\frac{3+9 \beta +9 \beta ^2\pm\sqrt{81 \beta ^4+162 \beta ^3+63 \beta ^2+6 \beta +1}}{2 \left(9 \beta ^2+6 \beta +1\right)}\, .
\ee
Concentrating on the asymptotic solution with $n=3/2$, we find that \eqref{except1} and \eqref{except2} are satisfied when $\b=1$ and $\b=\pm{1}/{\sqrt{27}}$ respectively. For these values of $\b$, a leading power of $n={3}/{2}$ is not allowed and instead to find the leading behaviour in these cases we must treat $\b=1$ and $\b=\pm{1}/{\sqrt{27}}$ separately from the start. As we will see below in these cases we just recover the other three power-law solutions. 

\paragraph{$\boldsymbol{\beta=1:}$} the leading piece in the limit $r\rightarrow\infty$ on the left-hand side is still of course given by \eqref{lhs}. 
Indeed this will be the case for any $\beta$ as the left-hand side contains no coefficients $c_i,\tilde{c}_i$ and is therefore independent of $\beta$. 
Also, since $\beta=1$ does not correspond to one of the values at which the leading parts of the coefficients $c_i,\tilde{c}_i$ vanish, see \eqref{coeffs d}, the leading piece on the right-hand side will still be given by \eqref{rhs}, but now with $\beta$ set to 1:
\begin{equation}
\label{rhs beta 1}
-\frac{5 n r^2}{768 \pi ^2} - \frac{25 \left(14 n^3-37 n^2+24 n\right) r^2}{2304\pi ^2 \left(16 n^2-21 n+2\right)}\,.
\end{equation}
Note that even though $\beta=1$ has been identified as a value at which the $r^n$ terms in the first denominator of \eqref{rhs} vanish, this is only when $n=3/2$ and so here we still see the right-hand side scaling as $r^2$. The scaling of the right and left-hand sides is the same as that in section \ref{Leading behaviour} and so by the same reasoning we see that $n$ must be less than 2 and therefore \eqref{rhs beta 1} must vanish in order to satisfy the fixed point equation for large $r$. 
The right-hand side \eqref{rhs beta 1} vanishes when $n=(62 \pm\sqrt{127})/59$ and also trivially for $n=0$. Indeed these are the remaining power-law solutions, in particular the former pair are the values for $n$ given by \eqref{double root soln} with $\beta=1$ as expected, while the latter is the solution \eqref{n=0}.


\paragraph{$\boldsymbol{\beta=\pm 1/ \sqrt{27}:}$} the right-hand side of the fixed point equation again scales like $r^2$ in the large $r$ limit, as this $\beta$ is also not one of the exceptional values appearing in \eqref{coeffs d}. 
We see that again $n$ must be less than 2 and that the right-hand side must vanish. Once again, the values of $n$ just correspond to  \eqref{double root soln}  when $\beta=\pm{1}/{\sqrt{27}}$.
And we still also have the $n=0$ solution \eqref{n=0}.

\subsubsection{Exceptions from the numerators}
\label{sec:excep-numer}
Exceptions to the leading behaviour $n=\nfrac{3}{2}$ also arise from particular choices for $\beta$ reducing the powers of $r$ appearing in the coefficients $\tilde{c}_i$ and $c_i$ which could result in an overall decrease in the leading power on the right-hand side of the fixed point equation. The values of $\beta$ for which the leading power of $r$ in the coefficients vanishes can be easily found from \eqref{coeffs d}.
Notably however, both numerators in \eqref{rhs} are satisfied independently for $n=\nfrac{3}{2}$. This means that for the values $\beta=\nfrac{1}{2},\nfrac{5}{6},\nfrac{1}{6}$ for which only one of the fractions becomes sub-dominant the $\beta$-independent solution $n=\nfrac{3}{2}$ remains valid. 

\paragraph{$\boldsymbol{\beta=1/6:}$} although we are concentrating on the $n=\nfrac32$ solution, for completeness we note that $\beta=\nfrac16$ does present an exception for the general power solutions \eqref{double root soln}. From \eqref{double root soln}, we would expect to find asymptotic series with leading powers $n= (19\pm\sqrt{73})/18 = 1.530, 0.5809$. However when $\beta=\nfrac16$, we see from \eqref{coeffs d} that $d_1$, $d_2$ and $d_4$ all vanish. Then from \eqref{rhs} we see that in this case the only solutions left for $n$ are the $n=0$ and $n=\nfrac32$ cases established in section \ref{Leading behaviour}.

\paragraph{$\boldsymbol{\beta=0:}$} in this case the leading powers of $r$ in all the coefficients vanish, cf.\ \eqref{coeffs d}.
(We see the implications of having $\b=0$ in section \ref{Sub-leading behaviour} where it represents a pole of all but one of the coefficients in the asymptotic expansion given in appendix \ref{AppendixA}.)
As a result, the right-hand side of the fixed point equation \eqref{fp} in general no longer scales as $r^2$ but instead goes like $r$. This apparently implies two solutions: either the asymptotic series is $\vp = A r^2 + k_1 r +\cdots$, since $r^2$ satisfies the left-hand side on its own, or  the asymptotic series takes the form $\vp = Ar+\cdots$, with the $r$ term then balancing both sides of the equation. However substituting $\vp=Ar^n$ into the right-hand side of \eqref{fp} (with $\beta=0$) we find the leading term is:
\be 
\label{beta0leading}
-{\frac {5\,n \left( 58\,{n}^{3}-389\,{n}^{2}+792\,n-477 \right) }{
4608\,{\pi }^{2} \left( n+3 \right)  \left( n-1 \right)  \left( n-2
 \right) }}\, r\,,
\ee
which thus presents an exception for both of these cases! The reason for this is that $n=1,2$ happen to be precisely the two powers that reduce the leading power in second denominator in this case, as can be seen from \eqref{except2}. Thus actually when $n=1$ or $n=2$, the second term on the right-hand side of \eqref{fp} contributes $\sim r^2$. Since it does so now with no free parameters ($\beta$ and $n$ having been fixed), neither suggested asymptotic solution will work: for $n=1$ because the correction is larger than the supposed leading term, while for $n=2$ there is nothing to balance it since the left-hand side vanishes identically. Finally, we consider general $n$. For $n>1$, the left-hand side dominates and we require that it vanishes for the fixed point equation to be satisfied, but this only gives us the already excluded $n=2$ solution. For Re$(n)<1$ the right hand side dominates and \eqref{beta0leading} must vanish on its own. The cubic in the numerator has no roots in this region and thus we are left with only an $n=0$ solution, namely \eqref{n=0no2}.
Note that this differs from \eqref{n=0}, in particular the first subleading power is now $1/r$.

\paragraph{} 
To summarise for the $n=3/2$ asymptotic series in particular, we have seen that this fails to exist at $\beta = 0$, 1, and $\pm 1/\sqrt{27}$. However as we will see, in general these exceptions do not obstruct the construction of the subleading terms or the subsequent determination of the missing parameters.

\subsection{Sub-leading behaviour}
\label{Sub-leading behaviour}
In this section we present the method for determining the subsequent terms in the asymptotic series \eqref{ansatz}. As already stated, we will concentrate on the $n=3/2$ solution.
We have seen in the previous section that for generic $\beta$ the fixed point equation scales like $r^2$ for large $r$. Choosing $\vp\sim A\, r^{\nfrac{3}{2}}$ causes the $r^2$ contribution to disappear, therefore satisfying the equation in the large $r$ limit. This is precisely because a leading power of $n=\nfrac{3}{2}$ is what is required to make the coefficient multiplying the $r^2$ term be identically zero.

Once the $r^2$ terms have vanished, the new leading large $r$ behaviour of the fixed point equation is $r^{\nfrac{3}{2}}$, coming from the undifferentiated $\vp$ on the left-hand side of \eqref{fp}. The next term in the solution should be such that it now cancels the pieces contributing to the new leading behaviour. It is with this in mind that we proceed to build the sub-leading terms of the solution.

We denote the next term in the solution by a function $\zeta(r)$ such that
\be
\vp(r)= A\, r^{3/2} + \zeta(r) +\cdots\,,
\ee
 where $\zeta$ grows more slowly than the leading term.
This implies that we can find leading corrections to the solution algorithmically, by taking large enough $r$ to allow linearising the fixed point equation in $\zeta$.
This will give us a linear differential equation for $\zeta$, where we keep only the leading parts in a large $r$ expansion of its coefficients. This equation is then set equal to the new leading piece, namely the $r^{\nfrac{3}{2}}$ piece discussed above, and is straightforward to solve for $\zeta$ since we only want the leading part of the particular integral.
In fact inspection of the fixed point equation shows that this contribution can only come from the right-hand side and that it requires $\zeta(r)\propto r$. 
Indeed with this choice, the leading contribution from the numerators on the right-hand side are then terms which scale like $r^3$. The leading terms from the denominators scale like $r^{\nfrac{3}{2}}$ (providing no exceptional cases arise). Together these give an overall contribution of $r^{\nfrac{3}{2}}$ to the right-hand side as required. Including the next term in the solution we now have
\be
\vp(r)= A\, r^{3/2} + k_1 r +\cdots \,.
\ee
To find  the coefficient $k_1$, we substitute the solution as given above into the fixed point equation and take the large $r$ limit. Collecting all terms on one side of the equation, the leading terms go like $r^{\nfrac{3}{2}}$ as expected, multiplied by a coefficient containing $k_1$. We require this coefficient to vanish in order to satisfy the fixed point equation and so $k_1$ must take the following form
\be
k_1 = \frac{3456 \pi ^2 A^2 (\beta -1) \left(27 \beta ^2-1\right)}{\beta  \left(1620 \beta ^4-2376 \beta ^3+903 \beta ^2+2 \beta -19\right)}\,.\label{Fval}
\ee
The next terms in the series are found by repeating this procedure.
After five iterations the solution becomes \eqref{solution}
where $k_1$ is given in \eqref{Fval} and the more lengthy expressions for the other coefficients are given in appendix \ref{AppendixA}. Note that a second constant $b$, independent of $A$, is found as a result, through particular integrals containing logs. At this point our solution therefore contains two independent parameters in total.

\subsection{Exceptions}
\label{exceptions}
The solution \eqref{solution} will break down at values of $\beta$ corresponding to poles of the coefficients $k_i$. As can be seen straight-away, $\beta=0$ is one such value. This has already been flagged-up as problematic in section \ref{Exceptions} where the trouble was traced back to the fact that when $\beta=0$, the leading part of each of the coefficients in the fixed point equation \eqref{coeffs d} vanishes. This means that \eqref{rhs} no longer represents the true asymptotic scaling behaviour of the fixed point equation and should not be used to derive the leading behaviour of the solution.

There are other poles in the coefficients besides at $\beta=0$, as can be readily seen from the full form of the coefficients given in appendix \ref{AppendixA}. We find that as we build the asymptotic series, new coefficients contain new poles, not featured in earlier terms. In this way we will build a countable, but apparently infinite, set of exceptional values of $\beta$. 
It is not clear how these exceptional values are distributed (for example whether they lie within some bounded region or not), but
since the real numbers are uncountable, we are always guaranteed values of 
$\beta$ for which there are no poles in the series.

If we do happen to choose a value for $\beta$ that gives rise to a pole then this signals that the term in the solution containing the pole does not have the correct scaling behaviour in order to satisfy the fixed point equation in this instance.
Take the first coefficient $k_1$ as an example. At a pole of $k_1$ (all except $\beta=0$), adding a piece on to the solution that goes like $r$ does not result in an $r^{\nfrac{3}{2}}$ contribution on the right-hand side of the fixed point equation as required, because the coefficient automatically vanishes in this case.

Instead we must look for a different sub-leading term. A less simple choice but one that works nonetheless is $r\log(r)$. The reason for this is that if all the derivatives hit the $r$ factor and not the $\log(r)$ factor then again the $r^{\nfrac{3}{2}}$ piece on the right-hand side must vanish identically, since it is as though the $\log(r)$ is just a constant multiplier for these pieces. We therefore know that in the asymptotic expansion at this order, the only terms that survive have the $\log(r)$ term differentiated. But this then maps $r\log(r)\mapsto1$ which is the power-law dependence we desired for the differentiated term. We will apply the same strategy in section \ref{sec:non-pow}.

Taking this as the sub-leading term, such that the solution now goes $\vp(r)= Ar^{\nfrac{3}{2}} + k_1r\log(r)$, gives rise to the desired $r^{\nfrac{3}{2}}$ term on the right-hand side, but now without the pole (i.e.\ a different $k_1$) and we can continue to build the series solution from there. This suggests that for a solution with leading behaviour $r^{\nfrac{3}{2}}$, each new set of poles associated with a new coefficient gives rise to further appearances of $\log(r)$ in that sub-leading term and therefore a plethora of different possible solutions dependent on these exceptional values for $\beta$.

\subsection{Finding the missing parameters}
\label{Missing parameters}
The asymptotic solution \eqref{solution} contains only two parameters but we are solving a third order ordinary differential equation and we know that local to a generic value of $r$ there is a three parameter set of solutions. 
In this section we linearise about the leading solution \eqref{solution} to uncover the missing parameters  \cite{Morris:1994ki,Morris:1994ie,Morris:1994jc,Dietz:2012ic,Dietz:2016gzg}. We do so by writing $A \mapsto A + \epsilon\eta(r)$ such that the solution becomes
\be
\label{soln change A}
\vp(r) = (A+\epsilon\eta(r))\, r^{\frac{3}{2}} + \cdots\,,
\ee
where $\epsilon\ll1$ and $\eta$ is some arbitrary function of $r$. Since the constant $A$ introduced in \eqref{ansatz} can take any value, we are permitted to change it by any constant amount. 
Thus a constant $\eta(r)$ should be a solution
in the asymptotic limit $r\rightarrow \infty$. We use this reasoning to help find the missing parameters. In section \ref{sec:non-pow} we will follow a related but different strategy.

We insert \eqref{soln change A}, complete with all modified sub-leading terms, into the fixed point equation \eqref{fp} and expand about $\epsilon=0$. We know already that at $\mathcal{O}(\epsilon^0)$ the fixed point equation is satisfied for large $r$, since at this order \eqref{soln change A} is equivalent to the original solution \eqref{solution}.
At $\mathcal{O}(\epsilon)$ in the large $r$ limit we obtain a third order ODE for $\eta(r)$ :
\be
\label{diff eta}
h_3\, r^5 \,\eta'''(r) + h_2\, r^4\,\eta''(r) + h_1\,r^3\eta'(r)=0\,,
\ee
where
\begin{align}
\label{coeffs f}
h_1&=\frac{5 \beta \left(-6156 \beta ^4+7020 \beta ^3-699 \beta ^2-508 \beta +83\right)}{6912 \pi ^2 A (\beta -1) \left(27 \beta ^2-1\right)}\,,\nonumber\\
h_2&= \frac{\beta  \left(-3240 \beta ^4+3078 \beta ^3+471 \beta ^2-421 \beta +47\right)}{864 \pi ^2 A (\beta -1) \left(27 \beta ^2-1\right)}\,,\nonumber\\
h_3 &= -\frac{\beta  (9 \beta +5) \left(6 \beta-1\right)^2}{576\pi ^2 A \left(27 \beta ^2-1\right)}\,.
\end{align}
Initially we would expect to have another term on the left-hand side of \eqref{diff eta} that looks like $h_0\eta(r)$ times $r$ to some power. However the coefficient $h_0$ vanishes up to the order of approximation of the solution we are working to. In fact this had to be so, since otherwise $\eta(r) =\text{constant}$ would not satisfy the equation.
This means that what would have been be a third order ODE is instead a second order equation in $\eta'$. This idea is analogous to the Wronskian method for  differential equations. The solution $\vp$ contains two independent parameters, $A$ and $b$, and so we already know two independent solutions of fixed point equation. These solutions can be used to build a Wronskian that satisfies a first order differential equation and which can then be used to find the unknown solution. We will not need this full machinery however.

The differential equation \eqref{diff eta} is invariant under changes of scale $r\mapsto s r$ and thus has power law solutions. Setting $\eta\propto r^{p}$ we find three solutions for $p$: the trivial solution $p_1=0$ as required for consistency with the possibility of $\eta=\text{constant}$, $p_2=-3/2$ and
\be
\label{p3}
p_3=\frac{-4212 \beta ^4+5508 \beta ^3-1437 \beta ^2-172 \beta +53}{6 (6 \beta -1)^2 \left(\beta-1\right)\left(9\beta+5\right)}\,.
\ee
The complete solution for $\eta$  is then given by a linear combination of these powers:
\be
\eta(r) = \delta A + \delta B \,r^{-\frac{3}{2}} + \delta C\, r^{p_3}\,,
\ee
where we have introduced infinitesimal parameters $\delta A$, $\delta B$ and $\delta C$. 
Finally, inserting $\eta$ back into the solution \eqref{soln change A} we find the change in the asymptotic series complete with the change in the missing parameter:
\be
\label{delta-pow}
\delta \varphi(r) \sim \delta A\, r^{\frac{3}{2}} + \delta B + \delta C\, r^{p_3 +\frac{3}{2}} \,.
\ee
The first parameter $\delta A$ resulted from perturbing the constant $A$. We see that the solution $p_2=-\nfrac{3}{2}$ was to be expected since $\delta B$ corresponds to perturbing the $b$ parameter. We have also uncovered one new parameter through $\delta C$.

Whether or not the $\delta C$ perturbation is kept in the series depends on the size of $p_3$: if $p_3>0$ then the perturbation grows faster than the leading series, invalidating it,  and thus will be excluded from the solution. If $p_3< 0$ it is kept, which happens when $\beta\in(-0.1809,0.1931)\cup(0.4042,0.8913)$. However then we notice that by increasing $r$, the $r^{p_3 +\nfrac{3}{2}}$ can be made arbitrarily smaller than the leading term $Ar^{\nfrac{3}{2}}$. Therefore the full asymptotic series is developed by adding 
\be 
C\, r^{p_3 +\frac{3}{2}}
\ee
to \eqref{solution}, i.e.\ with a now arbitrary size constant $C$. There are subleading terms to this which will look similar to those in \eqref{solution} but with an $r^{p_3}$ factor. As we develop the asymptotic series further, we will also find terms containing powers of $r^{p_3}$ coming from the non-linearity of the fixed point equation \eqref{fp}. This development  is similar to the development of asymptotic expansions in reference \cite{Dietz:2016gzg}, where sinusoidal and log terms are also involved,  and in reference \cite{Dietz:2012ic} where also special powers arise. 
In the present case we see that the asymptotic series takes the form of a triple expansion in $1/r$, $\log(r)$ and $r^{p_3}$ for large $r$.
The value of $p_3<0$ will determine the relative importance of all these terms. 

\begin{figure}
\centering
\includegraphics[scale=0.7]{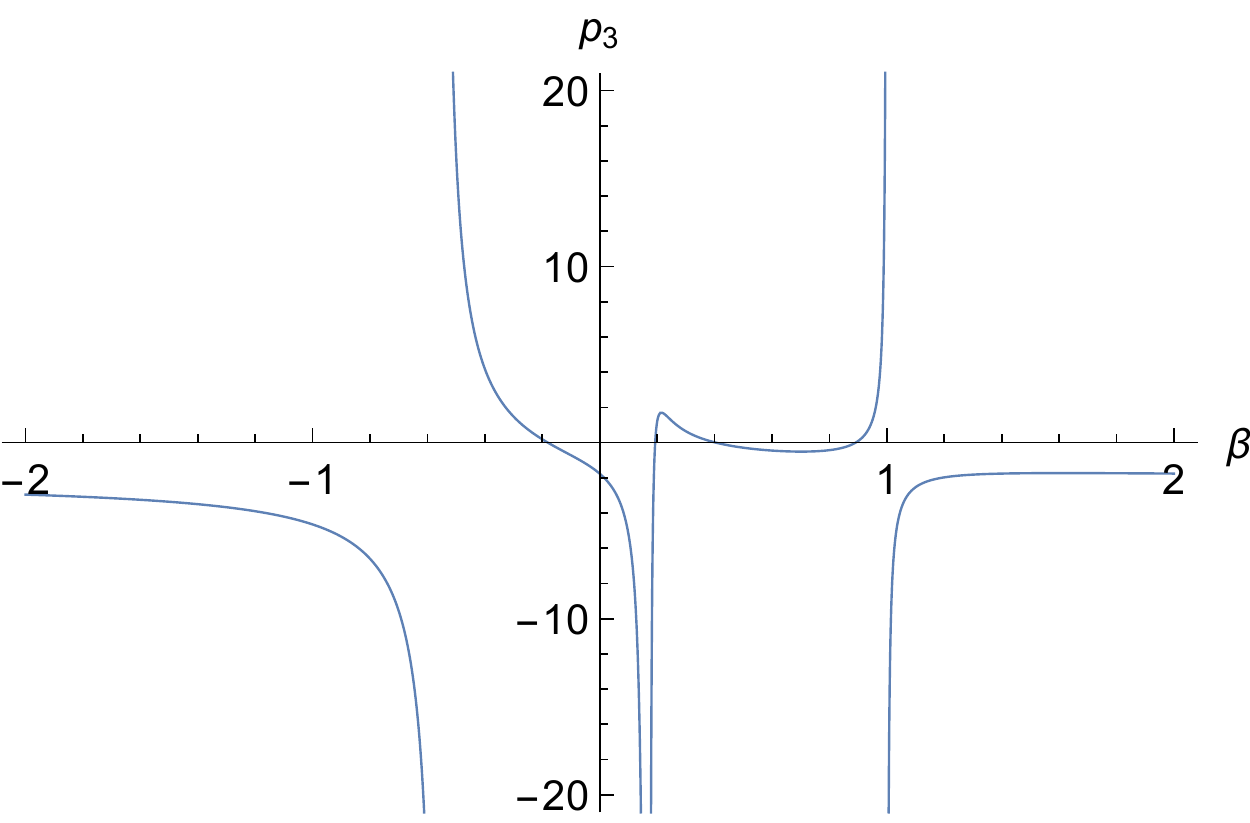}
\caption{A plot of $p_3$ against $\beta$. $p_3=0$ marks the line above which the perturbation $\delta C$ grows more quickly than the leading solution.}
\label{fig p3}
\end{figure}

The case $p_3=0$ needs careful examination: it corresponds to a solution $\eta'\propto 1/r$ in \eqref{diff eta}. Therefore in this case the last term in \eqref{delta-pow} actually appears as $\delta C\, r^{\nfrac32}\log(r)$ which rules it out, since this grows faster than the leading term.
%
The behaviour of $p_3$ is shown in figure \ref{fig p3}.

Knowing whether of not a missing parameter is excluded is crucial as the balance between the number of parameters and the number of constraints has important consequences for the nature of fixed point solutions.

There are three values of $\beta$ at which $p_3$ develops a pole, $\beta=-\nfrac{5}{9},\nfrac{1}{6}, 1$, as can be seen from \eqref{p3}, (also see figure \ref{fig p3}). The first two of these correspond to zeros of the coefficient $h_3$ meaning that at these values the differential equation \eqref{diff eta} is no longer the correct one and we must go to the next order in the large $r$ expansion of the $\eta'''$ coefficient. Doing this for $\beta=\nfrac{1}{6}$, we obtain the alternative equation for $\eta$:
\be
\label{diff eta 1/6}
\frac{13\,r^3}{48 A \pi^2}  \eta'''(r)+ \frac{r^4}{648 A \pi^2}  \eta''(r) + \frac{5 \,r^3 }{1296 A\pi^2}\eta'(r) = 0\,,
\ee
where the second and third coefficients are just $h_2$ and $h_1$ respectively with $\beta=\nfrac{1}{6}$ but where now the first coefficient is not the $\beta=\nfrac{1}{6}$ equivalent of $h_3$. Again there is no $\eta$ term because of the arguments given below \eqref{coeffs f} and this enables us to turn \eqref{diff eta 1/6} into a second order ODE by writing $\eta'(r)=\rho(r)$. We can then solve for $\rho(r)$ and integrate up to get $\eta(r)$. We find that for large $r$
\be
\eta(r)= \delta A + \delta B\, r^{-\frac{3}{2}} + \delta C \sqrt{r}\, e^{-\frac{r^2}{351}}\,.
\ee
Substituting $\eta$ back into \eqref{soln change A} we find
\be
\label{phi beta 1/6}
\delta \vp(r) \sim \delta A r^{\frac{3}{2}} + \delta B + \delta C\, r^2 e^{-\frac{r^2}{351}}\,.
\ee
Again, $\delta B$ results from perturbing the constant term, but now  $\delta C$  comes paired with an exponentially decaying piece. Since the exponential decays more rapidly than any power, the $\delta C$ term grows more slowly than any term we have found so far in the asymptotic solution \eqref{solution}. Therefore again we can replace $\delta C$ by $C$ and add this to the solution \eqref{solution}. Again the full series will involve powers of this term together with powers of $r$ and $\log(r)$, however even just the linear term will always be less important than any term in \eqref{solution} for sufficiently large $r$, and thus in practice one only need keep this linear term. In conclusion, when $\beta=\nfrac{1}{6}$,  the solution contains three independent parameters, $A, B$ and $C$, and takes the form of \eqref{full-pow-beta1/6}.
%

Following the same procedure for $\beta=-\nfrac{5}{9}$, the differential equation for $\eta$ is given by
\be
\frac{845\, r^4}{38016 \pi ^2 A} \eta'''(r) +\frac{166115\, r^4}{7185024 \pi ^2 A}\eta''(r) + \frac{830575\,r^3}{14370048 \pi ^2 A}\eta'(r) = 0\,.
\ee
Upon substituting the solution $\eta$ back into $\vp$ we obtain
\be
\delta \vp(r) \sim \delta A r^{\frac{3}{2}} + \delta B+ \delta C\, r^4 e^{-\frac{33223}{31941}r}\,.
\ee
We see that for $\beta=-\nfrac{5}{9}$ the solution also contains three independent parameters.

The remaining pole of $p_3$ at $\beta=1$ is the result of the coefficients $h_1$ and $h_2$ diverging. These coefficients also diverge at $\beta=\pm\nfrac{1}{\sqrt{27}}$, but since $h_3$ does as well, this behaviour is not captured by $p_3$: we are able to multiply through by $(27\beta^2 - 1)$ in \eqref{diff eta} thereby removing this pole from the differential equation.
Nonetheless, the value $\beta=\pm\nfrac{1}{\sqrt{27}}$ is still problematic. In fact both $\beta=\pm\nfrac{1}{\sqrt{27}}$ and $\beta=1$ correspond to values at which the leading solution \eqref{solution} already breaks down.
The issue can be traced back to zeros occurring in the denominators on the right-hand side of the fixed point equation as discussed in section \ref{Exceptions}.

There are further values of $\beta$, corresponding to the zeros of the coefficients \eqref{coeffs f}, for which the differential equation \eqref{diff eta} is no longer correct. One example which can be seen straightaway from \eqref{coeffs f} is $\beta=0$. This has been already flagged up as a troublesome value in section \ref{Exceptions} and is actually another value for which the leading solution is already not valid.

\subsection{Numerical comparison}
\label{sec:numer-pow}
The authors of reference \cite{Demmel2015b} tried matching their $\beta=1/6$ numerical solution to asymptotic behaviour given by:
\be 
\label{eq:fitansatz}
\vp(r) = A_2 \, r^2 \left(1 + u_1 \, r^{-1} + u_2 \, r^{-2}+\cdots \right) \, .
\ee
We have seen that this is not a valid asymptotic series for the fixed point equation \eqref{fp}, except at the special values $\beta_\pm$ given in \eqref{beta n=2} as in the case of the asymptotic solution (d) \eqref{n=2sol}. They also found no analytic match except at these special values, and  concluded that the asymptotic behaviour should be the result of a ``balanced regime'' which is taken to be $Ar^2$ but with logarithmic corrections. This bears some similarity to the asymptotic series we investigate in section \ref{sec:non-pow}, which however we will see in section \ref{sec:numer-non-pow} cannot provide the asymptotic solution because it does not have enough asymptotic parameters. Since the authors chose a value of $\beta$ that provides already $n_s=n_{ODE}=3$ constraints on the fixed point solution space through the fixed singularities, any number of asymptotic parameters less than the maximum $n_{ODE}=3$, will rule out a global solution. 

In this sense the authors struck lucky because we find a suitable power-law asymptotic solution with the maximum three parameters, namely \eqref{full-pow-beta1/6}.
The authors determined a fit of \eqref{eq:fitansatz} over the range $r\in[6,9]$. We can use this fit to
see how well our power-law asymptotic asymptotic solution \eqref{full-pow-beta1/6} does in matching their large $r$ behaviour as far as it was taken. As we will see, despite the very different leading behaviour at large $r$ we can find equally acceptable fits. 
\begin{figure}
\centering
\includegraphics[scale=0.4]{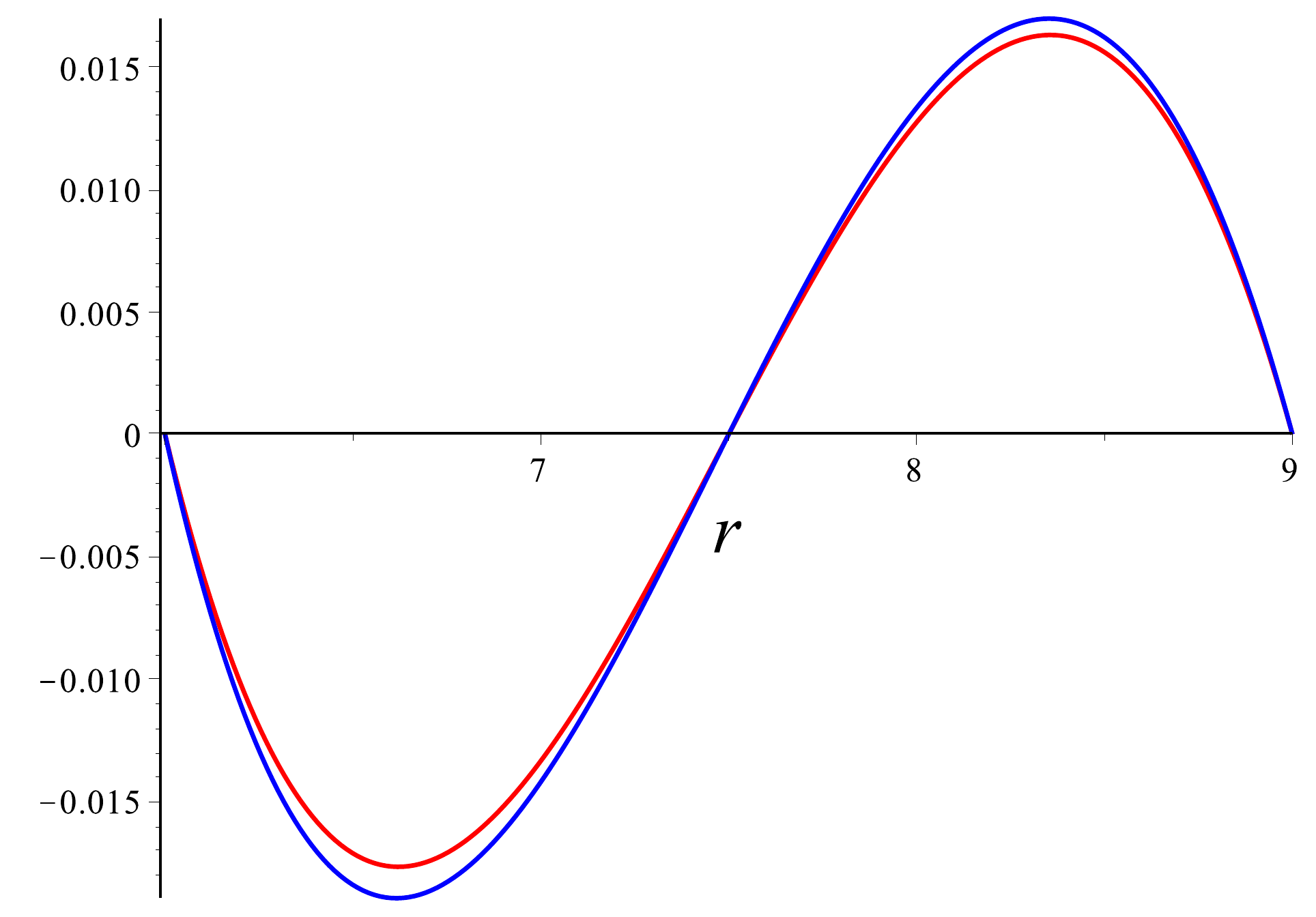}
\caption{Plot of the difference between \eqref{full-pow-beta1/6} and \eqref{eq:fitansatz}. The red curve uses \eqref{pow-param-sols1} and the blue curve uses \eqref{pow-param-sols2}.}
\label{fig:fit-pow}
\end{figure}

Their fit gave the solutions:
\be\label{eq:fitparamters}
\begin{split}
A^{\rm fit}_2 = & \, \phantom{-}\ 0.07705 \pm 0.00032 \, , \\
u_1^{\rm fit} = & \, -2.07514 \pm 0.05399 \, , \\
u_2^{\rm fit} = & \, -6.36855 \pm 0.25897 \, . 
\end{split}
\ee
Note that our asymptotic expansion \eqref{full-pow-beta1/6} to the level taken, is actually linear in $C$ and $\log(b)$, so it is straightforward to solve for these. Determining also $A$ by insisting that 
\eqref{full-pow-beta1/6} agrees with \eqref{eq:fitansatz} at $r=6$, 7.5 and 9, we find two solutions:
\begin{alignat}{2}
\label{pow-param-sols1}
A &= -5.6498\cdot10^{-5}\,, &\quad \log(b) =-4932.4\,, &\quad C =0.13864\,;\\
\label{pow-param-sols2}
A &=5.0025\cdot10^{-4}\,, &\quad\log(b) =3.6538\,,\ \ &\quad C=0.12571\,;
\end{alignat}
where the second seems more believable. On the other hand we note that the asymptotic expansion \eqref{full-pow-beta1/6}  suggests, but does not require,\footnote{for example we could rewrite the expansion in terms of $\log(b)$} that we apply it only to the region $r>b$, which would favour the first solution.
\begin{figure}
\centering
\includegraphics[scale=0.4]{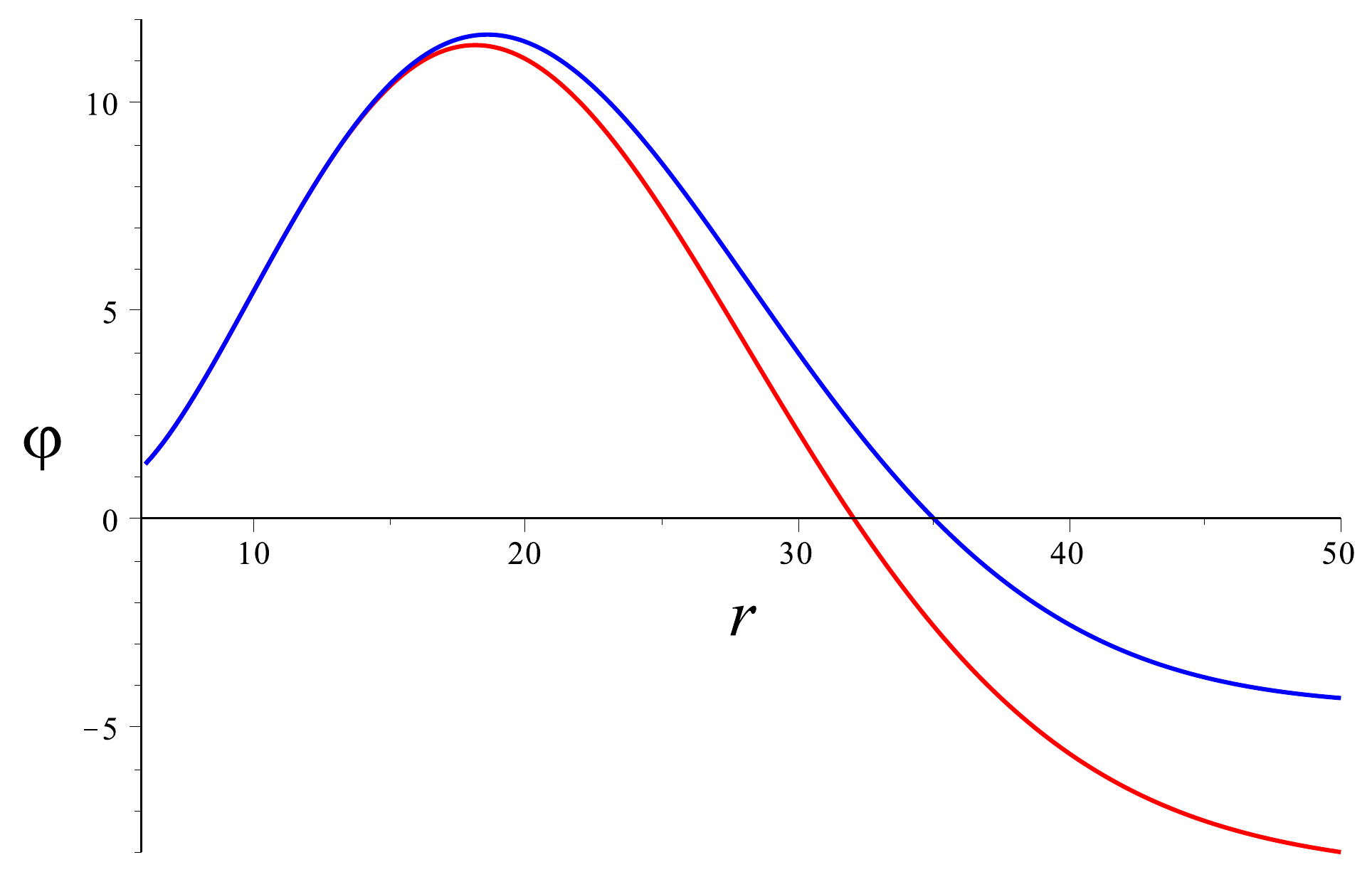}
\caption{The predicted large $r$ behaviour from the two fits. The red curve uses \eqref{pow-param-sols1} and the blue curve uses \eqref{pow-param-sols2}.}
\label{fig:fit-pow-asymptotic}
\end{figure}
It is not possible to distinguish by eye the solutions and the (fitted) numerical solution over the range $r\in[6,9]$, so instead we plot their difference in figure \ref{fig:fit-pow} for the two possibilities \eqref{pow-param-sols1} and \eqref{pow-param-sols2}. As can be seen the error is almost the same in both cases and competitive with the error implied by the spreads in \eqref{eq:fitparamters}. Clearly the two possibilities   \eqref{pow-param-sols1} and \eqref{pow-param-sols2} are not both correct. Determining which, if either, is correct, would require computing the numerical solution to larger $r$. In particular note that the asymptotic solution \eqref{full-pow-beta1/6} fits  because the final term dominates in the fitted region, where it provides the $r^2$-like behaviour necessary to fit the data. Its exponential decay only becomes significant once $r>\sqrt{351}=18.73$ after which our fitted solutions peak and then turn negative, with the leading asymptotic parts of \eqref{full-pow-beta1/6} finally taking over, as can be seen in figure \ref{fig:fit-pow-asymptotic}.

\section{Asymptotic expansion of a non-power law solution}
\label{sec:non-pow}

Power counting for $\vp(r)\sim r^n$ suggests that $\vp(r)\sim r^2$ should be the leading solution, since then the two sides of the fixed point equation, \eqref{lhs} and \eqref{rhs}, balance. 
However, as discussed just above section \ref{Exceptions},
this fails to be the case in general because it also happens that the left-hand side vanishes identically. Then the leading term on the right-hand side must also vanish, which is only true for specific values of $\beta$. The way out of this is analogous to that discussed in section \ref{exceptions}: since $r^2$ is annihilated by the left-hand side, we know that $r^2\log(r)$ will survive and furthermore give us a pure power $r^2$, which is what we will need in order to balance the $r^2$ power coming from taking ratios on the right-hand side. In this section we explain how to develop the non-power law asymptotic solution and find its missing parameters.

In fact for good measure we also tried the general ansatz $\vp(r)\sim r^n(\log r)^p$. Then one finds that on the right-hand side of \eqref{fp}, the first ratio $\sim r^2$ while the second ratio $\sim r^0$. Therefore balance is achieved only for $n=2$ and $p=1$, since then as we have just seen the left-hand side also behaves as $\sim r^2$.


\subsection{Leading behaviour}
\label{sec:non-pow-leading}
We begin by studying the leading behaviour for the $r^2\log r$ ansatz. In full, the above argument implies that this leading term has to be of the form
\be\vp(r)\sim k_1 r^2 \log\left(\tfrac{r}{b}\right)\,.\label{ansatzlog}\ee 
Demanding that these agree, we find that $k_1$ is determined:
\be\label{non-pow k1} k_1=\dfrac{1-72\b+156\b^2}{9216\pi^2}\,,\ee
while $b$ is left undetermined.

\subsection{Sub-leading behaviour}
\label{sec:non-pow-subleading}
For the next  terms, it is easier to first write
$\vp(r)=r^2f(r)$ for some function $f$ and then make the change of variables $\nfrac{r}{b}\mapsto e^x$ such that $\vp(r)\mapsto e^{2x}f(x)$. Finally we divide by $r^2\equiv e^{2x}$ to simplify the fixed point equation \eqref{fp}. Once we do this, we can expand the fixed point equation in small exponentials in the following way
\be 0=f_a+f_b e^{-x}+f_c e^{-2x}+\cdots \,.\label{expapp}\ee
Here $f_b,f_c\dots$ give corrections that are smaller than any power so for the moment we can discard these pieces and concentrate on 
$f_a$, which is given by the following expression
\beal f_a&=2 f'(x)-\frac{30 \beta  (2 \beta -1) (6 \beta -5) \left(2 f''(x)+5 f'(x)+2 f(x)\right)}{4608\pi ^2 \left((3 \beta -1) f'(x)+(6 \beta -5) f(x)\right)}-\nonumber\\
&-\frac{\beta  (1-6 \beta )^2 \left(2 (9 \beta +5) f'''(x)+(63 \beta +31) f''(x)+(63 \beta +25) f'(x)+6 (3 \beta +1) f(x)\right)}{4608\pi ^2 \left((3 \beta +1)^2 f''(x)+(3 \beta  (9 \beta +5)+1) f'(x)+6 \beta  (3 \beta +1) f(x)\right)}\,.\label{EqaA}\eeal
In order to find the subleading terms, the procedure is as follows.
From \eqref{ansatzlog} we know that for large $x$ (equivalent to large $r$), $f(x)=k_1 x$. Plugging this into \eqref{EqaA} we see that the leading piece in \eqref{EqaA} in the large $x$ limit is given by a constant at large $x$, as already indicated by the first term in $f_a$. We then recover the coefficient $k_1$ by demanding that the constant part cancels.
Once this is done, we find that the new leading behaviour of \eqref{EqaA} goes like $x^{-1}$ so next we must as a term $k_2 \log\left({x}/{c}\right)$ to the solution, where $c$ is an arbitrary constant, in order for the solution to satisfy the fixed point equation.
This term can be determined again from the first term in $f_a$.
Demanding that the coefficient of $x^{-1}$ in \eqref{EqaA} vanish, we find the value of $k_2$ (this and the rest of the coefficients are listed in appendix \ref{kivalues}).
To find the next term we substitute $f(x)$ as found so far i.e.\
\be f(x)=k_1 x+k_2 \log\left(\frac{x}{c}\right)\,,\ee
into \eqref{EqaA}, with the already known coefficients $k_1$ and $k_2$.
In doing this we find that the leading piece now behaves like ${\log(x)}/{x^2}$. This implies that we must add $k_3{\log\left({x}/{c}\right)}/{x}$ to our ansatz and find the value of $k_3$ that cancels this term. After more iterations of this procedure, we find  $f(x)$ is given by
\beal f(x)&=k_1 x+k_2\log\left(\dfrac{x}{c}\right)+k_3\dfrac{\log\left(\dfrac{x}{c}\right)}{x}+\dfrac{k_4}{x}+k_5\dfrac{\log^2\left(\dfrac{x}{c}\right)}{x^2}+\cdots \,, \label{f}
\eeal
where coefficients $k_i$ are given in the Appendix \ref{kivalues}.\par

It is worth noticing that there is a constant, $b$, in \eqref{ansatzlog} and another one, $c$, in \eqref{f}.
The constant $b$ is also captured in $f$ by using the translation invariance of $f_a$. Thus if $f(x)$ is a solution, so is $f(x+x_0)$, where $x_0\equiv-\log(b)$. One might think that we already have two free parameters in the solution. However, this is not the case: it is easy to see, with the values of appendix \ref{kivalues}, that
\be \dfrac{\partial f(x+x_0)}{\partial x_0}=-\dfrac{k_1 c}{k_2}\dfrac{\partial f(x+x_0)}{\partial c}\, ,\ee
which implies that the two constants can be combined into one, and therefore there is actually only one independent parameter. Since we already know we can dispense with $b$ and then recover it by $x$-translation invariance, in the following we set $b=c=1$ when working with $f(x)$ and instead fold them into a parameter $A$ in the end for $\vp(r)$. Indeed, changing variables back to $r$, we see that the whole solution is then given in \eqref{vptotal}.

\subsection{Exceptions}\label{exceptionss}

By looking at the coefficients $k_i$ in Appendix \ref{kivalues} it can be seen that there are some values of $\b$ for which \eqref{vptotal} is not an acceptable solution. These are listed below.

\paragraph{$\boldsymbol{\beta=0}$ and $\boldsymbol{\beta=\beta_\pm}:$}
 for these values $\vp(r)=r^2\log\left({r}/{b}\right)$ is not a solution of the fixed point equation. The latter choice $\beta=\beta_\pm$ was expected, since we have seen in \eqref{beta n=2}  that for these values $\vp(r)\sim r^2$ is a solution, without the need to add the $\log$ term. Actually, what happens  in both cases is that  the leading power on the right-hand side decreases, so the asymptotic behaviour is dictated by the left-hand side, which vanishes for $\vp(r)\sim r^2$ but not for $\vp(r)\sim r^2\log r$. However for $\beta=0$ we saw that this exceptional behaviour left us then with only the $n=0$ solution \eqref{n=0no2}.

\paragraph{$\boldsymbol{\beta=-\nfrac{1}{3}}$ and $\boldsymbol{\b=\nfrac{5}{6}}:$}
for these values the asymptotic solution is still of the form \eqref{vptotal}, or equivalently \eqref{f}, but because these coefficients cancel leading contributions in \eqref{fp} the coefficients $k_i$ have different values that do not correspond with just  substituting the above values of $\b$ into \eqref{kigeneral}. Following the same procedure, but with the right leading contributions for these cases, we find the new coefficients given in \eqref{ki13} for $\b=-\nfrac{1}{3}$, and \eqref{ki56} for $\b=\nfrac{5}{6}$. 

\subsection{Finding the missing parameters}
\label{missing params 2}

In order to find the two missing parameters, we linearise \eqref{EqaA} about \eqref{vptotal} by writing $f(x)\mapsto f(x)+\e\eta(x)$. This results in a differential equation for$\eta$:
\be
h_3x\eta'''(x)+h_2x\eta''(x)+ 2x^2\eta'(x)+h_0\eta(x)=0\label{eqa}\,,
\ee
where the $h_i$ are the following functions of $\b$ ($h_1=2$ so this simple value is already explicitly written in the above expression):
\beal
h_0&=\frac{4 \beta  \left(3 \beta  \left(9 \beta  \left(312 \beta ^2-334 \beta -5\right)+406\right)-29\right)+5}{3 \beta  (3 \beta +1) (6 \beta -5) (12 \beta  (13 \beta -6)+1)}\,,\label{h0nonpow}\\
h_2&=\frac{-4104 \beta ^4+108 \beta ^3+642 \beta ^2-37 \beta +1}{3 \beta  (3 \beta +1) \left(156 \beta ^2-72 \beta +1\right) }\,,\label{h2nonpow}\\
h_3&=-\frac{2 (1-6 \beta )^2 (9 \beta +5)}{3 (3 \beta +1) \left(156 \beta ^2-72 \beta +1\right) }\,.\label{h3nonpow}
\eeal
We already know one solution to this equation,
\be 
\label{f-prime}
\eta=\frac{\partial f}{\partial x} = k_1 + \frac{k_2}x -k_3 \frac{\log(x)}{x^2}+\cdots\,,
\ee
by translation invariance.  The other two solutions cannot go like a power for large $x$ since this would make the $\eta'''$ and $\eta''$ terms subleading already. In other words for power-law solutions, \eqref{eqa} behaves like a linear first order differential equation, with thus (up to a scale) only one solution. We need to ansatz a solution that can make $\eta''$ and/or $\eta'''$ as important as the $\eta$ or $\eta'$ terms. This motivates trying 
\be 
\label{Lp-ansatz}
\eta= e^{L x^p}\,,
\ee 
with $L\ne0$ and $p>1$ .
In that case, we have
\beal
\eta=& e^{L x^p}\,,\\
\eta'=& L p x^{p-1} e^{L x^p}\,,\\
\eta''=& L^2 p^2 x^{2 p-2} e^{L x^p}+L (p-1) p x^{p-2} e^{L x^p}\sim L^2 p^2 x^{2 p-2} e^{L x^p}\,,\\
\eta'''=& L^3 p^3 x^{3 p-3} e^{L x^p}+L^2 (p-1) p^2 x^{2 p-3} e^{L x^p}+L^2 p^2 (2 p-2) x^{2 p-3} e^{L x^p}\nonumber\\
&\quad\quad+L (p-2) (p-1) p x^{p-3} e^{L x^p}\sim L^3 p^3 x^{3 p-3} e^{L x^p}\,,
\eeal
where we are keeping only the (asymptotically) leading terms.
Therefore, from \eqref{eqa} we have
\be e^{L x^p}\left(h_0+2 L p x^{p+1}+h_2 L^2 p^2 x^{2 p-1}+h_3 L^3 p^3 x^{3 p-2}\right)=0\,.
\label{leadeq}\ee
Actually we can discard the $h_0$ term in this expression since for $p>1$ it will never be leading. Of the remaining terms, we want the leading ones to cancel one another and so there are three options to explore.

\paragraph*{Option 1:} The second and third terms are leading. If this is true then we require
\be
 p+1=2p-1\Rightarrow p=2\,,
 \ee
but for this value of $p$, the last term will become a leading term also and therefore this option is excluded.

\paragraph*{Option 2:} The last two terms are leading, meaning that
\be
2p-1=3p-2\Rightarrow p=1\,.
\ee
But for this value of $p$ the second term becomes leading and so this option is also excluded.

\paragraph*{Option 3:} The second and the last terms are leading such that
\be
p+1=3p-2\Rightarrow p=\dfrac{3}{2}\,.
\ee
This option is allowed, since the third term is now sub-leading.

Now demanding that the leading terms vanish, implies that $L$ has to fulfil the condition $2  +h_3 p^2 L^2=0$,
i.e.\
\be L=\pm\dfrac{2 }{3}\sqrt{\dfrac{- 2}{h_3}}\label{Bsol}\,.\ee
We see that $L$ will take a real or imaginary value depending on the sign of $h_3$.
For $\b<-\nfrac{5}{9},\; -\nfrac{1}{3}<\b<\b_-$ and $\b>\b_+$, $L$ is real, where $\beta_\pm$ is defined in \eqref{beta n=2}.
In this case we only keep the negative root in \eqref{Bsol} since the other one gives a $\eta$ which grows exponentially. Therefore we find for $L$ real there are only two parameters in the asymptotic solution, one coming from the leading solution and one in the form of a missing parameter.

In the case that $L$ takes an imaginary value, \eqref{Lp-ansatz} has unit modulus, so to see whether it is still an allowed ansatz we need to go to next order in the perturbation and compare its behaviour with the leading $k_1 x$ term in \eqref{vptotal}. To this end we substitute
\be 
\label{Lp-ansatz-corrected}
\eta= e^{L x^{\frac{3}{2}} +\zeta(x)}\,,
\ee
into \eqref{eqa} where $\zeta$ grows slower than $x^{\nfrac32}$. In this way we find that 
\be 
\zeta= 4\frac{h_2}{h_3}x\,.
\ee
Whether or not this perturbation is kept then depends on the sign of $h_2/h_3$. We find that for $-5/9<\beta<-1/3$ the sign is negative and therefore the perturbation is allowed.
Together with the two parameters, the two solutions in \eqref{Bsol} get combined into real oscillatory combinations with an exponential tail provided by $\zeta$.
This gives us a three-parameter asymptotic solution.
For the other region of imaginary $L$, namely $\b_-<\b<\b_+$, we find that $h_2/h_3>0$. Thus in this region \eqref{Lp-ansatz-corrected} is an exponentially growing perturbation and is excluded.
It follows that in this case the asymptotic solution only contains the one parameter $A$ in \eqref{vptotal}.

%
%

As in section \ref{Missing parameters}, we note that where these perturbations are allowed we can replace their linearised coefficients with full coefficients, since the perturbations can already be made as small as we like compared to the leading terms by increasing $x$.
We can summarise the full asymptotic solutions we have found so far as \eqref{non-pow-only}--\eqref{non-powOsc32}.

%

\subsection{Exceptions}
\label{sec:missing2exceptions}
There are several values of $\b$ for which the differential equation \eqref{eqa} is not valid. The first two given below relate to exceptions already considered in section \ref{exceptionss}, where we saw that the expansion coefficients in \eqref{vptotal} get altered. The remaining cases are caused  by the vanishing of one of the $h_i$ coefficients, \eqref{h0nonpow}--\eqref{h3nonpow}. In this case, that the corresponding term in \eqref{eqa} gets replaced by one which grows more slowly at large $x$. Since $h_0$ played no role in the above analysis, exceptions arise only from the vanishing of $h_3$ and $h_2$.



\subsubsection{Altered coefficients}
\label{sec:altered}

\paragraph{$\boldsymbol{\beta=-1/3}:$}
for this value, we instead use the coefficients in \eqref{ki13} and follow the procedure used previously to derive a differential equation for $\eta$:
\be\label{eqa3}
-\frac{9x^2 }{17}\eta'''(x)-\frac{45 x^2 }{34}\eta''(x)+2x^2\eta'(x)+\frac{775 }{238 }\eta(x)=0\,.\ee
We already know one solution is \eqref{f-prime}, however now with the coefficients $k_i$ from \eqref{ki13}. Then the leading behaviour of the other two solutions does not involve the undifferentiated $\eta$. Indeed, dividing by $x^2$, the other three terms on their own give a differential equation with constant coefficients which is therefore solved with $\eta = e^{Lx}$, while the undifferentiated $\eta$ term contributes $\sim e^{Lx}/x^2$, which can be neglected at leading order. $L$ thus solves a cubic. 
Discounting the $L=0$ solution (which is \eqref{f-prime} in disguise), we are left with a quadratic whose roots are $L=-L_\pm$, where
\be 
\label{Lpm}
L_\pm = \frac54\pm\frac{\sqrt {769}}{12}\,.
\ee
Since $-L_->0$ and $-L_+<0$, we discard the $-L_-$ solution and are  left with the two-parameter asymptotic solution \eqref{non-powExpL}.

%
%

\paragraph{$\boldsymbol{\beta={5}/{6}}:$}
as before, but using now \eqref{ki56} the equation reads
\be-\frac{50x }{21 }\eta'''(x)-\frac{688x }{105 }\eta''(x)+2x^2 \eta'(x)+ \frac{388 }{105 }\eta(x)=0\,.\ee
This has the same form as \eqref{eqa} so trying the same ansatz it has the same solution for $p=\nfrac{3}{2}$ and thus we find the two-parameter asymptotic solution \eqref{non-powExp32alt}.
\paragraph{}The remaining possible exceptional values can be arranged according to which coefficient of \eqref{eqa} they cause to vanish.

\subsubsection{Third derivative}
\label{sec:third derivative}

There are two values of $\beta$ that make $h_3$ vanish. For both of them, it is not that we must go to the next order in \eqref{EqaA} to get the leading term, but that the third derivative term vanishes identically there. Instead we need to go to higher order in the exponential expansion.

\paragraph{$\boldsymbol{\beta=1/6}:$} In this case, in the exponential expansion of the fixed point equation,
\be f_a+f_b e^{-x}+f_c e^{-2x}+\cdots\,,\ee
not only does the third derivative term vanish in $f_a$ but also in $f_b$, so in order to find the coefficient $h_3$ we need to consider $f_c$. The resulting differential equation for $\eta$ is
\be
\frac{78 x e^{-2 x}}{5 }\eta'''(x)-2x\eta''(x)+2x^2\eta'(x)+\frac{19 }{4}\eta(x)=0\,.
\ee
An ansatz of the form $e^{Le^{2x}}$ provides the perturbation that involves the third derivative, by balancing against the second derivative part, with the rest subleading. But we find that $L=5/78$, which being positive, rules this out of the asymptotic series. The other perturbation is found by neglecting the third derivative term. In this case we get,
\be-2x \eta''(x)+2 x^2 \eta'(x)+\frac{19 }{4}\eta(x)=0\,.\ee
With an ansatz $e^{L x^p}$, 
one finds the asymptotic solution $p=2,\, L={1}/{2}$. Again, this growing perturbation is ruled out in the asymptotic series, so we end up with only the one-parameter solution $f_{asy}(x-\log A)$, i.e.\ $\vp(r)$ as in \eqref{vptotal}.

\paragraph{$\boldsymbol{\beta=-5/9}:$}
now the coefficient for $\eta'''$ appears in $f_b$ and we get
\be-\frac{4563xe^{-x}}{2407 }  \eta'''(x)-\frac{23056x}{12035 }\eta''(x)+2x2 \eta'(x)+\dfrac{103627 }{24070}\eta(x)=0\,.\ee
Thus similar to the previous case, an ansatz of the form $e^{Le^{x}}$ provides the perturbation that involves the third derivative. Since then $L=-\nfrac{23056}{22815}<0$, this rapidly decaying perturbation provides one of the missing parameters. Neglecting the third derivative term we get a similar equation to the previous case, for which the missing perturbation is $e^{L x^p}$, with again  $p=2$ but now $L=\nfrac{12035}{23056 }$. This is therefore still an exponentially growing perturbation and is thus ruled out. Therefore in this case we have the two-parameter asymptotic solution \eqref{non-powExpExp}.

\subsubsection{Second derivative}\label{k2sec}

The coefficient $h_2$ vanishes for the two real roots of the quartic in \eqref{h2nonpow}, cf.\ table \ref{table2}.
\begin{table}
\centering
\begin{tabular}{c|c|c|c|c|c}
 $\b$ & $h_0$ 
 & $h_2$ & $h_3$& $L$ & $q=-\frac34-\frac{h_2}{2h_3}$ \\ 
\hline
$-0.4111$ &$4.1866 $
&$2.1950$ & $0.7840$ & $\pm  1.0648 i$  & -2.1499\\
$0.3800$ &$5.8825$
& $-7.6628$& $ 1.1209$ & $\pm 0.8905 i$ & 2.6681\\
\noalign{\smallskip}
\end{tabular} 
\caption{Parameters for the differential equation and solutions, in the case that \eqref{h2nonpow} vanishes.}
\label{table2}
\end{table}
The differential equation now reads (with a new $\eta''$ term and coefficient $h_2$):
\be h_3 x \eta'''(x)+h_2 \eta''(x)+ 2 x^2 \eta'(x)+ h_0 \eta(x)=0\,,\ee
where the coefficients are also given in the table. 
Comparing to the general case \eqref{eqa}, we see that the only structural difference is that the $\eta''$ is now even more subleading. Since it actually played no role in the general case in determining the (formally) leading behaviour, the same ansatz \eqref{Lp-ansatz} solves this case and thus we find $L$ is given by \eqref{Bsol} but with $h_3$ as given in table \ref{table2}, and thus $L$ takes the imaginary values also listed in the table. Therefore as we saw in the general case, to determine whether this perturbation survives we need to go to the next order. Substituting \eqref{Lp-ansatz-corrected} we find that this time it is solved to leading order by
$e^\zeta = x^q$, where $q$ is also given in the table. Since overall the perturbation must grow slower than $k_1x$, the leading term in \eqref{vptotal}, we see that the two perturbations are excluded for $\beta=0.3800$ and thus we have only the one-parameter solution \eqref{non-pow-only},
while for $\beta=-0.4111$ we have the three-parameter solution
\eqref{non-powOsc}.

\subsection{Numerical comparison}
\label{sec:numer-non-pow}

From section \ref{sec:dimensionality-ef}, we already know that the relevant solution for $\beta=1/6$, namely the unadorned \eqref{vptotal}, cannot be the asymptotic limit of the numerical solution found in reference \cite{Demmel2015b}, since we saw that its one free parameter is overconstrained. We can also see directly that the numerical solution, equivalently \eqref{eq:fitansatz} with \eqref{eq:fitparamters}, cannot match.
Using \eqref{kigeneral} we find at $\beta=1/6$:
\be 
k_1=-{\frac {5}{6912\,{\pi }^{2}}}, \
k_2=-{\frac {95}{55296\,{\pi }^{2}}}, \
k_3=-{\frac {1805}{442368\,{\pi }^{2}}}, \
k_4=-{\frac {95}{442368\,{\pi }^{2}}}, \
k_5={\frac {34295}{7077888\,{\pi }^{2}}}\,.
\ee
Since the expansion only makes sense for $r\gg A$, we see the asymptotic solution implies that at large $r$, we have $\vp<0$ with $|\vp|$ growing faster than $r^2$, which is qualitatively different from the numerical solution.

\section{Conclusions}
\label{sec:conclusions}

Despite the complicated nature of the fixed point equations resulting from functional truncations of the effective average action, in particular for the $f(R)$ approximation which then leads to a non-linear second or third order ODE for the corresponding scaled quantity $\vp(r)$, we have seen that by adopting techniques first developed in \cite{Morris:1994ki,Morris:1994ie,Morris:1994jc} and applied to this area in  \cite{Dietz:2012ic,Dietz:2016gzg}, it is reasonably straightforward to extract general key properties of the solutions, through an asymptotic analysis. The corresponding asymptotic solutions are set out as a summarised list in section \ref{sec:asymptotics-overview}, where also links are provided to the subsections where these are derived.

In particular, before resorting to a laborious numerical treatment, one can map out the dimensionality of the fixed point solution spaces using the counting formula \eqref{counting}. These spaces divide into sets depending on the number of free parameters, $n_{asy}$, in the corresponding  asymptotic  solution. We saw examples of this in section \ref{sec:dimensionality-ef}.
Finding the asymptotic solutions together with their complete set of free parameters is thus key to this, as it is in fact for validating any numerical solution (as discussed in section \ref{sec:validate}) since without matching to an asymptotic solution one can never be sure that the hoped-for global numerical solution does not end in a moveable singularity at some large $r$. Moreover, a full knowledge of the asymptotic behaviour provides insight and guidance for developing the numerical solution.
We provide an example of this in section \ref{sec:numer-pow} where we match the relevant asymptotic solution to the numerical solution found in reference \cite{Demmel2015b}, see also section \ref{sec:numer-non-pow}.

In the original applications \cite{Morris:1994ki,Morris:1994ie,Morris:1994jc,Morris:1995he,Morris:1997xj}, one immediately found the (unique) leading behaviour of the asymptotic solution since this was simply given by scaling dimensions (see \eqref{Vscale}), neglecting the complicated part of the fixed point equation that describes the quantum corrections. In functional truncations for quantum gravity, it is now clear that this is typically no longer the case, as discussed in section \ref{sec:quantum}. Instead the quantum corrections remain important no matter how large the curvature $R$ is taken, for readily identifiable physical reasons.\footnote{The same was found to be true for metric in the conformal truncation of reference \cite{Dietz:2016gzg}.} 

Thus a little more effort is required to find all possible leading terms for an asymptotic solution in functional truncations to quantum gravity. The strategy, as set out in section \ref{Leading behaviour}, is to start with a general ansatz, figure out which terms in the fixed point equation are then the most important at large $r$ and then require that these terms balance, i.e.\ that these leading pieces cancel amongst themselves. The possible ans\"atze are actually quite limited because most of any function $\vp(r)$ can be neglected in the large $r$ limit. In the example fixed point equation we chose, namely the ODE \eqref{fp} from reference \cite{Demmel2015b}, we tried a power law $\vp(r)\sim r^n$ as explained in section \ref{Leading behaviour}, resulting in solutions $n=0,$ $3/2$ and $n_\pm(\b)$ as summarised in cases (a) to (e) in section \ref{sec:asymptotics-overview}.
We also tried $\vp(r)\sim r^n (\log r)^p$, finding just the one solution, $n=2$ with $p=1$, that is presented in section \ref{sec:non-pow} and summarised as case (f) in section \ref{sec:asymptotics-overview}. Already this more complicated leading asymptotic solves the equation only through special circumstances, as explained at the beginning of  section \ref{sec:non-pow}. 

Carefully considering exceptions that appear in various regions, and at various special points of the endomorphism parameter $\beta$, including in sub-leading terms that we are about to discuss, we supply a total of 15 different asymptotic series in section \ref{sec:asymptotics-overview}. In fact as shown in section \ref{exceptions}, there are further modifications of \eqref{solution} at discrete values of $\beta$ signalled by divergences in one of the subleading coefficients, potentially countably infinite in number.

Developing the leading asymptotic into a series $\vp_{asy}(r)$, complete with sub-leading corrections, is  the most straightforward part of the procedure, cf.\ sections \ref{Sub-leading behaviour} and \ref{sec:non-pow-subleading}. However if the asymptotic series has $n_{asy}<n_{ODE}$ free parameters ($n_{ODE}$ being the order of the ODE),  we cannot be sure we have found the full asymptotic series until we have understood where the ``missing" parameters have gone. This is where we see another huge difference \cite{Dietz:2012ic,Dietz:2016gzg} from the early applications \cite{Morris:1994ki,Morris:1994ie,Morris:1994jc,Morris:1995he,Morris:1997xj}. There it was always the case that $n_{asy}=1$ while $n_{ODE}=2$. The missing parameter always corresponded to a perturbation that grew rapidly, faster than the leading term in the asymptotic series, and thus could not be added without invalidating it. This perturbation could be understood to be the linearised expression of moveable singularities in the ODE. On the contrary here it is typically the case that the full asymptotic series contains further free parameters. It is clear that this is another expression of the fact that the quantum corrections do not decouple in the large $r$ limit. 

Finding these parameters, or proving that they are legitimately excluded, can be straightforwardly achieved through the following strategy. We perturb the asymptotic solution, writing $\vp(r)=\vp_{asy}(r)+\zeta(r)$, and keep only terms linear in $\zeta$. The result is a linear ODE, which is typically simple, since in the coefficients we only need the leading terms at large $r$. The task is further simplified since we are only looking for the leading behaviour of the solutions $\zeta$, and since for every parameter $a$ in $\vp_{asy}(r)$ we already know that: 
\be 
\zeta(r) = \zeta_{a}(r) := \frac{\partial}{\partial a}\vp_{asy}(r)
\ee
is a solution.
To find the  solutions, $\zeta=\zeta_m(r)$ corresponding to the missing parameters, the easiest way is to find an ansatz which can balance different terms in the, now linear, ODE. With a little thought it is always possible to find all $n_{ODE}$ solutions. A helpful hint is provided by noting that the highest derivatives must  have a role to play in at least one of the solutions. Once we have found $n_{ODE}$ linearly independent solutions, we are ready to classify them. If they grow faster than the leading term in $\vp_{asy}(r)$, they have to be discarded, as explained above. On the other hand if they grow slower than this leading term, we can add them to the asymptotic series with a \emph{finite} coefficient. This is because we can always take $r$ large enough to make the linearisation step valid, whatever size of coefficient we take. In this paper we provide numerous examples of this procedure in sections 
\ref{Missing parameters}, \ref{missing params 2} and \ref{sec:missing2exceptions}, culminating in 11 different full asymptotic series in cases (e) and (f) in section \ref{sec:asymptotics-overview}, and a zoo of different $\zeta_m$, including powers of $r$, exponentials of $-r$ or $-r^2$, and $\sin\log r$ type terms. Needless to say, finding these missing terms is also important for matching to numerical solutions  \cite{Dietz:2012ic,Dietz:2016gzg}. We saw in section \ref{sec:numer-pow} that matching to the numerical solution found in reference \cite{Demmel2015b}, crucially relied on the $C\, r^2 e^{-\nfrac{r^2}{351}}$ term in \eqref{full-pow-beta1/6}. Matching at high accuracy these full asymptotic solutions to numerical solutions, requires developing the asymptotic series, complete with the new parameters, to higher order. We do not do this in the work presented in this chapter, but examples can be found in references. \cite{Dietz:2012ic,Dietz:2016gzg}, where we see that the non-linear parts of the ODE then generate sub-leading terms involving all the parameters. 

Although the eigenoperator spectrum was not addressed in this work, asymptotic techniques were developed for them also  \cite{Morris:1994ie,Morris:1994jc,Morris:1996xq,Morris:1996nx,Bridle:2016nsu} and applied to asymptotic safety in  \cite{Dietz:2012ic,Dietz:2016gzg,Benedetti:2013jk}.

In section \ref{sec:dimensionality-ef} we used the above full asymptotic series to map out the dimensions of the solution spaces for different values of the endomorphism parameter $\beta$. This endomorphism parameter, together with the other one $\alpha$ which was ultimately set to $\beta-2/3$, was introduced to provide extra flexibility in designing the way modes are integrated out in the flow equations \cite{Demmel2015b}, in particular with the aim of ensuring that for some value of this parameter there is an isolated fixed point solution suitable for building an asymptotically safe theory of quantum gravity. 
Much the same strategy has also been followed in references. \cite{Demmel:2014hla,Ohta:2015efa,Ohta2016,Falls:2016msz}. Such a freedom would indeed appear to be inherent in exact RG descriptions of quantum gravity, so it is certainly important to explore its consequences. However as we have seen in section \ref{sec:dimensionality-ef}, the freedom to change this parameter opens a Pandora's box. Depending on the value of $\beta$ and the asymptotic behaviour, there are no solutions, discrete fixed points, lines, or planar regions of fixed points. We discussed these briefly in section \ref{sec:which} in the light of results elsewhere in the literature. As we saw in section \ref{sec:intro-phys}, $\vp_{asy}(r)$ provides the fixed point equation of state through the limit in equation \eqref{phys}. In section \ref{sec:phys}, we saw this led to several possible scenarios.

Since quantum fluctuations remain strongly coupled at large $r$, it is not surprising that the results are sensitive to the formulation.  However ultimately we would want to see universality expressed as qualitatively the same behaviour for the fixed point and the corresponding equation of state, independent of the details of the regularisation, providing the regularisation is not singular in some way. Clearly, further research is required to improve the approximations. Fortunately the asymptotic techniques explained in this paper, are sufficiently powerful to allow the solution of much more sophisticated approximations, for example cases where the right-hand side of the flow equation is awkward or impossible to evaluate exactly \cite{Dietz:2016gzg}.
Finally, applying the techniques we have described here to other formulations that have already been developed \cite{Percacci:2015wwa,Labus:2015ska,Eichhorn:2015bna,Demmel:2014fk,Demmel:2012ub,Demmel:2013myx,Benedetti:2013jk,Demmel:2014hla,Demmel2015b,Ohta:2015efa,Ohta2016,Percacci:2016arh,Falls:2016msz,Ohta:2017dsq}, will no doubt further elucidate the situation.
\appsection{}
\section{Power law solution coefficients $k_i$}
\label{AppendixA}

\begin{align}
k_2 =& \frac{1}{\beta ^2 (\beta  (3 \beta  (36 \beta  (15 \beta\! -\! 22)\! +301)\! +2)\! -19)^2 (\beta  (3 \beta  (36 \beta  (21 \beta\!  -\! 37)\! +725)\! -164)\! -23)}\nonumber\\
&\times\Big(9 A\!  \left(589824 \pi ^4 A^2 (\beta\!  -\! 1)\!  \left(27 \beta ^2-1\right) (\beta  (3 \beta  (3 \beta  (9 \beta  (12 \beta  (9 \beta  (60 \beta \! -\! 151)\! +1103)\!\right.\nonumber\\
& -\! 3175)\left.-4000)+4370)+908)-271)+\beta  (\beta  (3 \beta  (36 \beta  (15 \beta -22)+301)+2)-19)^2\right.\nonumber\\
&\left.\times(\beta  (3 \beta  (54 \beta  (28 \beta -33)+473)+76)-35)\right)\,,\\
k_3 =&\frac{1}{\beta ^3 \left(324 \beta ^4-810 \beta ^3+636 \beta ^2-83 \beta -2\right) \left(1620 \beta ^4-2376 \beta ^3+903 \beta ^2+2 \beta -19\right)^3}\nonumber\\
&\!\times\!\!\dfrac{1}{ \left(2268 \beta ^4\! -\! 3996 \beta ^3\! +\! 2175 \beta ^2-\! 164 \beta \! -\! 23\right)}\nonumber\\
&\times\Big(\! 1152 \pi ^2 A^2\! \left(294912 \pi ^4 A^2\! \left(21664553744880 \beta ^{17}\right.\right.\nonumber\\
&-131103093477600 \beta ^{16}+335182432132080 \beta ^{15}-465992520928740 \beta ^{14}\nonumber\\
&+373012915696569 \beta ^{13}-160032473858853 \beta ^{12}+20341799162595 \beta ^{11}\nonumber\\
&+10879448697531 \beta ^{10}-3992311992294 \beta ^9-184358900772 \beta ^8+235681642062 \beta ^7\nonumber\\
&-3342432654 \beta ^6-9333036891 \beta ^5+381546579 \beta ^4+240424223 \beta ^3-15717769 \beta ^2\nonumber\\
&-\left.2976296 \beta +275350\right)+\beta  \left(1620 \beta ^4-2376 \beta ^3+903 \beta ^2+2 \beta -19\right)^2\nonumber\\
&\times  \left(3670485840 \beta ^{11}-13593079800 \beta ^{10}+19462865328 \beta ^9-13229473554 \beta ^8\nonumber\right.\\
&\left.\left.+3888160137 \beta ^7\! -\! 11847951 \beta ^6-212450220 \beta ^5+13245732 \beta ^4\! +\! 8143979 \beta ^3\! -\! 752317 \beta ^2\! \! \right.\right.\nonumber\\
&\left.\left.-148144 \beta +17570\right)\right)\Big)\, ,\\
k_4=&-\frac{1}{\beta ^4 \left(324 \beta ^4-2484 \beta ^3+2913 \beta ^2-500 \beta +7\right) \left(324 \beta ^4-810 \beta ^3+636 \beta ^2-83 \beta -2\right)}\nonumber\\
&\times\dfrac{1}{ \left(1620 \beta ^4-2376 \beta ^3+903 \beta ^2+2 \beta -19\right)^4 \left(2268 \beta ^4-3996 \beta ^3+2175 \beta ^2-164 \beta -23\right)}\nonumber\\
&\times\Big(2654208 \pi ^4 A^3\!\! \left(262440 \beta ^7-691092 \beta ^6+579798 \beta ^5-139563 \beta ^4-21348 \beta ^3+5382\right.\nonumber\\
&\left.+\beta ^2+1214 \beta -211\right) \left(294912 \pi ^4 A^2 \left(21664553744880 \beta ^{17}-131103093477600 \beta ^{16}\right.\right.\nonumber\\
& +335182432132080 \beta ^{15}-465992520928740 \beta ^{14}+373012915696569 \beta ^{13}\nonumber\\
&\! -\! 160032473858853 \beta^{12}\! +\! 20341799162595 \beta^{11}\! +\! 10879448697531 \beta ^{10}\!\!\nonumber\\
&\! -\! 3992311992294 \beta ^9 -\! 184358900772 \beta^8\! +\! 235681642062 \beta ^7\! -\! 3342432654 \beta ^6 \nonumber\\
&-9333036891 \beta ^5+\! 381546579 \beta ^4\!+240424223 \beta ^3-15717769 \beta ^2-2976296 \beta +275350\nonumber\\
&+\beta (1620 \beta^4-2376 \beta ^3+903 \beta ^2+2 \beta -19)^2 (3670485840 \beta^{11}-13593079800 \beta ^{10}\nonumber\\
&+19462865328 \beta ^9-13229473554 \beta ^8+3888160137 \beta ^7-11847951 \beta ^6\nonumber\\
&-212450220 \beta^5+13245732 \beta^4+8143979 \beta ^3-752317 \beta ^2-148144 \beta +17570))\Big).
\end{align}

Unfortunately the $k_5$ expression is too long to include in the paper. We also list the values of the $k_i$  for the special case $\beta=1/6$:
\beal 
k_1 &=-1296\,{\pi }^{2}{a}^{2}\,,\nonumber\\
k_2 &={\frac {27\,a \left( 3649536\,{\pi }^{4}{a}^
{2}+25 \right) }{20}}\,,\nonumber\\
k_3 &=-{\frac {1944\,{\pi }^{2}{a}^{2} \left( 
123254784\,{\pi }^{4}{a}^{2}+865 \right) }{25}}\,,\\
k_4 &=-{\frac {30233088\,{
\pi }^{4}{a}^{3} \left( 123254784\,{\pi }^{4}{a}^{2}+865 \right) }{125
}}\,,\nonumber \\
k_5 &=-{\frac {81\,a \left( 4034150189236224\,{\pi }^{8}{a}^{4}+
29839933440\,{\pi }^{4}{a}^{2}+18125 \right) }{4000}}\,.\nonumber
\eeal

\section{Non-power law solution coefficients $k_i$}
\label{kivalues}

Coefficients $k_i$ for general $\b$:
\bea
k_1\!\!\!  &\!\! =&\!\!\! \frac{156 \beta ^2-72 \beta +1}{9216 \pi ^2}\nonumber\,,\\
k_2\!\! \! &\!\! =&\!\!\! \frac{33696 \beta ^5-36072 \beta ^4-540 \beta ^3+4872 \beta ^2-116 \beta +5}{55296 \pi ^2 \beta  (3 \beta +1) (6 \beta -5)}\nonumber\,,\\
k_3\!\! \!&\!\! =&\!\! \!\frac{\left(33696 \beta ^5-36072 \beta ^4-540 \beta ^3+4872 \beta ^2-116 \beta +5\right)^2}{331776 \pi ^2 (5-6 \beta )^2 \beta ^2 (3 \beta +1)^2 \left(156 \beta ^2-72 \beta +1\right)}\nonumber\,,\\
k_4\!\! \!&\!\! =&\!\!\!\!\! \frac{2659392\beta ^8\! -\! 3044304 \beta ^7\! -\! 449064 \beta ^6\! +\! 971352 \beta ^5\! +\! 8748 \beta ^4\!\! -\! 67518 \beta ^3\! +\! 4119 \beta ^2\! +\! 235 \beta\! -\! 25}{331776 \pi ^2 (5-6 \beta )^2 \beta ^2 (3 \beta +1)^2}\nonumber ,\\
k_5 \!\!\! &\!\! =&\!\!\! -\frac{\left(33696 \beta ^5-36072 \beta ^4-540 \beta ^3+4872 \beta ^2-116 \beta +5\right)^3}{3981312 \pi ^2 \beta ^3 (3 \beta +1)^3 (6 \beta -5)^3 \left(156 \beta ^2-72 \beta +1\right)^2}\label{kigeneral}\,.
\eea

Coefficients for $\b=-{1}/{3}$:
\be
\label{ki13}
k_1=\frac{17}{3456 \pi ^2}, \,
k_2=\frac{775}{96768 \pi ^2},\,
k_3=\frac{600625}{46061568 \pi ^2},\,
k_4=\frac{349525}{46061568 \pi ^2},\,
k_5=-\frac{465484375}{43850612736 \pi ^2}
\ee
and for $\b=5/6$:
\be 
k_1=\frac{1}{576 \pi ^2},\,
k_2=\frac{97}{30240 \pi ^2},\,
k_3=\frac{9409}{1587600 \pi ^2},\,
k_4=\frac{4171}{705600 \pi ^2},\,
k_5=-\frac{912673}{166698000 \pi ^2}.\label{ki56}
\ee

\chapter{Outlook}
\label{cha:outlook}


In this thesis we have presented research considering different fundamental issues within asymptotic safety.

In chapter \ref{cha:recon} we provided two solutions to the reconstruction problem.
The first consisted of constructing a suitable bare action from the effective average action $\Gamma_k$ by using the corresponding Wilsonian action $S^k$ computed through the duality relation \eqref{duality-con}.
The second solution was a map from $\Gamma_k$ to a pair $\{\mathcal{S}^\Lambda,\Gamma^\Lambda_k\}$ as summarized in figure \ref{fig:map}.
There we also proved a remarkable duality relation between two effective average actions computed with different overall cutoff profiles but whose corresponding Wilsonian actions coincide. We note that this result is significant in general, not just within the context of the RG.
And although all relations were phrased in terms of a single scalar field, generalising the discussion to full quantum gravity would be straightforward and a useful extension of this work.

In chapter \ref{cha:back ind} we investigated background independence and understood the underlying reasons why msWIs generically forbid fixed points in the derivative expansion of conformally reduced gravity.
There we argued that no solutions to the flow equation exist if the msWIs are incompatible.
Even in the compatible case, we found out why msWIs can still forbid fixed points through the parameterisation of the conformal factor $f(\phi)$.
For example we see that fixed points are forbidden for exponential parametrisations $f(\phi)= \exp(\phi)$ (as long as the field grows a non-zero anomalous dimension). We note that the reasons for this conflict are general and not solely tied to the derivative expansion of the conformally truncated model. Therefore this issue could provide important constraints for example on the exponential parametrisations recently advocated in the literature \cite{Demmel:2015zfa,Eichhorn:2013xr,Eichhorn:2015bna,Nink:2014yya,Percacci:2015wwa,Labus:2015ska,Ohta:2015efa,Gies:2015tca,Dona:2015tnf}.

We reiterate that it surely seems significant that whenever the msWIs are compatible with the flow equations, it is possible to combine them to uncover a background independent description of the flow, including a background independent notion of the RG scale.
Of course this has only been shown in the $\mathcal{O}(\partial^2)$ approximation and in conformally reduced gravity so one might doubt whether this trick of combining equations would work for full quantum gravity.
Owing to this we also investigated how these issues might appear in polynomial truncations. From the polynomial expansion of the potential we saw that if the equations are to be compatible then in terms of vertices, the information they contain becomes highly redundant at sufficiently high order.
Again, these findings surely hint at the existence of simpler description.
Indeed, in \cite{Morris:2016nda} an alternative approach has been initiated which avoids these issues entirely since background independence is never broken.

Finally, in chapter \ref{cha:asymp} we studied the asymptotic behaviour of fixed point solutions in the $f(R)$ approximation.
We gave a detailed recipe of how to construct such solutions, including how to uncover the missing parameters. 
We found that quantum fluctuations do not decouple at large $R$, typically leading to elaborate asymptotic solutions containing several free parameters.
Depending on the value of the endomorphism parameter $\beta$, fixed point solution spaces of differing dimension were found e.g.\ there were no solutions, discrete fixed points, lines or planar regions.
However we would ultimately want the qualitative behaviour of the fixed point to be independent of the details of the regularisation and the very fact that this freedom exists suggests that the fixed singularities induced by the form of the cutoff are unphysical artifacts and should be eliminated wherever possible.

Even when these singularities are sufficiently eliminated \cite{Benedetti:2012}, the $f(R)$ approximation yields a continuum of fixed point solutions which furthermore support a continuous spectra of eigenoperators \cite{Dietz:2012ic}, the lack of constraints coming from the large field behaviour being responsible for this.
The structure of these solutions is governed by the conformal factor sector \cite{Benedetti:2012, Demmel2015b} and these findings are in fact a reflection of the conformal factor problem previously mentioned in section \ref{sec:approxs} and studied in reference \cite{Dietz:2016gzg}.
More work is needed to understand how these issues might be overcome in full quantum gravity and possible routes to pursue are discussed at the end of \cite{Dietz:2016gzg}.

Throughout this thesis our discussions have only been concerned with pure gravity, but for an asymptotically safe theory to be a viable description of gravitational dynamics within our universe it must be compatible with matter i.e.\ a gravitational fixed point must persist upon the inclusion of matter fields.
Incorporating matter fields into the formalism is straightforward. The structure of the flow equation \eqref{FRGE} remains the same, $\varphi$ then just represents the set of all fields and the Hessian and cutoff operator $R_k$ become block matrices labelled by the different fields.
It is unlikely that the standard model is asymptotically safe by itself and so coupling to gravity is conjectured to induce a suitable fixed point for all matter fields \cite{Shaposhnikov:2009pv}.
For evidence supporting this see e.g.\ \cite{Eichhorn:2011pc, Eichhorn:2012va, Eichhorn:2015woa, Eichhorn:2016esv, Eichhorn:2017ylw, Eichhorn:2017eht, Eichhorn:2016vvy, Christiansen:2017qca, Zanusso:2009bs, Harst:2011zx, Oda:2015sma, Dona:2015tnf, Dona:2013qba, Dona:2014pla}.
A further motivation for investigating matter-gravity interactions is that they could become relevant for tests of quantum gravity. For example, the asymptotic safety scenario could be ruled out if new matter discovered at particle accelerators failed to be compatible with the existence of a fixed point.
Understanding the compatibility of matter and asymptotically safe quantum gravity continues to be an active area of research within the community.


Indeed, even if asymptotic safety proves to be internally consistent, it will be experimental evidence that ultimately decides if it provides the correct description of nature.
As noted at the very beginning of this thesis, strong gravitational effects manifest themselves at the Planck scale.
This makes cosmology, and the early-universe cosmology in particular, a promising domain for testing asymptotic safety.
Efforts towards explaining inflation from an asymptotic safety point of view are being made, with predictions on cosmic inflation given in \cite{Bonanno:2010bt, Bonanno:2015fga, Saltas:2015vsc, Nielsen:2015una, Kofinas:2016lcz, Falls:2016wsa}.
In terms of astrophysical processes, predictions for the final stages of black hole evaporation as a result of running couplings have also been made, see e.g.\ \cite{Bonanno:1998ye, Becker:2012js, Becker:2012jx, Koch:2013owa}.
However, it may also be possible to put asymptotic safety to the test at energies accessible at earth bound experiments.
If extra dimensions exist, this could lead to a lowering of the Planck scale to the TeV scale.
This then opens the door for directly probing quantum gravitational effects in the near future, for example by using photon-photon scattering as argued in \cite{Dobrich:2012nv}.

Whilst it is imperative to make contact with experiment through asymptotic safety, we end this final chapter by re-emphasizing that it is also crucial to continue working on the fundamental aspects of the approach and that this remains an important avenue of research to pursue.




\cleardoublepage{}
\phantomsection{}
\addcontentsline{toc}{chapter}{Bibliography}

\bibliography{refs/refs}
\bibliographystyle{JHEP}

\end{document}